\documentclass[pdflatex,sn-mathphys-num]{sn-jnl}

\usepackage{graphicx}%
\usepackage{multirow}%
\usepackage{amsmath,amssymb,amsfonts}%
\usepackage{amsthm}%
\usepackage{mathrsfs}%
\usepackage[title]{appendix}%
\usepackage{xcolor}%
\usepackage{textcomp}%
\usepackage{manyfoot}%
\usepackage{booktabs}%
\usepackage{algorithm}%
\usepackage{algorithmicx}%
\usepackage{algpseudocode}%
\usepackage{listings}%
\usepackage{amsmath}
\usepackage{array}
\usepackage{ amssymb }
\usepackage{braket}
\usepackage{qcircuit}
\usepackage{soul}
\usepackage{braket} 
\usepackage{relsize}

\usepackage{amsmath}
\usepackage{ amssymb }
\usepackage{braket}
\usepackage{qcircuit}
\usepackage{soul}
\usepackage{braket}

\theoremstyle{thmstyleone}%
\theoremstyle{thmstyletwo}%

\theoremstyle{thmstylethree}%

\begin{document}

\title{{\LARGE Poisson structure and Integrability of a Hamiltonian flow for the inhomogeneous six-vertex model}}
\author*{\fnm{Pete} \sur{Rigas}}\email{pbr43@cornell.edu}

\affil*{Newport Beach, CA 92625, United States, ORCID: 0009-0003-1053-9720}

\date{}

\maketitle

Abstract{
  . We confirm a conjecture related to a Hamiltonian flow on inhomogeneous limit shapes of the 6-vertex model that was recently formulated in a 2022 work due to Keating, Reshetikhin, and Sridhar, [8], hence confirming a conjecture of the authors as to whether the flow is integrable. To demonstrate that such an integrability property of the Hamiltonian holds from the action-angle variables, we make use of previous methods for studying Hamiltonian systems, implemented by Faddeev and Takhtajan, in which it was shown that integrability of a Hamiltonian system holds for the nonlinear Schroedinger equation by computing action-angle variables from the Poisson bracket, which is connected to the analysis of entries of the monodromy and transfer matrices. For the inhomogeneous six-vertex model, an approach for computing the action-angle variables is possible through formulating several relations between entries of the Quantum monodromy, and transfer, matrices. }


\bigskip

\textbf{MSC Class}: 82B23; 82C10; 81U02; 34L25; 60K35; 82B20; 82B27

\section{Introduction}

\subsection{Overview}

\noindent The six-vertex model, originally introduced by chemists in [13], has emerged as a model of several studies in Probability theory, with recent efforts devoted towards rigorous descriptions of the localization, and delocalization, phases of the height function [4], the free energy [5], transition between disordered and antiferroelectric phases [7], and other related aspects pertaining to Russo-Seymour-Welsh results and crossing probabilities of the height function over strips of the square lattice, and interior of the cylinder, [14], in addition to classical work at the beginning of the seventies for computing the residual free entropy of the model on the torus [11]. To answer one conjecture raised in [8] regarding the integrability of the Hamiltonian flow for the six-vertex model, we make use of approaches described in [6] for determining the action-angle variables of the Hamiltonian that is introduced in the next section below. Generally, this pursuit fits within the efforts of previous works in Mathematical Physics from our ability in being able to determine, explicitly, action angle coordinates for which the dynamics of a system is linear. In the case of the six-vertex model with inhomogeneities, we exhibit that such an integrability property holds, from integrability of inhomogeneous limit shapes, by performing several computations with the Poisson bracket. Such computations are reliant upon the structure of L-operators, which in the case of domain wall boundary conditions have been previously examined within the Quantum inverse scattering framework that is discussed further in the next section, for Hamiltonian systems. Under domain wall boundary conditions for the inhomogeneous six-vertex model, the structure of the corresponding L-operators, namely that they are expressed in terms of Pauli basis elements, admits for several explicit computations that can be used to approximate asymptotic behaviors of the Quantum monodromy matrix, from those of the transfer matrix.

The properties of such matrices are reliant upon the Poisson bracket which we define below. Given the fact that the Poisson bracket is a bilinear operation on operators (equivalently, which is described in this work as entries of the transfer and Quantum monodromy matrices obtained from product representations of L-operators for the 6-vertex model under domain-wall boundary conditions), satisfies the following set of properties, given test functions $f$, $g$ and $h$:

{\small \begin{itemize}
    \item [$\bullet$] \textit{Anticommutativity}. $\big\{ f, g \big\}  =  - \big\{ g , f \big\} $,

    \item[$\bullet$] \textit{Bilinearity}. For real $a,b$, $\big\{ af + bg , h \big\} = a \big\{ f ,h \big\} + b \big\{ g , h \big\},$ and $\big\{ h , af + bg \big\} = a \big\{ h , f \big\} + b \big\{ h , g \big\} $,

    \item[$\bullet$] \textit{Leibniz' rule}. $\big\{ fg , h \big\} = \big\{ f , h \big\} g + f \big\{ g , h \big\}$,

    \item[$\bullet$] \textit{Jacobi identity}. $\big\{ f , \big\{ g , h \big\} \big\} + \big\{ g , \big\{ h , f \big\} \big\} + \big\{ h , \big\{ f , g \big\} \big\} = 0$.  \end{itemize} }

Throughout the remainder of the manuscript, to perform computations with respect to the above Poisson bracket we fix an underlying manifold and operators that are obtained from the transfer and Quantum monodromy matices of the 6-vertex model. Given the choice of an underlying manifold, straightforwardly one can also fix a Poisson manifold, namely a manifold of the aforementioned form endowed with the Poisson bracket. To demonstrate how conserved quantities underlie integrability of limit shapes introduced by Keating, Reshetikhin and Sridhar in [8], we describe how infinitely many conserved quantities associated with solutions to the Euler-Lagrange equations are obtained. We state the equations of interest, in addition to the accompanying conserved quantities, below. 

In comparison to transfer, and Quantum monodromy, matrices for other models of Statistical Physics, where it be spin chains, or related systems over the lattice, transfer, and Quantum monodromy, matrices for the six-vertex model take on very specific relations with R-matrices that can be thought of in future circumstances, whether through the Quantum R, or combinatorial R, matrices. Equipped with an understanding of the asymptotic properties of the transfer matrix, one can then study the Quantum monodromy matrix to establish connections with a collection of sixteen Poisson brackets along with the connection of such a family of brackets to integrability and exact solvability. In [8], to quantify the role of a particular Hamiltonian flow on integrability of inhomogeneous limit shapes Keating, Reshetikhin and Sridhar demonstrated that conserved quantities of the form,

{\small

\begin{align*}
  {\tiny \big[} \textbf{T} {\tiny\big( u ,H  , V )} , \textbf{T} {\tiny \big( u^{\prime} ,H ,  V ) \big] } = 0  , 
\end{align*}

}

\noindent from the commutation of transfer matrices for the 6-vertex model with respect to the Lie bracket ${\tiny \big[ \cdot , \cdot  \big] }$ and spectral parameters $u \neq u^{\prime}$ under the external fields $H,V$ can be obtained from,
 
 {\small
 
 \begin{align*}
 \sigma {\tiny \big( s , t \big)} \equiv \underset{H,V}{\mathrm{max}}  \big\{  s \cdot H + t \cdot V - f_{\textit{torus}} {\tiny \big( H , V )  }           \big\}        ,
 \end{align*}

 }

 \noindent corresponding to the Legendre transform of the free energy of the 6-vertex model over the torus for the free-energy functional $f$, the two external fields $H,V$ and $s,t \in {\tiny \big[}   - 1 \big/ 2   , 1 \big/ 2    {\tiny \big]}$. Given an $R \equiv R_{ab}$ matrix for the 6-vertex model, associated with the conserved quantities from the above statement of commutativity of $\textbf{T}{\tiny \big(} u {\tiny\big)}$ and $\textbf{T}{\tiny \big(} u^{\prime} {\tiny\big)}$, the so-called RTT relation states,

  {\small
  
  \begin{align*}
    \textbf{T} {\tiny\big( u , 0  , V )}        \textbf{T} {\tiny\big( u , 0   , V )}   R {\tiny \big(}   u   {\tiny \big)}   \equiv \textbf{T}_a {\tiny\big( u , 0   , V )} \textbf{T}_b {\tiny\big( u , 0   , V )}  R_{ab} {\tiny \big(}   u   {\tiny \big)}  =  R_{ab} {\tiny \big(}   u   {\tiny \big)} \textbf{T}_b {\tiny\big( u , 0   , V )} \textbf{T}_a {\tiny\big( u , 0   , V )} \equiv   R {\tiny \big(}   u   {\tiny \big)} \textbf{T} {\tiny\big( u , 0   , V )} \textbf{T} {\tiny\big( u , 0   , V )}  .
  \end{align*} 
  
  }

\noindent For either a planar domain or cylinder $M$ denote $\chi : \partial M \longrightarrow \textbf{R}$. The above transform that is dependent upon the free energy of the 6-vertex model over the torus can be used to define the Euler-Lagrange equation, 

{\small

\begin{align*}
      \partial^2_1  \sigma {\tiny \big(}  \partial_x h , \partial_y h       {\tiny \big)} \cdot  \partial^2_x h + \partial_1 \partial_2 \sigma {\tiny \big(}   \partial_x h , \partial_y h       {\tiny \big)}  \cdot \partial_x \partial_y h      +    \partial^2_2 \sigma {\tiny \big(} \partial_x h , \partial_y h     {\tiny \big)}  \cdot \partial^2_y h =     0    ,
\end{align*} 

}

\noindent where $h$ denotes the height function of the 6-vertex model (ie, a graph homomorphism supported) which takes the form,

{\small

\begin{align*}
h : F {\tiny \big( } \textbf{Z}^2 {\tiny \big)} \longrightarrow \textbf{Z} , 
\end{align*} 

}

\noindent subject to the conditions $h {\tiny \big(} 0 , y {\tiny \big)} = \chi_1  {\tiny \big(} y {\tiny \big)}$, $h {\tiny \big(} T , y {\tiny \big)} = \chi_2  {\tiny \big(} y {\tiny \big)} + C$ for some $T>0$, $C \in \textbf{R}$, with the partial derivatives,

 {\small
 
 \begin{align*}
\partial^2_1 \cdot  \equiv \bigg( \frac{\partial \cdot }{\partial \chi_1} \bigg)^2  ,  \\ \\    \partial_x h \equiv \frac{\partial h}{\partial x } , \\ \\ \partial_y h \equiv  \frac{\partial h}{\partial y }  , \\ \\ \partial^2_x \cdot \equiv \bigg( \frac{\partial \cdot }{\partial x} \bigg)^2   , \\ \\ \partial_1 \partial_2 \cdot  \equiv \frac{\partial}{\partial \chi_1 } \bigg( \frac{\partial \cdot }{\partial \chi_2 } \bigg)  , \\ \\  \partial_x \partial_y h \equiv \frac{\partial^2 h}{\partial x \partial y}  , \\ \\ \partial^2_2 \cdot \equiv \bigg( \frac{\partial \cdot }{\partial \chi_2} \bigg)^2  , \\ \\    \partial^2_y h \equiv     \frac{\partial^2 h}{\partial y^2}    .
 \end{align*}

 }

 \noindent The Hamiltonian flow associated with the above statement of the Euler-Lagrange equation takes the form,

 {\small
 
 \begin{align*}
   H_u {\tiny \big(} p , h {\tiny \big)} \equiv     \int_0^L \underset{\nu  \in \textbf{R}}{\mathrm{max}}   \big\{      p \nu - \sigma_u {\tiny \big(} \nu , \xi {\tiny \big)}  + V \nu    \big\} \mathrm{d}y        ,
 \end{align*}

 }

 \noindent where $u$ denotes a spectral parameter and $p,h : {\tiny \big[} 0 , L {\tiny \big]} \longrightarrow \textbf{R}$ and,
 
 {\small
 
 \begin{align*} 
       \xi \equiv \theta + x \big/ 2 + y \big/ 2   , 
 \end{align*} 
 
 }
 
 \noindent for $\theta \in {\tiny \big[} 0 , 2 \pi {\tiny \big]}$. In this paper we are interested in characterizing how integrability of the 6-vertex model can be formulated, and analyzed, from the flow $H$. Given the constraint $\partial_x h , \partial_y h \in {\tiny \big[} - 1 \big/ 2 , 1 \big/ 2 {\tiny \big]}$, the limiting free energy of the 6-vertex model,

 {\small

 \begin{align*}
   f_{D,\chi} \equiv \underset{h \in h ( \chi)}{\mathrm{max}} \int_D \sigma {\tiny \big(} \partial_x h , \partial_y h  {\tiny \big)}  \text{ } \mathrm{d} x \text{ } \mathrm{d} y   , 
 \end{align*}
 
 }

 \noindent supported over a simply connected domain $D \subset \textbf{Z}^2$ has the minimizer,

 {\small
 
 \begin{align*}
 S {\tiny \big[ } h {\tiny \big]} \equiv \int_D \sigma {\tiny ( }    \partial_x h , \partial_y h     {\tiny \big)} \text{ } \mathrm{d} x \text{ } \mathrm{d} y .
 \end{align*} 
 
 }

 \noindent The above minimizer over a simply connected domain over the cylinder takes the form,

  {\small
 
 \begin{align*}
 S {\tiny \big[ } h {\tiny \big]} \equiv \int_D \big\{ \sigma {\tiny ( }    \partial_x h , \partial_y h     {\tiny \big)} + V \partial_x h  \big\}  \text{ } \mathrm{d} x \text{ } \mathrm{d} y .
 \end{align*} 
 
 } 

 \bigskip 
 
A collection of brackets, constituting the two-dimensional Poisson structure described explicitly below is mathematically related to the objects,
 
 {\small \begin{align*}
      { \big\{}   h {\tiny \big( y \big)}  ,   p {\tiny \big( y^{\prime} \big)}  {\big\}}  = \delta {\tiny \big( y - y^{\prime} \big)} ,  \\ \\  { \big\{}   p {\tiny \big( y \big)}  ,   p {\tiny \big( y^{\prime} \big)}  { \big\}}   = 0  ,  \\ \\   { \big\{}  h {\tiny \big( y \big)}  ,   h {\tiny \big( y^{\prime} \big)}  {\big\}}   = 0  ,  
 \end{align*}  } 
 
 \noindent where $p$ denote coordinate charts over the tangent bundle to the height function and $\delta$ is a function that equals $1$ iff $y \equiv y^{\prime}$ and is otherwise zero. The limit shape associated with solutions to the Euler-Lagrange equation stated above is obtained from the minimzer $h_0$ of the functional,

 {\small
 
 \begin{align*}
     S {\tiny \big[ } h {\tiny \big]} \equiv \int_0^L \int_0^T  \sigma_{u (y ) - v ( x ) } {\tiny ( }    \partial_x h , \partial_y h     {\tiny \big)}   \text{ } \mathrm{d} x \text{ } \mathrm{d} y       .
 \end{align*} 
 
 }

 \noindent Altogether, the following adaptation of the Quantum inverse scattering method allows a resolution of one conjecture raised by the authors of [8], as to whether the Hamiltonian flow for the six-vertex model that they introduce is integrable. Besides the forthcoming two-dimensional integrability statement for the six-vertex model, further adaptations to other models of Statistical Physics remains of great interest. Under other circumstances, for three-dimensional vertex models, correlations, in addition to the lattice over which interactions occur, can be expected to have a significant impact on the integrable, and exactly solvable, structure of the model and closely related settings, [1,2,3,6,8,9,15].

In the presence of inhomogeneities for the 6-vertex model, an underlying Hamiltonian for the six-vertex model can be shown to satisfy integrability properties, through  a set of sixteen relations, which can be derived from entries of the monodromy and transfer matrices, which is not only intimately connected to the R-matrices satisfying the Yang-Baxter equation, but also to infinite volume limit of the transfer matrices as $N \longrightarrow + \infty$, from which standard properties of the Poisson bracket can be used to asymptotically evaluate entries of the $N$ th power of the monodromy matrix (the complete set of such relations that the monodromy matrix satisfies are provided in \textit{2.4}). In comparison to other studies that have quantified crossing probabilities, [4], and even the free energy landscape, [5], of the six-vertex model, the following application of the Quantum inverse scattering framework predominantly relies upon integrable, rather than discrete, probabilistic structures. In higher dimensional vertex models, besides the questions of interest mentioned at the end of the previous paragraph, it would be intriguing to explore connections between such discrete and integrable structures. From the discrete perspective, the fact that the six-vertex model has a height function representation that satisfies the Fortuin-Kestelyn-Ginibre lattice condition, as well as a few other properties, is of tremendous utility for obtaining crossing probability estimates of the height function under a wide variety of boundary conditions. From the integrable perspective, to demonstrate that the same vertex model satisfies relations that can be leveraged for exact solvability, under similar classes of boundary conditions computations with the Poisson bracket parametrize a two-dimensional family that can be approximated near the weak finite volume limit. With computations that are provided for asymptotic approximations of nine Poisson brackets, computations for the eighth Poisson bracket are the most complicated, from several cross terms arising from entries of the Quantum monodromy matrix.

\subsection{This paper's contributions}

\noindent This paper characterizes properties of action-angle coordinates for the 6-vertex model in the presence of inhomogeneities. Properties of such coordinates are paramount in determining whether a model of interest is integrable. While integrability and exact solvability can be characterized through objects that are of more algebraic nature, examining these properties with L-operators under domain wall boundaries remains of interest to explore. Under domain walls, L-operators that have been previously identified in the literature have been used to compute two-dimensional correlation functions. For our purposes, while having explicit contour integral representations for the $6$-vertex model is informative, it does not depend upon action-angle coordinates. 

To characterize integrability we examine the behavior of the system,

\[
\left\{\!\begin{array}{ll@{}>{{}}l} \underline{(1)}:      \big\{  A {\tiny \big( u \big)  }       , A {\tiny \big( u^{\prime} \big) }   \big\} 
\text{ , } \\  \vdots \\ \underline{(16)}: \big\{  D {\tiny\big( u \big) }  ,       D {\tiny \big( u^{\prime} \big) } \big\} \text{, }
\end{array}\right.
\]

\noindent of Poisson brackets, where,

{\small \begin{align*}
 \textit{Finite volume expansion of the first operator of the 6-vertex transfer matrix} \equiv    A {\tiny\big( u \big) } \end{align*}
 
 \begin{align*} \equiv \underset{N \longrightarrow + \infty}{\mathrm{lim}}    A_N {\tiny\big( u \big)  }       ,  \end{align*}
 
 \begin{align*}  \vdots \end{align*}

 \begin{align*}\textit{Finite volume expansion of the fourth operator of the 6-vertex transfer matrix} \equiv    D {\tiny \big( u \big) }  \end{align*}
 
 \begin{align*} \equiv \underset{N \longrightarrow + \infty}{\mathrm{lim}}    D_N {\tiny\big( u \big)}      .
\end{align*} }

\noindent We provide an overview of the results for approximating entries from L-operators, in addition to accompanying Poisson brackets, in \textit{Table 1} and \textit{Table 2} below. As a bilinear mapping, several standard properties of the Poisson bracket are provided in the next section. That is, to deduce that an approximation, '$\approx$' is said to hold, where,

{\small \begin{align*}
 \big\{  f \approx g \big\} \Longleftrightarrow \big\{   f \equiv g + \mathrm{o} {\tiny \big( N \big) }  \overset{N \longrightarrow + \infty}{\longrightarrow}             f \equiv g      \big\} ,       
\end{align*}
 }

\noindent for two test functions $f$ and $g$, of real and compact support, namely,

{\small \begin{align*}
  f , g \in \underset{\textit{test functions}}{\bigcup} \big\{ \textit{test functions F} : F        \textit{ can be represented with a Poisson bracket of A, B, C, or D} \end{align*}

 \begin{align*} \textit{ operators}                                           \big\}   .
\end{align*} }

\noindent Under the Poisson bracket, integrability of a model of interest can be related to the existence of suitable action-angle coordinates. From a weak volume expansion of $B \big( u \big)$, that is, the second block operator in the transfer matrix, through,

{\small \begin{align*}
  B {\tiny\big( u \big)}  \equiv \underset{N\longrightarrow + \infty}{\mathrm{lim}}  \underset{1 \leq j \leq N}{\prod} B_j {\tiny\big( u \big) }   .
\end{align*} }

\bigskip

\begin{tabular}{|l|l|}
\hline\parbox[t]{0.25\textwidth}{
\begin{itemize}
\item \textbf{Lemma} \textit{1}
\item \textbf{Lemma} \textit{2}
\item \textbf{Lemma} \textit{3}
\item \textbf{Lemma} \textit{4}
\end{itemize}}& 
\parbox[t]{0.73\textwidth}{
\begin{itemize}
\item First entry of the product of three 6-vertex L-operators
\item Second entry of the product of three 6-vertex L-operators
\item Third entry of the product of three 6-vertex L-operators
\item Fourth entry of the product of three 6-vertex L-operators
\end{itemize}}\\
\hline
\end{tabular}
\noindent \textit{Table 1}. An overview of the results, with accompanying Proofs in the Appendix section, for computing closed form representations of product representations for L-operators.

\bigskip

\noindent As $N \longrightarrow + \infty$ the above product of each $B_j {\tiny\big( u \big)}$ operator is related to the expansion,

\[\begin{bmatrix}
 \bigg\{       A_{i-3} {\tiny\big( \lambda_{\alpha} \big)} + B_{i-3} {\tiny \big( \lambda_{\alpha} \big) } \bigg\} \cdot   \bigg\{    \big( \mathrm{sin} {\tiny\big( 2 \eta \big)} \big)^{n-(i-3)} \mathscr{A}^{\prime}_1  + \mathscr{A}^{\prime}_2 +  \mathscr{A}^{\prime}_3 \bigg\}   &  \big( \mathcal{E}_1 \big)^{\prime\prime\prime }   \\ \bigg\{         C_{i-3} {\tiny\big( \lambda_{\alpha} \big) } + D_{i-3} {\tiny\big( \lambda_{\alpha} \big)}       \bigg\} \cdot   \bigg\{   \big( \mathrm{sin} {\tiny\big( 2 \eta \big)}  \big)^{n-(i-3)} \mathscr{C}^{\prime}_1 + \mathscr{C}^{\prime}_2 + \mathscr{C}^{\prime}_3  \bigg\}   &  \big( \mathcal{E}_2 \big)^{\prime\prime\prime }  \end{bmatrix} \cdot 
    \prod_{i \leq i^{\prime} \leq n}    \begin{bmatrix}
             \textbf{1}^{i^{\prime}}   &             \textbf{2}^{i^{\prime}}      \\
           \textbf{3}^{i^{\prime}}   &  \textbf{4}^{i^{\prime}}    
  \end{bmatrix}   \text{, } 
 \]

\begin{tabular}{|l|l|}
\hline\parbox[t]{0.25\textwidth}{
\begin{itemize}
\item \textbf{Lemma} \textit{6}
\item \textbf{Lemma} \textit{7}
\item \textbf{Lemma} \textit{8}
\item \textbf{Lemma} \textit{9}
\item \textbf{Lemma} \textit{10}
\item \textbf{Lemma} \textit{11}
\item \textbf{Lemma} \textit{12}
\item \textbf{Lemma} \textit{13}
\item \textbf{Lemma} \textit{14}
\end{itemize}}& 
\parbox[t]{0.73\textwidth}{
\begin{itemize}
\item First Poisson bracket, $\mathcal{P}_1$, for $ \big\{ A \big( u \big) , A \big( u^{\prime} \big) \big\} $  \item Second Poisson bracket, $\mathcal{P}_2$, for $ \big\{ A \big( u \big) , A \big( u^{\prime} \big) \big\} $ \item Third Poisson bracket, $\mathcal{P}_3$, for $ \big\{ A \big( u \big) , A \big( u^{\prime} \big) \big\} $ \item Fourth Poisson bracket, $\mathcal{P}_4$, for $ \big\{ A \big( u \big) , A \big( u^{\prime} \big) \big\} $ \item Fifth Poisson bracket, $\mathcal{P}_5$, for $ \big\{ A \big( u \big) , A \big( u^{\prime} \big) \big\} $ \item Sixth Poisson bracket, $\mathcal{P}_6$, for $ \big\{ A \big( u \big) , A \big( u^{\prime} \big) \big\} $ \item Seventh Poisson bracket, $\mathcal{P}_7$, for $ \big\{ A \big( u \big) , A \big( u^{\prime} \big) \big\} $ \item Eighth Poisson bracket, $\mathcal{P}_8$, for $ \big\{ A \big( u \big) , A \big( u^{\prime} \big) \big\} $   \item Ninth Poisson bracket, $\mathcal{P}_9$, for $ \big\{ A \big( u \big) , A \big( u^{\prime} \big) \big\} $ 
\end{itemize}}\\
\hline
\end{tabular}
\noindent \textit{Table 2}. An overview of the results provided in \textit{2.4} for approximating nine Poisson brackets of interest.

 \bigskip

 \noindent where,

{\small \begin{align*}
\big( \mathcal{E}_1 \big)^{\prime\prime\prime }  \equiv \big( \mathcal{E}_1 \big)^{\prime\prime\prime }  {\tiny\big(       \lambda , \alpha , i , n \big)}                , \\ \\ \big( \mathcal{E}_2 \big)^{\prime\prime\prime }  \equiv       \big( \mathcal{E}_2 \big)^{\prime\prime\prime }   {\tiny\big(       \lambda , \alpha , i , n \big)  }    , \\ \\ 
    \begin{bmatrix}
             \textbf{1}^{i^{\prime}}   &             \textbf{2}^{i^{\prime}}      \\
           \textbf{3}^{i^{\prime}}   &  \textbf{4}^{i^{\prime}}    
  \end{bmatrix}    \equiv   \begin{bmatrix}
   \text{First entry}_i     & \text{Second entry}_i \\ \text{Third entry}_i & \text{Fourth entry}_i
  \end{bmatrix} \end{align*}

  \begin{align*} \equiv  \bigg[ \begin{smallmatrix}
   \text{First entry  of the domain-wall 6-vertex transfer matrix }_i     & \text{Second entry  of the domain-wall 6-vertex transfer matrix }_i \\ \text{Third entry  of the domain-wall 6-vertex transfer matrix }_i & \text{Fourth entry  of the domain-wall 6-vertex transfer matrix }_i
  \end{smallmatrix} \bigg] .
\end{align*} }

\noindent The closed form representation of the above quantities, $\big( \mathcal{E}_1 \big)^{\prime\prime\prime } $ and $\big( \mathcal{E}_2 \big)^{\prime\prime\prime }$, are provided in the series of statements for $\textbf{Lemma}$ \textit{1}-\textit{6} provided in \textit{2.2}.

\subsection{Description of Quantum inverse scattering methods for the $6$-vertex model, from the $20$-vertex model}

\noindent We first provide a general description of  the scattering method below.

\begin{itemize}
    \item[$\bullet$] The Quantum Inverse Scattering Method (QISM) approach, is based upon seminal work from Faddeev and Takhatajan, [6], which asserts that Integrability of Hamiltonian systems are integrable, or equivalently, relating to exactly solvable structure, [15]. The method includes:

\begin{itemize}
\item[$\bullet$] Formulating well posed solutions to the Nonlinear Schrodinger equation,
Introducing expressions for the Poisson brackets,

\item[$\bullet$]
Formulating sets of conditions for action-angle variables, which approximate dynamics that are approximately linear,

\item[$\bullet$]
Characterizing integrability from approximating a collection of Poisson brackets.
\end{itemize}

\item[$\bullet$] The QISM approach has wide appeal within Mathematical and Statistical Physics, particularly for characterizing integrability of closely related systems. Of the systems that can be of interest, and amenable to the QISM approach, previous work of the author in this area includes:

\begin{itemize}
\item[$\bullet$] Characterizing exact solvability of the 6-vertex model, from integrability of inhomogeneous limit shapes and a Hamiltonian flow,
\item[$\bullet$]
Characterizing exact integrability, and solvability, of the 4-vertex and 20-vertex models,
\item[$\bullet$]
Formulating contour integral representations for the 20-vertex model under domain-wall boundary conditions.
\end{itemize}
\end{itemize}

\subsubsection{QISM for the 20-vertex model}

For boundary conditions $\xi$, either those with sufficiently flat slope, ie, flat, or domain walls, such as those introduced over the pentagonal lattice, {[15]}, the 20-vertex probability measure supported over the triangular lattice, $\textbf{T}$, takes the form,

{\small \begin{align*}
   \textbf{P}^{20V , \xi}_{\textbf{T}} {\tiny \big[ \cdot \big]}  \equiv  \textbf{P}^{20V}_{\textbf{T}} {\tiny\big[ \cdot \big] }   \text{, }
\end{align*} }

\noindent which is explicitly given by the ratio of the vertex weight function and the partition function,

{\small \begin{align*}
\textbf{P}^{20V}_{\textbf{T}} {\tiny \big[      \omega         \big] }    \equiv \textbf{P}^{20V} {\tiny\big[   \omega     \big] }     =  \frac{w_{20V}{\tiny \big(\omega \big) } }{Z^{20V}_{\textbf{T}}} \equiv \frac{w {\tiny \big(\omega \big)}}{Z_{\textbf{T}}} \text{, }
\end{align*} }

\noindent for some vertex configuration $\omega \in \Omega^{20V}$ - the 20-vertex sample space, and weights similar to those introduced in the previous section for the 6-vertex model, namely, {[15]},

{\small \begin{align*}
     w_0 \equiv   a_1 a_2 a_3   \text{, } 
\\
    w_1 \equiv   b_1 a_2 b_3   \text{, } \\    w_2    \equiv   b_1 a_2 c_3    \text{, } \\
   w_3 \equiv       a_1 b_2 b_3 + c_1 c_2 c_3  \text{, } \\     w_4 \equiv    c_1 a_2 a_3        \text{, } \\               w_5 \equiv   b_1 c_2 a_3   \text{, } \\  w_6 \equiv  b_1 b_2 a_3    \text{, } 
\end{align*} }

\noindent and the partition function,

{\small \begin{align*}
 Z_{\textbf{T}} \equiv \underset{\omega \in \Omega^{20V}}{\sum}  w {\tiny\big( \omega \big)}  \text{. }
\end{align*} }

\noindent As was the case for the two-dimensional vertex model, the 6-vertex model, introduced in the previous subsection, one can also introduce finite volume approximations to the transfer matrix before taking the weak infinite volume limit, from spectral parameters $u,v,w$, with, [15],

{\small \[
T  {\tiny \big(   \underline{M} , N ,  \underline{\lambda_{\alpha}} , u , v , w \big) }  \equiv    \overset{\underline{M}}{\underset{\underline{j}=0}{\prod}}  \text{ }  \overset{-N}{\underset{i=0}{\prod}} \bigg\{   \mathrm{exp} \big( \lambda_3 {\tiny\big( q^{-2} \xi^{s_i} \big)} \big) \bigg[     \begin{smallmatrix}      q^{D_i}       &    q^{-2} a_i q^{-D_i-D_j} \xi^{s-s_i}        &   a_i a_{j} q^{-D_i - 3D_j} \xi^{s - s_i - s_j}  \\ a^{\dagger}_i q^{D_i} \xi^{s_i} 
             &      q^{-D_i + D_j} - q^{-2} q^{D_i -D_j} \xi^{s}     &     - a_j q^{D_i - 3D_j} \xi^{s-s_j}  \\ 0  &    a^{\dagger}_j q^{D_j} \xi^{s_j} &  q^{-D_j} \\    \end{smallmatrix} \bigg]    \bigg\}  \text{. } 
\] } 

\noindent As $M \longrightarrow + \infty$,  $N \longrightarrow - \infty$, the finite volume approximation above, in weak infinite volume, takes the form,

{\small \begin{align*}
T {\tiny\big(   \underline{\lambda}  \big) }  \equiv T {\tiny\bigg( + \infty , - \infty ,   \underline{\lambda_{\alpha}} , {\tiny\big\{ u_i \big\} }, {\tiny\big\{ v_j  \big\} }, {\tiny\big\{ w_k \big\}}  \bigg) }=   \underset{\underline{M} \longrightarrow + \infty}{\mathrm{lim}} \text{ }  \underset{N \longrightarrow - \infty}{\mathrm{lim}}     T  {\tiny\big(   M , N , \lambda_{\alpha} ,  v  ,  u    \big) }    \text{, } 
\end{align*} } 

\noindent with, [15],

{\small \begin{align*}
 \textbf{T}^{3D} {\tiny \big( \underline{\lambda} \big)  }  \equiv    \underset{ N \longrightarrow -  \infty}{\underset{\underline{M} \longrightarrow + \infty}{\mathrm{lim}}} \mathrm{tr} \bigg\{     \overset{\underline{M}}{\underset{j=0}{\prod}}  \text{ }  \overset{-N}{\underset{k=0}{\prod}} \mathrm{exp} \big( \lambda_3 {\tiny \big( q^{-2} \xi^{s^j_k} \big) }  \big)    \bigg[ \begin{smallmatrix}     q^{D^j_k}       &    q^{-2} a^j_k q^{-D^j_k -D^j_{k+1}} \xi^{s-s^k_j}        &  *_2\\ \big( a^j_k \big)^{\dagger} q^{D^j_k} \xi^{s^j_k} 
             &     *_1   &     - a^j_k q^{D^j_k - 3D^j_{k+1}} \xi^{s-s^j_k}  \\ 0  &    a^{\dagger}_j q^{D^j_k} \xi^{s^j_k} &  q^{-D^j_k} \\    \end{smallmatrix} \bigg]      \bigg\}  
\text{, }
\end{align*} }

\noindent for,

{\small \begin{align*}
  *_1 \equiv  q^{-D^j_k + D^j_{k+1}} - q^{-2} q^{D^j_k -D^j_{k+1}} \xi^{s}    \text{, } \\   *_2 \equiv a^j_k a^j_{k+1} q^{-D^j_k - 3D^j_{k+1}} \xi^{s - s^j_k - s^j_{k+1}}   \text{. }
\end{align*} }

\subsubsection{QISM for the 6-vertex, 20-vertex and 4-vertex models}

\noindent The Quantum inverse scattering method, beyond the methods seminally introduced in [6], is widely applicable to several vertex models. In previous work of the author, in the highest possible three-dimensional setting for the 20-vertex model under domain-wall boundary conditions asymptotic expressions for the transfer and Quantum monodromy matrices have been obtained. Before stating how integrability, and action-angle coordinates, can be formulated for the two-dimensional state space over which the 6-vertex model is supported, we draw the attention of the reader to the 4-vertex and 20-vertex models. Under the fixed boundary conditions (through scalar boundary conditions) for the 4-vertex model, for determining whether integrable, or hybrid integrable, properties of the model holds, one manipulates products of quantities of the form,

\begin{align*}
    L^{4V} {\tiny \big(  n | u \big)} \equiv L {\tiny\big( n | u \big) }    \equiv  \begin{bmatrix}
         L_{11} {\tiny \big( n | u \big)}  & L_{12} {\tiny \big( n | u \big) }  \\ L_{21} {\tiny\big( n | u \big)} & L_{22} {\tiny\big( n | u \big) }
    \end{bmatrix}  = \begin{bmatrix}
       - u e_n & \sigma^{-}_n \\ \sigma^{+}_n & u^{-1} e_n 
    \end{bmatrix}
    \text{, }
\end{align*}

\noindent corresponding to the 4-vertex L-operator, with degrees of freedom,

\begin{align*}
 \bigg\{ \mathrm{DOF\text{ }  1} \equiv \underset{n \in \textbf{Z}}{\bigcup}  \sigma^-_n     \text{, }  \mathrm{DOF\text{ }  2} \equiv  \underset{n \in \textbf{Z}}{\bigcup}  \sigma^+_n    \text{, }  \mathrm{DOF\text{ }  3} \equiv  \underset{n \in \textbf{Z}}{\bigcup}  e_n  \bigg\} \text{, }
\end{align*}

\noindent respectively spanned by the two standard Pauli basis elements, and the basis element $e_n$ of $\textbf{Z}$. For each vertex model, determining integrability under the presence of different boundary conditions (in addition to the existence of other quantities, as examined in [10]) amounts to determining closed form approximations to operators,

{\small \[
\left\{\!\begin{array}{ll@{}>{{}}l} 
  A^{6V} , B^{6V} , C^{6V} , D^{6V}  \text{, } \\ A^{20V}, B^{20V}, C^{20V}, D^{20V}, E^{20V}, F^{20V}, G^{20V}, H^{20V}, I^{20V}  \text{, } \\  A^{4V} , B^{4V} , C^{4V} , D^{4V}  \text{, } 
\end{array}\right.
\] }

\noindent appearing in,

\begin{align*}
\begin{bmatrix}
       A^{6V} {\tiny \big( \lambda_{\alpha} , v_k \big)  } & B^{6V} {\tiny \big( \lambda_{\alpha} , v_k \big)  } \\
    C^{6V} {\tiny \big( \lambda_{\alpha}, v_k \big) } & D^{6V} {\tiny \big( \lambda_{\alpha} , v_k \big)}   \text{ }  
  \end{bmatrix} \text{, }
\begin{bmatrix}
 A^{20V} {\tiny \big( \underline{u} 
 \big) }  & D^{20V} {\tiny\big( \underline{u} \big) } & G^{20V} {\tiny\big( \underline{u}  \big)} \\ B^{20V} {\tiny\big( \underline{u}  \big)} & E^{20V} {\tiny\big( \underline{u}  \big)} & H^{20V} {\tiny\big( \underline{u}  \big) } \\ C^{20V} {\tiny\big( \underline{u}  \big) } &  F^{20V} {\tiny\big( \underline{u}  \big)} & I^{20V} {\tiny\big( \underline{u}  \big) }
\end{bmatrix} \text{, }
    \begin{bmatrix}
        A^{4V} {\tiny\big( \lambda_{\alpha} , v_k  \big)}  & B^{4V} {\tiny\big(  \lambda_{\alpha} , v_k  \big)}  \\ C^{4V} {\tiny\big(  \lambda_{\alpha} , v_k  \big)} & D^{4V} {\tiny\big(  \lambda_{\alpha} , v_k  \big) } 
    \end{bmatrix}     \text{, }
\end{align*}

\noindent respectively corresponding to finite volume approximations of the 6-vertex, 20-vertex, and 4-vertex, models. From finite volume representations of transfer matrices for the 6-vertex, 20-vertex, and 4-vertex, models, one can study tensor products of Poisson brackets of transfer matrices, each of which take the form,

\begin{align*}
     \bigg\{  \begin{bmatrix}
        A^{6V} {\tiny\big( \lambda_{\alpha} , v_k  \big)} & B^{6V} {\tiny\big(  \lambda_{\alpha} , v_k  \big)} \\ C^{6V} {\tiny\big(  \lambda_{\alpha} , v_k  \big)} & D^{6V} {\tiny\big(  \lambda_{\alpha} , v_k  \big) }
    \end{bmatrix}   \overset{\bigotimes}{,} \begin{bmatrix}
        A^{6V} {\tiny\big( \lambda^{\prime}_{\alpha} , v^{\prime}_k  \big)} & B^{6V} {\tiny\big( \lambda^{\prime}_{\alpha} , v^{\prime}_k    \big)} \\ C^{6V} {\tiny\big(  \lambda^{\prime}_{\alpha} , v^{\prime}_k    \big) } & D^{6V} {\tiny\big( \lambda^{\prime}_{\alpha} , v^{\prime}_k   \big) } 
    \end{bmatrix}   \bigg\}  \text{, }
\bigg\{    \begin{bmatrix}
 A {\tiny\big( \underline{u} 
 \big)} & D {\tiny\big( \underline{u} \big)}  & G {\tiny\big( \underline{u}  \big)} \\ B {\tiny\big( \underline{u}  \big) } & E {\tiny\big( \underline{u}  \big) } & H {\tiny\big( \underline{u}  \big)  } \\ C {\tiny\big( \underline{u}  \big)}  &  F {\tiny\big( \underline{u}  \big)}  & I {\tiny\big( \underline{u}  \big) } 
\end{bmatrix} \end{align*}

\begin{align*} \overset{\bigotimes}{,}\begin{bmatrix}
 A {\tiny \big( \underline{u^{\prime}} \big) } & D {\tiny \big( \underline{u^{\prime}} \big) }  & G {\tiny \big( \underline{u^{\prime}} \big) } \\ B {\tiny\big( \underline{u^{\prime}}\big) } & E {\tiny\big( \underline{u^{\prime}} \big)}  & H {\tiny\big( \underline{u^{\prime}} \big)}   \\ C {\tiny\big( \underline{u^{\prime}} \big) } &  F {\tiny\big( \underline{u^{\prime}} \big)} & I {\tiny \big( \underline{u^{\prime}} \big) }
\end{bmatrix} \bigg\} \text{, }
     \bigg\{  \begin{bmatrix}
        A^{4V} {\tiny\big( \lambda_{\alpha} , v_k  \big)}  & B^{4V} {\tiny \big(  \lambda_{\alpha} , v_k  \big) } \\ C^{4V} {\tiny\big(  \lambda_{\alpha} , v_k  \big) }  & D^{4V} {\tiny\big(  \lambda_{\alpha} , v_k  \big) }
    \end{bmatrix}  \overset{\bigotimes}{,} \begin{bmatrix}
        A^{4V} {\tiny\big( \lambda^{\prime}_{\alpha} , v^{\prime}_k   \big)} & B^{4V} {\tiny\big( \lambda^{\prime}_{\alpha} , v^{\prime}_k    \big)}  \\ C^{4V} {\tiny\big( \lambda^{\prime}_{\alpha} , v^{\prime}_k   \big) } & D^{4V} {\tiny\big( \lambda^{\prime}_{\alpha} , v^{\prime}_k   \big) }
    \end{bmatrix}   \bigg\}      \text{, }
\end{align*}

\noindent respectively. To quantify asymptotic properties of the transfer and Quantum monodromy matrices, for,

\begin{align*}
  E^{\mathrm{6V}} {\tiny \big( x - v_k  , x \big)}  \equiv  \mathrm{exp} \big[             \mathrm{coth} {\tiny \big( {\eta} \big/ {2} + i \alpha_j - v_k \big) }            \big]    \text{, } \end{align*}
  
  \begin{align*} E^{3D,\mathrm{6V}} {\tiny\big(   \underline{u^{\prime}}  ,  v^{\prime}_k - v_k ,  u^{\prime}_k - u_k  , w^{\prime}_k - w_k \big)}   \equiv    \mathrm{exp} \bigg\{   {1} \big/ {2i} \cdot \begin{bmatrix} 1 & 0 & 0 \\ 0 & -1 & 0 \\ 0 & 0 & 0 \end{bmatrix}   +  \begin{bmatrix} 0 & 0 & \psi  \\ 0 & \bar{\psi}   & 0 \\ 0 & 0 & 0 \end{bmatrix}        \bigg\}        \text{, } \end{align*}

     \begin{align*} E^{\mathrm{4V}} {\tiny \big( x - v_k  , x \big) }  \equiv       \mathrm{exp} \bigg\{   {1} \big/ {2i} \cdot  \begin{bmatrix} 1 & 0  \\ 0 & -1  \end{bmatrix}   +  \begin{bmatrix} 0 & 0   \\ 0 & \bar{\psi}      \end{bmatrix}        \bigg\}                 \text{, }
\end{align*}

\noindent with respect to the standard tensor product operation $ \bigotimes \cdot$ one has the collection of relations,

{\small \begin{align*}
      \big\{  T^{6V}_{-} {\tiny \big( x , \underline{\lambda} \big) } \overset{\bigotimes}{,} T^{6V}_{-} {\tiny \big( x , \underline{\mu} \big)  }          \big\} = r^{6V} {\tiny \big( \underline{\lambda} - \underline{\mu} \big) } \cdot T^{6V}_{-} {\tiny \big( x , \underline{\lambda} \big) } \bigotimes T^{6V}_{-} {\tiny \big( x , \underline{\mu} \big) }  - T^{6V}_{-} {\tiny \big( x , \underline{\lambda} \big) } \bigotimes T^{6V}_{-} {\tiny \big( x , \underline{\mu} \big) }  \end{align*}

      \begin{align*}    \times r^{6V}_{-} {\tiny \big( \underline{\lambda} - \underline{\mu} \big) }         \text{, } \end{align*}

      \begin{align*}        \big\{  T^{6V}_{+} {\tiny \big( x , \underline{\lambda} \big) } \overset{\bigotimes}{,} T^{6V}_{+} {\tiny \big( x , \underline{\mu} \big) }           \big\} = T^{6V}_{+} {\tiny \big( x , \underline{\lambda} \big) } \bigotimes T^{6V}_{+} {\tiny \big( x , \underline{\mu} \big) } \cdot r^{6V}_{+} {\tiny \big( \underline{\lambda} - \underline{\mu} \big) }  - r^{6V} {\tiny \big( \underline{\lambda} - \underline{\mu} \big) } \cdot T^{6V}_{+} {\tiny \big( x , \underline{\lambda} \big) }\end{align*}

      \begin{align*}    \bigotimes T^{6V}_{+} {\tiny \big( x , \underline{\mu} \big)  }   \end{align*}

      \begin{align*}    \big\{  T^{3D}_{-} {\tiny \big( x , \underline{\lambda} \big) } \overset{\bigotimes}{,} T^{3D}_{-} {\tiny \big( x , \underline{\mu} \big)  }          \big\} = r {\tiny \big( \underline{\lambda} - \underline{\mu} \big) } \cdot T^{3D}_{-} {\tiny \big( x , \underline{\lambda} \big)} \bigotimes T^{3D}_{-} {\tiny \big( x , \underline{\mu} \big)}  - T^{3D}_{-} {\tiny \big( x , \underline{\lambda} \big) } \bigotimes T^{3D}_{-} {\tiny \big( x , \underline{\mu} \big)}  \end{align*}

      \begin{align*}    \times r^{3D}_{-} {\tiny \big( \underline{\lambda} - \underline{\mu} \big)   }       \text{, } \end{align*}

      \begin{align*}        \big\{  T^{3D}_{+} {\tiny \big( x , \underline{\lambda} \big) } \overset{\bigotimes}{,} T^{3D}_{+} {\tiny \big( x , \underline{\mu} \big) }          \big\} = T^{3D}_{+} {\tiny \big( x , \underline{\lambda} \big) } \bigotimes T^{3D}_{+} {\tiny \big( x , \underline{\mu} \big)} \cdot  r^{3D}_{+} {\tiny \big( \underline{\lambda} - \underline{\mu} \big)}  - r^{3D} {\tiny \big( \underline{\lambda} - \underline{\mu} \big) } \cdot T^{3D}_{+} {\tiny \big( x , \underline{\lambda} \big) } \end{align*}

      \begin{align*}    \bigotimes T^{3D}_{+} {\tiny \big( x , \underline{\mu} \big)  }  \text{, }  \end{align*}

      \begin{align*}      \big\{  T^{4V}_{-} {\tiny \big( x , \underline{\lambda} \big) } \overset{\bigotimes}{,} T^{4V}_{-} {\tiny \big( x , \underline{\mu} \big)  }          \big\} = r^{4V} {\tiny \big( \underline{\lambda} - \underline{\mu} \big) } \cdot T^{4V}_{-} {\tiny \big( x , \underline{\lambda} \big)} \bigotimes T^{4V}_{-} {\tiny \big( x , \underline{\mu} \big)}  - T^{4V}_{-} {\tiny\big( x , \underline{\lambda} \big) } \bigotimes T^{4V}_{-} {\tiny \big( x , \underline{\mu} \big)}  \end{align*}

      \begin{align*}    \times r^{4V}_{-} {\tiny \big( \underline{\lambda} - \underline{\mu} \big)   }       \text{, } \end{align*}

      \begin{align*}      \big\{  T^{4V}_{+} {\tiny \big( x , \underline{\lambda} \big) } \overset{\bigotimes}{,} T^{4V}_{+} {\tiny \big( x , \underline{\mu} \big)   }         \big\} = T^{4V}_{+} {\tiny \big( x , \underline{\lambda} \big) } \bigotimes T^{4V}_{+} {\tiny \big( x , \underline{\mu} \big) } \cdot r^{4V}_{+} {\tiny \big( \underline{\lambda} - \underline{\mu} \big) }  - r^{4V} {\tiny \big( \underline{\lambda} - \underline{\mu} \big) } \cdot T^{4V}_{+} {\tiny \big( x , \underline{\lambda} \big) } \end{align*}

      \begin{align*}    \bigotimes T^{4V}_{+} {\tiny \big( x , \underline{\mu} \big)   } \text{, }
\end{align*} }

\noindent for the transfer matrices of the 6-vertex, 20-vertex, and 4-vertex, models satisfy,

\begin{align*}
   T^{6V}_{\pm} {\tiny \big( x , {\lambda} \big) } = \underset{y \longrightarrow \pm \infty}{\mathrm{lim}} T {\tiny \big( x , y , {\lambda} \big)} \cdot  E^{6V} {\tiny \big( y , {\lambda} \big) } \text{, } \\ 
 T^{3D}_{\pm} {\tiny \big( x , \underline{\lambda} \big) } = \underset{y \longrightarrow \pm \infty}{\mathrm{lim}} T {\tiny \big( x , y , \underline{\lambda} \big)} \cdot  E^{20V} {\tiny \big( y , \underline{\lambda} \big)   } 
 \text{, } \\   T^{4V}_{\pm} {\tiny \big( x , {\lambda} \big) } = \underset{y \longrightarrow \pm \infty}{\mathrm{lim}} T {\tiny \big( x , y , {\lambda} \big) } \cdot  E^{4V} {\tiny \big( y , {\lambda} \big)  } \text{, }
\end{align*}

\noindent While each transfer matrix defined through the weak finite volume procedure above (through the choice of suitably chosen spectral parameters $\lambda$ over $\textbf{Z}^2$, and $\underline{\lambda}$ over $\textbf{T}$) is respectively given by, for $j,k \in \textbf{N}$,

\begin{align*}
      T^{6V}_a  {\tiny \big( u , \big\{ v_k \big\} , H , V \big) }  \equiv 
  \begin{bmatrix}
       A {\tiny \big( u \big) }  & B {\tiny \big( u \big)}   \\
    C {\tiny \big( u \big) } & D {\tiny \big( u \big) } \text{ }  
  \end{bmatrix}   \text{, } \end{align*}

  \begin{align*} \textbf{T}^{3D} {\tiny \big( \underline{\lambda} \big) }  \equiv    \underset{ N \longrightarrow -  \infty}{\underset{\underline{M} \longrightarrow + \infty}{\mathrm{lim}}} \mathrm{tr} \bigg\{     \overset{\underline{M}}{\underset{j=0}{\prod}}  \text{ }  \overset{-N}{\underset{k=0}{\prod}} \mathrm{exp} \big( \lambda_3 {\tiny \big( q^{-2} \xi^{s^j_k} \big) }  \big)     \bigg[ \begin{smallmatrix}     q^{D^j_k}       &    q^{-2} a^j_k q^{-D^j_k -D^j_{k+1}} \xi^{s-s^k_j}        & *_1 \\ \big( a^j_k \big)^{\dagger} q^{D^j_k} \xi^{s^j_k} 
             &     *_2 &   - a^j_k q^{D^j_k - 3D^j_{k+1}} \xi^{s-s^j_k}  \\ 0  &    a^{\dagger}_j q^{D^j_k} \xi^{s^j_k} &  q^{-D^j_k} \\    \end{smallmatrix} \bigg]      \bigg\} \text{, } \end{align*}

             \begin{align*} T^{4V} {\tiny \big( \underline{u} \big) } \equiv     \underset{0 \leq j \leq M}{\prod}  L^{4V} {\tiny \big( j | u \big) } \equiv    \begin{bmatrix}
        A {\tiny \big( \underline{u} \big)} & B {\tiny \big( \underline{u} \big)} \\ C {\tiny\big( \underline{u} \big)} & D {\tiny\big( \underline{u} \big)} 
    \end{bmatrix}     \text{, }
\end{align*}

\noindent for,

\begin{align*}
  *_1 \equiv  a^j_k a^j_{k+1} q^{-D^j_k - 3D^j_{k+1}} \xi^{s - s^j_k - s^j_{k+1}}    \text{, }\\  \\  *_2 \equiv  q^{-D^j_k + D^j_{k+1}} - q^{-2} q^{D^j_k -D^j_{k+1}} \xi^{s}       \text{. }
\end{align*}

\noindent An adaptation of such an integrability property was also demonstrated by the author is being formulated for the 20-vertex model, {[15]}, with the L-operators, integrable structures of vertex models is ultimately dependent upon whether Poisson brackets,

{\small \begin{align*}
    \big\{  \Phi^{6V} {\tiny \big( {\lambda} \big)  } ,   \bar{\Phi^{6V} {\tiny \big( {\lambda} \big) }  } \big\}      \text{, } \big\{  \Phi^{20V} {\tiny \big( \underline{\lambda} \big)}  ,   \bar{\Phi^{20V} {\tiny \big( \underline{\lambda} \big)}  } \big\}       \text{, }  \big\{  \Phi^{4V} {\tiny \big( {\lambda} \big) } ,   \bar{\Phi^{4V} {\tiny \big( {\lambda} \big) } } \big\}       \text{, } 
\end{align*} }

\noindent of action-angle variables, with respect to their complex conjugates, vanish. The existence of such a vanishing Poisson bracket would imply that, with respect to the action-angle coordinates, the associated, potentially time-dependent, dynamics of the system would be linear. In comparison to other vertex models described in this article, the 20-vertex model, as the highest-dimensional counterpart, exhibits a Poisson structure with the most number of brackets. That is, in comparison to the following collection,

\[
\left\{\!\begin{array}{ll@{}>{{}}l} \underline{(1)}:      \big\{  A {\tiny \big( u \big)     }   , A {\tiny \big( u^{\prime} \big) }   \big\} 
\text{ , } \\  \vdots  \\ \underline{(16)}: \big\{  D {\tiny \big( u \big) } ,       D {\tiny \big( u^{\prime} \big) }  \big\} \text{, }
\end{array}\right.
\]



\noindent of 16 Poisson brackets from block representations of the transfer matrix for the 6-vertex model, the transfer for the 20-vertex model having more block entry representations, from the fact that,

\[
\begin{bmatrix}
 A {\tiny \big( \underline{u} \big) }  & D {\tiny \big( \underline{u} \big)}   & G {\tiny \big( \underline{u} \big) }  \\ B {\tiny \big( \underline{u} \big) }  & E {\tiny \big( \underline{u} \big) }  & H {\tiny \big( \underline{u} \big) }  \\ C {\tiny \big( \underline{u} \big) }  &  F {\tiny \big( \underline{u} \big) } & I {\tiny \big( \underline{u} \big) }
\end{bmatrix}  \text{, } 
\]

\noindent implies that the Poisson structure has $81$ brackets, which take the form, [15],

{\small \[
\left\{\!\begin{array}{ll@{}>{{}}l} \boxed{(1)}:      \big\{  A {\tiny\big( \underline{u} \big) }       , A {\tiny\big( \underline{u^{\prime}} \big)   } \big\} 
\text{, } &  \boxed{(22)}: \big\{ C {\tiny\big( \underline{u} \big) } , D {\tiny \big( \underline{u^{\prime}} \big) } \big\} \text{, }  &  \boxed{(43)}: \big\{ E {\tiny \big( \underline{u} \big)}  , G {\tiny\big( \underline{u^{\prime}} \big)} \big\} \text{, }  \\  \boxed{(2)}:  \big\{         A {\tiny \big( \underline{u} \big)  }      ,        B {\tiny \big( \underline{u^{\prime}} \big) }     \big\}  \text{
, } &  \boxed{(23)}:   \big\{  C {\tiny \big( \underline{u} \big) } , E {\tiny \big( \underline{u^{\prime}} \big) }  \big\}  \text{, } &  \boxed{(44)}: \big\{ E {\tiny \big( \underline{u} \big)} , H {\tiny\big( \underline{u^{\prime}} \big)}  \big\} \text{, }   \\  \boxed{(3)}:   \big\{   A {\tiny \big( \underline{u} \big)  }     ,  C {\tiny \big( \underline{u^{\prime}} \big)} \big\} 
 \text{, } &  \boxed{(24)}: \big\{ C {\tiny\big( \underline{u} \big)} , F {\tiny\big( \underline{u^{\prime}} \big) }\big\} \text{, } & \boxed{(45)}: \big\{ E {\tiny\big( \underline{u} \big)} , I {\tiny\big( \underline{u^{\prime}} \big)} \big\} \text{, } \\ \boxed{ (4)}: \big\{   A {\tiny\big( \underline{u} \big)  }     ,  D {\tiny\big( \underline{u^{\prime}} \big)}  \big\}   \text{, } & \boxed{(25)}: \big\{ C {\tiny \big( \underline{u} \big)}  , G {\tiny\big( \underline{u^{\prime} } \big)}  \big\} \text{, } &  \boxed{(46)}: \big\{ F {\tiny\big( \underline{u} \big) } ,     A {\tiny \big( \underline{u^{\prime }} \big)}  \big\} \text{, } \\    \boxed{(5)}: \big\{ A {\tiny\big( \underline{u} \big)} , E {\tiny\big( \underline{u^{\prime}} \big)} \big\}  \text{,  } &   \boxed{(26)}: \big\{ C {\tiny\big( \underline{u} \big)} , H {\tiny\big( \underline{u^{\prime}} \big)}  \big\} \text{, } & \boxed{(47)}:  \big\{ F {\tiny\big( \underline{u} \big)} , B {\tiny\big( \underline{u^{\prime}}  \big)}  \big\} \text{, }  \\   \boxed{(6)} : \big\{ A {\tiny\big( \underline{u} \big)} , F {\tiny\big( \underline{u^{\prime} } \big)} \big\}  \text{ , }   & \boxed{(27)}: \big\{ C {\tiny\big( \underline{u} \big)} , I {\tiny\big( \underline{u^{\prime}}  \big)} \big\} \text{, } &     \boxed{(48)}: \big\{ F {\tiny\big( \underline{u} \big)} , C {\tiny\big( \underline{u^{\prime}} \big)}  \big\} \text{, }     \\   \boxed{(7)}: \big\{ A {\tiny\big( \underline{u} \big)} , G {\tiny\big( \underline{u^{\prime}} \big) } \big\} \text{, }  &    \boxed{(28)}: \big\{ D {\tiny\big( \underline{u} \big) } , A {\tiny\big(\underline{u^{\prime}} \big)}   \big\} \text{, } & \boxed{(49)}:   \big\{ F {\tiny\big( \underline{u} \big)} , D {\tiny\big( \underline{u^{\prime}} \big)} \big\} \text{, }  \\     \boxed{(8)}: \big\{ A {\tiny\big( \underline{u} \big) }, H {\tiny\big( \underline{u^{\prime}} \big)} \big\} &  \boxed{(29)}: \big\{  D {\tiny\big( \underline{u} \big) } , B {\tiny\big( \underline{u^{\prime}} \big)  } \big\} \text{, } &   \boxed{(50)}: \big\{ F {\tiny\big( \underline{u} \big)} , E {\tiny\big( \underline{u^{\prime}} \big)}  \big\} \text{, }  \\ \boxed{(9)}: \big\{ A {\tiny\big( \underline{u} \big)} , I {\tiny\big( \underline{u^{\prime}} \big) }\big\}  \text{, } &    \boxed{(30)}: \big\{  D {\tiny\big( \underline{u} \big)}  , C {\tiny\big( \underline{u^{\prime}} \big)} \big\} \text{, } &  \boxed{(51)}:   \big\{ F {\tiny\big( \underline{u} \big) } , F {\tiny\big( \underline{u^{\prime}} \big)} \big\} \text{, }  \\  \boxed{(10)} : \big\{ B {\tiny\big( \underline{u} \big)} , A {\tiny\big( \underline{u^{\prime}} \big)} \big\} \text{, } &    \boxed{(31)}: \big\{  D {\tiny\big( \underline{u} \big) } ,       D {\tiny\big( \underline{u^{\prime}} \big)  \big\} } \text{, } &  \boxed{(52)}: \big\{ F {\tiny\big( \underline{u} \big) }, G {\tiny\big( \underline{u^{\prime}} \big)}  \big\} \text{, }
\\     \boxed{(11)}: \big\{ B {\tiny\big( \underline{u} \big)} , B {\tiny\big( \underline{u^{\prime}} \big)} \big\} \text{, } &   \boxed{(32)} : \big\{ D {\tiny\big( \underline{u} \big)} , E {\tiny\big( \underline{u^{\prime}}\big)} \big\} \text{, } &   \boxed{(53)}: \big\{ F {\tiny\big( \underline{u} \big) } , H {\tiny\big( \underline{u^{\prime}} \big)} \big\} \text{, } \\  \boxed{(12)}: \big\{ B {\tiny\big( \underline{u} \big) }, C {\tiny\big( \underline{u^{\prime}} \big)} \big\} \text{, }  &  \boxed{(33)}: \big\{ D {\tiny\big( \underline{u} \big)} , F {\tiny\big( \underline{u^{\prime}} \big)} \big\} \text{, } &   \boxed{(54)}:  \big\{ F {\tiny\big( \underline{u} \big)} , I {\tiny\big( \underline{u^{\prime}} \big)} \big\} \text{, }  
\\     \boxed{(13)}: \big\{ B {\tiny\big( \underline{u} \big)}  , D {\tiny\big( \underline{u^{\prime}} \big)}   \big\} \text{, } & \boxed{(34)}: \big\{ D {\tiny\big( \underline{u} \big) }, G {\tiny\big( \underline{u^{\prime}} \big)} \big\} \text{, }  &  \boxed{(55)}: \big\{ G {\tiny\big( \underline{u} \big)} , A {\tiny\big( \underline{u^{\prime}} \big) } \big\} \text{, }  \end{array}\right.
\]

 \[\left\{\!\begin{array}{ll@{}>{{}}l}     \boxed{(14)}: \big\{ B {\tiny\big( \underline{u} \big)} , E {\tiny\big( \underline{u^{\prime}} \big)} \big\}  \text{, } &   \boxed{(35)}: \big\{ D {\tiny\big( \underline{u} \big)} , H {\tiny\big( \underline{u^{\prime}} \big)} \big\} \text{, } &   \boxed{(56)}: \big\{ G {\tiny\big( \underline{u} \big)} , B {\tiny\big( \underline{u^{\prime}} \big)} \big\} \text{, } \\        \boxed{(15)}: \big\{ B {\tiny\big( \underline{u} \big)} , F {\tiny\big( \underline{u^{\prime}} \big)} \big\}  \text{, } &   \boxed{(36)}: \big\{ D {\tiny\big( \underline{u} \big)} , I {\tiny\big( \underline{u^{\prime}} \big)} \big\} \text{, } &  \boxed{(57)}:  \big\{ G {\tiny\big( \underline{u} \big)} , C {\tiny\big( \underline{u^{\prime}} \big) }  \big\} \text{, } \\    \boxed{(16)}: \big\{ B {\tiny\big( \underline{u} \big)} , G {\tiny\big( \underline{u^{\prime}} \big)} \big\} \text{, } & \boxed{(37)}: \big\{ E {\tiny\big( \underline{u} \big) }, A {\tiny\big( \underline{u^{\prime}} \big)} \big\}  \text{, } &  \boxed{(58)}: \big\{ G {\tiny\big( \underline{u} \big)}, D {\tiny\big( \underline{u^{\prime} }\big) } \big\} \text{, } \\    \boxed{(17)}: \big\{ B {\tiny\big( \underline{u} \big)} , H {\tiny\big( \underline{u^{\prime}} \big) } \big\} \text{, } & \boxed{(38)}: \big\{ E {\tiny\big( \underline{u} \big) } , B {\tiny\big( \underline{u^{\prime}} \big) }\big\} \text{, } & \boxed{(59)} : \big\{ G {\tiny\big( \underline{u} \big)}  , E {\tiny\big( \underline{u^{\prime}} \big) } \big\} \text{, }  
 \\ \boxed{(18)}: \big\{ B {\tiny\big( \underline{u} \big)} , I {\tiny\big( \underline{u^{\prime}} \big) } \big\}   \text{, }  &    \boxed{(39)}: \big\{ E {\tiny\big( \underline{u} \big)} , C {\tiny\big( \underline{u^{\prime}} \big)}  \big\} \text{, } & \boxed{(60)}:  \big\{ G {\tiny\big( \underline{u} \big)} , F {\tiny\big( \underline{u^{\prime}} \big) } \big\} \text{, }  \\ \boxed{(19)}:  \big\{ C {\tiny\big( \underline{u} \big)}  , A {\tiny\big( \underline{u^{\prime}} \big) }  \big\} \text{, } &    \boxed{(40)}:   \big\{ E {\tiny\big( \underline{u} \big)} , D {\tiny\big( \underline{u^{\prime}} \big) } \big\} \text{, } & \boxed{(61)}: \big\{ G {\tiny\big( \underline{u} \big) } , G {\tiny\big( \underline{u^{\prime}} \big)}  \big\} \text{, } \\ \boxed{(20)}: \big\{ C {\tiny\big( \underline{u} \big)} , B {\tiny\big( \underline{u^{\prime}} \big) } \big\} \text{, } & \boxed{(41)}: \big\{ E {\tiny\big( \underline{u} \big) }, E {\tiny\big( \underline{u^{\prime}} \big)} \big\} \text{, } & \boxed{(62)}: \big\{ G {\tiny\big( \underline{u} \big)} , H {\tiny\big( \underline{u^{\prime}} \big)}  \big\} \text{, } \\ \boxed{(21)}: \big\{ C {\tiny\big( \underline{u} \big) }, C {\tiny\big( \underline{u^{\prime}} \big)} \big\} \text{, }  &  \boxed{(42)}: \big\{ E {\tiny\big( \underline{u} \big) } , F {\tiny\big( \underline{u^{\prime}} \big)}  \big\} \text{, } &  \boxed{(63)}: \big\{ G {\tiny\big( \underline{u} \big)} , I {\tiny\big( \underline{u^{\prime}} \big)}  \big\} \text{, }
 \end{array}\right.
\]  }

\noindent corresponding to the first $63$ brackets within the structure, and,

{\small \[\left\{\!\begin{array}{ll@{}>{{}}l}   \boxed{(64)}: \big\{ H {\tiny\big( \underline{u} \big)} , A {\tiny\big( \underline{u^{\prime}} \big)}  \big\} \text{, } & \boxed{(70)}: \big\{ H {\tiny\big(\underline{u} \big) }, G {\tiny\big( \underline{u^{\prime}}  \big) } \big\} \text{, } & \boxed{(76)}: \big\{ I {\tiny\big( \underline{u} \big)} , D {\tiny\big( \underline{u^{\prime}} \big)} \big\} \text{, }  \\ \boxed{(65)}: \big\{ H {\tiny\big( \underline{u} \big)} , B {\tiny\big( \underline{u^{\prime}} \big)} \big\} \text{, } & \boxed{(71)}:  \big\{ H {\tiny\big( \underline{u} \big)} , H {\tiny\big( \underline{u^{\prime} }\big)} \big\} \text{, }      &   \boxed{(77)}: \big\{ I {\tiny\big( \underline{u} \big)} , E {\tiny\big( \underline{u^{\prime}} \big)}  \big\} \text{, }   \\ \boxed{(66)}: \big\{ H {\tiny\big( \underline{u} \big) }, C {\tiny\big( \underline{u^{\prime}} \big) }\big\} \text{, } &  \boxed{(72)}: \big\{ H {\tiny\big( \underline{u} \big)} , I {\tiny\big( \underline{u^{\prime}} \big)} \big\} \text{, }         &       \boxed{(78)}: \big\{ I {\tiny\big( \underline{u} \big)} , F {\tiny\big( \underline{u^{\prime}} \big)} \big\} \text{, }    \\ \boxed{(67)} : \big\{ H {\tiny\big( \underline{u} \big)} , D {\tiny\big( \underline{u^{\prime}} \big)} \big\} \text{ , }  &  \boxed{(73)}: \big\{ I {\tiny\big( \underline{u} \big)} , A {\tiny\big( \underline{u^{\prime}} \big)} \big\} \text{, }        &   \boxed{(79)}: \big\{ I {\tiny\big( \underline{u} \big)} , G {\tiny\big( \underline{u^{\prime}} \big)} \big\} \text{, }  \\ \boxed{(68)}: \big\{ H {\tiny\big( \underline{u} \big) }, E {\tiny\big( \underline{u^{\prime}} \big)}  \big\} \text{, }   &    \boxed{(74)}: \big\{ I {\tiny\big( \underline{u} \big)} , B {\tiny\big( \underline{u^{\prime}} \big)} \big\} \text{, }        &         \boxed{(80)}:  \big\{ I {\tiny\big( \underline{u} \big)} , H {\tiny\big( \underline{u^{\prime}} \big)} \big\} \text{, }          \\ \boxed{(69)}: \big\{ H {\tiny\big( \underline{u} \big)} , F {\tiny\big( \underline{u^{\prime}} \big)} \big\} \text{, }  &  \boxed{(75)}: \big\{ I {\tiny\big( \underline{u} \big)} , C {\tiny\big( \underline{u^{\prime}} \big)} \big\} \text{, }     & \boxed{(81)}: \big\{ I {\tiny\big( \underline{u} \big)} , I {\tiny\big( \underline{u^{\prime}} \big)} \big\} \text{, }   
\end{array}\right.
\] }

\noindent corresponding to the remaining $17$ brackets within the structure. To quantify asymptotic properties of the limit,

\[ \textbf{T}^{3D} {\tiny\big( \underline{\lambda} \big) }   \equiv    \underset{ N \longrightarrow -  \infty}{\underset{\underline{M} \longrightarrow + \infty}{\mathrm{lim}}} \mathrm{tr} \bigg\{      \overset{\underline{M}}{\underset{j=0}{\prod}}  \text{ }  \overset{-N}{\underset{k=0}{\prod}} \mathrm{exp} \big( \lambda_3 {\tiny\big( q^{-2} \xi^{s^j_k} \big)}  \big)     \bigg[ \begin{smallmatrix}     q^{D^j_k}       &    q^{-2} a^j_k q^{-D^j_k -D^j_{k+1}} \xi^{s-s^k_j}        & *_1 \\ \big( a^j_k \big)^{\dagger} q^{D^j_k} \xi^{s^j_k} 
             &     *_2 &   - a^j_k q^{D^j_k - 3D^j_{k+1}} \xi^{s-s^j_k}  \\ 0  &    a^{\dagger}_j q^{D^j_k} \xi^{s^j_k} &  q^{-D^j_k} \\    \end{smallmatrix} \bigg]      \bigg\}  \text{, }
\]

\noindent over $\textbf{T}$, we compute products of L-operators by taking into account contributions from the following groups of terms. The various terms associated with entries of the three-dimensional product representation of L-operators provided for the transfer matrix of the 20-vertex model provided above correspond to powers of $q$, images of the unital associative mapping $\xi$, in addition to cross term products between $q$ and $\xi$.

\bigskip

\noindent Products of blocks representations in transfer matrices for the 20-vertex model satisfy,

{\small \[
\left\{\!\begin{array}{ll@{}>{{}}l} 
   \underline{G {\tiny\big( \underline{u} \big) } E {\tiny\big( \underline{u^{\prime}} \big)}  C {\tiny\big( \underline{u^{\prime\prime}} \big)} } =       f {\tiny\big( \lambda_{\alpha} , \lambda_r , \lambda_{r^{\prime}} \big)}    f {\tiny\big( \lambda , \lambda^{\prime} \big)}     C {\tiny\big( \underline{u^{\prime\prime}} \big)} E {\tiny\big( \underline{u^{\prime}} \big)}  G {\tiny\big( \underline{u} \big) }    +    f {\tiny\big( \lambda_{\alpha} , \lambda_r , \lambda_{r^{\prime}} \big) } g {\tiny\big( \lambda^{\prime} , \lambda \big) }   \\ \times   C {\tiny\big( \underline{u^{\prime}}    \big)} E {\tiny\big( \underline{u^{\prime\prime}} \big) }   G {\tiny\big( \underline{u} \big)                 }           +        g {\tiny\big( \lambda_{\alpha} , \lambda_r , \lambda_{r^{\prime}} \big)   }   f {\tiny\big(   \lambda , \lambda^{\prime} \big)}  C {\tiny\big( \underline{u^{\prime}} \big) E \big( \underline{u} \big)    }     G {\tiny\big( \underline{u^{\prime\prime} } \big)     }         +  g {\tiny\big( \lambda_{\alpha} , \lambda_r , \lambda_{r^{\prime}} \big)      } g {\tiny \big( \lambda^{\prime} , \lambda \big)} \\ \times   C {\tiny\big(  \underline{u}     \big)} E {\tiny\big( \underline{u^{\prime}}       \big)  }                    G {\tiny\big( \underline{u^{\prime\prime} } \big) }                \text{, }  \\  \vdots  \\ 
         \underline{A {\tiny\big( \underline{u } \big)}  E {\tiny\big( \underline{u^{\prime}} \big) } I {\tiny\big( \underline{u^{\prime\prime}} \big) } }       =           f {\tiny\big( \lambda_{\alpha}  , \lambda_r , \lambda_{r^{\prime}} \big)    }    f {\tiny\big( \lambda , \lambda^{\prime} \big)  }   I {\tiny\big( \underline{u^{\prime\prime}} \big) } E {\tiny\big( \underline{u^{\prime}} \big) } A {\tiny\big( \underline{u} \big)}    +    f {\tiny\big( \lambda_{\alpha}  , \lambda_r , \lambda_{r^{\prime}} \big)  }   g {\tiny\big( \lambda^{\prime} , \lambda \big)  }        \\  \times      I {\tiny\big( \underline{u^{\prime}} \big)} E {\tiny\big( \underline{u^{\prime\prime}} \big)  }         A {\tiny\big( \underline{u} \big) }    +   g {\tiny\big( \lambda_{\alpha} , \lambda_r , \lambda_{r^{\prime}} \big)    }    f {\tiny\big( \lambda , \lambda^{\prime} \big)  }   I {\tiny\big( \underline{u}  \big)} E {\tiny\big( \underline{u^{\prime\prime}} \big) }              A {\tiny\big( \underline{u^{\prime}} \big)  }    + g {\tiny\big( \lambda_{\alpha} , \lambda_r , \lambda_{r^{\prime}} \big)  }g {\tiny\big( \lambda^{\prime} , \lambda \big)  }     \\ \times               I {\tiny\big( \underline{u^{\prime\prime}} \big)   }        E {\tiny\big( \underline{u}  \big)  }    A {\tiny\big( \underline{u^{\prime}} \big)  }            \text{, } 
\end{array}\right.
\] }

\noindent where,

{\small \begin{align*}
   f {\tiny\big( \lambda_{\alpha} , \lambda_{r} , \lambda_{r^{\prime}} \big) } \equiv     \frac{\mathrm{sin} {\tiny\big( \lambda_{r^{\prime}} - \lambda_r - \lambda_{\alpha} + 2 \eta  \big)} }{\mathrm{sin} {\tiny\big( \lambda_{r^{\prime}} - \lambda_r - \lambda_{\alpha } \big)} }                    \text{, } \end{align*}

   \begin{align*} 
g {\tiny\big( \lambda_{\alpha} , \lambda_{r} , \lambda_{r^{\prime}} \big)}  \equiv   \frac{\mathrm{sin} {\tiny\big( 2 \eta \big)}}{\mathrm{sin} {\tiny\big( \lambda_{r^{\prime}} - \lambda_r - \lambda_{\alpha} \big)} }                  \text{, } 
\end{align*} }

\noindent similar to the functions used to formulate correlation functions of the 6-vertex model under domain-wall boundary conditions, [3]. The above expressions for the product of block representations of the 20-vertex transfer matrix is generated from the embedded Poisson bracket, [15],

\[\
\bigg\{    \begin{bmatrix}
 A {\tiny\big( \underline{u} 
 \big)} & D {\tiny\big( \underline{u} \big) } & G {\tiny\big( \underline{u}  \big)} \\ B {\tiny\big( \underline{u}  \big)} & E {\tiny\big( \underline{u}  \big)} & H {\tiny\big( \underline{u}  \big)}  \\ C {\tiny\big( \underline{u}  \big) } &  F {\tiny\big( \underline{u}  \big)} & I {\tiny\big( \underline{u}  \big) }
\end{bmatrix}\overset{\bigotimes}{,} \bigg\{ \begin{bmatrix}
 A {\tiny\big( \underline{u^{\prime}} \big) }& D {\tiny\big( \underline{u^{\prime}} \big) } & G {\tiny\big( \underline{u^{\prime}} \big) }\\ B {\tiny\big( \underline{u^{\prime}}\big)}  & E {\tiny\big( \underline{u^{\prime}} \big)} & H {\tiny\big( \underline{u^{\prime}} \big)}  \\ C {\tiny\big( \underline{u^{\prime}} \big) } &  F {\tiny\big( \underline{u^{\prime}} \big) }& I {\tiny\big( \underline{u^{\prime}} \big) } 
\end{bmatrix}  \overset{\bigotimes}{,} \begin{bmatrix}
 A {\tiny\big( \underline{u^{\prime\prime}} \big)} & D {\tiny\big( \underline{u^{\prime\prime}} \big)  }& G {\tiny\big( \underline{u^{\prime\prime}} \big) }\\ B {\tiny\big( \underline{u^{\prime\prime}}\big) }& E {\tiny\big( \underline{u^{\prime\prime }} \big)} & H {\tiny\big( \underline{u^{\prime\prime}} \big)  } \\ C {\tiny\big( \underline{u^{\prime\prime}} \big)  }&  F {\tiny\big( \underline{u^{\prime\prime}} \big) }& I {\tiny\big( \underline{u^{\prime\prime}} \big)} 
\end{bmatrix}  \bigg\}  \text{ } \bigg\} \text{, } 
\]

\noindent where each transfer matrix is dependent upon a different collection of spectral parameters, whether it be $u, u^{\prime}$, or $u^{\prime\prime}$.

\subsection{Objects for the $6$-vertex model}

\noindent Over the torus $\textbf{T}_{N} \equiv \big( V {\tiny\big(} \textbf{T}_N {\tiny \big)} , E {\tiny\big(} \textbf{T}_N {\tiny\big)} \big)$, the six-vertex model can be defined through the probability measure,

{\small \begin{align*}
 \textbf{P}_{\textbf{T}_N} {\tiny\big[ \omega \big]} \equiv \textbf{P} {\tiny\big[ \omega \big] } \equiv     \frac{w {\tiny\big( \omega \big)}}{Z_{\textbf{T}_N}}           \text{, } 
\end{align*} } 

\noindent where $\omega$ is a \textit{six-vertex configuration} determined by the six possible configurations, with the weight in the numerator of the probability measure taking the form,

{\small \begin{align*}
  w_{\mathrm{6V}} {\tiny\big( \omega \big) }\equiv w {\tiny\big( \omega \big)} \equiv a_1^{n_1} a_2^{n_2} b_1^{n_3} b_2^{n_4} c_1^{n_5} c_2^{n_6}  \text{, }
\end{align*} } 

\noindent for $a_1 , a_2 , b_1 , b_2 , c_1 , c_2 \geq 0$, with the partition function,

{\small \begin{align*}
 Z_{\textbf{T}_N} {\tiny\big( \omega , \Omega \big)}  \equiv Z_{\textbf{T}_N} = \underset{\omega \in \Omega ( \textbf{T}_N ) }{\sum} w {\tiny\big( \omega \big) }  \text{. } 
\end{align*}  } 

\noindent Besides $\textbf{P}_{\textbf{T}_N} {\tiny \big[ \cdot \big]}$, the disorder parameter of the six-vertex model is of the form,

{\small \begin{align*}
    \Delta \equiv \frac{a_1 a_2 + a_3 a_4 - a_5 a_6}{2 \sqrt{a_1 a_2 a_3 a_4}}    \text{. } 
\end{align*} }

\noindent For non-symmetric weights, the weights of the six-vertex model admit the parametrization,

{\small \begin{align*}
   a_1 \equiv      a \text{ }  \mathrm{exp} {\tiny\big(  H + V \big)  } \text{, } \\ 
   a_2 \equiv   a  \text{ }  \mathrm{exp} {\tiny\big( - H - V \big) }  \text{, } \\  b_1 \equiv  \text{ }  \mathrm{exp} {\tiny\big( H - V \big) }   \text{, } \\ b_2 \equiv \text{ }  \mathrm{exp} {\tiny\big( - H + V \big) }  \text{, } \\ c_1 \equiv  c \lambda  \text{, } \\ c_2 \equiv c \lambda^{-1} \text{, } 
\end{align*} } 

\noindent for $a_1 \equiv a_2 \equiv a$, $b_1 \equiv b_2 \equiv b$, $c_1 \equiv c_2 \equiv c$, and $\lambda \geq 1$, and external fields $H,V$. From such a parametrization of the weights as given above, one can form the so-called $R$-matrix, for the standard basis of $\textbf{C}^2$, with,

{\small \[
R \equiv R {\tiny\big( u , H , V \big)}  \equiv 
  \begin{bmatrix}
      a \text{ }  \mathrm{exp} {\tiny\big(  H + V \big) }   & 0 & 0 & 0  \\
    0 & b \text{ } \mathrm{exp} {\tiny\big( H - V \big)} & c & 0  \\0 & c & b \text{ }  \mathrm{exp} {\tiny\big( - H + V \big)} & 0 \\ 0 & 0 & 0 & a \text{ }  \mathrm{exp} {\tiny\big( - H - V \big) } \\ 
  \end{bmatrix} \text{, }
\] }

\noindent in the tensor product basis $e_1 \otimes e_1$, $e_1 \otimes e_2$, $e_2 \otimes e_1$, $e_2 \otimes e_2$, for $e_1 \equiv {\tiny\big[} 1 \text{  } 0 {\tiny\big]^{\mathrm{T}}}$ and $e_1 \equiv {\tiny\big[} 0 \text{  } 1 {\tiny\big]^{\mathrm{T}}}$. For vanishing external fields $H \equiv V \equiv 0$, if we denote $R {\tiny\big( u \big)} \equiv R {\tiny\big( u , 0 , 0 \big)}$, the matrix above admits the identity,

{\small \begin{align*}
     R {\tiny\big( u , H , V \big) } =    \bigg[    \mathrm{diag} \big[ \mathrm{exp} {\tiny\big(} \frac{H}{2} {\tiny\big)} , \mathrm{exp} {\tiny\big(} - \frac{H}{2} {\tiny\big)}  \big] \otimes      \mathrm{diag} \big[ \mathrm{exp} {\tiny\big( \frac{V}{2} \big)} , \mathrm{exp} {\tiny\big( - \frac{V}{2} \big)}  \big]    \bigg] \cdot R {\tiny\big( u \big)} \cdot  \bigg[     \mathrm{diag} \big[ \mathrm{exp} {\tiny\big( }\frac{H}{2} {\tiny\big) }, \mathrm{exp} {\tiny\big(} - \frac{H}{2} {\tiny\big) } \big] \\ \otimes    \mathrm{diag} \big[ \mathrm{exp} {\tiny\big( } \frac{V}{2} {\tiny\big)} , \mathrm{exp} {\tiny\big(} - \frac{V}{2} {\tiny\big)}   \big]          \bigg]   \equiv  {\tiny\big(} D^H \otimes D^V {\tiny\big)} \cdot  R {\tiny\big(} u {\tiny\big) } \cdot {\tiny\big( } D^H \otimes D^V {\tiny\big)  }     \text{, } 
\end{align*} }

\noindent over $\textbf{C}^2 \otimes \textbf{C}^2 \otimes \textbf{C}^2$. In widely celebrated work, [1], Baxter demonstrated that the $R$ matrix satisfies the Yang-Baxter equation,

{\small \begin{align*}
R_{12} {\tiny\big( u \big) } R_{13} {\tiny\big( u + v \big) }R_{23} {\tiny\big( v \big) }=  R_{23} {\tiny\big( v \big)} R_{13} {\tiny\big( u + v \big)} R_{12} {\tiny\big( u \big)  }   \text{. } 
\end{align*} }

\noindent For $\Delta < -1$, Baxter's parametrization, for the weights, given $0 < u < \eta$,

{\small \begin{align*}
        a \equiv   \mathrm{sinh} {\tiny\big( \eta - u \big) }\text{, } b \equiv \mathrm{sinh} {\tiny\big( u \big) } \text{, }     c     \equiv \mathrm{sinh} {\tiny\big( \eta \big) }  \text{, } 
\end{align*} } 

\noindent allows one to classify the eigenvalues of the transfer matrix which is related to the Bethe equations. Besides this fact, equipped with $u$, the horizontal lines of the square lattice, $H$ and $V$, two external fields, the partition function over the $N \times M$ torus, $\textbf{T}_{NM}$, can be expressed with the summation,

{\small \begin{align*}
      Z_{\textbf{T}_{MN}}   {\tiny\big( u , H \big) } \equiv    Z_{\textbf{T}_{MN}}  = \overset{N}{\underset{n=0}{\sum}}  \mathrm{exp} {\tiny\big( M \big( N - 2 n \big) V \big) }  \cdot  Z_{\textbf{T}_{MN}}^n {\tiny\big( u , H \big)  }        \text{, } 
\end{align*} }

\noindent for the semigrand canonical partition function,

{\small \begin{align*}
 Z_{\textbf{T}_{MN}}^n {\tiny\big( u , H \big) }  \equiv   \underset{\{ \alpha_i \}}{\underset{ i \geq 1}{\sum}}   \big( \Lambda_{\{ \alpha_i \}} {\tiny\big( u , H \big) } \big)^M            \text{, } 
\end{align*} } 

\noindent for countably many solutions $\big\{ \alpha_i \big\}$ to the Bethe equation, 

{\small \begin{align*}
     \overset{N}{\underset{k=1}{\prod}}   \frac{\mathrm{sinh} {\tiny\big(   \frac{\eta}{2} + i \alpha_j - v_k  \big)}}{\mathrm{sinh} {\tiny\big(     \frac{\eta}{2} - i \alpha_j + v_k    \big)}}       = \mathrm{exp} {\tiny\big( 2 H N          \big)} \cdot \overset{n}{\underset{m=1, m \neq j}{\prod} }            \frac{\mathrm{sinh} {\tiny\big(   i \big( \alpha_j - \alpha_m \big)+ \eta    \big)} }{\mathrm{sinh} {\tiny\big(   i \big( \alpha_j - \alpha_m \big) - \eta      \big)} }      \text{, } 
\end{align*} } 

\noindent and for the eigenvalue $\Lambda_{\{ \alpha_i \}}$,

{\small \begin{align*}
\Lambda \equiv  \Lambda_{\{ \alpha_i \}} \equiv \mathrm{exp} {\tiny\big( N H \big)} \cdot  \overset{N}{\underset{k=1}{\prod}}  \mathrm{sinh} {\tiny\big( \eta - u + v_k \big)} \overset{n}{\underset{j=1}{\prod}} \frac{\mathrm{sinh}{\tiny\big( \frac{\eta}{2} + u - i \alpha_j \big)} }{\mathrm{sinh} {\tiny\big( \frac{\eta}{2} - u + i \alpha_j \big) }} + \mathrm{exp} {\tiny\big( - NH \big)}  \cdot  \overset{n}{\underset{k=1}{\prod}} \mathrm{sinh} {\tiny\big( u - v_k \big) } \end{align*}

 \begin{align*} \times  \overset{n}{\underset{j=1}{\prod}}  \frac{\mathrm{sinh}{\tiny\big( \frac{3 \eta}{2} - u + i \alpha_j  \big) } }{\mathrm{sinh} {\tiny\big( u - \frac{\eta}{2} - i \alpha_j  \big)}  } \text{, }
\end{align*} } 

\noindent with corresponding eigenvector,

{\small \begin{align*}
\overset{n}{\underset{i=1}{\prod}}              B {\tiny\big( \alpha_i \big)}  \ket{\Downarrow}
  \equiv      \overset{n}{\underset{i=1}{\prod}}              B {\tiny\big( \alpha_i \big)}  \bigg[ \overset{n}{\underset{i=1}{\bigotimes}} \ket{\downarrow}
 \bigg]            \equiv   \overset{n}{\underset{i=1}{\prod}}              B {\tiny\big( \alpha_i \big) }  \bigg[  \ket{\downarrow}
 {\otimes}   \overset{n-2}{\cdots} \otimes  \ket{\downarrow}
 \bigg]     \equiv  \overset{n}{\underset{i=1}{\prod}}              B {\tiny\big( \alpha_i \big) } \bigg[  {{0}\choose{1}} 
 {\otimes}   \overset{n-2}{\cdots} \otimes  {{0}\choose{1} }
 \bigg] \text{, } 
\end{align*} } 

\noindent of the transfer matrix,

{\small \begin{align*}
   t \big( u , {\tiny\big\{ v_k \big\}}  , H , V \big) : \big( \textbf{C}^2 \big)^{\otimes N} \longrightarrow \big( \textbf{C}^2 \big)^{\otimes N}    \mapsto \bigg\{ \overset{N}{\underset{i=1}{\prod}}         D_i^{2V}  \cdot   \mathrm{Tr}_a \big[       T_a \big( u , {\tiny\big\{ v_k \big\} } , H , 0 \big)       \big]        \bigg\}     \text{, } 
\end{align*} } 

\noindent which, explicitly, is proportional to the trace of the Quantum monodromy matrix, which is given by,

{\small \begin{align*}
 T_a \big( u , {\tiny\big\{ v_k \big\}} , H , 0 \big) : \textbf{C}^2 \otimes \big( \textbf{C}^2 \big)^{\otimes N} \longrightarrow \textbf{C}^2 \otimes \big( \textbf{C}^2 \big)^{\otimes N}      \mapsto    \bigg\{ \overset{N}{\underset{i=1}{\prod}}   \mathrm{diag} \big( \mathrm{exp} {\tiny\big( 2H \big)} ,  \mathrm{exp} {\tiny\big( 2 H \big)}  \big)  \cdot      R_{ia} {\tiny\big( u - v_i \big) }    \bigg\}   \text{, } 
\end{align*} }

\noindent where each $v_i$ is chosen so that each $u-v_i$ is a spectral parameter given at site $i$. In the next section, to apply the formalism first introduced for Hamiltonian systems in the nonlinear Schroedinger equation, [6], to the Quantum monodromy and transfer matrices of the six-vertex model, observe that the product of diagonal matrices with each $R_{ia}$ is equivalent to, [8],

{\small \begin{align*}
     \mathrm{diag} \big( \mathrm{exp} {\tiny\big( 2H \big)} , \mathrm{exp} {\tiny\big( 2 H \big)}  \big)  \bigg[    R_{1a} {\tiny\big( u - v_1 \big) }  \cdots  \times  R_{(N-1)a} {\tiny\big( u - v_{N-1} \big) } \bigg]   \mathrm{diag} \big( \mathrm{exp} {\tiny\big( 2H \big) }, \mathrm{exp} {\tiny\big( 2 H \big)}   \big)  \cdot   R_{Na} {\tiny\big( u - v_N \big) } \text{. } 
\end{align*} } 

\noindent To formulate the Hamiltonian flow for the six-vertex model, from the statement of the Bethe equations, and their eigenvalues, introduce the functions $\psi^{\pm}_u {\tiny\big( \alpha + i u \big)}  \equiv  \psi_{\pm} {\tiny\big( \alpha + i u\big)}$, which are given by, [8],

{\small \begin{align*}
     \psi_{+} {\tiny\big( \alpha + i u\big)} = \mathrm{log} \bigg\{  \big|       \frac{\mathrm{exp} {\tiny\big(  \eta + 2 u   \big) } - \mathrm{exp} {\tiny\big(    2 i \alpha    \big) }}{\mathrm{exp} {\tiny\big( \eta - 2 i \alpha \big) } - \mathrm{exp} {\tiny\big(   2u      \big)} } \big| \bigg\}  = \mathrm{log}\bigg\{   \big| \frac{\mathrm{sinh} {\tiny\big(    \frac{\eta}{2} + u - i \alpha    \big)}  }{\mathrm{sinh} {\tiny\big(    \frac{\eta}{2} - u + i \alpha    \big)}  }  \big| \bigg\}  \equiv \mathrm{log}\bigg\{    \frac{\big| \mathrm{sinh}  {\tiny\big(    \frac{\eta}{2} + u - i \alpha    \big) } \big| }{ \big|\mathrm{sinh} {\tiny\big(    \frac{\eta}{2} - u + i \alpha    \big) } \big| }   \bigg\}   \end{align*}
      
      \begin{align*} =   \mathrm{log}\bigg\{   \frac{\mathrm{sinh} {\tiny\big(    \frac{\eta}{2} + u - i \alpha    \big) } }{\mathrm{sinh} {\tiny\big(    \frac{\eta}{2} - u + i \alpha    \big)}  }  \bigg\}     \text{, }  \end{align*}
      
      \begin{align*}     \psi_{-} {\tiny\big( \alpha + i u\big)  } = \mathrm{log} \bigg\{    \big|   \frac{\mathrm{exp} {\tiny\big(  2 \eta + 2 i \alpha \big) } - \mathrm{exp} {\tiny\big( 2 u - \eta \big)}}{\mathrm{exp} {\tiny\big( 2 u  \big)  }- \mathrm{exp} {\tiny\big( \eta + 2 i \alpha  \big)}} \big|  \bigg\}   \equiv  \mathrm{log} \bigg\{      \frac{\big| \mathrm{exp} {\tiny\big(  2 \eta + 2 i \alpha \big) } - \mathrm{exp} {\tiny\big( 2 u - \eta \big) }\big| }{ \big| \mathrm{exp} {\tiny\big( 2 u  \big) } - \mathrm{exp} {\tiny\big( \eta + 2 i \alpha  \big)}  \big| }   \bigg\}  \end{align*}
      
      \begin{align*}   \equiv \mathrm{log} \bigg\{   \frac{ \big| \mathrm{sinh} {\tiny\big(     \frac{3 \eta}{2} - u + i \alpha     \big) }\big| }{ \big| \mathrm{sinh} {\tiny\big(    u - \frac{\eta}{2} - i \alpha     \big) } \big| }     \bigg\}     =  \mathrm{log} \bigg\{   \frac{\mathrm{sinh} {\tiny\big(     \frac{3 \eta}{2} - u + i \alpha     \big) } }{\mathrm{sinh} {\tiny\big(    u - \frac{\eta}{2} - i \alpha     \big) }  }     \bigg\}     \text{, } 
\end{align*} } 

\noindent as well as the relation,

{\small \begin{align*}
    \Theta {\tiny\big( \alpha - \beta \big)}  \equiv \frac{1+\mathrm{exp} \big(   i p {\tiny\big( \alpha \big)} + i p {\tiny\big( \beta \big) }       \big) - 2 \Delta \cdot  \mathrm{exp} \big(     i p {\tiny\big( \alpha \big)  }           \big) }{1+\mathrm{exp} \big(     i p {\tiny\big( \alpha \big)}  + i p {\tiny\big( \beta \big)  }              \big) - 2 \Delta \cdot \mathrm{exp} \big(             i p {\tiny\big( \beta \big)}    \big)} = - \frac{\mathrm{sinh} {\tiny\big( i \alpha - i \beta + \eta \big)} }{\mathrm{sinh} {\tiny\big( i \alpha - i \beta - \eta \big)} }   \text{, } 
\end{align*} }

\noindent for,

{\small \begin{align*}
    p {\tiny\big( \alpha \big)}  = \mathrm{log} \bigg\{   \big|       \frac{\mathrm{sinh} {\tiny\big(   \frac{\eta}{2} + i \alpha      \big) }}{\mathrm{sinh} {\tiny\big(  \frac{\eta}{2} - i \alpha   \big)} }    \big|    \bigg\}   \equiv \mathrm{log} \bigg\{          \frac{\mathrm{sinh} {\tiny\big(   \frac{\eta}{2} + i \alpha      \big)} }{\mathrm{sinh} {\tiny\big(  \frac{\eta}{2} - i \alpha   \big)}  }       \bigg\}     \text{, } 
\end{align*} } 

\noindent from which the statement of the Bethe equations is equivalent to, in terms of a parameterization of $\psi_{-} {\tiny\big( \alpha + v_k \big)}$, and $\mathrm{exp} \big( \Theta {\tiny\big( \alpha_j - \alpha_m \big)}\big)$, [8], 

{\small \begin{align*}
       \overset{N}{\underset{k=1}{\prod}} {\mathrm{exp} \big(   \psi_{+} {\tiny\big(  \alpha + v_k      \big)}  \big) }  = \frac{1}{\mathrm{exp} {\tiny\big( 2 H N          \big)} }  \cdot        {\underset{m=1, m \neq j}{\prod} }           \frac{{\tiny\big( -1 \big)^m} }{\mathrm{exp} \big( \Theta {\tiny\big( \alpha_j - \alpha_m \big)} \big) }            = \frac{1}{\mathrm{exp} {\tiny\big( 2 H N          \big)} } \cdot 
 \overset{n}{\underset{m=1, m \neq j}{\prod} }            \frac{\mathrm{sinh} {\tiny\big(   i \big( \alpha_j - \alpha_m \big) - \eta      \big)}}{\mathrm{sinh} {\tiny\big(   i \big( \alpha_j - \alpha_m \big)+ \eta    \big) } }           \text{, } 
\end{align*} }

\noindent while the statement for the eigenvalue of the Bethe equations is equivalent to, given some ${\tiny\big\{ \alpha_i \big\}_{i \in \textbf{N}}}$, in terms of a parameterization of $\psi_{\pm} {\tiny\big( \alpha + v_k \big)}$, 

{\small \begin{align*}
     \Lambda \big( \psi_{\pm} {\tiny\big( \alpha + i u \big)}  \big)  \equiv  \Lambda_{\{ \alpha_i \}} \equiv \mathrm{exp} {\tiny\big( N H \big) } \cdot \overset{N}{\underset{k=1}{\prod}}  \mathrm{sinh} {\tiny\big( \eta - u + v_k \big) } \end{align*}
      
      \begin{align*}   \times \overset{n}{\underset{j=1}{\prod}}   \mathrm{exp} \big(  \psi_{+} {\tiny\big( \alpha_j + i u\big) } \big)           + \mathrm{exp} {\tiny\big( - NH \big)}  \cdot \overset{n}{\underset{k=1}{\prod}} \mathrm{sinh} {\tiny\big( u - v_k \big) }   \cdot  \overset{n}{\underset{j=1}{\prod}}  \mathrm{exp} \big(   \psi_{-} {\tiny\big( \alpha_j + i u\big)  }    \big)  \text{. } 
\end{align*} }

\noindent To introduce the Hamiltonian formulation of the six-vertex model, in which given a height function representation, from the set of all possible asymptotic height functions $\mathcal{H}_{L,q}$, for Lipchitz $h \sim \mathcal{H}_{L,q}$ and $x , x^{\prime} \in {\tiny\big[ 0 , L \big]}$, with,

{\small \begin{align*}
    {\tiny\big|} h{\tiny\big(x \big) }-   h   {\tiny\big(  x^{\prime}      \big) \big|} < {\tiny\big|} x - x^{\prime} {\tiny\big|  } \text{, }
\end{align*} } 

\noindent and periodic, with,

{\small \begin{align*}
   h {\tiny\big( L , y \big)} = h {\tiny\big( 0 , y \big)}  + q   \text{, } 
\end{align*} }

\noindent for $h : {\tiny\big[ 0 , L \big]} \longrightarrow \textbf{R}$, and fixed $0 < q < 1$. From such a sampling of the height function, as well as another periodic function $\pi {\tiny\big( x \big)}$ in $x$, the pair $\big(  \pi  {\tiny\big(} x {\tiny\big)} , h{\tiny\big(} x {\tiny\big) }\big)$ can be identified with the cotangent space $T^{*} \mathcal{H}_{L,q}$, while the flow of the Hamiltonian can be identified with the pair $\big( \pi {\tiny\big( x , y \big)} , h {\tiny\big( x , y \big) } \big)$, in which,

{\small \begin{align*}
        H_u \big( \pi {\tiny\big( x , y \big) }, h {\tiny\big( x , y \big)} \big) \equiv H_u {\tiny\big( \pi , h \big) }  = {\int}_{[0,L]}   \mathcal{H}_{u - v ( x) } \big[  \partial_x h {\tiny\big( x \big) } ,    \pi {\tiny\big( x \big) }   \big]  \mathrm{d} x    \text{, } 
\end{align*} } 

\noindent over $T^{*} \mathcal{H}_{L,q}$, for a solution $h \big( x , y \big)$ to the Euler Lagrange equations. Under the integral over $x$ provided above, $\mathcal{H}_u$ is the semigrand canonical free energy, [8],

{\small \begin{align*}
  \mathcal{H}_u {\tiny\big( q , H \big)}  \equiv  \mathcal{H}_u = \mathrm{log} \big[   Z_{\textbf{T}_{MN}}^n {\tiny\big( u , H \big)     }  \big]               \text{, } 
\end{align*} } 

\noindent which can alternatively be expressed as the maximum over $\pm$, with,

{\small \begin{align*}
  \mathcal{H}_u {\tiny\big( q , H \big)}  \equiv \underset{\pm}{\mathrm{max}}  \big\{  \mathcal{H}^{\pm}_u {\tiny\big( q , H \big)}  \big\} \equiv \underset{\pm}{\mathrm{max}}  \big\{     \pm H + l_{\pm} + \int_C  \psi^{\pm}_u {\tiny\big( \alpha \big) }  \rho {\tiny\big( \alpha \big) }  \text{ } \mathrm{d} \alpha   \big\}     \text{, } \tag{$\mathrm{H}$}
\end{align*} }

\noindent with $l_{-} \equiv \mathrm{log} \text{ } \mathrm{sinh} {\tiny\big( \eta - u \big)}$, $l_{+} \equiv \mathrm{log} \text{ } \mathrm{sinh} \text{ }  u$, and the density,

{\small \begin{align*}
  \rho {\tiny\big( \alpha \big) }\equiv       \# \big\{    \text{Bethe roots along contours } C        \big\}   \equiv \underset{\alpha > 0 }{\bigcup}  \big\{ \alpha : \alpha \cap C \neq \emptyset \big\} \equiv  \bigg|  \big\{ \alpha : \alpha \cap C \neq \emptyset \big\}      \bigg| \text{. } 
\end{align*} }

\noindent To connect the cotangent space $T^{*}_{\phi_1} \mathcal{H}_{q, l}$ with $T^{*}_{\phi_2} \mathcal{H}_{q, l}$ at time $T$ given the initial flow line $ {\tiny\big( \pi_0 , h_0 \big)}$, one determines the unique critical point of the functional,

{\small \begin{align*}
   S  {\tiny\big( \pi , h \big) }  \equiv     S    =   {\int}_{[0,L]} \text{ }  {\int}_{[0,T]}    \big[                \pi {\tiny\big( x , y \big) } \partial_y h {\tiny\big( x , y \big)} - \mathcal{H}_{u - v ( x ) }\big( \partial_x h  {\tiny\big( x , y \big) }, \pi {\tiny\big( x , y \big)}  \big)             \big]    \mathrm{d} y \text{ }  \mathrm{d} x          \text{, } 
\end{align*} } 

\noindent which is ${\tiny\big( \pi_0 , h_0 \big)}$. Altogether,

{\small \begin{align*}
        Z^n_{\textbf{T}_{MN}}  = \mathrm{exp} \big( NM \cdot \mathcal{H}_u {\tiny\big( q , H \big) } \big) \big( 1 + \mathrm{o} {\tiny\big( 1 \big)} \big)           \text{, } 
\end{align*} } 

\noindent asymptotically, for $M >> N$, $n>0$, and $q = {n} \big/ {N}$ as $N \longrightarrow + \infty$.

\subsection{Hamiltonian flows and integrability from the transfer and quantum monodromy matrices of the 6-vertex model}

\noindent One application of the Hamiltonian approach from [8], which demonstrates that the Euler-Lagrange equations have infinitely many conserved quantities, is discussed below.

\subsubsection{Previous application of the Hamiltonian formulation for the nonlinear Schroedinger equation}

\noindent With  $T_a \big( u , {\tiny\big\{ v_k \big\} } , H , V \big)$, which besides the previous diagonal representation can also be expressed as,

{\small \[
T_a  \big( u , {\tiny\big\{ v_k \big\}}  , H , V \big)  \equiv 
  \begin{bmatrix}
       A {\tiny\big( u \big) }  & B {\tiny\big( u \big)  } \\
    C {\tiny\big( u \big) } & D {\tiny\big( u \big) }  \text{ }  
  \end{bmatrix} \text{, }
\] }

\noindent one may proceed to determine the angle-action variables of the Hamiltonian system, as formulated in [6].

\bigskip

\noindent To implement the computations for this approach for the nonlinear Schroedinger equation, introduce, under slightly different parameters, the reduced monodromy matrix, from the limit as $x \longrightarrow + \infty$, and as $y \longrightarrow - \infty$, (\textit{Chapter III}, [6]),

{\small \begin{align*}
\underset{N \longrightarrow + \infty}{\mathrm{lim}} T_N {\tiny\big( \lambda \big)}  \equiv   T {\tiny\big( \lambda \big) }   = \underset{y \longrightarrow - \infty}{\underset{x \longrightarrow + \infty}{\mathrm{lim}} } E {\tiny\big( -x , \lambda \big) } \cdot T {\tiny\big( x , y , \lambda \big)} \cdot  E {\tiny\big( y , \lambda \big) }   \text{, } 
\end{align*} } 

\noindent for the matrix exponential,

{\small \begin{align*}
  E {\tiny\big( x , \lambda \big)} \equiv \mathrm{exp} {\tiny\big( \lambda x \cdot U_1 
 \big)} \text{, } 
\end{align*} } 

\noindent and also from the clockwise exponential action,

{\small \begin{align*}
     T_L {\tiny\big( \lambda \big)} \equiv {\mathrm{exp}}  \bigg[ \underset{\mathrm{clockwise\text{ }  action}}{\int_{[-L,L]}} U {\tiny\big( x , \lambda \big)}  \mathrm{d} x  \bigg]    \text{, } 
\end{align*} }

\noindent for a real parameter $\lambda$, with the integrand $U$, a $2 \times 2$ matrix, and $N \geq L$, which satisfies the first order PDE, [6],

{\small \begin{align*}
   \frac{\partial F}{\partial x} =   U {\tiny\big( x , \lambda \big) } \cdot    F    \text{, } 
\end{align*} }

\noindent for the vector valued function $F {\tiny\big( x , t \big)} \equiv F \equiv {\tiny\big[ f_1 , f_2 \big]^{\mathrm{T}}}$. The nonconstant prefactor for $F$ takes the form of the linear combination,

{\small \begin{align*}
  U = U_0 + \lambda \cdot  U_1  \text{, } 
\end{align*} }

\noindent with,

{\small \[
\bigg\{ U_0  \equiv \sqrt{\chi}
  \begin{bmatrix}
       0  &  \psi \\
    \bar{\psi}  & 0 \text{ }  
  \end{bmatrix} \text{, }   U_1  \equiv 
 \frac{1}{2i} \begin{bmatrix}
   1     &  0  \\
   0  &  -1 \text{ }  
  \end{bmatrix}    \bigg\} . 
\] }

\noindent The coordinates appearing in the definition of $U_0$ satisfy the second order PDE, the nonlinear Schroedinger equation, with solutions $\psi$, [6],

{\small \begin{align*}
   i \cdot \frac{\partial \psi}{\partial t} = - \frac{\partial^2 \psi}{\partial x^2} + 2 \chi \big| \psi \big|^2 \cdot  \psi      \text{, } 
\end{align*} }

\noindent for real $\chi$, a coupling constant, with $\big| \psi \big|^2 = \psi \bar{\psi}$. In the following infinite limit, the solutions satisfy,

{\small \begin{align*}
\underset{x \longrightarrow \pm \infty}{\mathrm{lim}} {\tiny\psi \big( x \big)} = \rho^{(\pm)}_1 \text{, } \\  \underset{x \longrightarrow \pm \infty}{\mathrm{lim}} \psi {\tiny\big( x \big) }
 = \rho^{(\pm)}_2 \text{, } 
\end{align*} }

\noindent for real $\rho^{(\pm)}_1, \rho^{(\pm)}_2$ with $\rho^{(\pm)}_1\neq \rho^{(\pm)}_2$. Equipped with solutions to the nonlinear PDE above, the Hamiltonian admits the following integral representation over the real line, (\textit{Chapter III}, [6]),

{\small \begin{align*}
 H {\tiny\big( x , \psi , \bar{\psi} , \chi \big)  } \equiv H \equiv \int_{\textbf{R}} \bigg[ \frac{\partial \psi}{\partial x} \frac{\partial \bar{\psi}}{\partial x} + \chi \cdot  \bar{\psi}^2 \psi^2 \bigg]    \text{ } \mathrm{d} x \text{, } 
\end{align*} }

\noindent Under the matrix representation for the reduced monodromy matrix, (\textit{Chapter I}, [6]),

{\small \[
T_L {\tiny\big( \lambda \big) } \equiv T  {\tiny\big( \lambda \big) } \equiv 
  \begin{bmatrix}
     a {\tiny\big( \lambda \big) }  & \epsilon \cdot \bar{b}{\tiny \big( \lambda \big) }  \\
   b{\tiny \big( \lambda \big)  } & \bar{a} {\tiny\big( \lambda \big) } \text{ }  
  \end{bmatrix} \text{, }
\] }

\noindent the transition coefficients for discrete spectra occur iff $\epsilon \equiv -1$, and are given by,

{\small \begin{align*}
    T^{(\mathrm{1})}_{-} {\tiny\big( x , \lambda_j \big)} = \gamma_j \cdot T^{(\mathrm{2})}_{+} {\tiny\big( x , \lambda_j \big) }          \text{, } 
\end{align*} }

\noindent for $1 \leq j \leq n$, where the $\lambda_j$ parameter appearing in $T^{(\mathrm{1})}_{-} {\tiny\big( x , \cdot \big)}$, and in $T^{(\mathrm{2})}_{+} {\tiny\big( x , \cdot \big)}$, is the collection of zeros, ${\tiny\big\{ \lambda_j \big\}_{j \in \textbf{N}}}$, of $a {\tiny\big( \lambda \big)}$ in the upper half plane, and $\gamma_j$ denotes another real parameter, the \textit{transition amplitude}, between $T^{(\mathrm{1})}_{-} {\tiny\big( x , \cdot \big)}$ and $T^{(\mathrm{2})}_{+} {\tiny\big( x , \cdot \big)}$. For another real parameter $\mu$, the identity

{\small \begin{align*}
  \big\{     T {\tiny\big( x , y , \lambda \big) }\overset{\bigotimes}{,}     T {\tiny\big( x , y , \mu \big) }  \big\}  =  \big\{    r {\tiny\big( \lambda - \mu \big) }     ,  T {\tiny\big( x , y , \lambda \big)} \bigotimes  T {\tiny\big( x , y , \mu \big) }  \big\}   \text{, } 
\end{align*} } 

\noindent of the Poisson bracket $\big\{ \cdot , \cdot \big\}$, which satisfies the anticommutativity, and bilinearity properties, Leibniz' rule, and the Jacobi identity, holds for $y < x$, where,

{\small \begin{align*}
  r_{\pm} {\tiny\big( \lambda - \mu \big) } = \underset{y \longrightarrow \pm \infty}{\mathrm{lim}}   E {\tiny\big( y , \mu - \lambda \big)  } \bigotimes E {\tiny\big( y , \lambda - \mu \big)} \cdot  r {\tiny\big( \lambda - \mu \big)    }      \text{, } 
\end{align*} } 

\noindent under the convention for the tensor product, between $A$ and $B$, with respect to the Poisson bracket, (\textit{Chapter III}, [6]),

{\small \begin{align*}
 \big\{     A \overset{\otimes}{,}   B \big\}_{jk,mn} \equiv \big\{   A_{jm} , B_{kn}     \big\}   \text{. } 
\end{align*} }

\subsubsection{Poisson bracket and the action-angle variables}

\noindent The action of the Poisson bracket that is implemented for demonstrating that the Hamiltonian flow for the nonlinear Schroedinger equation is integrable depends upon whether the arguments of the Poisson bracket are functionals, or matrix functionals. In the first case, the bracket takes the form, (\textit{Chapter III}, [6]),

{\small \begin{align*}
 \big\{ F ,  G \big\}   \equiv    i \int_{[-L,L]} \bigg[           \frac{\delta F}{\delta \psi} \cdot \frac{\delta G}{\delta \bar{\psi}} - \frac{\delta F}{\delta \bar{\psi}} \cdot \frac{\delta G}{\delta \psi }             \bigg] \text{ } \mathrm{d} x  \text{, } 
\end{align*} }

\noindent for functionals $F$ and $G$, while for the second case, the bracket takes the form,

{\small \begin{align*}
 \big\{  A \overset{\bigotimes}{,} B \big\} \equiv     i \int_{[-L,L]} \bigg[  \frac{\delta A}{\delta \psi} \bigotimes \frac{\delta B}{\delta \bar{\psi}} - \frac{\delta A}{\delta \bar{\psi}} \bigotimes  \frac{\delta B}{\delta \psi }                     \bigg]    \text{ } \mathrm{d} x      \text{, } 
\end{align*} }

\noindent for matrix functionals $A$ and $B$, and with $\psi \equiv \psi {\tiny \big( x \big)}$ and $\bar{\psi} \equiv \bar{\psi} {\tiny \big( x \big)} $ having support over ${\tiny\big( - L , L \big)}$. Similarly, from $T {\tiny\big( x , y , \lambda \big)}$,

{\small \begin{align*}
 T_{\pm} {\tiny\big( x , \lambda \big) }= \underset{y \longrightarrow \pm \infty}{\mathrm{lim}} T {\tiny\big( x , y , \lambda \big)} \cdot E {\tiny\big( y , \lambda \big) }  \text{. } 
\end{align*} } 

\noindent From the first identity involving the Poisson bracket, one similarly has, as either $y \longrightarrow - \infty$, or as $y \longrightarrow + \infty$, independently of $y$,

{\small \begin{align*}
      \big\{  T_{-} {\tiny \big( x , \lambda \big)} \overset{\bigotimes}{,} T_{-} {\tiny\big( x , \mu \big)     }      \big\} = r {\tiny\big( \lambda - \mu \big)}  \cdot  T_{-} {\tiny\big( x , \lambda \big)} \bigotimes T_{-} {\tiny\big( x , \mu \big)} - T_{-} {\tiny\big( x , \lambda \big)} \bigotimes T_{-} {\tiny\big( x , \mu \big)} \cdot r_{-} {\tiny\big( \lambda - \mu \big)    }    \text{, } \\     \big\{  T_{+} {\tiny\big( x , \lambda \big)} \overset{\bigotimes}{,} T_{+} {\tiny\big( x , \mu \big) }          \big\} = T_{+} {\tiny\big( x , \lambda \big) }\bigotimes T_{+} {\tiny\big( x , \mu \big) }\cdot r_{+} {\tiny\big( \lambda - \mu \big) }- r {\tiny\big( \lambda - \mu \big) }\cdot T_{+} {\tiny\big( x , \lambda \big) }\bigotimes T_{+} {\tiny\big( x , \mu \big) }       \text{, } 
\end{align*} } 

\noindent while, between $T_{-} {\tiny \big( x , \lambda \big)}$ and $T_{+} {\tiny\big( x , \lambda \big)}$,

{\small \begin{align*}
   \big\{  T_{-} {\tiny\big( x , \lambda \big) }\overset{\bigotimes}{,} T_{+} {\tiny\big( x , \lambda \big)}      \big\} = 0 \text{. } 
\end{align*} } 

\noindent From the quantities introduced thus far, the action angle variables, given a Hamiltonian flow, are determined by whether the Poisson bracket vanishes or not, in which the only terms with nonvanishing Poisson bracket are,

{\small \begin{align*}
        \big\{  \Phi {\tiny\big( \lambda \big) } ,   \bar{\Phi {\tiny\big( \lambda \big)}  } \big\} = 0          \text{, } 
\end{align*} }

\noindent for the function,

{\small \begin{align*}
  \Phi {\tiny\big( \lambda \big) }\equiv \sqrt{\rho {\tiny\big( \lambda \big)}} \cdot \mathrm{exp} \big( - i \cdot  \phi {\tiny\big( \lambda \big)} \big) \text{, } 
\end{align*} } 

\noindent and its complex conjugate, $\bar{\Phi {\tiny\big( \lambda \big)}}$, as well as,

{\small \begin{align*}
      \rho {\tiny\big( \lambda \big)} \equiv \frac{1}{2 \pi \chi} \cdot \mathrm{log} \big[      1 + \epsilon \big| b {\tiny\big( \lambda \big)} \big|^2 \big]  \text{, }  \end{align*}

 \begin{align*} \phi {\tiny\big( \lambda \big)} \equiv - \mathrm{arg} \big( b {\tiny\big( \lambda \big) } \big)   \text{. } 
\end{align*} }

\noindent With the monodromy, and transfer, matrices of the previous section which were respectively denoted with $t \big( u , {\tiny\big\{ v_k \big\} }, H , V \big) $ and $ T_a \big( u , {\tiny\big\{ v_k \big\} }, H , 0 \big) $, in the next section we state the key property that the action-angle variables satisfy.

\subsection{Action-angle variables for the Hamiltonian flow in the $\Delta < - 1$ regime of the six-vertex model}

\noindent To make use of the Poisson bracket for determining which variables with respect to it vanish, it suffices to demonstrate that the Poisson bracket vanishes for certain entries of the Quantum monodromy matrix. We recall the representation presented in \textit{1.4.1} with, $T_a \big( u , {\tiny\big\{ v_k \big\}} , H , V \big)$ of the Quantum monodromy matrix, which was introduced alongside the definition of the transfer matrix, $t \big( u , {\tiny\big\{ v_k \big\} } , H , V \big)$.

\subsubsection{Inhomogeneities in the horizontal direction}

\noindent In the setting of the six-vertex model, the action-angle variables are determined by the structure of the Quantum monodromy matrix, and hence of the transfer matrix.

\bigskip

\noindent From the background provided in the previous section on the Hamiltonian structure of the nonlinear Schroedinger equation, instead of sending $x \longrightarrow \pm \infty$, and $y \longrightarrow - \infty$, we send the spectral parameters at each site $i$ to $-\infty$, in which,

{\small \begin{align*}
  u - v_1 \longrightarrow - \infty , \cdots , u - v_N \longrightarrow - \infty   \text{, } 
\end{align*} }

\noindent from each term of the $\mathrm{R}$-matrix appearing in the Quantum monodromy matrix, 

{\small \begin{align*}
   T_a {\tiny \big( u , H , V \big)}   =   \underset{v_k \longrightarrow - \infty} {\mathrm{lim}}  T_a \big(    u ,  {\tiny\big\{ v_k \big\}}  , H , V \big)  =    \underset{v_i \longrightarrow - \infty} {\mathrm{lim}} \text{ } \bigg[  \overset{N}{\underset{i=1}{\prod}}   \big( 
    \mathrm{diag} \big( \mathrm{exp} {\tiny\big( 2H \big)} ,  \mathrm{exp} {\tiny\big( 2 H \big)}  \big)   \big)  \cdot    R_{ia} {\tiny\big( u - v_i \big)}  \bigg]   \text{. } 
   \end{align*} }

\noindent As $v_i \longrightarrow - \infty, \cdots , v_N \longrightarrow - \infty$, the above limit is equivalent to, 
   
   {\small \begin{align*}
   \underset{v_N \longrightarrow - \infty}{\underset{\vdots}{\underset{v_1 \longrightarrow - \infty}{\mathrm{lim}}}}  \bigg[ \big( 
       \mathrm{diag} \big( \mathrm{exp} {\tiny\big( 2H \big)} , \mathrm{exp} {\tiny\big( 2 H \big) } \big) \big) \cdot  R_{1a} {\tiny\big( u - v_1 \big) } \times \cdots \times  \big( \mathrm{diag} \big( \mathrm{exp} {\tiny\big( 2H \big)}  , \mathrm{exp} {\tiny\big( 2 H \big)}  \big) \big)   \cdot    R_{Na}  {\tiny\big( u - v_N  \big)  }   \bigg]     \text{. } 
\end{align*} }

\noindent From $T_a$ above, in comparison to the monodromy matrix $T{\tiny\big( \lambda \big)}$ for the nonlinear Schroedinger equation, the terms dependent on $u$, $B {\tiny\big( u \big)}$ and $C {\tiny\big( u \big)}$, are not complex conjugates of one another as $\bar{b {\tiny\big( \lambda \big)}}$ and $b {\tiny\big( \lambda \big)}$ are, as provided in the definition of $T {\tiny\big( \lambda \big)}$. 

\bigskip

\noindent In order to demonstrate that the six-vertex model is integrable from properties of the Hamiltonian flow, and to also make sense of the multi dimensional limit show above, it suffices to show that the Poisson bracket of the generator of motion integrals vanish. First, recall, from ($\mathrm{H}$), 

{\small \begin{align*}
  \mathcal{H}_u {\tiny\big( q , H \big)} \equiv \underset{\pm}{\mathrm{max}}    \text{ } \mathcal{H}^{\pm}_u {\tiny\big( q , H \big)}  \equiv \underset{\pm}{\mathrm{max}}  \big\{     \pm H + l_{\pm} + \int_C  \psi^{\pm}_u {\tiny\big( \alpha \big)} \rho {\tiny\big( \alpha \big)}    \text{ } \mathrm{d} \alpha   \big\}   \text{, } 
\end{align*} } 

\noindent which, explicitly in terms of $\psi$, is equivalent to,

{\small \[
\mathcal{H}^{\pm}_u {\tiny\big( q , H \big)}  \equiv  \text{ } 
\left\{\!\begin{array}{ll@{}>{{}}l}       + \Rightarrow    \underset{+}{\mathrm{max}}  \big\{     H + \mathrm{log} \text{ } \mathrm{sinh} \text{ }  u + \int_C  \mathrm{log}\big[  \frac{\mathrm{sinh} {\tiny\big(    \frac{\eta}{2} + u - i \alpha  - iu  \big) }}{\mathrm{sinh} {\tiny\big(    \frac{\eta}{2} - u + i \alpha - iu    \big)} }  \big]    \rho {\tiny\big( \alpha \big) }   \text{ } \mathrm{d} \alpha   \big\}    
\text{, } \\  \\ 
 - \Rightarrow   \underset{-}{\mathrm{max}}  \big\{  -   H + \mathrm{log} \text{ } \mathrm{sinh} {\tiny\big( \eta - u \big)} + \int_C  \mathrm{log} \big[  \frac{\mathrm{sinh} {\tiny\big(     \frac{3 \eta}{2} - u + i \alpha - iu     \big)} }{\mathrm{sinh} {\tiny\big(    u - \frac{\eta}{2} - i \alpha  - iu    \big)} }     \big]      \rho {\tiny\big( \alpha \big) }  \text{ } \mathrm{d} \alpha   \big\}     \text{. }  
\end{array}\right.
\] } 

\noindent With each motion integral introduced above for $+$ and for $-$, we proceed to identify the action-angle variables for the Hamiltonian flow of the six-vertex model from the motion integrals above.

\bigskip

\noindent To make such an identification of which entries of the monodromy matrix appear in expressions for the generating functions of the the motion integrals, observe that the Quantum monodromy matrix, denoted earlier with $T_a \big( u , {\tiny\big\{ v_k \big\} }, H , V \big)$, upon setting $u \equiv \lambda_{\alpha}$ for the eigenvalue of the Quantum monodromy matrix, can be expressed as,

{\small \[
\begin{bmatrix}
       A {\tiny\big( \lambda_{\alpha} \big)  } & B {\tiny\big( \lambda_{\alpha} \big)   }\\
    C {\tiny\big( \lambda_{\alpha} \big)}  & D {\tiny\big( \lambda_{\alpha} \big) }  \text{ }  
  \end{bmatrix} \text{, }
\] } 

\noindent which admits a decomposition in terms of $L$ operators, from the product,

{\small \begin{align*}
     \overset{N-1}{\underset{i=0}{\prod}}   L_{\alpha , N - i  } {\tiny\big( \lambda_{\alpha} , v_{N-i} \big)  }       \text{, } 
\end{align*} } 

\noindent where the $L$ operator is defined as, [2],

{\small \[
       L_{\alpha , k   } {\tiny\big( \lambda_{\alpha} , v_{k} \big) }    \equiv 
  \begin{bmatrix}
     \mathrm{sin} {\tiny\big( \lambda_{\alpha} - v_k + \eta \sigma^z_k \big)  }     &    \mathrm{sin} {\tiny\big( 2 \eta \big) }\sigma^{-}_k    \\
      \mathrm{sin} {\tiny\big( 2 \eta \big) }\sigma^{+}_k     &   \mathrm{sin}  {\tiny\big( \lambda_{\alpha} - v_k - \eta \sigma^z_k \big)}      
  \end{bmatrix}  \text{, } 
\] } 

\noindent corresponding to the $k$ th horizontal line, and $\alpha$ th vertical line, with $L_{\alpha,k} \curvearrowright \bigg(  \text{vertical space} \bigotimes \text{horizontal space} \bigg) $. From the block decomposition above in terms of $A$, $B$, $C$ and $D$, one also has that,

{\small \begin{align*}
  \mathrm{tr} \big(  T_a \big( \lambda_a , {\tiny\big\{ v_k \big\}}  , H , V \big) \big)  =    A {\tiny\big( \lambda_{\alpha} \big)} + D {\tiny\big( \lambda_{\alpha} \big) }   \text{. } 
\end{align*} }

\subsubsection{Paper organization}

\noindent In the remaining subsection before computing L-operators in the $\Delta < -1$ regime of the six-vertex model \textit{2}, we provide an overview of the Hamiltonian formulation discussed in [6], from which computations are performed using the Poisson bracket, either on functionals or on matrix functionals, to determine the action-angle variables which are required for the statement of integrability of the Hamiltonian flow that are provided in the statement of \textbf{Theorem}. Such variables coincide with ones which have nonzero Poisson bracket, which can be simplified from the set of sixteen relations obtained from the $N$ th power of the monodromy matrix in \textit{2.4}. The quantities that are included in the Poisson bracket are taken from entries of the monodromy matrix as $x \longrightarrow \pm \infty$, $y \longrightarrow \pm \infty$, or as $x \longrightarrow \pm \infty$, $y \longrightarrow \pm \infty$ simultaneously. In order to make use of a similar approach for the Hamiltonian formulation described in the previous section from [7] in the context of the Hamiltonian flow for the six-vertex model in the presence of vertical and/or horizontal inhomogeneities (similar to those considered by the author for the 20-vertex model under a three-dimensional encoding of domain-wall boundary conditions, [15]), one must identify manners in which the Hamiltonian formulation of the six-vertex model depends on the structure of the L-operator. Such structures are not only dependent on the statement of the Bethe equations, but also on the eigenvalues of the Bethe equation itself, and Pauli operators $\sigma^{-}$ and $\sigma^{+}$. In \textit{2} we analyze such properties of the Quantum mondromy, and transfer, matrices, from L-operators. The particular choice of L-operator that we take allows for us to obtain expressions for entries of the monodromy matrix that is introduced for the Hamiltonian flow of the six-vertex model, thereby allowing for us to deduce the desired action-angle variables of the Hamiltonian flow, hence concluding that the Hamiltonian flow in the presence of inhomogeneities satisfies an integrability property. The forthcoming expression for action-angle variables, and related expressions involving product representations of L-operators, is intimately related to the behavior of the entropy, and several other characteristics of vertex models that have previously been examined in seminal works, [1,2,6,11,12,13,16,17].

\section{Statement of Main Results: Approximating Poisson brackets in the finite weak volume limit, action-angle coordinates and their complex conjugate, integrability of ${\tiny H_u}$}

\subsection{Overview}

\noindent In the current section, \textit{2}, to further describe inhomogeneities of the six-vertex model that influence constructions of the transfer and Quantum monodromy matrices, along with connections for computing action-angle variables, we obtain a system of recursive relations for each entry of the monodromy matrix which can provide expressions that we asymptotically approximate in the large $N$ limit from nine Poisson brackets. Equipped with such a background, in \textit{2.2}, we introduce the L-operator, and implement several computations, as suggested in the last section of [8], for computing each entry of the monodromy matrix raised to an arbitrary positive power. These relations are of upmost importance in performing computations in subsequent sections with the Poisson bracket, in which in [6] it was shown, for the nonlinear Schroedinger equation, that there are sixteen relations that can be asymptotically approximated from properties of the Poisson bracket. For the inhomogeneous six-vertex model, each such relation relation is obtained from the fact that the Poisson bracket of the tensor product of two reduced monodromy matrices can be expressed in terms of a Poisson bracket between entries of reduced monodromy matrices. Asymptotically as $N \longrightarrow + \infty$, from the monodromy matrix defined in [8] for studying inhomogeneities of the six-vertex model, sixteen relations from the Poisson bracket of the tensor product of reduced monodromy matrices can also be formulated in the context of the Hamiltonian flow and the Quantum monodromy matrix, thereby establishing that desired integrable and exactly solvable properties hold.

From the expressions given in the item below for each Poisson bracket, there exists a corresponding summation over operators from the block representation of the the two-dimensional product representation,

{\small \[ \begin{bmatrix}
       A  {\tiny\big( u \big) }  & B  {\tiny\big( u \big) }  \\
    C  {\tiny\big( u \big)}  & D  {\tiny\big( u \big) } \text{ }  
  \end{bmatrix}  , 
\]  }

\noindent for the transfer matrix introduced in the next section. Such a product representation for the two-dimensional transfer matrix is obtained from manipulating L-operators that have been previously obtained in the presence of domain-wall boundary conditions, {[3]}. As a fundamental object that plays many roles within the Quantum inverse scattering framework, the transfer matrix is used to obtain information on the Quantum monodromy matrix in the large finite volume limit. Despite the fact that lower dimensional versions of the transfer, and Quantum monodromy, matrices are investigated in this work, there not only remains potential to further investigate more algebraic characteristics of objects that are used in the Quantum inverse scattering framework, but also to establish representation theoretic, and combinatorially minded, observations. To simplify some of the notation appearing for each Poisson bracket, denote,

{\small \begin{align*}
    \underset{m,n^{\prime}: m + n^{\prime} = n-3 }{\sum} *   \equiv    \underset{m,n^{\prime}: m + n^{\prime} = n-3 }{\underset{m, n^{\prime} \in \textbf{Z}}{\underset{1 \leq i^{\prime} \leq n^{\prime}}{\sum}}}   *        \text{, }
\end{align*} }

\noindent where,

{\small \[
* \in \left\{\!\begin{array}{ll@{}>{{}}l}      \big\{  A  {\tiny\big( u \big)   }     , A  {\tiny\big( u^{\prime} \big) }  \big\} 
\text{, } \\ \vdots  \\  \big\{  D  {\tiny\big( u \big) } ,       D  {\tiny\big( u^{\prime} \big) } \big\} \text{. }
\end{array}\right.
\] }

\noindent To approximate the first Poisson bracket appearing in the collection of above relations,

{\small \begin{align*}
     \big\{  A  {\tiny\big( u \big)       } , A  {\tiny\big( u^{\prime} \big)  }  \big\}   ,
\end{align*} }

\noindent To demonstrate that the two below conditions stated in \textbf{Theorem} hold we approximate the first above Poisson bracket, depending upon $A \big( u \big)$ and $A \big( u^{\prime} \big)$, from products of the domain-wall L-operator, [3],

{\small \[
       L_{\alpha , k   }  {\tiny\big( \lambda_{\alpha} , v_{k} \big)  }   \equiv 
  \begin{bmatrix}
     \mathrm{sin}  {\tiny\big( \lambda_{\alpha} - v_k + \eta \sigma^z_k \big)  }     &    \mathrm{sin}  {\tiny\big( 2 \eta \big) } \sigma^{-}_k    \\
      \mathrm{sin}  {\tiny\big( 2 \eta \big)} \sigma^{+}_k     &   \mathrm{sin}   {\tiny\big( \lambda_{\alpha} - v_k - \eta \sigma^z_k \big) }     
  \end{bmatrix}  \text{, } 
\] } 

\noindent where,

{\small \[
 \left\{\!\begin{array}{ll@{}>{{}}l}  
  \lambda_{\alpha} \equiv \textit{Spectral parameter along the horizontal line } \alpha \textit{ of } \textbf{Z}^2   , \\ \\  \nu_k \equiv \textit{Spectral parameter along the vertical line } k \textit{ of } \textbf{Z}^2      , \\ \\ \sigma^z_k \equiv \textit{First matrix basis element for the Pauli group}  ,  \\ \\ \sigma^-_k \equiv    \textit{First matrix basis element for the Pauli group}      , \\ \\ \eta \equiv   \mathrm{arcsin}   {\tiny\big[ {c} \big/ {2} \big] }    ,           
\end{array}\right.
\] }

\noindent for the third isotropic weight, $c$, of the inhomogeneous $6$-vertex model.

\bigskip 

\noindent To demonstrate that the transfer matrices $\textbf{T} {\tiny \big( u \big)}$ and $\textbf{T} {\tiny \big( u^{\prime} \big)}$ imply,

{\small

\begin{align*}
   \big\{    \mathrm{Tr} \big[ \textbf{T} {\tiny \big( u \big)}  \big] ,    \mathrm{Tr} \big[ \textbf{T} {\tiny \big( u^{\prime} \big)}  \big]  \big\} = 0       , 
\end{align*}

}

\noindent from the standard identity,

 {\small
 
 \begin{align*}
     \big\{    \textbf{T} {\tiny \big( u \big)} ,  \textbf{T} {\tiny \big( v \big))}     \big\} = \big[  r {\tiny \big( u - v \big)}  , \textbf{T} {\tiny \big( u \big)} \cdot  \textbf{T} {\tiny \big( v \big))}  \big]    , 
 \end{align*}

 }

 \noindent for the $R$-matrix, $r$, in addition to the fact that the transfer matrix given a choice of the spectral parameter can be expressed as a double commutator,

{\small

\begin{align*}
    \textbf{T} {\tiny \big( u \big)} =   \big[        \big[  A  {\tiny \big( u \big)} ,  B   {\tiny \big( u \big)} \big] , \big[ C  {\tiny \big( u \big)} , D  {\tiny \big( u \big)}  \big] \big]    \equiv         \big[        \big[  A ,  B \big] , \big[ C , D  \big] \big]    , 
\end{align*}

}

\noindent we demonstrate how product expansions of L-operators under domain-wall boundary conditions can be obtained from the following result.



\subsection{Proposition}

\noindent \textbf{Proposition} (\textit{a decomposition of the first Poisson bracket, } $\big\{ A  {\tiny\big( u \big) } , A  {\tiny\big( u^{\prime} \big) } \big\} $ \textit{into the nine below brackets for L-operators of the 6-vertex model under domain-wall boundary conditions}). Fix $n,n^{\prime}, m >0$. As $N \longrightarrow + \infty$, the collection of Poisson brackets, $\mathcal{P}_1 , \cdots , \mathcal{P}_9$, given by,

{\small \[
\left\{\!\begin{array}{ll@{}>{{}}l}   
    \mathcal{P}_1  \equiv    \big\{    \mathscr{P}_1 \big( \mathrm{sin} {\tiny \big( 2 \eta \big)} \big)^{n-3} \mathscr{A}_1   ,   \mathscr{P}_2      \big( \mathrm{sin} {\tiny\big( 2 \eta \big)} \big)^{n-3}\mathscr{A}^{\prime}_1     \big\}
 \text{, } \\ 
 \mathcal{P}_2 \equiv    \big\{  \mathscr{P}_1   \big( \mathrm{sin} {\tiny\big( 2 \eta \big)} \big)^{n-3}    \mathscr{A}_1     ,                  \mathscr{P}_2 \mathscr{A}^{\prime}_2   \big\}  \text{, }  \\ 
 \mathcal{P}_3 \equiv  \big\{ \mathscr{P}_1 \big( \mathrm{sin} {\tiny\big( 2 \eta \big)} \big)^{n-3} \mathscr{A}_1 , \mathscr{P}_2 \mathscr{A}^{\prime}_3 \big\} \text{, }  \\  \mathcal{P}_4  \equiv \big\{ \mathscr{P}_1 \mathscr{A}_2 , \mathscr{P}_2 \big( \mathrm{sin}  {\tiny\big( 2 \eta \big)} \big)^{n-3} \mathscr{A}^{\prime}_1 \big\}   \text{, } \\  \mathcal{P}_5 \equiv   \big\{     \mathscr{P}_1 \mathscr{A}_2 , \mathscr{P}_2 \mathscr{A}^{\prime}_2        \big\} \text{, } 
\\   \mathcal{P}_6 \equiv     \big\{ \mathscr{P}_1 \mathscr{A}_2 , \mathscr{P}_2 \mathscr{A}^{\prime}_3 \big\}  \text{, } \\    \mathcal{P}_7 \equiv  \big\{ \mathscr{P}_1 \mathscr{A}_3  ,  \mathscr{P}_2 \big( \mathrm{sin} {\tiny\big( 2 \eta \big)} \big)^{n-3} \mathscr{A}^{\prime}_1  \big\} \text{, } \\ 
\mathcal{P}_8 \equiv \big\{ \mathscr{P}_1 \mathscr{A}_3 , \mathscr{P}_2 \mathscr{A}^{\prime}_2 \big\} 
   \text{, }  \\ 
    \mathcal{P}_9  \equiv    \big\{       \mathscr{P}_1      \mathscr{A}_3       ,     \mathscr{P}_2            \mathscr{A}^{\prime}_3         \big\} 
 \text{, } 
 \end{array}\right.
\]

\noindent which satisfies,

{\small \begin{align*}
  \big\{ A {\tiny\big( u \big)} , A {\tiny\big( u^{\prime} \big)} \big\}  \approx   \mathcal{P}_1 + \mathcal{P}_2 +     \mathcal{P}_3 + \mathcal{P}_4 +  \mathcal{P}_5 + \mathcal{P}_6 +   \mathcal{P}_7 + \mathcal{P}_8 + \mathcal{P}_9      ,
\end{align*} }

\noindent are given by the a sequence of approximations provided in the Appendix (specifically, \textit{3.1.1}-\textit{3.1.9}).

\bigskip 

\noindent Taking the summation of the collection of Poisson brackets $\mathscr{C} \equiv \underset{1 \leq i \leq 9}{\bigcup} \mathcal{P}_i$ yields,

{\small \begin{align*}
    \underset{1 \leq i \leq 9}{\sum}  \mathcal{P}_i       \text{, } 
\end{align*} }

\noindent equals,

{\small \begin{align*}
  \bigg\{   \bigg( \bigg\{      A_3 {\tiny\big( u \big)} + B_3 {\tiny\big( u\big)} \bigg\} \cdot  \bigg\{    \big( \mathrm{sin} {\tiny\big( 2 \eta \big)}  \big)^{n-3} \mathscr{A}_1  + \mathscr{A}_2 +  \mathscr{A}_3 \bigg\}    \bigg)     , \bigg(   \bigg\{       A_3 {\tiny\big( u^{\prime} \big)} + B_3 {\tiny\big( u^{\prime} \big)} \bigg\} \cdot   \bigg\{    \big( \mathrm{sin} {\tiny\big( 2 \eta \big) } \big)^{n-3} \mathscr{A}^{\prime}_1  + \mathscr{A}^{\prime}_2 \end{align*}
  
  \begin{align*} +  \mathscr{A}^{\prime}_3 \bigg\}  \bigg)   \bigg\}         \text{. } 
\end{align*} }

\noindent To demonstrate that the desired conclusion above holds for the main result, which can be extended, through very similar manipulations of the Poisson bracket for fifteen other relations that are also parametrized in entries of the monodromy matrix, we perform several computations with L-operators in \textit{2}. From performing computations with the L-operators, we obtain, explicitly, expressions for $\mathscr{A}_1$, $\mathscr{A}_2$, $\mathscr{A}_3$, $\mathscr{A}^{\prime}_1$, $\mathscr{A}^{\prime}_2$, and $\mathscr{A}^{\prime}_3$. Below, we introduce the action angle variables and the result that the action-angle variables, in canonical coordinates, vanish.

\bigskip

\noindent \textit{Proof of Proposition 1}. The desired representation of the above Poisson brackets for L-operators of the 6-vertex model under domain-wall boundary conditions is shown to hold from a series of six Lemmas established in the Appendix, from which we conclude the argument. \boxed{}

\bigskip

\noindent As a general matter of fact, it is not expected that action-angle variables will have a vanishing Poisson bracket, such as the one below, for other models besides the 6-vertex model. For the 6-vertex model, the primary notion underlying integrability of a Hamiltonian flow is obtained from properties relating to integrability of inhomogeneous limit shapes. For higher-dimensional vertex models, along with other vertex models that have the same dimensionality as the state space of the 6-vertex model, it remains of future interest to explore additional adaptations of the Quantum inverse scattering method.

\bigskip

\noindent \textbf{Definition} (\textit{action-angle variables for the inhomogeneous six-vertex model}). For,

{\small \begin{align*}
  B {\tiny\big( u \big)} \equiv \underset{N\longrightarrow + \infty}{\mathrm{lim}}  \underset{1 \leq j \leq N}{\prod} B_j {\tiny\big( u \big) }   ,
\end{align*} } 

\noindent define,

{\small \begin{align*}
      \varphi {\tiny\big( u \big)} = - \mathrm{arg} \big( B {\tiny\big( u \big)}  \big)     \text{, } \end{align*}

  \begin{align*} \rho {\tiny\big( u \big) }  =   \frac{1}{\pi} \cdot \mathrm{log} \big[ 1 + \epsilon \big|  B {\tiny\big( u \big)}  \big|^2 \big]                  \text{, } 
\end{align*} }

\noindent corresponding to action-angle variables for the inhomogeneous six-vertex model.

\subsection{Theorem $1$}

\noindent \textbf{Theorem} \textit{1} (\textit{vanishing of the Poisson bracket of the action-angle variables in canonical coordinates}). The following condition holds: 

\begin{itemize}

\item[$\bullet$] \textit{Vanishing of the Poisson brackets with respect to the canonical coordinates}. One has that,

{\small \begin{align*}
      \big\{ \varphi {\tiny\big( u_1 \big)} , \varphi {\tiny\big( u_2 \big)} \big\}    =   \big\{   \rho {\tiny\big( u_1 \big)    }      ,    \rho {\tiny\big( u_2 \big) }   \big\} = 0       \text{. } 
\end{align*} } 
\end{itemize}

\noindent \textit{Proof of Theorem 1}. By direct computation, for the above transcriptions of the action-angle coordinates and its complex conjugate to vanish with respect to $\big\{\cdot , \cdot \big\}$, one has that, 
 
 {\small
 
 \begin{align*}
    \bigg\{ \underset{N\longrightarrow + \infty}{\mathrm{lim}}  \underset{1 \leq j \leq N}{\prod} B_j {\tiny\big( u \big) }   , \underset{N\longrightarrow + \infty}{\mathrm{lim}}  \underset{1 \leq j \leq N}{\prod} B_j {\tiny\big( u^{\prime} \big) }   \bigg\} =   \underset{N\longrightarrow + \infty}{\mathrm{lim}}  \bigg\{  \underset{1 \leq j \leq N}{\prod} B_j {\tiny\big( u \big) }   ,   \underset{1 \leq j \leq N}{\prod} B_j {\tiny\big( u^{\prime} \big) }   \bigg\}  , 
 \end{align*} 
 
 }

 \noindent can be rearranged as,

 {\small
 
 \begin{align*}
\underset{N\longrightarrow + \infty}{\mathrm{lim}}     \big\{   \mathrm{sin} {\tiny\big( 2 \eta \big) }\sigma^{+}_N {\tiny \big( u \big))}  ,  \mathrm{sin} {\tiny\big( 2 \eta \big) }\sigma^{+}_N  {\tiny \big( u^{\prime} \big))}         \big\}           =  \mathrm{sin} {\tiny\big( 2 \eta \big) }    \times \underset{N\longrightarrow + \infty}{\mathrm{lim}}     \big\{   \sigma^{+}_N {\tiny \big( u \big))}  ,  \sigma^{+}_N  {\tiny \big( u^{\prime} \big))}         \big\}          =   \mathrm{sin} {\tiny\big( 2 \eta \big) }    \times \delta {\tiny \big( u - u^{\prime} \big)}                   . 
 \end{align*} 
 
 }

 \noindent Altogether given the analog of the Poisson bracket of $B {\tiny \big( u \big))}$ with $B {\tiny \big( u^{\prime} \big))}$ presented above a direct adaption of the computation of action-angle coordinates presented in [6] yields the desired result, from which we conclude the argument. \boxed{}

\subsection{Theorem $2$}

\noindent \textbf{Theorem} \textit{2} (\textit{integrability of the Hamiltonian flow previously identified by Keating, Reshetikhin and Sridhar for inhomogeneous limit shapes}). The following condition holds:

\begin{itemize}
\item[$\bullet$] \textit{Integrability of the flow generated by $H_u$ through the object $\mathcal{H}_u$}. Recall,

 {\small \[
\mathcal{H}^{\pm}_u {\tiny\big( q , H \big)}  \equiv  \text{ } 
\left\{\!\begin{array}{ll@{}>{{}}l}       + \Rightarrow    \underset{+}{\mathrm{max}}  \big\{     H + \mathrm{log} \text{ } \mathrm{sinh} \text{ }  u + \int_C  \mathrm{log}\big[  \frac{\mathrm{sinh} {\tiny\big(    \frac{\eta}{2} + u - i \alpha  - iu  \big) }}{\mathrm{sinh} {\tiny\big(    \frac{\eta}{2} - u + i \alpha - iu    \big)} }  \big]    \rho {\tiny\big( \alpha \big) }   \text{ } \mathrm{d} \alpha   \big\}    
\text{, } \\  \\ 
 - \Rightarrow   \underset{-}{\mathrm{max}}  \big\{  -   H + \mathrm{log} \text{ } \mathrm{sinh} {\tiny\big( \eta - u \big)} + \int_C  \mathrm{log} \big[  \frac{\mathrm{sinh} {\tiny\big(     \frac{3 \eta}{2} - u + i \alpha - iu     \big)} }{\mathrm{sinh} {\tiny\big(    u - \frac{\eta}{2} - i \alpha  - iu    \big)} }     \big]      \rho {\tiny\big( \alpha \big) }  \text{ } \mathrm{d} \alpha   \big\}     \text{. }  
\end{array}\right.
\] }

\noindent from which one has that,

 {\small

 \begin{align*}
     \big\{   \mathcal{H}^{\pm}_u {\tiny\big( q , H \big)}      , \mathcal{H}^{\pm}_v  {\tiny\big( q , H \big)}   \big\} = 0         ,
 \end{align*}

 }

 \noindent hence implying that the desired integrability property of the Hamiltonian flow holds, given the existence of infinitely many conserved quantities.

 \bigskip

\noindent \textit{Proof of Theorem 2}. Consider the underlying phase space to be that of $\textbf{Z}^2$. To demonstrate that the desired integrability property of the Hamiltonian flow from limit shapes that are solutions to the Euler-Lagrange equation,

{\small

\begin{align*}
      \partial^2_1  \sigma {\tiny \big(}  \partial_x h , \partial_y h       {\tiny \big)} \cdot  \partial^2_x h + \partial_1 \partial_2 \sigma {\tiny \big(}   \partial_x h , \partial_y h       {\tiny \big)}  \cdot \partial_x \partial_y h      +    \partial^2_2 \sigma {\tiny \big(} \partial_x h , \partial_y h     {\tiny \big)}  \cdot \partial^2_y h =     0    ,
\end{align*} 

}

\noindent from the constraints on the slope, and boundary conditions of the height function,

{\small 

\begin{align*} h {\tiny \big(} 0 , y {\tiny \big)} = \chi_1  {\tiny \big(} y {\tiny \big)}, \end{align*}

 \begin{align*} h {\tiny \big(} T , y {\tiny \big)} = \chi_2  {\tiny \big(} y {\tiny \big)} + C,   \end{align*}

 \begin{align*} \textit{Slope of the height function along the $x$-coordinate axis}  \equiv  \partial_x h  \in {\tiny \big[} - 1 \big/ 2 , 1 \big/ 2 {\tiny \big]}   , \end{align*}

 \begin{align*} \textit{Slope of the height function along the $y$-coordinate axis}  \equiv   \partial_y h  \in {\tiny \big[} - 1 \big/ 2 , 1 \big/ 2 {\tiny \big]}    , 
 \end{align*}
 }

 \noindent for some $T>0$, $C \in \textbf{R}$, holds, we first recall the following. Namely, under the Hamiltonian flow,

 {\small
 
 \begin{align*}
 \textit{Hamiltonian flow} \equiv  H_u {\tiny \big(} p , h {\tiny \big)} \equiv     \int_0^L \underset{\nu  \in \textbf{R}}{\mathrm{max}}   \big\{      p \nu - \sigma_u {\tiny \big(} \nu , \xi {\tiny \big)}  + V \nu    \big\} \mathrm{d}y        , 
 \end{align*}

 \begin{align*}
       \xi \equiv \theta + x \big/ 2 + y \big/ 2      , 
 \end{align*}

 \begin{align*}
  \textit{Surface tension} \equiv   \sigma {\tiny \big( s , t \big)} \equiv \underset{H,V}{\mathrm{max}}  \big\{  s \cdot H + t \cdot V - f_{\textit{torus}} {\tiny \big( H , V )  }           \big\}            , 
 \end{align*}

 \begin{align*}
  \textit{Cylindrical free energy}  \equiv   f_{D,\chi} \equiv \underset{h \in h ( \chi)}{\mathrm{max}} \int_D \sigma {\tiny \big(} \partial_x h , \partial_y h  {\tiny \big)}  \text{ } \mathrm{d} x \text{ } \mathrm{d} y     , 
 \end{align*}

 }

\noindent for $p,h : {\tiny \big[} 0 , L {\tiny \big]} \longrightarrow \textbf{R}$, $\theta \in {\tiny\big[ 0 , 2 \pi \big]}$, ${\tiny \big( x , y \big)} \in \textbf{R}^2$, $s,t \in {\tiny\big[ - 1 \big/ 2 , 1 \big/ 2  \big]}$, to demonstrate that integrability of the $H_u$ displayed above holds from the existence of infinitely many conserved quantities due to commutativity,

{\small

\begin{align*}
  {\tiny \big[} \textbf{T} {\tiny\big( u ,H  , V )} , \textbf{T} {\tiny \big( u^{\prime} ,H ,  V ) \big] } = 0  , 
\end{align*}

}

\noindent of two transfer matrices for the 6-vertex model one performs the following computations with respect to the symplectic form,

{\small

\begin{align*}
\big\{ \cdot , \cdot \big\},
\end{align*}

}

\noindent satisfying the properties,

{\small \begin{itemize}
    \item [$\bullet$] \textit{Anticommutativity}. $\big\{ f, g \big\}  =  - \big\{ g , f \big\} $,

    \item[$\bullet$] \textit{Bilinearity}. For real $a,b$, $\big\{ af + bg , h \big\} = a \big\{ f ,h \big\} + b \big\{ g , h \big\},$ and $\big\{ h , af + bg \big\} = a \big\{ h , f \big\} + b \big\{ h , g \big\} $,

    \item[$\bullet$] \textit{Leibniz' rule}. $\big\{ fg , h \big\} = \big\{ f , h \big\} g + f \big\{ g , h \big\}$,

    \item[$\bullet$] \textit{Jacobi identity}. $\big\{ f , \big\{ g , h \big\} \big\} + \big\{ g , \big\{ h , f \big\} \big\} + \big\{ h , \big\{ f , g \big\} \big\} = 0$,  \end{itemize} }

\noindent given the test functions $f,g,h$ (we abbreviate the above list of properties for the Poisson bracket with $(\mathrm{AC})$, $(\mathrm{BL})$, $(\mathrm{LR})$ and $(\mathrm{JI})$, respectively, for short should they appear in the following computation for demonstrating that $\mathcal{H}_u$ is in involution $\mathcal{H}_v$ for $u \neq v$ through the stated integrability property of conserved quantities of $\mathcal{H}$). Given the previously identified phase space, denote $X_f$ and $X_g$ corresponding to the vector fields for $f$ and $g$, respectively. The symplectic manifold (ie, a pair of a manifold $M$ over the previously identified phase space endowed with the operation,

{\small \begin{align*}
     \omega \equiv {\tiny \big[ X_F, X_G \big]}    ) ,
\end{align*} }

 \noindent is in the correspondence,

 {\small

 \begin{align*}
   \big\{ \cdot , \cdot \big\}    \Longleftrightarrow  \big[ \cdot , \cdot  \big]  ,
 \end{align*}

 }
 
 \noindent where the Poisson bracket on the LHS takes as input operator (which are obtained from product expansions of L-operators for the 6-vertex model under domain-wall boundary conditions provided in the \textbf{Proposition} above), while the symplectic form on the RHS takes as input vector fields corresponding to the test functions $f,g$. Given such a correspondence as the one displayed above underlying the symplectic operation $\omega$ to argue that the desired integrability property of $H_u$ holds it suffices to verify that the integrals corresponding to solutions of the Euler-Lagrange equation are in involution (ie, that the Poisson bracket between any two of the integrals vanishes). Furthermore, besides confirming that such integrals are in involution one must also count the number of conserved quantities against the dimension of the phase space. To this end by direct computation from the identity,

{\small

\begin{align*}
   \bigg[  \overset{N}{\underset{i=1}{\prod}}   \mathrm{diag} \big( \mathrm{exp} {\tiny\big( 2H \big)} ,  \mathrm{exp} {\tiny\big( 2 H \big)}  \big)  \cdot      R_{ia} {\tiny\big( u - v_i \big) }   ,   \overset{N}{\underset{i=1}{\prod}}   \mathrm{diag} \big( \mathrm{exp} {\tiny\big( 2H \big)} ,  \mathrm{exp} {\tiny\big( 2 H \big)}  \big)  \cdot      R_{ia} {\tiny\big( u^{\prime} - v^{\prime}_i \big) }    \bigg] = 0      ,
\end{align*} 

}

 \noindent the conserved quantities associated with the identity,

{\small

\begin{align*}
        {\tiny \big[} \textbf{T} {\tiny\big( u ,H  , V )} , \textbf{T} {\tiny \big( v ,H ,  V ) \big] } = 0        ,
\end{align*} 

}

\noindent for the spectral parameters $u \neq v$ can be shown to be in involution from the computation,

{\small

\begin{align*}
  \big\{   \mathcal{H}^{\pm}_u {\tiny\big( q , H \big)}      , \mathcal{H}^{\pm}_v  {\tiny\big( q , H \big)}   \big\}  , 
\end{align*} 

}

\noindent which we decompose into the following four possible cases,

{\small

\begin{align*}
  \big\{   \mathcal{H}^{+}_u {\tiny\big( q , H \big)}      , \mathcal{H}^{+}_v  {\tiny\big( q , H \big)}   \big\}  ,  \\ \\      \big\{   \mathcal{H}^{+}_u {\tiny\big( q , H \big)}      , \mathcal{H}^{-}_v  {\tiny\big( q , H \big)}   \big\}  , \\ \\  \big\{   \mathcal{H}^{-}_u {\tiny\big( q , H \big)}      , \mathcal{H}^{+}_v  {\tiny\big( q , H \big)}   \big\}  , \\ \\   \big\{   \mathcal{H}^{-}_u {\tiny\big( q , H \big)}      , \mathcal{H}^{-}_v  {\tiny\big( q , H \big)}   \big\}    , 
\end{align*} 

}

\noindent under the encoding of $+$, or $-$ boundary conditions for $H$. Indeed proceeding with the first possible case from the list of Poisson brackets provided above implies,

{\small

\begin{align*}
    \big\{   \mathcal{H}^{+}_u {\tiny\big( q , H \big)}      , \mathcal{H}^{+}_v  {\tiny\big( q , H \big)}   \big\} \equiv         \bigg\{     {\mathrm{max}}  \big\{     H + \mathrm{log} \text{ } \mathrm{sinh} \text{ }  u + \int_C  \mathrm{log}\big[  \frac{\mathrm{sinh} (    \frac{\eta}{2} + u - i \alpha  - iu  ) }{\mathrm{sinh} (    \frac{\eta}{2} - u + i \alpha - iu    ) }  \big]    \rho \big( \alpha \big)   \text{ } \mathrm{d} \alpha   \big\}        \\     ,    {\mathrm{max}}  \big\{     H + \mathrm{log} \text{ } \mathrm{sinh} \text{ }  v + \int_C  \mathrm{log}\big[  \frac{\mathrm{sinh} (    \frac{\eta}{2} + v - i \alpha  - i v   ) }{\mathrm{sinh} (    \frac{\eta}{2} - v  + i \alpha - i v     ) }  \big]    \rho \big( \alpha \big)   \text{ } \mathrm{d} \alpha   \big\}         \bigg\} \end{align*}
    
    \begin{align*} \overset{(\mathrm{BL})}{=} \bigg\{         \underset{+}{\mathrm{max}} \text{ }  H     ,    \underset{+}{\mathrm{max}} \text{ }  \big\{     H + \mathrm{log} \text{ } \mathrm{sinh} \text{ }  v + \int_C  \mathrm{log}\big[  \frac{\mathrm{sinh} (    \frac{\eta}{2} + v - i \alpha  - i v   ) }{\mathrm{sinh} (    \frac{\eta}{2} - v  + i \alpha - i v     ) }  \big]    \rho \big( \alpha \big)   \text{ } \mathrm{d} \alpha   \big\}          \bigg\}  \end{align*}
    
    \begin{align*}               +     \bigg\{      \underset{+}{\mathrm{max}} \text{ }          \big\{  \mathrm{log} \text{ } \mathrm{sinh} \text{ }  u      \big\}    ,    \underset{+}{\mathrm{max}} \text{ }  \big\{     H + \mathrm{log} \text{ } \mathrm{sinh} \text{ }  v + \int_C  \mathrm{log}\big[  \frac{\mathrm{sinh} (    \frac{\eta}{2} + v - i \alpha  - i v   ) }{\mathrm{sinh} (    \frac{\eta}{2} - v  + i \alpha - i v     ) }  \big]    \rho \big( \alpha \big)   \text{ } \mathrm{d} \alpha   \big\}         \bigg\} \end{align*}

 \begin{align*} +         \bigg\{     \underset{+}{\mathrm{max}} \text{ }  \big\{     \int_C  \mathrm{log}\big[  \frac{\mathrm{sinh} (    \frac{\eta}{2} + u - i \alpha  - iu  ) }{\mathrm{sinh} (    \frac{\eta}{2} - u + i \alpha - iu    ) }  \big]    \rho \big( \alpha \big)   \text{ } \mathrm{d} \alpha       \big\}            ,      \underset{+}{\mathrm{max}} \text{ }   \big\{     H + \mathrm{log} \text{ } \mathrm{sinh} \text{ }  v + \int_C  \mathrm{log}\big[  \frac{\mathrm{sinh} (    \frac{\eta}{2} + v - i \alpha  - i v   ) }{\mathrm{sinh} (    \frac{\eta}{2} - v  + i \alpha - i v     ) }  \big]  \end{align*}

  \begin{align*} \times \rho \big( \alpha \big)   \text{ } \mathrm{d} \alpha   \big\}             \bigg\}   \\ 
\end{align*} 

\begin{align*}
{=} \text{ } \underset{+}{\mathrm{max}} \bigg[  \bigg\{           H     ,      \big\{     H + \mathrm{log} \text{ } \mathrm{sinh} \text{ }  v + \int_C  \mathrm{log}\big[  \frac{\mathrm{sinh} (    \frac{\eta}{2} + v - i \alpha  - i v   ) }{\mathrm{sinh} (    \frac{\eta}{2} - v  + i \alpha - i v     ) }  \big]    \rho \big( \alpha \big)   \text{ } \mathrm{d} \alpha   \big\}          \bigg\}   \end{align*}
    
    \begin{align*}               +     \bigg\{                \mathrm{log} \text{ } \mathrm{sinh} \text{ }  u          ,     \big\{     H + \mathrm{log} \text{ } \mathrm{sinh} \text{ }  v + \int_C  \mathrm{log}\big[  \frac{\mathrm{sinh} (    \frac{\eta}{2} + v - i \alpha  - i v   ) }{\mathrm{sinh} (    \frac{\eta}{2} - v  + i \alpha - i v     ) }  \big]    \rho \big( \alpha \big)   \text{ } \mathrm{d} \alpha   \big\}         \bigg\} \end{align*}

 \begin{align*} +         \bigg\{          \int_C  \mathrm{log}\big[  \frac{\mathrm{sinh} (    \frac{\eta}{2} + u - i \alpha  - iu  ) }{\mathrm{sinh} (    \frac{\eta}{2} - u + i \alpha - iu    ) }  \big]    \rho \big( \alpha \big)   \text{ } \mathrm{d} \alpha               ,       \big\{     H + \mathrm{log} \text{ } \mathrm{sinh} \text{ }  v + \int_C  \mathrm{log}\big[  \frac{\mathrm{sinh} (    \frac{\eta}{2} + v - i \alpha  - i v   ) }{\mathrm{sinh} (    \frac{\eta}{2} - v  + i \alpha - i v     ) }  \big]    \rho \big( \alpha \big)   \text{ } \mathrm{d} \alpha   \big\}             \bigg\}   \bigg] \\ 
\end{align*}

\begin{align*}
\overset{(\mathrm{BL})}{=}    \underset{+}{\mathrm{max}} \bigg[  \bigg\{           \int_{\textbf{R}} \bigg[ \frac{\partial \psi}{\partial x} \frac{\partial \bar{\psi}}{\partial x} + \chi \cdot  \bar{\psi}^2 \psi^2 \bigg]    \text{ } \mathrm{d} x       , \int_{\textbf{R}} \bigg[ \frac{\partial \psi}{\partial x^{\prime}} \frac{\partial \bar{\psi}}{\partial x^{\prime}} + \chi \cdot  \bar{\psi}^2 \psi^2 \bigg]    \text{ } \mathrm{d} x^{\prime}   \bigg\}  + \bigg\{ \int_{\textbf{R}} \bigg[ \frac{\partial \psi}{\partial x} \frac{\partial \bar{\psi}}{\partial x}  + \chi \cdot  \bar{\psi}^2 \psi^2 \bigg]    \text{ } \mathrm{d} x , \end{align*}
    
    \begin{align*}    \mathrm{log} \text{ } \mathrm{sinh} \text{ }  v      \bigg\} + \bigg\{   \int_{\textbf{R}} \bigg[ \frac{\partial \psi}{\partial x} \frac{\partial \bar{\psi}}{\partial x} + \chi \cdot  \bar{\psi}^2 \psi^2 \bigg]    \text{ } \mathrm{d} x     ,   \int_C  \mathrm{log}\big[  \frac{\mathrm{sinh} (    \frac{\eta}{2} + v - i \alpha  - i v   ) }{\mathrm{sinh} (    \frac{\eta}{2} - v  + i \alpha - i v     ) }  \big]   \rho \big( \alpha \big)   \text{ } \mathrm{d} \alpha   \big\}             \bigg\}    \end{align*}
    
    \begin{align*}    +   \bigg\{ \mathrm{log} \text{ } \mathrm{sinh} \text{ }  u     , \int_{\textbf{R}} \bigg[ \frac{\partial \psi}{\partial x^{\prime}} \frac{\partial \bar{\psi}}{\partial x^{\prime}} + \chi \cdot  \bar{\psi}^2 \psi^2 \bigg]    \text{ } \mathrm{d} x^{\prime}   \bigg\} + \big\{      \mathrm{log} \text{ } \mathrm{sinh} \text{ }  u             ,  \mathrm{log} \text{ } \mathrm{sinh} \text{ }  v     \big\} \end{align*}
    
    \begin{align*}   + \bigg\{ \mathrm{log} \text{ } \mathrm{sinh} \text{ }  u  ,   \int_C  \mathrm{log}\big[  \frac{\mathrm{sinh} (    \frac{\eta}{2} + v - i \alpha  - i v   ) }{\mathrm{sinh} (    \frac{\eta}{2} - v  + i \alpha - i v     ) }  \big]    \rho \big( \alpha \big)   \text{ } \mathrm{d} \alpha   \bigg\} \end{align*}
    
    \begin{align*}   +     \bigg\{         \int_C  \mathrm{log}\big[  \frac{\mathrm{sinh} (    \frac{\eta}{2} + u - i \alpha  - iu  ) }{\mathrm{sinh} (    \frac{\eta}{2} - u + i \alpha - iu    ) }  \big]    \rho \big( \alpha \big)   \text{ } \mathrm{d} \alpha            ,                    \int_{\textbf{R}} \bigg[ \frac{\partial \psi}{\partial x^{\prime}} \frac{\partial \bar{\psi}}{\partial x^{\prime}} + \chi \cdot  \bar{\psi}^2 \psi^2 \bigg]    \text{ } \mathrm{d} x^{\prime}        \bigg\}      \end{align*}
    
    \begin{align*}   +     \bigg\{         \int_C  \mathrm{log}\big[  \frac{\mathrm{sinh} (    \frac{\eta}{2} + u - i \alpha  - iu  ) }{\mathrm{sinh} (    \frac{\eta}{2} - u + i \alpha - iu    ) }  \big]    \rho \big( \alpha \big)   \text{ } \mathrm{d} \alpha            ,              \mathrm{log} \text{ } \mathrm{sinh} \text{ } v   \bigg\}   \end{align*}
    
    \begin{align*}   +   \bigg\{         \int_C  \mathrm{log}\big[  \frac{\mathrm{sinh} (    \frac{\eta}{2} + u - i \alpha  - iu  ) }{\mathrm{sinh} (    \frac{\eta}{2} - u + i \alpha - iu    ) }  \big]    \rho \big( \alpha \big)   \text{ } \mathrm{d} \alpha            ,    \int_C  \mathrm{log}\big[  \frac{\mathrm{sinh} (    \frac{\eta}{2} + v - i \alpha  - i v   ) }{\mathrm{sinh} (    \frac{\eta}{2} - v  + i \alpha - i v     ) }  \big]    \rho \big( \alpha \big)   \text{ } \mathrm{d} \alpha              \bigg\}   \bigg]  .  \\ \end{align*}   
}

\noindent To evaluate the above superposition, it suffices to repeatedly apply $(\mathrm{BL})$ wherever necessary. Importantly, by making repeated use of the identity,

{\small

\begin{align*}
   \int_C  \mathrm{log}\big[  \frac{\mathrm{sinh} (    \frac{\eta}{2} + u - i \alpha  - iu  ) }{\mathrm{sinh} (    \frac{\eta}{2} - u + i \alpha - iu    ) }  \big]    \rho \big( \alpha \big)   \text{ } \mathrm{d} \alpha           =       \int_C      \mathrm{log}          \frac{\underset{1 \leq n \leq + \infty}{\sum}          \frac{2 \big( \frac{\eta}{2} + u - i \alpha  - iu   \big)^n}{n! }         }{    \underset{1 \leq n \leq + \infty}{\sum} \frac{2 \big(   \frac{\eta}{2} - u + i \alpha - iu      \big)^n }{n!} }   \cdot              \rho \big( \alpha \big)   \text{ } \mathrm{d} \alpha  \end{align*}
      
      \begin{align*}         =      \int_C      \mathrm{log} \bigg[              \underset{1 \leq n \leq + \infty}{\sum}       \bigg( \frac{\sqrt{2}}{n!} \bigg)^2 \bigg(   \frac{   \frac{\eta}{2} - u + i \alpha - iu  }{ \frac{\eta}{2} - u + i \alpha - iu    }      \bigg)^n           \bigg]       \rho \big( \alpha \big)   \text{ } \mathrm{d} \alpha                    
\end{align*} 

  \begin{align*}         =                \int_C     \underset{1 \leq n \leq + \infty}{\prod}      \mathrm{log} \bigg[                 \bigg( \frac{\sqrt{2}}{n!} \bigg)^2 \bigg(   \frac{   \frac{\eta}{2} - u + i \alpha - iu  }{ \frac{\eta}{2} - u + i \alpha - iu    }      \bigg)^n           \bigg]       \rho \big( \alpha \big)   \text{ } \mathrm{d} \alpha                    
\end{align*} 

\begin{align*}
    =     \int_C   \bigg\{   \underset{1 \leq n \leq + \infty}{\prod}      \mathrm{log} \bigg[                 \bigg( \frac{\sqrt{2}}{n!} \bigg)^2 \bigg(   \frac{   \frac{\eta}{2} - u + i \alpha - iu  }{ \frac{\eta}{2} - u + i \alpha - iu    }      \bigg)^n           \bigg]   \cdot     \mathrm{log} \bigg[    \bigg( \frac{\sqrt{2}}{n!} \bigg)^2     \bigg(        \frac{\frac{\eta}{2} + i \alpha}{\frac{\eta}{2} - i \alpha }                     \bigg)^n              \bigg]   \bigg\}    \text{ } \mathrm{d} \alpha                            
\end{align*}

\begin{align*}
    \approx      \int_C   \bigg\{   \underset{0 \leq n \leq + \infty}{\prod}      \mathrm{log} \bigg[                 \bigg( \frac{\sqrt{2}}{n!} \bigg)^2 \bigg(   \frac{   \frac{\eta}{2} - u + i \alpha - iu  }{ \frac{\eta}{2} - u + i \alpha - iu    }      \bigg)^n           \bigg]   \cdot     \mathrm{log} \bigg[    \bigg( \frac{\sqrt{2}}{n!} \bigg)^2     \bigg(        \frac{\frac{\eta}{2} + i \alpha}{\frac{\eta}{2} - i \alpha }                     \bigg)^n              \bigg]   \bigg\}    \text{ } \mathrm{d} \alpha                            
\end{align*} 

\begin{align*}
=     \int_C   \bigg\{            \bigg( \frac{\sqrt{2}}{n!} \bigg)^4   \bigg(         \frac{   \frac{\eta}{2} - u + i \alpha - iu  }{ \frac{\eta}{2} - u + i \alpha - iu    }      \times \frac{\frac{\eta}{2} + i \alpha}{\frac{\eta}{2} - i \alpha }                     \bigg)^n                  \bigg\}    \text{ } \mathrm{d} \alpha     ,
\end{align*}

}

 \noindent for the integral with support over the density of Bethe roots. The final expression obtained above, through the polynomial approximation in the spectral parameter $u$ and additional variable $\alpha$, is of the form,

 {\small
 
\begin{align*}
   \int_C   \bigg\{            \bigg( \frac{\sqrt{2}}{n!} \bigg)^4   \bigg(         \frac{   \frac{\eta}{2} - u + i \alpha - iu  }{ \frac{\eta}{2} - u + i \alpha - iu    }      \times \frac{\frac{\eta}{2} + i \alpha}{\frac{\eta}{2} - i \alpha }                     \bigg)^n                  \bigg\}    \text{ } \mathrm{d} \alpha  =     \bigg( \frac{\sqrt{2}}{n!} \bigg)^4   \cdot  \int_C   \bigg\{          \bigg(         \frac{   \frac{\eta}{2} - u + i \alpha - iu  }{ \frac{\eta}{2} - u + i \alpha - iu    }      \times \frac{\frac{\eta}{2} + i \alpha}{\frac{\eta}{2} - i \alpha }                     \bigg)^n                  \bigg\}    \text{ } \mathrm{d} \alpha   \\    \\               =     \bigg( \frac{\sqrt{2}}{n!} \bigg)^4   \cdot  \int_C   \bigg\{                      \frac{\big( \frac{\eta}{2} \big)^2}{ \big( \frac{\eta}{2} - u + i \alpha - iu \big) \big(   \frac{\eta}{2} - i \alpha   \big)   }    +   \frac{\frac{\eta}{2}  \cdot i \alpha }{ \big( \frac{\eta}{2} - u + i \alpha - iu \big) \big(   \frac{\eta}{2} - i \alpha   \big)   }     -   \frac{u \cdot \frac{\eta}{2}  }{ \big( \frac{\eta}{2} - u + i \alpha - iu \big) \big(   \frac{\eta}{2} - i \alpha   \big)   }   \end{align*}

   \begin{align*} - \frac{ u \cdot i \alpha}{ \big( \frac{\eta}{2} - u + i \alpha - iu \big) \big(   \frac{\eta}{2} - i \alpha   \big)}    + \frac{i \alpha \cdot \frac{\eta}{2}}{ \big( \frac{\eta}{2} - u + i \alpha - iu \big) \big(   \frac{\eta}{2} - i \alpha   \big)}           + \frac{\big( i \alpha \big)^2}{\big( \frac{\eta}{2} - u + i \alpha - iu \big) \big(   \frac{\eta}{2} - i \alpha   \big)}   \end{align*}

   \begin{align*}   - \frac{iu \cdot \frac{\eta}{2}}{\big( \frac{\eta}{2} - u + i \alpha - iu \big) \big(   \frac{\eta}{2} - i \alpha   \big)}    + \frac{ u \cdot \alpha }{\big( \frac{\eta}{2} - u + i \alpha - iu \big) \big(   \frac{\eta}{2} - i \alpha   \big)}     \bigg\}^n     \text{ } \mathrm{d} \alpha    \\    
\end{align*}

\begin{align*}
=      \bigg( \frac{\sqrt{2}}{n!} \bigg)^4   \cdot  \int_C      \bigg\{   \underset{0 \leq k \leq n}{\sum}        \binom{n}{k}      \bigg[      \frac{\big( \frac{\eta}{2} \big)^2}{ \big( \frac{\eta}{2} - u + i \alpha - iu \big) \big(   \frac{\eta}{2} - i \alpha   \big)   }           \bigg]^{n-k}   \cdot \bigg\{                 \underset{0 \leq k^{\prime} \leq k}{\sum}              \binom{k}{k^{\prime}}  \bigg[       \frac{\frac{\eta}{2}  \cdot i \alpha }{ \big( \frac{\eta}{2} - u + i \alpha - iu \big) \big(   \frac{\eta}{2} - i \alpha   \big)   }      \bigg]^{k-k^{\prime}}   \end{align*}

   \begin{align*}  \cdot \bigg\{   \underset{0 \leq k^{\prime\prime} \leq k^{\prime}}{\sum}        \binom{k^{\prime}}{k^{\prime\prime}}   \bigg[                  -   \frac{u \cdot \frac{\eta}{2}  }{ \big( \frac{\eta}{2} - u + i \alpha - iu \big) \big(   \frac{\eta}{2} - i \alpha   \big)   }      \bigg]^{k^{\prime} - k^{\prime\prime}} \cdot \bigg\{      \underset{0 \leq k^{\prime\prime\prime} \leq k^{\prime\prime}}{\sum}   \binom{k^{\prime\prime}}{k^{\prime\prime\prime}}    \bigg[   - \frac{ u \cdot i \alpha}{ \big( \frac{\eta}{2} - u + i \alpha - iu \big) \big(   \frac{\eta}{2} - i \alpha   \big)}   \bigg]^{k^{\prime\prime} - k^{\prime\prime\prime}}   \end{align*}

   \begin{align*}   \cdot \bigg\{     \underset{0 \leq k^{\prime\prime\prime\prime} \leq k^{\prime\prime\prime}}{\sum}       \binom{k^{\prime\prime\prime}}{k^{\prime\prime\prime\prime}}        \bigg[    \frac{i \alpha \cdot \frac{\eta}{2}}{ \big( \frac{\eta}{2} - u + i \alpha - iu \big) \big(   \frac{\eta}{2} - i \alpha   \big)}                   \bigg]^{k^{\prime\prime\prime\prime} - k^{\prime\prime\prime}} \cdot \bigg\{                          \underset{0 \leq k^{\prime\prime\prime\prime\prime} \leq k^{\prime\prime\prime\prime} }{\sum}      \binom{k^{\prime\prime\prime\prime}}{k^{\prime\prime\prime\prime\prime}}   \end{align*}

 \begin{align*}  \times    \bigg[     \frac{\big( i \alpha \big)^2}{\big( \frac{\eta}{2} - u + i \alpha - iu \big) \big(   \frac{\eta}{2} - i \alpha   \big)}                \bigg]^{k^{\prime\prime\prime\prime} - k^{\prime\prime\prime\prime\prime} }    \cdot \bigg\{          \underset{0 \leq   k^{\prime\prime\prime\prime\prime\prime} \leq k^{\prime\prime\prime\prime\prime} }{\sum}        \binom{k^{\prime\prime\prime\prime\prime}}{k^{\prime\prime\prime\prime\prime\prime}}    \bigg[               - \frac{iu \cdot \frac{\eta}{2}}{\big( \frac{\eta}{2} - u + i \alpha - iu \big) \big(   \frac{\eta}{2} - i \alpha   \big)} \bigg]^{k^{\prime\prime\prime\prime\prime}-k^{\prime\prime\prime\prime\prime\prime}}            \end{align*}

   \begin{align*}     \times    \bigg[  \frac{ u \cdot \alpha }{\big( \frac{\eta}{2} - u + i \alpha - iu \big) \big(   \frac{\eta}{2} - i \alpha   \big)}       \bigg]^{k^{\prime\prime\prime\prime\prime\prime}}   \bigg\}       \bigg\}   \bigg\}                     \bigg\}       \bigg\}                   \bigg\} \bigg\}   \text{ } \mathrm{d} \alpha                      ,  \\ 
\end{align*}

\begin{align*}
\equiv \mathcal{I} {\tiny \big( u ,  \alpha \big)} , 
\end{align*}

 }

 \noindent where the above identity is shown to hold from a direct computation in the last section of the Appendix $\bigg(\text{which implies } 
   \int_C  \mathrm{log}\big[  \frac{\mathrm{sinh} (    \frac{\eta}{2} + u - i \alpha  - iu  ) }{\mathrm{sinh} (    \frac{\eta}{2} - u + i \alpha - iu    ) }  \big]    \rho \big( \alpha \big)   \text{ } \mathrm{d} \alpha        \approx \mathcal{I} {\tiny \big( u ,  \alpha \big)}\bigg)$. As a result, $\big\{   \mathcal{H}^{+}_u {\tiny\big( q , H \big)}      , \mathcal{H}^{+}_v  {\tiny\big( q , H \big)}   \big\}$, corresponding to the first possible bracket under $+$ boundary conditions, term by term is equal to:

\begin{itemize}
\item[$\bullet$] \textbf{1}: \textit{Self-interacting term between the Hamiltonian terms}. One has that,

{\small \begin{align*}
    \bigg\{           \int_{\textbf{R}} \bigg[ \frac{\partial \psi}{\partial x} \frac{\partial \bar{\psi}}{\partial x} + \chi \cdot  \bar{\psi}^2 \psi^2 \bigg]    \text{ } \mathrm{d} x       , \int_{\textbf{R}} \bigg[ \frac{\partial \psi}{\partial x^{\prime}} \frac{\partial \bar{\psi}}{\partial x^{\prime}} + \chi \cdot  \bar{\psi}^2 \psi^2 \bigg]    \text{ } \mathrm{d} x^{\prime}   \bigg\}   \end{align*}

   \noindent is equal to, by ($\mathrm{BL}$),
   
   \begin{align*}    \bigg\{       \int_{\textbf{R}}    \frac{\partial \psi}{\partial x} \frac{\partial \bar{\psi}}{\partial x}     \text{ } \mathrm{d} x  ,      \int_{\textbf{R}}   \bigg[ \frac{\partial \psi}{\partial x^{\prime}} \frac{\partial \bar{\psi}}{\partial x^{\prime}} + \chi \cdot  \bar{\psi}^2 \psi^2 \bigg]    \mathrm{d} x^{\prime}          \bigg\}    +           \bigg\{       \int_{\textbf{R}} \chi \cdot  \bar{\psi}^2 \psi^2    \text{ } \mathrm{d} x   ,      \int_{\textbf{R}}   \bigg[ \frac{\partial \psi}{\partial x^{\prime}} \frac{\partial \bar{\psi}}{\partial x^{\prime}} \end{align*}

 \begin{align*}  + \chi \cdot  \bar{\psi}^2 \psi^2 \bigg]    \text{ } \mathrm{d} x^{\prime}          \bigg\}                       =                \bigg\{       \int_{\textbf{R}}    \frac{\partial \psi}{\partial x} \frac{\partial \bar{\psi}}{\partial x}     \text{ } \mathrm{d} x  ,      \int_{\textbf{R}}    \frac{\partial \psi}{\partial x^{\prime}}       \frac{\partial \bar{\psi}}{\partial x^{\prime}} \mathrm{d} x^{\prime}          \bigg\}  +   \bigg\{       \int_{\textbf{R}}    \frac{\partial \psi}{\partial x} \frac{\partial \bar{\psi}}{\partial x}     \text{ } \mathrm{d} x  , \end{align*}

 \begin{align*}       \int_{\textbf{R}}  \chi \cdot  \bar{\psi}^2 \psi^2   \mathrm{d} x^{\prime}          \bigg\}   +    \bigg\{       \int_{\textbf{R}} \chi \cdot  \bar{\psi}^2 \psi^2    \text{ } \mathrm{d} x    ,      \int_{\textbf{R}}    \frac{\partial \psi}{\partial x^{\prime}} \frac{\partial \bar{\psi}}{\partial x^{\prime}}   \text{ } \mathrm{d} x^{\prime}          \bigg\}      \end{align*}

         \begin{align*} 
          +    \bigg\{       \int_{\textbf{R}} \chi \cdot  \bar{\psi}^2 \psi^2    \text{ } \mathrm{d} x    ,      \int_{\textbf{R}}   \chi \cdot  \bar{\psi}^2 \psi^2    \text{ } \mathrm{d} x^{\prime}          \bigg\}  
\end{align*} 

\item[$\bullet$] \textbf{2}: \textit{Hamiltonian term with the natural logarithm of the hyperbolic sine of the spectral parameter $v$}. One has that, 

\begin{align*}
     \bigg\{ \int_{\textbf{R}} \bigg[ \frac{\partial \psi}{\partial x} \frac{\partial \bar{\psi}}{\partial x}  + \chi \cdot  \bar{\psi}^2 \psi^2 \bigg]    \text{ } \mathrm{d} x ,    \mathrm{log} \text{ } \mathrm{sinh} \text{ }  v      \bigg\}                \\ 
\end{align*} 

   \noindent is equal to, by ($\mathrm{BL}$),

\begin{align*}
    \bigg\{ \int_{\textbf{R}}  \frac{\partial \psi}{\partial x} \frac{\partial \bar{\psi}}{\partial x}     \text{ } \mathrm{d} x ,    \mathrm{log} \text{ } \mathrm{sinh} \text{ }  v      \bigg\}              +  \bigg\{ \int_{\textbf{R}}    \chi \cdot  \bar{\psi}^2 \psi^2    \text{ } \mathrm{d} x ,    \mathrm{log} \text{ } \mathrm{sinh} \text{ }  v      \bigg\}                   ,
\end{align*}

\item[$\bullet$] \textbf{3}: \textit{Hamiltonian term with the integral over the density of Bethe roots}. One has that, 

\begin{align*}
        \bigg\{   \int_{\textbf{R}} \bigg[ \frac{\partial \psi}{\partial x} \frac{\partial \bar{\psi}}{\partial x} + \chi \cdot  \bar{\psi}^2 \psi^2 \bigg]    \text{ } \mathrm{d} x     ,   \int_C  \mathrm{log}\big[  \frac{\mathrm{sinh} (    \frac{\eta}{2} + v - i \alpha  - i v   ) }{\mathrm{sinh} (    \frac{\eta}{2} - v  + i \alpha - i v     ) }  \big]   \rho \big( \alpha \big)   \text{ } \mathrm{d} \alpha   \big\}             \bigg\}                      \\  
\end{align*}

\noindent is equal to, by ($\mathrm{BL}$),

\begin{align*}
    \bigg\{   \int_{\textbf{R}}  \frac{\partial \psi}{\partial x} \frac{\partial \bar{\psi}}{\partial x}  \text{ } \mathrm{d} x     ,   \int_C  \mathrm{log}\big[  \frac{\mathrm{sinh} (    \frac{\eta}{2} + v - i \alpha  - i v   ) }{\mathrm{sinh} (    \frac{\eta}{2} - v  + i \alpha - i v     ) }  \big]   \rho \big( \alpha \big)   \text{ } \mathrm{d} \alpha   \big\}             \bigg\}                     +  \bigg\{   \int_{\textbf{R}}  \chi \cdot  \bar{\psi}^2 \psi^2   \text{ } \mathrm{d} x     ,  \\  \int_C  \mathrm{log}\big[  \frac{\mathrm{sinh} (    \frac{\eta}{2} + v - i \alpha  - i v   ) }{\mathrm{sinh} (    \frac{\eta}{2} - v  + i \alpha - i v     ) }  \big]   \rho \big( \alpha \big)   \text{ } \mathrm{d} \alpha   \big\}             \bigg\}     \\                  
\end{align*}

\begin{align*}
    \approx    \bigg\{   \int_{\textbf{R}}  \frac{\partial \psi}{\partial x} \frac{\partial \bar{\psi}}{\partial x}  \text{ } \mathrm{d} x     ,   \mathcal{I} {\tiny \big( u ,  \alpha \big)}    \big\}             \bigg\}                     +  \bigg\{   \int_{\textbf{R}}  \chi \cdot  \bar{\psi}^2 \psi^2   \text{ } \mathrm{d} x     ,    \mathcal{I} {\tiny \big( u ,  \alpha \big)}       \bigg\}  \overset{(\mathrm{LR})}{\approx} 0  . 
\end{align*}

\item[$\bullet$] \textbf{4}: \textit{Nautral logarithm of the hyperbolic sine of the spectral parameter $u$ with the Hamiltonian term}. One has that, 

\begin{align*}
         \bigg\{ \mathrm{log} \text{ } \mathrm{sinh} \text{ }  u     , \int_{\textbf{R}} \bigg[ \frac{\partial \psi}{\partial x^{\prime}} \frac{\partial \bar{\psi}}{\partial x^{\prime}} + \chi \cdot  \bar{\psi}^2 \psi^2 \bigg]    \text{ } \mathrm{d} x^{\prime}   \bigg\}                                                                            \\  
\end{align*}

\noindent is equal to, by ($\mathrm{BL}$),

\begin{align*}
   \bigg\{ \mathrm{log} \text{ } \mathrm{sinh} \text{ }  u     , \int_{\textbf{R}} \frac{\partial \psi}{\partial x^{\prime}} \frac{\partial \bar{\psi}}{\partial x^{\prime}}   \text{ } \mathrm{d} x^{\prime}   \bigg\}    +      \bigg\{ \mathrm{log} \text{ } \mathrm{sinh} \text{ }  u     , \int_{\textbf{R}} \chi \cdot  \bar{\psi}^2 \psi^2    \text{ } \mathrm{d} x^{\prime}   \bigg\}        ,
\end{align*}

\item[$\bullet$] \textbf{5}: \textit{Self-interacting term between the natural logarithm of the hyperbolic sine of the spectral parameters $u$ and $v$}. One has that,

\begin{align*}
          \big\{      \mathrm{log} \text{ } \mathrm{sinh} \text{ }  u             ,  \mathrm{log} \text{ } \mathrm{sinh} \text{ }  v     \big\}   =  \bigg\{    \mathrm{log} \text{ }                  \frac{\mathrm{exp} {\tiny \big(  u \big) } + \mathrm{exp} {\tiny \big( - u \big) }}{2}      ,      \mathrm{log} \text{ }      \frac{\mathrm{exp} {\tiny \big(  v \big) } + \mathrm{exp} {\tiny \big( - v \big) }}{2}               \bigg\}                                           \approx  \frac{1}{2} \big\{      u                     ,   v                         \big\}                 \\  
\end{align*}

\item[$\bullet$] \textbf{6}: \textit{Natural logarithm of the hyperbolic sine of the spectral parameter $u$ with the integral over the density of Bethe roots}. One has that,

\begin{align*}
      \bigg\{ \mathrm{log} \text{ } \mathrm{sinh} \text{ }  u  ,   \int_C  \mathrm{log}\big[  \frac{\mathrm{sinh} (    \frac{\eta}{2} + v - i \alpha  - i v   ) }{\mathrm{sinh} (    \frac{\eta}{2} - v  + i \alpha - i v     ) }  \big]    \rho \big( \alpha \big)   \text{ } \mathrm{d} \alpha   \bigg\} \approx         \big\{ \mathrm{log} \text{ } \mathrm{sinh} \text{ }  u  ,      \mathcal{I} {\tiny \big( u ,  \alpha \big)}           \big\}   \overset{(\mathrm{LR})}{\approx} 0                                            \\  
\end{align*}

\item[$\bullet$] \textbf{7}: \textit{The integral over the density of Bethe roots with the Hamiltonian term}. One has that,

\begin{align*}
      \bigg\{         \int_C  \mathrm{log}\big[  \frac{\mathrm{sinh} (    \frac{\eta}{2} + u - i \alpha  - iu  ) }{\mathrm{sinh} (    \frac{\eta}{2} - u + i \alpha - iu    ) }  \big]    \rho \big( \alpha \big)   \text{ } \mathrm{d} \alpha            ,                    \int_{\textbf{R}} \bigg[ \frac{\partial \psi}{\partial x^{\prime}} \frac{\partial \bar{\psi}}{\partial x^{\prime}} + \chi \cdot  \bar{\psi}^2 \psi^2 \bigg]    \text{ } \mathrm{d} x^{\prime}        \bigg\} \approx         \bigg\{  \mathcal{I} {\tiny \big( u ,  \alpha \big)} ,  \\       \int_{\textbf{R}} \bigg[ \frac{\partial \psi}{\partial x^{\prime}} \frac{\partial \bar{\psi}}{\partial x^{\prime}} + \chi \cdot  \bar{\psi}^2 \psi^2 \bigg]    \text{ } \mathrm{d} x^{\prime}                \bigg\}      \overset{(\mathrm{LR})}{\approx} 0                                                                 \\  
\end{align*} 

\item[$\bullet$] \textbf{8}: \textit{The integral over the density of Bethe roots with the natural logarithm of the hyperbolic sine of the spectral parameter $v$}. One has that,

\begin{align*}
     \bigg\{         \int_C  \mathrm{log}\big[  \frac{\mathrm{sinh} (    \frac{\eta}{2} + u - i \alpha  - iu  ) }{\mathrm{sinh} (    \frac{\eta}{2} - u + i \alpha - iu    ) }  \big]    \rho \big( \alpha \big)   \text{ } \mathrm{d} \alpha            ,              \mathrm{log} \text{ } \mathrm{sinh} \text{ } v   \bigg\} \approx \big\{  \mathcal{I} {\tiny \big( u ,  \alpha \big)}    ,         \mathrm{log} \text{ } \mathrm{sinh} \text{ } v       \big\}  \overset{(\mathrm{LR})}{\approx} 0    \\        \end{align*}

\item[$\bullet$] \textbf{9}: \textit{Self-interacting term of the integrals over the density of Bethe roots}. One has that,

\begin{align*}
          \bigg\{         \int_C  \mathrm{log}\big[  \frac{\mathrm{sinh} (    \frac{\eta}{2} + u - i \alpha  - iu  ) }{\mathrm{sinh} (    \frac{\eta}{2} - u + i \alpha - iu    ) }  \big]    \rho \big( \alpha \big)   \text{ } \mathrm{d} \alpha            ,    \int_C  \mathrm{log}\big[  \frac{\mathrm{sinh} (    \frac{\eta}{2} + v - i \alpha  - i v   ) }{\mathrm{sinh} (    \frac{\eta}{2} - v  + i \alpha - i v     ) }  \big]    \rho \big( \alpha \big)   \text{ } \mathrm{d} \alpha              \bigg\} \\   \end{align*}

    \begin{align*} \approx    \big\{   \mathcal{I} {\tiny \big( u ,  \alpha \big)}  , \mathcal{I} {\tiny \big( v ,  \alpha \big)}        \big\} \overset{(\mathrm{LR})}{\approx} 0     \\      
\end{align*}

}

\end{itemize}

\noindent Besides the above case in which each Hamiltonian flow $\mathcal{H}$ has $+$ boundary conditions one can straightforwardly adapt the above computation to similarly demonstrate that $\big\{   \mathcal{H}^{+}_u {\tiny\big( q , H \big)}   \approx  \mathcal{H}^{-}_v  {\tiny\big( q , H \big)}   \big\} \approx  \big\{   \mathcal{H}^{-}_u {\tiny\big( q , H \big)}   \approx  \mathcal{H}^{+}_v  {\tiny\big( q , H \big)}   \big\},    \big\{   \mathcal{H}^{-}_u {\tiny\big( q , H \big)}      , \mathcal{H}^{-}_v  {\tiny\big( q , H \big)}   \big\} \approx 0$. As a result, by the Liouville-Arnold Theorem, if there exists a manifold with symplectic structure $\omega$ that takes as input two vector fields, the above computations for demonstrating that the integrals from limit-shapes of the inhomogeneous 6-vertex model that are solutions to the Euler-Lagrange equation. Furthermore, given the above decomposition through $\mathcal{I}$ that was analytically obtained to approximate the integral against the density of Bethe roots counting the number of spectral parameters $u \neq v$ for which,

{\small

\begin{align*}
    \mathrm{dim} \big\{  u \neq v :       \big\{   \mathcal{H}^{\pm}_u {\tiny\big( q , H \big)}      , \mathcal{H}^{\pm}_v  {\tiny\big( q , H \big)} \big\}  \equiv 0     \big\}   =   n {\tiny \big( n - 1 \big)}     ,
\end{align*}

}

\noindent against the dimension of the sympletic manifold $M$ implies,

{\small

\begin{align*}
     \frac{ \mathrm{dim} \big\{  u \neq v :       \big\{   \mathcal{H}^{\pm}_u {\tiny\big( q , H \big)}      , \mathcal{H}^{\pm}_v  {\tiny\big( q , H \big)} \big\}  \equiv 0     \big\} }{\mathrm{dim} M } =   \frac{  n {\tiny \big( n - 1 \big)}    }{\mathrm{dim} M }    ,
\end{align*}

}

from which we conclude the argument. \boxed{}

\end{itemize}

\section{Returning to the Quantum monodromy matrix}

\subsection{Overview}

\noindent From expressions obtained for entries of the monodromy matrix,

{\small \[
\begin{bmatrix}
       A {\tiny\big( \lambda_{\alpha} \big) }  & B {\tiny\big( \lambda_{\alpha} \big) }  \\
    C {\tiny\big( \lambda_{\alpha} \big)}  & D {\tiny\big( \lambda_{\alpha} \big)}  \text{ }  
  \end{bmatrix} ,
\] }

\noindent in terms of $A_{3} {\tiny\big( \lambda_{\alpha} \big)}$, $B_3 {\tiny\big( \lambda_{\alpha} \big)}$, $C_3 {\tiny\big( \lambda_{\alpha} \big)}$ and $D_3 {\tiny\big( \lambda_{\alpha} \big)}$, below we perform the following computations with respect to the Poisson bracket. Related to the main result of the paper, below one must argue the following result pertaining to the Bethe equations.

\bigskip

\noindent \textbf{Lemma} \textit{BE 1} (\textit{mapping of the Bethe equations into a higher-dimensional space}). Fix solutions to the Bethe equations, $\alpha_j$. For each $j$, there exists functions $\mathscr{U}^j_1 , \mathscr{U}^j_2 , \mathscr{U}^j_3$ and $\mathscr{U}^j_4$, satisfying the following conditions, for functions,

{\small \begin{align*}
 F_1 \equiv F_1 {\tiny\big(  \frac{1}{2} ,  \alpha_j  , - v_k \big)}   \equiv \mathrm{sinh} {\tiny\big(     \frac{\eta}{2} + i \alpha_j - v_k           \big)} \text{, } 
\end{align*} \begin{align*}
  F_2 \equiv   F_2  {\tiny\big(  \frac{1}{2} , -  \alpha_j  ,  v_k \big) }  \equiv            \mathrm{sinh} {\tiny\big( \frac{\eta}{2} - i \alpha_j + v_k \big)  }   \text{, } 
\end{align*} }

\noindent from which one obtains the system,

{\small \begin{align*}
      \frac{\partial  F_1}{\partial  v_k  }            =  \mathscr{U}^j_1 {\tiny\bigg[ \frac{\eta}{2} , \alpha_j , - v_k \bigg] } \cdot  F_1       \text{, } \end{align*}

 \begin{align*} \frac{\partial F_2 }{\partial   v_k}  =  \mathscr{U}^j_2 {\tiny\bigg[ \frac{\eta}{2} , - \alpha_j , v_k \bigg]  } \cdot   F_2     \text{, } 
 \end{align*} }

\noindent for terms on the left hand side of the Bethe equations, under the product from $k=1$ to $k=N$, while, for terms on the right hand side of the Bethe equations, introduce,

{\small \begin{align*}
     F_3 \equiv F_3 {\tiny\big( 1 , \alpha_j - \alpha_m , 0  \big) }  \equiv \mathrm{sinh} \big(  i {\tiny\big( \alpha_j - \alpha_m \big) + \eta \big)   }    \big)   \text{, } 
\end{align*}  \begin{align*}
     F_4 \equiv F_4 {\tiny\big( - 1, \alpha_j - \alpha_m  , 0 \big)} \equiv  \mathrm{sinh} \big(  {\tiny i \big( \alpha_j - \alpha_m \big) - \eta \big) }      \big)   \text{, } 
\end{align*} }

\noindent from which one obtains the system,

{\small \begin{align*}
   \frac{\partial F_3}{\partial \alpha_m} = \mathscr{U}^j_3 {\tiny\bigg[   1 , \alpha_j - \alpha_m , 0    \bigg] } \cdot F_3 {\tiny\big( 1 , \alpha_j - \alpha_m , 0 \big) }  \text{, }   \end{align*}

  \begin{align*} \frac{\partial F_4}{\partial \alpha_m } = \mathscr{U}^j_4 {\tiny\bigg[   - 1 , \alpha_j - \alpha_m , 0       \bigg]}   \cdot     F_4 {\tiny\big( - 1 , \alpha_j - \alpha_m , 0 \big)  } \text{, } 
\end{align*} }

\noindent under the product from $m = 1$ to $m=n$, except for $m = j$.

\bigskip

\noindent For a collection of solutions ${\tiny\big\{ \alpha_j \big\}_{j \in \textbf{N}}}$,

{\small \begin{align*}
             \mathscr{U}^{\textbf{N}}_1 {\tiny\bigg[     \frac{\eta}{2} ,  \alpha_j , - v_k \bigg]} =       \underset{j \in \textbf{N}}{\bigcup}  \mathscr{U}^j_1 {\tiny\bigg[     \frac{\eta}{2} ,  \alpha_j , - v_k \bigg]} \text{, } \end{align*}

  \begin{align*}   \mathscr{U}^{\textbf{N}}_2 {\tiny\bigg[ \frac{\eta}{2} , - \alpha_j ,  v_k \bigg]} =  \underset{j \in \textbf{N}}{\bigcup}  \mathscr{U}^j_2 {\tiny\bigg[     \frac{\eta}{2} ,  - \alpha_j ,  v_k \bigg]    }        \text{, } \end{align*}

  \begin{align*}   \mathscr{U}^{\textbf{N}}_3 {\tiny\bigg[  1 , \alpha_j - \alpha_m , 0 \bigg]  } =  \underset{j \in \textbf{N}}{\bigcup}  \mathscr{U}^j_3  {\tiny\bigg[ 1, \alpha_j - \alpha_m , 0\bigg]    }   \text{, } \end{align*}

  \begin{align*}    \mathscr{U}^{\textbf{N}}_4 {\tiny\bigg[  -1 , \alpha_j - \alpha_m , 0 \bigg] } =       \underset{j \in \textbf{N}}{\bigcup}   \mathscr{U}^j_4 {\tiny\bigg[ -1, \alpha_j - \alpha_m , 0\bigg] }   \text{. }
\end{align*} }

\noindent \textbf{Lemma} \textit{BE 2} (\textit{mapping the Bethe equations to the four functions obtained in the previous result}). From the functions obtained in the previous result, the Bethe equations can be mapped to the relation,

{\small \begin{align*}
   \overset{N}{\underset{k=1}{\prod}}  \frac{\mathscr{U}^j_1 {\tiny\big( \frac{\eta}{2} , \alpha_j , - v_k  \big)} \cdot  F_1  {\tiny\big( \frac{1}{2} , \alpha_j , - v_k \big)     }   }{ \mathscr{U}^j_2 {\tiny\big( \frac{\eta}{2} , - \alpha_j , v_k \big)} \cdot     F_2  {\tiny\big( \frac{1}{2} , - \alpha_j , v_k \big)}  }  = \mathrm{exp} {\tiny\big( 2 H N \big) } \cdot    \overset{n}{\underset{m=1 , m \neq j }{\prod} }   
 \frac{\mathscr{U}^j_3 {\tiny\big(   1 , \alpha_j - \alpha_m , 0    \big)  \cdot F_3 {\tiny\big( 1 , \alpha_j - \alpha_m , 0 \big) } }  }{\mathscr{U}^j_4 {\tiny\big(   - 1 , \alpha_j - \alpha_m , 0       \big)   } \cdot     F_4 {\tiny\big( - 1 , \alpha_j - \alpha_m , 0 \big) }  }       \text{. }
\end{align*} }

\bigskip

\noindent From the previous two results, in the next result below we express $\frac{F_1}{F_2}$ and $\frac{F_3}{F_4}$, in terms of basis functions that are dependent on each solution $\alpha_k$ to the Bethe equations.

\bigskip

\noindent \textbf{Lemma} \textit{BE 3} (\textit{expressing} $F_1, F_2, F_3,$ \textit{and} $F_4$ \textit{in the solution space for the Bethe equation}). There exists functions $\mathscr{S}_1 \equiv \mathscr{S}_1 {\tiny\big( \eta , \alpha_j , v_k \big)}$ and $\mathscr{S}_2 \equiv \mathscr{S}_2 {\tiny\big( \eta , \alpha_j , v_k \big)}$, for which,

{\small \begin{align*}
  F_{1,2} = \mathscr{S}_1 \cdot  F_1 + \mathscr{S}_2 \cdot F_2   \text{, } 
\end{align*} } 

\noindent in the basis spanned by $F_1$ and $F_2$, as well as functions $\mathscr{S}_3 \equiv \mathscr{S}_3 {\tiny\big( \eta , \alpha_j , v_k \big)}$ and $\mathscr{S}_4 \equiv \mathscr{S}_4 {\tiny\big( \eta , \alpha_j , v_k \big)}$, for which,

{\small \begin{align*}
F_{3,4} = \mathscr{S}_3 \cdot  F_3 + \mathscr{S}_4 \cdot  F_4  \text{, } 
\end{align*} } 

\noindent in the basis spanned by $F_3$ and $F_4$.

\bigskip

\noindent \textbf{Lemma} \textit{BE 4} (\textit{spanning set of the entire solution space of the Bethe equations}). For all solutions to the Bethe equations, the spanning set is equivalent to the union of spanning sets for each $j$.

\subsection{Sixteen relations from the monodromy matrix}

\noindent With the action given by the exponential in the prefactor to the basis element $F_1$,

{\small \begin{align*}
  \frac{\mathrm{exp} \big( - {\tiny\big( \frac{\eta}{2} + i \alpha_j - v_k \big)}  \big) }{  \mathrm{exp} \big( - {\tiny\big( \frac{\eta}{2} + i \alpha_j - v_k \big) } 
 \big) - \mathrm{exp} \big( \frac{\eta}{2} + i \alpha_j - v_k \big)}   \text{, } 
\end{align*} }

\noindent write,

{\small \begin{align*}
 \bigg[ \frac{\mathrm{exp} \big(  \frac{\eta}{2}  + i \alpha_j - v_k  \big)}{\mathrm{exp} \big(  \frac{\eta}{2}  + i \alpha_j - v_k  \big) }  \bigg]  \cdot  \frac{\mathrm{exp} \big( - {\tiny\big( \frac{\eta}{2} + i \alpha_j - v_k \big)} \big) }{  \mathrm{exp} \big( - {\tiny \big( \frac{\eta}{2} + i \alpha_j - v_k \big)}  
 \big) - \mathrm{exp} \big( \frac{\eta}{2} + i \alpha_j - v_k \big)}    \text{, } 
\end{align*} }

\noindent so that rearranging terms gives,

{\small \begin{align*}
       \frac{1}{  1     - \mathrm{exp} \big( 2 {\tiny \big( \frac{\eta}{2} + i \alpha_j - v_k \big) }   \big) }    \text{. }
\end{align*}
 }

\noindent Hence, the set of linear combinations under the functions $F_1$ and $F_2$

{\small \begin{align*}
  \bigg[  \frac{1}{  1     - \mathrm{exp} \big( 2 {\tiny\big( \frac{\eta}{2} + i \alpha_j - v_k \big) }   \big) }     \bigg]  \cdot F_1 {\tiny \big( \frac{1}{2} , \alpha_j , - v_k \big)}   +  \bigg[  \frac{1}{ -  1     + \mathrm{exp} \big( 2 {\tiny\big( \frac{\eta}{2} + i \alpha_j - v_k \big)  } \big) }    \bigg] \cdot  F_2 {\tiny \big( \frac{1}{2} , \alpha_j , v_k \big) }  \text{, } 
\end{align*} }

\noindent with a similar set of relations holding for the basis spanned by $F_3$ and $F_4$.

\bigskip

\noindent Furthermore, from the Poisson bracket of the tensor product of $T_a \big( u , {\tiny \big\{ v_k \big\}} \big)$ with $T_a \big( u^{\prime} ,  {\tiny\big\{ v^{\prime}_k \big\}} \big)$,

{\small \begin{align*}
      \bigg[  r_{a,+}          {\tiny\big( v_k - v^{\prime}_k \big)}     \cdot    T_a \big( u ,  {\tiny\big\{ v_k \big\} }   \big) \bigg]   \bigotimes    T_a \big( u^{\prime} ,  {\tiny\big\{ v^{\prime}_k \big\} } \big) -   T_a \big( u ,  {\tiny\big\{ v_k \big\} }  \big) \bigotimes  \bigg[    T_a \big( u^{\prime} ,  {\tiny\big\{ v^{\prime}_k \big\} }  \big)   \cdot  r_{a,-}       {\tiny\big( v_k - v^{\prime}_k  \big)          }     \bigg]  \text{, } 
\end{align*} }

\noindent one can form a set of sixteen relations for the reduced monodromy matrices of the inhomogeneous six-vertex model, akin to the sixteen relations which are satisfied for the reduced monodromy matrices of the nonlinear Schroedinger equation, [5],

{\small \[
\left\{\!\begin{array}{ll@{}>{{}}l} (1):      \big\{  A  {\tiny\big( u \big) }       , A  {\tiny\big( u^{\prime} \big) }    \big\} 
\text{, } \\ \vdots   \\ (16): \big\{  D  {\tiny\big( u \big) } ,       D  {\tiny\big( u^{\prime} \big) }  \big\} \text{, }
\end{array}\right.
\] }

\noindent from the tensor product, which in the coordinates $u,u^{\prime}$ is equivalent to,

\[
\big\{ T_a \big( u ,  {\tiny\big\{ v_k \big\}}  \big) \overset{\bigotimes}{,}   T_a \big( u^{\prime} ,  {\tiny\big\{ v^{\prime}_k \big\}}  \big)   \big\} = \bigg\{    \begin{bmatrix} 
A  {\tiny\big( u \big)}  & B  {\tiny\big( u \big) } \\ C  {\tiny\big( u \big)} & D  {\tiny\big( u \big)  } 
\end{bmatrix}\overset{\bigotimes}{,}   \begin{bmatrix} 
A  {\tiny\big( u^{\prime} \big) } & B  {\tiny\big( u^{\prime} \big)}  \\ C  {\tiny\big( u^{\prime} \big)} & D  {\tiny\big( u^{\prime} \big) }   
\end{bmatrix}  \bigg\} \text{, } 
\]

\noindent which by the definition of Poisson bracket of the tensor product of the two reduced monodromy matrices, is,

\[
   \bigg\{    \bigg[  r_{a,+}          {\tiny\big( v_k - v^{\prime}_k \big) } \cdot \begin{bmatrix} 
A  {\tiny\big( u \big)}  & B  {\tiny\big( u \big)}  \\ C  {\tiny\big( u \big)} & D  {\tiny\big( u \big) }  
\end{bmatrix}  \bigg]   \bigotimes   \begin{bmatrix} 
A  {\tiny\big( u^{\prime} \big)}  & B  {\tiny\big( u^{\prime} \big) } \\ C  {\tiny\big( u^{\prime} \big)} & D  {\tiny\big( u^{\prime} \big)}    
\end{bmatrix}  \bigg\}  - \bigg\{   \begin{bmatrix} 
A  {\tiny\big( u \big)}  & B  {\tiny\big( u \big) } \\ C  {\tiny\big( u \big) } & D  {\tiny\big( u \big)   } 
\end{bmatrix} \bigotimes  \bigg[  \begin{bmatrix} 
A  {\tiny\big( u^{\prime} \big)}   & B  {\tiny\big( u^{\prime} \big) } \\ C  {\tiny\big( u^{\prime} \big)} & D  {\tiny\big( u^{\prime} \big) }  
\end{bmatrix}  \] \[ \times  r_{a,-}      {\tiny \big( v_k - v^{\prime}_k  \big)  }             \bigg]  \bigg\}   \text{. } 
\]

\noindent From the set of all possible relations, those which vanish with respect to the Poisson bracket constitute the action-angle variables. The next result determines which relations, from the sixteen listed above, vanish.

\bigskip

\noindent \textit{Proof of the Proposition providing a decomposition of the bracket $\big\{ {\tiny A \big( u \big) A \big( u^{\prime} \big)} \big\}$}. To demonstrate that the statement above holds, it suffices to compute the Poisson bracket for each of the sixteen relations. Beginning with the first relation, write,

{\small \begin{align*}
 ( 1 )  \equiv \bigg\{ \bigg[ A_3 {\tiny\big( u \big)} + B_3 {\tiny\big( u \big) } \bigg]  \big( \mathrm{sin} {\tiny\big( 2 \eta \big) }\big)^{n-3} \mathscr{A}_1       , \bigg(  \bigg\{      A_3 {\tiny\big( u^{\prime} \big)} + B_3 {\tiny\big( u^{\prime} \big)}  \bigg\} \cdot   \bigg\{    \big( \mathrm{sin} {\tiny\big( 2 \eta \big)}  \big)^{n-3} \mathscr{A}^{\prime}_1  + \mathscr{A}^{\prime}_2 +  \mathscr{A}^{\prime}_3 \bigg\} \bigg)      \bigg\} \end{align*}

  \begin{align*}  +  \bigg\{ \bigg[ A_3 {\tiny\big( u \big)} + B_3 {\tiny\big( u \big) }  \bigg]   \mathscr{A}_2  , \bigg(  \bigg\{       A_3 \big( u^{\prime} \big) + B_3 \big( u^{\prime} \big) \bigg\} \cdot   \bigg\{   \big( \mathrm{sin} {\tiny\big( 2 \eta \big)}  \big)^{n-3} \mathscr{A}^{\prime}_1  + \mathscr{A}^{\prime}_2 +  \mathscr{A}^{\prime}_3 \bigg\}   \bigg)  \bigg\}    \end{align*}

  \begin{align*}  +  \bigg\{  \bigg[ A_3 {\tiny\big( u \big)} + B_3 {\tiny\big( u \big) }\bigg]  \mathscr{A}_3    ,  \bigg( \bigg\{     A_3 {\tiny\big( u^{\prime} \big)} + B_3 {\tiny\big( u^{\prime} \big)}  \bigg\} \cdot  \bigg\{   \big( \mathrm{sin} {\tiny \big( 2 \eta \big)}  \big)^{n-3} \mathscr{A}^{\prime}_1  + \mathscr{A}^{\prime}_2 +  \mathscr{A}^{\prime}_3 \bigg\}     \bigg)   \bigg\} \text{, } 
\end{align*} }

\noindent which can be further rearranged, after a second application of bilinearity of the Poisson bracket, as,

{\small \begin{align*}
 \bigg\{ \bigg[ A_3 {\tiny\big( u \big)} + B_3 {\tiny\big( u \big)} \bigg]  \big( \mathrm{sin} {\tiny\big( 2 \eta \big) } \big)^{n-3} \mathscr{A}_1     , \bigg[ A_3 {\tiny\big( u^{\prime} \big)} + B_3 {\tiny\big( u^{\prime} \big) }\bigg] \big( \mathrm{sin} {\tiny\big( 2 \eta \big) }\big)^{n-3} \mathscr{A}^{\prime}_1    \bigg\}  
 +   \bigg\{    \bigg[ A_3 {\tiny\big( u \big)} + B_3 {\tiny\big( u \big)}  \bigg]  \end{align*}

  \begin{align*}  \times  \big( \mathrm{sin} {\tiny\big( 2 \eta \big)}  \big)^{n-3} \mathscr{A}_1  ,              \bigg[ A_3 {\tiny\big( u^{\prime} \big)} + B_3 {\tiny\big( u^{\prime} \big)} \bigg]  \mathscr{A}^{\prime}_2       \bigg\}   +    \bigg\{  \bigg[ A_3 {\tiny\big( u \big)} + B_3 {\tiny\big( u \big) }\bigg]  \big( \mathrm{sin} {\tiny\big( 2 \eta \big)} \big)^{n-3} \mathscr{A}_1  \end{align*}

  \begin{align*}     ,  \bigg[ A_3 {\tiny\big( u^{\prime} \big)}  + B_3 {\tiny\big( u^{\prime} \big)} \bigg] \mathscr{A}^{\prime}_3   \bigg\}   
+  \bigg\{    \bigg[ A_3 {\tiny\big( u \big)} + B_3 {\tiny\big( u \big)}  \bigg]  \mathscr{A}_2   \end{align*}

  \begin{align*}   ,      \bigg[ A_3 {\tiny\big( u^{\prime} \big) } + B_3 {\tiny\big( u^{\prime} \big)} \bigg] \big( \mathrm{sin} {\tiny\big( 2 \eta \big)} \big)^{n-3} \mathscr{A}^{\prime}_1              \bigg\}  +  \bigg\{  \bigg[ A_3 {\tiny\big( u \big)} + B_3 {\tiny\big( u \big)} \bigg] \mathscr{A}_2  ,      \bigg[ A_3 {\tiny\big( u^{\prime} \big)} \end{align*}

  \begin{align*}  + B_3 {\tiny\big( u^{\prime} \big) }\bigg] \mathscr{A}^{\prime}_2        \bigg\}  +  \bigg\{   \bigg[ A_3 {\tiny\big( u \big)} + B_3 {\tiny\big( u \big) }\bigg] \mathscr{A}_2  ,     \bigg[ A_3 {\tiny\big( u^{\prime} \big) } + B_3 {\tiny\big( u^{\prime} \big)} \bigg] \mathscr{A}^{\prime}_3   \bigg\}  \end{align*}

  \begin{align*}   +      \bigg\{  \bigg[ A_3 {\tiny\big( u \big)}  + B_3 {\tiny\big( u \big) }\bigg] \mathscr{A}_3  \end{align*}

  \begin{align*}   ,   \bigg[ A_3 {\tiny\big( u^{\prime} \big)} + B_3 {\tiny\big( u^{\prime} \big)} \bigg] \big( \mathrm{sin} {\tiny\big( 2 \eta \big)}  \big)^{n-3}  \mathscr{A}^{\prime}_1  \bigg\}   +      \bigg\{  \bigg[ A_3 {\tiny\big( u \big)} \end{align*}

  \begin{align*}  + B_3 {\tiny\big( u \big)} \bigg] \mathscr{A}_3   ,   \bigg[ A_3 {\tiny\big( u^{\prime} \big)} + B_3 {\tiny\big( u^{\prime} \big)} \bigg] \mathscr{A}^{\prime}_2  \bigg\}  +  \bigg\{  \bigg[ A_3 {\tiny\big( u \big)}  + B_3 {\tiny\big( u \big)}  \bigg] \mathscr{A}_3  \end{align*}

  \begin{align*}   ,   \bigg[ A_3 {\tiny\big( u^{\prime} \big)} + B_3 {\tiny\big( u^{\prime} \big)} \bigg] \mathscr{A}^{\prime}_3  \bigg\}  \text{. } 
\end{align*} } 

\noindent With two more applications of bilinearity of the Poisson bracket to each term in the superposition above, the previous expression is equivalent to,

{\small \begin{align*}
       \underset{(\mathscr{P}_1 , \mathscr{P}_2 ) \in ( B_3 ( u ) , A_3 ( u^{\prime}))}{\underset{(\mathscr{P}_1 , \mathscr{P}_2 ) \in ( A_3  ( u ) , B_3 (u^{\prime})) }{\underset{(\mathscr{P}_1 , \mathscr{P}_2 ) \in (A_3 ( u ) , A_3 ( u^{\prime})) }{\underset{(\mathscr{P}_1 , \mathscr{P}_2 ) \in (B_3 ( u ) , B_3 ( u^{\prime})) }{\sum}}}}   \big\{    \mathscr{P}_1 \big( \mathrm{sin} {\tiny \big( 2 \eta \big) }  \big)^{n-3} \mathscr{A}_1   ,   \mathscr{P}_2      \big( \mathrm{sin} {\tiny\big( 2 \eta \big) } \big)^{n-3}\mathscr{A}^{\prime}_1     \big\}    \end{align*}

   \begin{align*} 
       +    \underset{(\mathscr{P}_1 , \mathscr{P}_2 )  \in ( B_3 ( u ) , A_3 ( u^{\prime}))}{\underset{(\mathscr{P}_1 , \mathscr{P}_2 ) \in ( A_3  ( u ) , B_3 (u^{\prime})) }{\underset{(\mathscr{P}_1 , \mathscr{P}_2 ) \in (A_3 ( u ) , A_3 ( u^{\prime})) }{\underset{(\mathscr{P}_1  , \mathscr{P}_2   ) \in (B_3 ( u ) , B_3 ( u^{\prime})) }{\sum}}}}  \big\{  \mathscr{P}_1   \big( \mathrm{sin} {\tiny\big( 2 \eta \big)} \big)^{n-3}    \mathscr{A}_1     ,                  \mathscr{P}_2 \mathscr{A}^{\prime}_2   \big\}  +     \underset{(\mathscr{P}_1 , \mathscr{P}_2 ) \in ( B_3 ( u ) , A_3 ( u^{\prime}))}{\underset{(\mathscr{P}_1 , \mathscr{P}_2 ) \in ( A_3  ( u ) , B_3 (u^{\prime})) }{\underset{(\mathscr{P}_1 , \mathscr{P}_2 ) \in (A_3 ( u ) , A_3 ( u^{\prime})) }{\underset{(\mathscr{P}_1  , \mathscr{P}_2   ) \in (B_3 ( u ) , B_3 ( u^{\prime})) }{\sum}}}}       \big\{   \mathscr{P}_1      \big( \mathrm{sin} {\tiny\big( 2 \eta \big) } \big)^{n-3}   \mathscr{A}_1    \end{align*}

   \begin{align*}  ,     \mathscr{P}_2 \mathscr{A}^{\prime}_3       \big\}  \end{align*}

   \begin{align*}   
       +  \underset{(\mathscr{P}_1 , \mathscr{P}_2 )  \in ( B_3 ( u ) , A_3 ( u^{\prime}))}{\underset{(\mathscr{P}_1 , \mathscr{P}_2 ) \in ( A_3  ( u ) , B_3 (u^{\prime})) }{\underset{(\mathscr{P}_1 , \mathscr{P}_2 ) \in (A_3 ( u ) , A_3 ( u^{\prime})) }{\underset{(\mathscr{P}_1  , \mathscr{P}_2   ) \in (B_3 ( u ) , B_3 ( u^{\prime})) }{\sum}}}}       \big\{   \mathscr{P}_1 \mathscr{A}_2  ,  \mathscr{P}_2 \big( \mathrm{sin} {\tiny\big( 2 \eta \big) } \big)^{n-3} \mathscr{A}^{\prime}_1    \big\} + \underset{(\mathscr{P}_1 , \mathscr{P}_2 ) \in ( B_3 ( u ) , A_3 ( u^{\prime}))}{\underset{(\mathscr{P}_1 , \mathscr{P}_2 ) \in ( A_3  ( u ) , B_3 (u^{\prime})) }{\underset{(\mathscr{P}_1 , \mathscr{P}_2 ) \in (A_3 ( u ) , A_3 ( u^{\prime})) }{\underset{(\mathscr{P}_1  , \mathscr{P}_2   ) \in (B_3 ( u ) , B_3 ( u^{\prime})) }{\sum}}}}       \big\{               \mathscr{P}_1   \mathscr{A}_2     ,          \mathscr{P}_2        \mathscr{A}^{\prime}_2 \big\} \end{align*}
 
 \begin{align*} +  \underset{(\mathscr{P}_1 , \mathscr{P}_2 )  \in ( B_3 ( u ) , A_3 ( u^{\prime}))}{\underset{(\mathscr{P}_1 , \mathscr{P}_2 ) \in ( A_3  ( u ) , B_3 (u^{\prime})) }{\underset{(\mathscr{P}_1 , \mathscr{P}_2 ) \in (A_3 ( u ) , A_3 ( u^{\prime})) }{\underset{(\mathscr{P}_1  , \mathscr{P}_2   ) \in (B_3 ( u ) , B_3 ( u^{\prime})) }{\sum}}}}  \big\{ \mathscr{P}_1 \mathscr{A}_2 , \mathscr{P}_2  \mathscr{A}^{\prime}_3       \big\} +  \underset{(\mathscr{P}_1 , \mathscr{P}_2 ) \in ( B_3 ( u ) , A_3 ( u^{\prime}))}{\underset{(\mathscr{P}_1 , \mathscr{P}_2 ) \in ( A_3  ( u ) , B_3 (u^{\prime})) }{\underset{(\mathscr{P}_1 , \mathscr{P}_2 ) \in (A_3 ( u ) , A_3 ( u^{\prime})) }{\underset{(\mathscr{P}_1  , \mathscr{P}_2   ) \in (B_3 ( u ) , B_3 ( u^{\prime})) }{\sum}}}}  \big\{     \mathscr{P}_1 \mathscr{A}_3     ,      \mathscr{P}_2 \big( \mathrm{sin} {\tiny\big( 2 \eta \big)}  \big)^{n-3} \mathscr{A}^{\prime}_1       \big\}  \end{align*}

   \begin{align*}   +  \underset{(\mathscr{P}_1 , \mathscr{P}_2 )  \in ( B_3 ( u ) , A_3 ( u^{\prime}))}{\underset{(\mathscr{P}_1 , \mathscr{P}_2 ) \in ( A_3  ( u ) , B_3 (u^{\prime})) }{\underset{(\mathscr{P}_1 , \mathscr{P}_2 ) \in (A_3 ( u ) , A_3 ( u^{\prime})) }{\underset{(\mathscr{P}_1  , \mathscr{P}_2   ) \in (B_3 ( u ) , B_3 ( u^{\prime})) }{\sum}}}}  \big\{    \mathscr{P}_1 \mathscr{A}_3          ,    \mathscr{P}_2     \mathscr{A}^{\prime}_2    \big\}  + \underset{(\mathscr{P}_1 , \mathscr{P}_2 ) \in ( B_3 ( u ) , A_3 ( u^{\prime}))}{\underset{(\mathscr{P}_1 , \mathscr{P}_2 ) \in ( A_3  ( u ) , B_3 (u^{\prime})) }{\underset{(\mathscr{P}_1 , \mathscr{P}_2 ) \in (A_3 ( u ) , A_3 ( u^{\prime})) }{\underset{(\mathscr{P}_1  , \mathscr{P}_2   ) \in (B_3 ( u ) , B_3 ( u^{\prime})) }{\sum}}}}  \big\{       \mathscr{P}_1      \mathscr{A}_3         ,     \mathscr{P}_2            \mathscr{A}^{\prime}_3         \big\}  \text{, } 
\end{align*} } 

\noindent after isolating Poisson brackets together in groups of four, from which thirty six terms from the superposition on the previous page can be expressed with nine summations. To lighten the notation in the superposition above, denote,

\[
 \mathscr{P}   \equiv \left\{\!\begin{array}{ll@{}>{{}}l} (\mathscr{P}_1 , \mathscr{P}_2 ) \in ( B_3  {\tiny\big( u \big)} , B_3 {\tiny\big(u^{\prime} \big)} )   \text{, } \\ (\mathscr{P}_1 , \mathscr{P}_2 ) \in ( A_3  {\tiny\big( u \big) } , A_3 {\tiny\big(u^{\prime} \big)} )    \text{, }      \\   ( \mathscr{P}_1 , \mathscr{P}_2 ) \in ( A_3 {\tiny\big( u \big) }, B_3 {\tiny\big(u^{\prime} \big) })   \text{, }   \\ ( \mathscr{P}_1 , \mathscr{P}_2 ) \in ( B_3 {\tiny\big( u \big) }, A_3   
 {\tiny\big(u^{\prime} \big)}  ) \text{, }
\end{array}\right.
\]

\noindent For the Poisson bracket below, fix some $\big( \mathscr{P}^{\prime} {\tiny\big( u \big) } , \mathscr{P}^{\prime\prime} {\tiny\big( u \big)}  \big) \equiv \big( \mathscr{P}^{\prime} , \mathscr{P}^{\prime\prime} \big) \in \mathscr{P}$. The relation above will be used for the other fifteen relations to determine which ones vanish with respect to the Poisson bracket. We make use of the formula, by bilinearity of the Poisson bracket,

{\small \begin{align*}
 \underset{  \mathscr{P}   }{\sum} \big\{        \mathscr{P}^{\prime} {\tiny\big( u \big)} + \mathscr{P}^{\prime\prime} {\tiny\big( u \big)} , \mathscr{P}^{\prime} {\tiny\big( u^{\prime} \big)} + \mathscr{P}^{\prime\prime} {\tiny\big( u^{\prime} \big)}  \big\}  =   \underset{  \mathscr{P}   }{\sum}  \bigg[ \big\{   \mathscr{P}^{\prime} {\tiny\big( u \big)  }    ,  \mathscr{P}^{\prime} {\tiny\big( u^{\prime} \big) } \big\} +  \big\{ \mathscr{P}^{\prime} {\tiny\big( u \big) }  , \mathscr{P}^{\prime\prime} {\tiny\big( u \big)  } \big\} \end{align*}

 \begin{align*} + \big\{   \mathscr{P}^{\prime\prime} {\tiny\big( u \big) } ,    \mathscr{P}^{\prime} {\tiny\big( u^{\prime} \big) }  \big\}  +   \big\{  \mathscr{P}^{\prime\prime} {\tiny\big( u \big) }   ,   \mathscr{P}^{\prime\prime} {\tiny\big( u^{\prime} \big) }    \big\}   \bigg]                        \text{, }
\end{align*} }

\noindent The other fifteen relations are included after having obtained approximate expressions for each Poisson bracket from the first relation provided above. From each of the sixteen relations provided above as a summation over Poisson brackets, it is possible to determine which one of the relations vanishes by computing each one of the Poisson brackets individually. To this end, for the first relation, (1), rearrange terms from each one of the nine Poisson bracket terms. Before differentiating each expression which appears in the formula for the Poisson bracket observe,

{\small \begin{align*}
     \underline{\mathscr{P}_1 \big( \mathrm{sin}  {\tiny\big( 2 \eta \big) }  \big)^{n-3} \mathscr{A}_1 }    \equiv    \mathscr{P}_1 \big( \mathrm{sin}  {\tiny\big( 2 \eta \big) } \big)^{n-3} 
\cdot   \underset{1 \leq i \leq n-3}{\prod}   \sigma^{-,+}_{n-i}   \text{, } \end{align*}

 \begin{align*} 
 \underline{ \mathscr{P}_2 \big( \mathrm{sin}  {\tiny \big( 2 \eta \big) } \big)^{n-3} \mathscr{A}^{\prime}_1   }     \equiv     \mathscr{P}_2 \big( \mathrm{sin}  {\tiny\big( 2 \eta \big)}  \big)^{n-3} \cdot   \underset{1 \leq i \leq n-3}{\prod}   \sigma^{-,+}_{n-i}              \text{, } \end{align*}

 \begin{align*}         \underline{\mathscr{P}_1 \big( \mathrm{sin}  {\tiny\big( 2 \eta \big) }  \big)^{n-3} \mathscr{A}_1  }      \equiv   \mathscr{P}_1 \big( \mathrm{sin}  {\tiny\big( 2 \eta \big)} \big)^{n-3}  \cdot    \underset{1 \leq i \leq n-3}{\prod}   \sigma^{-,+}_{n-i}           \text{, }   \end{align*}

 \begin{align*}   \underline{\mathscr{P}_2 \mathscr{A}^{\prime}_2   }  \equiv   \mathscr{P}_2 \cdot      \underset{1 \leq i \leq n-3}{\prod}   \mathrm{sin}  {\tiny\big( u^{\prime} - v_{n-i} + \eta \sigma^z_{n-i} \big)  }   \text{, }    \end{align*}

 \begin{align*} 
 \underline{\mathscr{P}_1 \big( \mathrm{sin} \big( 2 \eta \big) \big)^{n-3} \mathscr{A}_1      }       \equiv   \mathscr{P}_1 \big( \mathrm{sin}  {\tiny \big( 2 \eta \big)}  \big)^{n-3} \cdot \underset{1 \leq i \leq n-3}{\prod}   \sigma^{-,+}_{n-i}        \text{, }  \end{align*}

 \begin{align*}
 \underline{\mathscr{P}_2 \mathscr{A}^{\prime}_3 }   \equiv    \mathscr{P}_2 \cdot   {\underset{1 \leq j \leq m}{\sum} }  \bigg[     \underset{1 \leq i \leq m}{\prod} \mathrm{sin} {\tiny\big( u^{\prime} - v_{n-i} \pm \eta \sigma^z_{n-j} \big) }  \cdot \big( \mathrm{sin} {\tiny\big( 2 \eta \big) } \big)^{n^{\prime}-1}  \cdot   \underset{1 \leq j \leq n^{\prime}}{ \prod}  \sigma^{-,+}_{n-j}             \bigg]         \text{, }\end{align*}

 \begin{align*} 
\underline{ \mathscr{P}_1 \mathscr{A}_2 }   \equiv   \mathscr{P}_1  \cdot  \underset{1 \leq i \leq n-3}{\prod}   \mathrm{sin} {\tiny\big( u - v_{n-i} + \eta \sigma^z_{n-j} \big) }   \text{, } \end{align*}

 \begin{align*} \underline{\mathscr{P}_2 \big( \mathrm{sin} \big( 2 \eta \big) \big)^{n-3} \mathscr{A}^{\prime}_1 }  \equiv \mathscr{P}_2 \big( \mathrm{sin} {\tiny\big( 2 \eta \big)}  \big)^{n-3} \cdot    \underset{1 \leq i \leq n-3}{\prod}  \sigma^{-,+}_{n-i}    \text{, } \end{align*}

 \begin{align*}     \underline{\mathscr{P}_1 \mathscr{A}_2}    \equiv  \mathscr{P}_1 
\cdot   \underset{1 \leq i \leq n-3}{\prod}   \mathrm{sin} {\tiny\big( u - v_{n-i} + \eta \sigma^z_{n-j} \big) }   \text{, } \end{align*}

 \begin{align*}  \underline{\mathscr{P}_2 \mathscr{A}^{\prime}_2    }    \equiv   \mathscr{P}_2 \cdot  \underset{1 \leq i \leq n-3}{\prod}   \mathrm{sin} {\tiny \big( u^{\prime} - v_{n-i} + \eta \sigma^z_{n-j} \big) }        \text{, } \end{align*}

 \begin{align*} \underline{\mathscr{P}_1 \mathscr{A}_2 }    \equiv  \mathscr{P}_1 
\cdot   \underset{1 \leq i \leq n-3}{\prod}    \mathrm{sin} {\tiny\big( u - v_{n-i} + \eta \sigma^z_{n-j} \big)  }       \text{, } \end{align*}

 \begin{align*}  \underline{\mathscr{P}_2 \mathscr{A}^{\prime}_3 }  \equiv  \mathscr{P}_2   \cdot    {\underset{1 \leq j \leq m}{\sum} }   \bigg[     \underset{1 \leq i \leq m}{\prod} \mathrm{sin} {\tiny\big( u^{\prime} - v_{n-i} \pm \eta \sigma^z_{n-j} \big) }   \cdot  \big( \mathrm{sin} {\tiny \big( 2 \eta \big) } \big)^{n^{\prime}-1}  \cdot  \underset{1 \leq j \leq n^{\prime}}{ \prod}  \sigma^{-,+}_{n-j}       \text{, } \end{align*}

 \begin{align*}   \underline{\mathscr{P}_1 \mathscr{A}_3 }   \equiv \mathscr{P}_1 \cdot     {\underset{1 \leq j \leq m}{\sum} }    \bigg[   \underset{1 \leq i \leq m}{\prod} \mathrm{sin} {\tiny\big( u - v_{n-i} \pm \eta \sigma^z_{n-j} \big)  } \cdot \big( \mathrm{sin} {\tiny\big( 2 \eta \big)}  \big)^{n^{\prime}-1}   \cdot  \underset{1 \leq j \leq n^{\prime}}{ \prod}  \sigma^{-,+}_{n-j}       \bigg]          \text{, } \end{align*}

 \begin{align*}  \underline{\mathscr{P}_2 \big( \mathrm{sin} \big( 2 \eta \big) \big)^{n-3} \mathscr{A}^{\prime}_1}  \equiv   \mathscr{P}_2 \big( \mathrm{sin} {\tiny \big( 2 \eta \big) }  \big)^{n-3}  \cdot   \underset{1 \leq i \leq n-3}{\prod}  \sigma^{-,+}_{n-i}   \text{, }  \end{align*}

 \begin{align*} \underline{\mathscr{P}_1 \mathscr{A}_3 }      \equiv  \mathscr{P}_1  \cdot     {\underset{1 \leq j \leq m}{\sum} }   \bigg[    \underset{1 \leq i \leq m}{\prod} \mathrm{sin} {\tiny\big( u - v_{n-i} \pm \eta \sigma^z_{n-j} \big) } \cdot  \big( \mathrm{sin} {\tiny\big( 2 \eta \big)}  \big)^{n^{\prime}-1}  \cdot   \underset{1 \leq j \leq n^{\prime}}{ \prod}  \sigma^{-,+}_{n-j}     \bigg]      \text{ }    \text{, } \end{align*}

 \begin{align*}
    \underline{\mathscr{P}_2 \mathscr{A}^{\prime}_2 }     \equiv   \mathscr{P}_2  \cdot    \underset{1 \leq i \leq n-3}{\prod}  \mathrm{sin} {\tiny\big( u^{\prime} - v_{n-i} + \eta \sigma^z_{n-j} \big)    }    \text{, } \end{align*}

 \begin{align*}     \underline{\mathscr{P}_1 \mathscr{A}_3}   \equiv   \mathscr{P}_1  \cdot    {\underset{1 \leq j \leq m}{\sum} }   \bigg[      \underset{1 \leq i \leq m}{\prod} \mathrm{sin} {\tiny \big( u - v_{n-i} \pm \eta \sigma^z_{n-j} \big)}    \cdot  \big( \mathrm{sin} {\tiny\big( 2 \eta \big) } \big)^{n^{\prime}-1}  \cdot    \underset{1 \leq j \leq n^{\prime}}{ \prod}  \sigma^{-,+}_{n-j}          \bigg]         \text{, } \end{align*}

 \begin{align*}  \underline{\mathscr{P}_2 \mathscr{A}^{\prime}_3 }  \equiv  \mathscr{P}_2   \cdot   {\underset{1 \leq j \leq m}{\sum} }   \bigg[      \underset{1 \leq i \leq m}{\prod} \mathrm{sin} {\tiny \big( u^{\prime} - v_{n-i} \pm \eta \sigma^z_{n-j} \big)} \cdot  \big( \mathrm{sin} {\tiny\big( 2 \eta \big) } \big)^{n^{\prime}-1}   \cdot    \underset{1 \leq j \leq n^{\prime}}{ \prod}  \sigma^{-,+}_{n-j}     \bigg]        \text{. } 
\end{align*} } 

\noindent The nine Poisson brackets appearing in the first relation, $\big\{ A {\tiny \big( u \big))} , A {\tiny \big( u^{\prime} \big))} \big\}$ are provided in \textit{3.11}-\textit{3.19} of the Appendix. Combining each of the nine obtained relations provided in the Appendix demonstrates that the first Poisson bracket along the diagonal of the transfer matrix for the 6-vertex model under domain-wall boundary conditions vanishes, from which we conclude the argument. \boxed{}

\subsubsection{Overview of extending the computations with the Poisson bracket to the remaining fifteen relations}

\noindent To exhibit how the computations performed for evaluating the nine Poisson brackets from the first relation can be extended to the remaining fifteen relations, we provide an outline of the terms involved in the second relation below. In particular, these terms are parametrized in different entries of the monodromy matrix, namely $\mathscr{B}^{\prime}_1$, $\mathscr{B}^{\prime}_2$, and $\mathscr{B}^{\prime}_3$ instead of $\mathscr{A}^{\prime}_1$, $\mathscr{A}^{\prime}_2$, and $\mathscr{A}^{\prime}_3$, and can be expressed in terms of a superposition of nine terms, obtained from a superposition of thirty six Poisson brackets as provided for the first relation. The terms, excluding those which are dependent on $\mathscr{A}_1$, $\mathscr{A}_2$, $\mathscr{A}_3$, $\mathscr{B}^{\prime}_1$, $\mathscr{B}^{\prime}_2$, or $\mathscr{B}^{\prime}_3$, appearing in the second relation are,

{\small \begin{align*}
              \underline{\mathscr{P}_2 \big( \mathrm{sin} {\tiny \big( 2 \eta \big) } \big)^{n-3} \mathscr{B}^{\prime}_1}     \equiv     \mathscr{P}_2 \big( \mathrm{sin} {\tiny  \big( 2 \eta \big) }  \big)^{n-3} \cdot  \underset{2 \leq i \leq n - (i-3)}{\prod}      \sigma^{-,+}_{n-i}      \text{, } \end{align*}

 \begin{align*} 
              \underline{\mathscr{P}_1 \mathscr{B}^{\prime}_2}    \equiv   \mathscr{P}_1 \cdot  \underset{2 \leq i \leq n- ( i -3 )}{\prod}  \mathrm{sin} {\tiny  \big( \lambda_{\alpha } - v_{n-i} + \eta \sigma^z_{n-j} \big) }       \text{, }  \end{align*}

 \begin{align*}    \underline{\mathscr{P}_2 \big( \mathrm{sin} {\tiny  \big( 2 \eta \big)  } \big)^{n-3} \mathscr{B}^{\prime}_3}  =    \mathscr{P}_2 \big( \mathrm{sin} {\tiny  \big( 2 \eta \big)  } \big)^{n-3}    \cdot       {\underset{1 \leq j \leq m}{\sum} }   \bigg[    \underset{2 \leq i \leq m}{\prod} \mathrm{sin} \big( \lambda_{\alpha} - v_{n-i} \pm \eta \sigma^z_{n-i} \big)  \cdot    \big( \mathrm{sin} \big( 2 \eta \big) \big)^{n^{\prime}-1} \end{align*}

 \begin{align*} \times   \underset{2 \leq j \leq n^{\prime}}{ \prod}  \sigma^{-,+}_{n-j}      \bigg]        \text{, } \end{align*}

 \begin{align*}
             \underline{ \mathscr{P}_2   \big( \mathrm{sin} {\tiny  \big( 2 \eta \big)  } \big)^{n-3} \mathscr{B}^{\prime}_1} \equiv      \mathscr{P}_2   \big( \mathrm{sin} {\tiny  \big( 2 \eta \big) }  \big)^{n-3}  \cdot \underset{2 \leq i \leq n - (i-3)}{\prod}        \sigma^{-,+}_{n-i}                        \text{, } \end{align*}

 \begin{align*}  \underline{\mathscr{P}_2 \mathscr{B}^{\prime}_2 } \equiv \mathscr{P}_2 \cdot   \underset{2 \leq i \leq n - ( i-3)}{\prod}    
 \mathrm{sin} {\tiny  \big( \lambda_{\alpha} - v_{n-i} + \eta \sigma^z_{n-i} \big)  } 
         \text{, } \end{align*}

 \begin{align*} \underline{\mathscr{P}_2 \mathscr{B}^{\prime}_3}   \equiv  \mathscr{P}_2   \cdot {\underset{1 \leq j \leq m}{\sum} }   \bigg[      \underset{2 \leq i \leq m}{\prod} \mathrm{sin} {\tiny  \big( \lambda_{\alpha} - v_{n-i} \pm \eta \sigma^z_{n-j} \big) }   \cdot  \big( \mathrm{sin} {\tiny  \big( 2 \eta \big)}  \big)^{n^{\prime}-1}   \cdot    \underset{2 \leq j \leq n^{\prime}}{ \prod}  \sigma^{-,+}_{n-j}      \bigg]        \text{, } \end{align*}

 \begin{align*}        \underline{\mathscr{P}_2 \big( \mathrm{sin} \big( 2 \eta \big) \big)^{n-3} \mathscr{B}^{\prime}_1}  \equiv    \mathscr{P}_2 \big( \mathrm{sin} {\tiny  \big( 2 \eta \big)}  \big)^{n-3} \cdot  \underset{2 \leq i \leq n-(i-3)}{\prod}     \sigma^{-,+}_{n-i}   \text{, } \end{align*}

 \begin{align*}   \underline{\mathscr{P}_2 \mathscr{B}^{\prime}_2}  \equiv \mathscr{P}_2 \cdot    \underset{2 \leq i \leq n-(i-3)}{\prod}            \mathrm{sin} {\tiny  \big( \lambda_{\alpha} - v_{n-i} + \eta \sigma^z_{n-j} \big)    }               \text{, }  \end{align*}

 \begin{align*}       \underline{\mathscr{P}_2 \mathscr{B}^{\prime}_3}  \equiv \mathscr{P}_2 \cdot {\underset{1 \leq j \leq m}{\sum} }   \bigg[       \underset{2 \leq i \leq m}{\prod} \mathrm{sin} {\tiny  \big( \lambda_{\alpha} - v_{n-i} \pm \eta \sigma^z_{n-j} \big)  }  \cdot  \big( \mathrm{sin} {\tiny  \big( 2 \eta \big) }  \big)^{n^{\prime}-1} \cdot   \underset{2 \leq j \leq n^{\prime}}{ \prod}  \sigma^{-,+}_{n-j}     \bigg]              \text{. } \end{align*} } 

\noindent To compute each Poisson bracket appearing in (2), the second relation, one can directly apply previous arguments for computing the Poisson brackets for the first relation. For sake of not repeating similar computations with the Poisson bracket than those which were provided earlier for the decomposition of $\big\{ {\tiny A \big( u \big) , A \big( u^{\prime} \big)} \big\}$. To obtain a similar decomposition for each of the Poisson brackets,

{\small

\begin{align*}
 \big\{ {\tiny B \big( u \big) , B \big( u^{\prime} \big)} \big\}  , \cdots  \big\{ {\tiny D \big( u \big) , D \big( u^{\prime} \big)} \big\}   , 
\end{align*} 

}

\noindent one can reformulate each bracket within the system of $16$ relations. Previously, in other work of the author, [15], such a decomposition was provided for the transfer and Quantum monodromy matrices of the 20-vertex models under domain-wall boundary conditions. In comparison to two-dimensional domain-wall boundary conditions, three-dimensional domain-wall boundary conditions for the 20-vertex model imply that a system of relations for the Poisson brackets of operators along the diagonal takes the form,

{\small \[
\left\{\!\begin{array}{ll@{}>{{}}l} \boxed{(1)}:      \big\{  A \big( \underline{u} \big)        , A \big( \underline{u^{\prime}} \big)   \big\} = \bigg\{     \bigg[   \underline{\mathcal{T}^1_{(1,1)}} + \underline{\mathcal{T}^2_{(1,1)}}  + \underline{\mathcal{T}^3_{(1,1)}}  + \underline{\mathcal{T}^4_{(1,1)}}  + \underline{\mathcal{T}^5_{(1,1)}} + \underline{\mathcal{T}^6_{(1,1)}}      \bigg]  \bigg[  \underline{A_1 \big( \underline{u} \big) } 
        + \underline{A_2 \big( \underline{u} \big) }  + \underline{A_3 \big( \underline{u} \big) }  + \underline{A_4  \big( \underline{u} \big) } \\  + \underline{A_5 \big( \underline{u} \big) }  \bigg]    ,      \bigg[   
 \big( \underline{\mathcal{T}^1_{(1,1)}}\big)^{\prime} + \big( \underline{\mathcal{T}^2_{(1,1)}} \big)^{\prime}  +  \big(  \underline{\mathcal{T}^3_{(1,1)}}\big)^{\prime}  + \big( \underline{\mathcal{T}^4_{(1,1)}}\big)^{\prime}  + \big( \underline{\mathcal{T}^5_{(1,1)}}\big)^{\prime} + \big( \underline{\mathcal{T}^6_{(1,1)}}   \big)^{\prime}    \bigg]  \bigg[  \underline{A_1 \big( \underline{u^{\prime}} \big) } 
        + \underline{A_2 \big( \underline{u^{\prime}} \big) }  + \underline{A_3 \big( \underline{u^{\prime}} \big) }  \\ + \underline{A_4  \big( \underline{u^{\prime}} \big) }  + \underline{A_5 \big( \underline{u^{\prime}} \big) }  \bigg]     \bigg\}
\text{ , } \\ \vdots 
\\  \boxed{(81)}: \big\{ I \big( \underline{u} \big) , I \big( \underline{u^{\prime}} \big) \big\} = \bigg\{         \bigg[  \underline{\mathcal{T}^1_{(3,3)}} + \underline{\mathcal{T}^2_{(3,3)} }+ \underline{\mathcal{T}^3_{(3,3)}}  + \underline{\mathcal{T}^4_{(3,3)}        }     \bigg]  \bigg[   \underline{I_1 \big( \underline{u} \big)}  + \underline{I_2 \big( \underline{u} \big)}   + \underline{I_3 \big( \underline{u} \big)}    + \underline{I_4 \big( \underline{u} \big)}  
            \bigg]   ,   \bigg[  \big( \underline{\mathcal{T}^1_{(3,3)}} \big)^{\prime} \\ + \big( \underline{\mathcal{T}^2_{(3,3)} } \big)^{\prime}  + \big( \underline{\mathcal{T}^3_{(3,3)}}  \big)^{\prime} + \big( \underline{\mathcal{T}^4_{(3,3)}} \big)^{\prime}            \bigg]  \bigg[   \underline{I_1 \big( \underline{u^{\prime}} \big)}  + \underline{I_2 \big( \underline{u^{\prime}} \big)}   + \underline{I_3 \big( \underline{u^{\prime}} \big)}    + \underline{I_4 \big( \underline{u^{\prime}} \big)}  
            \bigg]                    \bigg\} \text{, }
\end{array}\right.
\]  }

 \noindent where,

 {\small
 
 \begin{align*}
\bigg[   \underline{\mathcal{T}^1_{(1,1)}} + \underline{\mathcal{T}^2_{(1,1)}}  + \underline{\mathcal{T}^3_{(1,1)}}  + \underline{\mathcal{T}^4_{(1,1)}}  + \underline{\mathcal{T}^5_{(1,1)}} + \underline{\mathcal{T}^6_{(1,1)}}      \bigg]   {\tiny \big( u \big)} \equiv \bigg[   \underline{\mathcal{T}^1_{(1,1)}} + \underline{\mathcal{T}^2_{(1,1)}}  + \underline{\mathcal{T}^3_{(1,1)}}  + \underline{\mathcal{T}^4_{(1,1)}}  + \underline{\mathcal{T}^5_{(1,1)}} + \underline{\mathcal{T}^6_{(1,1)}}      \bigg]  ,  \end{align*}
   
   \begin{align*} \bigg[   
 \big( \underline{\mathcal{T}^1_{(1,1)}}\big)^{\prime} + \big( \underline{\mathcal{T}^2_{(1,1)}} \big)^{\prime}  +  \big(  \underline{\mathcal{T}^3_{(1,1)}}\big)^{\prime}  + \big( \underline{\mathcal{T}^4_{(1,1)}}\big)^{\prime}  + \big( \underline{\mathcal{T}^5_{(1,1)}}\big)^{\prime} + \big( \underline{\mathcal{T}^6_{(1,1)}}   \big)^{\prime}    \bigg]  {\tiny \big( u^{\prime} \big)} \equiv   \bigg[   
 \big( \underline{\mathcal{T}^1_{(1,1)}}\big)^{\prime} + \big( \underline{\mathcal{T}^2_{(1,1)}} \big)^{\prime}  +  \big(  \underline{\mathcal{T}^3_{(1,1)}}\big)^{\prime} \end{align*}

   \begin{align*}     + \big( \underline{\mathcal{T}^4_{(1,1)}}\big)^{\prime}  + \big( \underline{\mathcal{T}^5_{(1,1)}}\big)^{\prime} + \big( \underline{\mathcal{T}^6_{(1,1)}}   \big)^{\prime}    \bigg]   ,  \end{align*}
   
   \begin{align*}        \bigg[  \underline{\mathcal{T}^1_{(3,3)}} + \underline{\mathcal{T}^2_{(3,3)} }+ \underline{\mathcal{T}^3_{(3,3)}}  + \underline{\mathcal{T}^4_{(3,3)}        }     \bigg]  {\tiny \big( u \big)} \equiv     \bigg[  \underline{\mathcal{T}^1_{(3,3)}} + \underline{\mathcal{T}^2_{(3,3)} }+ \underline{\mathcal{T}^3_{(3,3)}}  + \underline{\mathcal{T}^4_{(3,3)}        }     \bigg]    ,  \end{align*}

\begin{align*}
\vdots
\end{align*}

   \begin{align*}    \bigg[  \big( \underline{\mathcal{T}^1_{(3,3)}} \big)^{\prime} + \big( \underline{\mathcal{T}^2_{(3,3)} } \big)^{\prime} + \big( \underline{\mathcal{T}^3_{(3,3)}}  \big)^{\prime} + \big( \underline{\mathcal{T}^4_{(3,3)}} \big)^{\prime}            \bigg]   {\tiny \big( u^{\prime} \big)}       \equiv  \bigg[  \big( \underline{\mathcal{T}^1_{(3,3)}} \big)^{\prime} + \big( \underline{\mathcal{T}^2_{(3,3)} } \big)^{\prime}  + \big( \underline{\mathcal{T}^3_{(3,3)}}  \big)^{\prime} + \big( \underline{\mathcal{T}^4_{(3,3)}} \big)^{\prime}            \bigg]   , 
 \end{align*} 
 
 }

 \noindent one can obtain the relationships satisfied by the remaining Poisson brackets for the 6-vertex model under domain-wall boundary conditions. Explicitly the remaining terms are read off from the superpositions,

{\small \begin{align*}
  \underset{\mathscr{P}}{\sum} \big\{    \mathscr{P}_1 \big( \mathrm{sin} {\tiny \big( 2 \eta \big)}  \big)^{n-3} \mathscr{A}_1   ,   \mathscr{P}_2      \big( \mathrm{sin}  {\tiny \big( 2 \eta \big)}  \big)^{n-3}\mathscr{B}^{\prime}_1     \big\}   +      \underset{\mathscr{P}}{\sum} \big\{  \mathscr{P}_1   \big( \mathrm{sin}  {\tiny \big( 2 \eta \big)}  \big)^{n-3}    \mathscr{A}_1     ,                  \mathscr{P}_2 \mathscr{B}^{\prime}_2   \big\} \end{align*}

  \begin{align*} 
  +  \underset{\mathscr{P}}{\sum}      \big\{   \mathscr{P}_1      \big( \mathrm{sin}  {\tiny  \big( 2 \eta \big) }  \big)^{n-3}   \mathscr{A}_1   ,     \mathscr{P}_2 \mathscr{B}^{\prime}_3       \big\}  +  \underset{\mathscr{P}}{\sum}     \big\{   \mathscr{P}_1 \mathscr{A}_2  ,  \mathscr{P}_2 \big( \mathrm{sin}  {\tiny \big( 2 \eta \big)}  \big)^{n-3} \mathscr{B}^{\prime}_1    \big\} \end{align*}

  \begin{align*} +  \underset{\mathscr{P}}{\sum}      \big\{               \mathscr{P}_1   \mathscr{A}_2  ,          \mathscr{P}_2        \mathscr{B}^{\prime}_2 \big\} + \underset{\mathscr{P}}{\sum} \big\{ \mathscr{P}_1 \mathscr{A}_2 , \mathscr{P}_2  \mathscr{B}^{\prime}_3       \big\} +  \underset{\mathscr{P}}{\sum} \big\{     \mathscr{P}_1 \mathscr{A}_3     ,      \mathscr{P}_2 \big( \mathrm{sin}  {\tiny \big( 2 \eta \big)}  \big)^{n-3} \mathscr{B}^{\prime}_1       \big\} \end{align*}

  \begin{align*} +  \underset{\mathscr{P}}{\sum} \big\{    \mathscr{P}_1 \mathscr{A}_3          ,    \mathscr{P}_2     \mathscr{B}^{\prime}_2    \big\}  +  \underset{\mathscr{P}}{\sum} \big\{       \mathscr{P}_1      \mathscr{A}_3       ,     \mathscr{P}_2            \mathscr{B}^{\prime}_3         \big\}         \text{, } 
\end{align*}

\noindent for (2) and similarly for $(3), \cdots, (16)$.

\subsection{Conclusion}

\noindent In this paper we presented an application of the Quantum inverse scattering method for concluding that a conjecture regarding integrability of a Hamiltonian flow formulated by Keating, Reshetikhin and Sridhar, [8], holds. Separate from previous adaptations of the Quantum inverse scattering method in this work we demonstrated how L-operators previously obtained for the 6-vertex model under domain-wall boundary conditions, [3], can be leveraged to conclude that integrability holds in the presence of the Hamiltonian flow identified by the three authors of the aforementioned work, which argued that infinitely many conserved quantities from integrals of motion of the Euler Lagrange equations hold. While seminal work on the 6-vertex model due to Baxter, [1], has long established that integrability of the 6-vertex model holds, demonstrating that this integrability property continues to hold under the commutativity of the transfer matrices for different spectral parameters,

{\small

\begin{align*}
\bigg\{ \textbf{T} {\tiny \big( u \big)} \textbf{T} {\tiny \big( u^{\prime} \big)} =  \textbf{T} {\tiny \big( u^{\prime} \big)} \textbf{T} {\tiny \big( u \big)} \bigg\} \Longleftrightarrow \bigg\{  {\tiny \big[} \textbf{T} {\tiny\big( u ,H , V , V )} , \textbf{T} {\tiny \big( u^{\prime} ,H , V , V ) \big] } = 0  \bigg\}  ,
\end{align*} 

}

\noindent under the Hamiltonian flow on limit shapes formulated by Keating, Reshetikhin and Sridhar. To further build upon seminal work from Baxter, [1], in addition to that of several other authors regarding exact solvability of vertex models from Statistical Physics, [2,3,5,6,9,10,11,12,17,18], we made use of the Arnold-Liouville Theorem which asserts that suitable action-angle coordinates exist. Related to integrability of the 6-vertex model from the above commutativity property of the transfer matrix in the Appendix we demonstrate, as a straightforward approach of the Quantum inverse scattering method, how the transfer and Quantum monodromy matrices can be constructed (the computations underlying several formulae demonstrating how product representation of L-operators of the 6-vertex model are provided in \textit{4.2}-\textit{4.6}). The construction of the transfer and Quantum monodromy matrices associated with domain-wall boundary conditions of the 6-vertex model is also intimately related to computations underlying the Bethe equations provided in \textit{4.7}-\textit{4.11}. 

\section{Appendix}

\subsection{Superposition for $\big\{ A {\tiny  \big( u \big) } , A {\tiny  \big( u^{\prime} \big) }  \big\}$}

\subsubsection{First Poisson bracket}

\begin{itemize}
    
\item[$\bullet$] \textit{First Poisson bracket}. One has,

{\small \begin{align*}
\mathcal{P}_1    \approx   \big( \mathcal{P}_1 \big)_1  -   \big( \mathcal{P}_1 \big)_2\text{,}   \end{align*}  }

\noindent where,

{\small \begin{align*}
\big( \mathcal{P}_1 \big)_1   \equiv    \big[ \big( \mathrm{sin} {\tiny  \big( 2 \eta \big) } \big)^{n-3} 
\mathscr{C}_1  \big]^2  \cdot     \frac{A_3 {\tiny \big( u \big) } B_3 {\tiny  \big( u^{\prime} \big) } }{ u - u^{\prime} }       \text{,} \end{align*}

   \begin{align*} \big( \mathcal{P}_1 \big)_2  \equiv   \big[ \big( \mathrm{sin} {\tiny \big( 2 \eta \big)} \big)^{n-3} 
\mathscr{C}_1  \big]^2  \cdot   \frac{A_3 {\tiny \big( u^{\prime} \big)}  B_3 {\tiny  \big( u \big) } }{u^{\prime} - u}      \text{. }
\end{align*} }

\end{itemize}

\subsubsection{Second Poisson bracket}

\begin{itemize}

\item[$\bullet$] \textit{Second Poisson bracket}. One has, 

{\small \begin{align*}
\mathcal{P}_2  \approx        - \big( \mathcal{P}_2 \big)_1  +  \big( \mathcal{P}_2 \big)_2\text{, }       
\end{align*} }

\noindent where,

{\small \begin{align*}
  \big( \mathcal{P}_2 \big)_1 \equiv \big[ \big( \mathrm{sin } {\tiny  \big( 2 \eta \big) }  \big)^{n-3} \mathscr{C}_1    \big]   \cdot    \frac{A_3 {\tiny \big( u \big) }  B_3 {\tiny \big( u^{\prime} \big) }  }{u - u^{\prime}}    \text{,} \end{align*}
  
  \begin{align*} \big( \mathcal{P}_2 \big)_2    \equiv  \big[ \big( \mathrm{sin } {\tiny\big( 2 \eta \big)  } \big)^{n-3} \mathscr{C}_1    \big]   \cdot         \frac{B_3 {\tiny \big( u \big)}  A_3 {\tiny \big( u^{\prime} \big)}  }{u^{\prime} - u }           \text{. }
\end{align*}  } 

\end{itemize}

\subsubsection{Third Poisson bracket}

\begin{itemize}

\item[$\bullet$] \textit{Third Poisson bracket}. One has,

    {\small \begin{align*}
 \mathcal{P}_3 \approx   \big( \mathcal{P}_3 \big)_1  +   \big( \mathcal{P}_3 \big)_2
 -  \big( \mathcal{P}_3 \big)_3 \text{,} 
\end{align*} }

\noindent where,

{\small \begin{align*}
  \big( \mathcal{P}_3 \big)_1  \equiv   \bigg[   {\underset{m,n^{\prime} : m + n^{\prime} = n -3}{\sum}}    \big( \mathrm{sin} {\tiny \big( 2 \eta \big) }  \big)^{n^{\prime}-1}   \bigg]  \bigg[  \frac{\partial  \big[ \underset{1 \leq i \leq m}{\prod} \big( \mathscr{C}_2 \big)_i  \big] }{\partial u} \frac{\partial B_3 {\tiny\big( u \big)  }  }{\partial u^{\prime}}        + \frac{\partial   \big[ \underset{1 \leq i \leq m}{\prod} \big( \mathscr{C}_2 \big)_i  \big]  }{\partial u} \frac{\partial A_3 {\tiny \big( u \big) }  }{\partial u^{\prime}} \bigg]   \text{,} \end{align*}
  
  \begin{align*} \big( \mathcal{P}_3 \big)_2 \equiv \bigg[    {\underset{m,n^{\prime} : m + n^{\prime} = n -3}{\sum}}   \big( \mathrm{sin} {\tiny\big( 2 \eta \big) } \big)^{n^{\prime}-1}      \bigg]\bigg[ \frac{\partial \big[      \underset{1 \leq i \leq m}{\prod} \big( \mathscr{C}_2 \big)_i     \big] }{\partial u^{\prime}}  \frac{\partial B_3 {\tiny\big( u \big) }}{\partial u}  + \frac{\partial \big[      \underset{1 \leq i \leq m}{\prod} \big( \mathscr{C}_2 \big)_i     \big]}{\partial u^{\prime}}    \frac{\partial A_3 {\tiny\big( u \big)} }{\partial u}   \bigg] \text{,} \end{align*}
  
  \begin{align*} \big( \mathcal{P}_3 \big)_3 \equiv   
  \bigg[   {\underset{1\leq j \leq  n^{\prime}}{\underset{1 \leq i \leq m}{\sum} }}   \big( \mathscr{C}_3 \big)_{i,j}             \bigg] \bigg[  \big( \mathrm{sin} {\tiny\big( 2 \eta \big)}  \big)^{n-3}    \frac{B_3 \big( u \big) A_3 {\tiny\big( u^{\prime} \big)} }{u - u^{\prime}}     +   \big( \mathrm{sin} {\tiny\big( 2 \eta \big)} \big)^{n-3}       
   \frac{A_3 {\tiny\big( u \big) }  B_3 {\tiny \big( u^{\prime} \big) } }{u - u^{\prime}}      \bigg] \text{. }
\end{align*} } 

 \end{itemize}

\subsubsection{Fourth Poisson bracket}

\begin{itemize}

\item[$\bullet$] \textit{Fourth Poisson bracket}. One has,

{\small \begin{align*}
 \mathcal{P}_4 \approx  - \big( \mathcal{P}_4 \big)_1 - \big( \mathcal{P}_4 \big)_2 + \big( \mathcal{P}_4 \big)_3 + \big( \mathcal{P}_4 \big)_4 + \big( \mathcal{P}_4 \big)_5 + \big( \mathcal{P}_4 \big)_6  \text{, }  
\end{align*} } 

\noindent where,

{\small \begin{align*}
    \big( \mathcal{P}_4 \big)_1  \equiv \big[ \big( \mathrm{sin} {\tiny \big( 2 \eta \big) }  \big)^{n-3}              \big( \mathscr{C}_1 \big)_i           \big]  \cdot   \frac{A_3 {\tiny \big( u \big)  } B_3 {\tiny\big( u^{\prime} \big) } }{u-u^{\prime}}  \text{, } \end{align*}
   
   \begin{align*} \big( \mathcal{P}_4 \big)_2  \equiv  \big[ \big( \mathrm{sin} {\tiny\big( 2 \eta \big)}  \big)^{n-3}              \big( \mathscr{C}_1 \big)_i           \big]    \cdot    \frac{B_3 {\tiny \big( u \big) } A_3 {\tiny \big( u^{\prime } \big)} }{u-u^{\prime}}        \text{, } \end{align*}
   
   \begin{align*}    \big( \mathcal{P}_4 \big)_3 \equiv     \mathscr{A}_2 A_3 {\tiny \big( u \big)}  \frac{B_3 {\tiny\big( u^{\prime} \big)}  }{\partial u^{\prime}}  \cdot  \underset{1 \leq i \leq n-3}{\sum}      \frac{\partial \big( \mathscr{C}_2 \big)_i}{\partial u}  \text{, } \end{align*}
   
   \begin{align*} \big( \mathcal{P}_4 \big)_4 \equiv     \mathscr{A}_2 B_3 {\tiny\big(u \big) } \frac{\partial B_3 {\tiny\big( u^{\prime} \big)}  }{\partial u} \cdot    \underset{1 \leq i \leq n-3}{\sum}  \frac{\partial \big( \mathscr{C}_2 \big)_i}{\partial u^{\prime}}    \text{, } \end{align*}
   
   \begin{align*}      \big( \mathcal{P}_4 \big)_5 \equiv    \mathscr{A}_2    B_3 {\tiny\big( u \big) }
 \frac{\partial   A_3 {\tiny\big( u^{\prime} \big) } }{\partial u}     \cdot   \underset{1 \leq i \leq n-3}{\sum}                \frac{\partial \big( \mathscr{C}_2 \big)_i}{\partial u^{\prime}}      \text{, } \end{align*}
   
   \begin{align*}    \big( \mathcal{P}_4 \big)_6      \equiv   \mathscr{A}_2    B_3 {\tiny \big( u \big) } 
 \frac{\partial   A_3 {\tiny \big( u^{\prime} \big) } }{\partial u}    \cdot   \underset{1 \leq i \leq n-3}{\sum}                \frac{\partial \big( \mathscr{C}_2 \big)_i}{\partial u^{\prime}}  \text{. }
\end{align*} }

\end{itemize}

\subsubsection{Fifth Poisson bracket}

\begin{itemize}

\item[$\bullet$] \textit{Fifth Poisson bracket}. One has,
  
{\small \begin{align*}
    \mathcal{P}_5  \approx   -   \big( \mathcal{P}_5 \big)_1      -    \big( \mathcal{P}_5 \big)_2          -  \big( \mathcal{P}_5 \big)_3       -   \big( \mathcal{P}_5 \big)_4            -  \big( \mathcal{P}_5 \big)_5  +   \big( \mathcal{P}_5 \big)_6          -  \big( \mathcal{P}_5 \big)_7       -   \big( \mathcal{P}_5 \big)_8  \text{, }             
\end{align*} } 

\noindent where,

{\small \begin{align*}
  \big( \mathcal{P}_5 \big)_1   \equiv    \frac{B_3 {\tiny \big(u^{\prime} \big)} A_3 {\tiny \big( u \big) }   B_3 {\tiny\big( u \big)  }  \mathscr{C}_2  }{u^{\prime} - u} \frac{\partial  B_3 {\tiny\big( u^{\prime} \big)}  }{\partial u^{\prime}}  \cdot        \underset{1 \leq i \leq n-3}{\sum}          \frac{\partial \big( \mathscr{C}_2 \big)_i}{\partial u} \cdot  \underset{1 \leq j  \leq n-3}{\prod} \big( \mathscr{C}_2 \big)_j         \text{, } \end{align*}
 
 \begin{align*}  
  \big( \mathcal{P}_5 \big)_2 \equiv \frac{B_3 {\tiny\big(u^{\prime} \big) } A_3 {\tiny\big( u \big) }  A_3 {\tiny\big( u \big) }\mathscr{C}_2  }{u^{\prime} - u}  \frac{\partial  B_3 {\tiny\big( u^{\prime} \big)} }{\partial u^{\prime}} 
\cdot      \underset{1 \leq k \leq n-3}{\sum}          \frac{\partial \big( \mathscr{C}_2 \big)_k}{\partial u}  \cdot \underset{1 \leq j  \leq n-3}{\prod} \big( \mathscr{C}_2 \big)_j    \text{, }  \end{align*}
 
 \begin{align*}    \big( \mathcal{P}_5 \big)_3 \equiv     
         \frac{B_3 {\tiny \big( u \big) } A_3 {\tiny\big( u^{\prime} \big) }  B_3 {\tiny\big( u \big) } \mathscr{C}_2 }{u - u^{\prime}}  \frac{\partial  B_3 {\tiny\big( u^{\prime} \big)}  }{\partial u^{\prime}}  \cdot     \underset{1 \leq k \leq n-3}{\sum}          \frac{\partial \big( \mathscr{C}_2 \big)_k}{\partial u} \cdot \underset{1 \leq j  \leq n-3}{\prod} \big( \mathscr{C}_2 \big)_j   \text{, }  \end{align*}
 
 \begin{align*}    \big( \mathcal{P}_5 \big)_4 \equiv  \frac{B_3 {\tiny\big( u \big) } A_3 {\tiny\big( u^{\prime} \big)}  A_3 {\tiny\big( u \big)  }  \mathscr{C}_2 }{u - u^{\prime}}  \frac{\partial  B_3 {\tiny\big( u^{\prime} \big) } }{\partial u^{\prime}}  \cdot        \underset{1 \leq k \leq n-3}{\sum}          \frac{\partial \big( \mathscr{C}_2 \big)_k}{\partial u} \cdot \underset{1 \leq j \leq n-3}{\prod} \big( \mathscr{C}_2 \big)_j      \text{, }  \end{align*}
 
 \begin{align*}    \big( \mathcal{P}_5 \big)_5   \equiv   B_3 {\tiny\big( u \big)   }  \frac{\partial  A_3 {\tiny\big( u^{\prime} \big) } }{\partial u^{\prime}}   \cdot      \underset{1 \leq k \leq n-3}{\sum}  \frac{\partial \big( \mathscr{C}_2 \big)_k}{\partial u} \cdot  \underset{1 \leq i \neq j \leq n-3}{\prod} \big( \mathscr{C}_2 \big)_i  \big( \mathscr{C}_2 \big)_j  
    \text{, } \end{align*}
 
 \begin{align*}   \big( \mathcal{P}_5 \big)_6 \equiv  B_3 {\tiny\big( u \big) } \frac{\partial A_3 {\tiny\big( u^{\prime} \big)}}{\partial u}  \cdot   \underset{1 \leq k \leq n-3}{\sum}   \frac{\partial \big( \mathscr{C}_2 \big)_k }{\partial u}   \cdot  \underset{1 \leq i \neq j \leq n-3}{\prod} \big( \mathscr{C}_2 \big)_i  \big( \mathscr{C}_2 \big)_j  
  \text{,}  \end{align*}
 
 \begin{align*}   \big( \mathcal{P}_5 \big)_7 \equiv  A_3 {\tiny \big( u \big) } \frac{\partial A_3 {\tiny\big( u^{\prime} \big)} }{\partial u}    \cdot   \underset{1 \leq k \leq n-3}{\sum} \frac{\partial \big( \mathscr{C}_2\big)_k}{\partial u}  \cdot \underset{1 \leq i \neq j \leq n-3}{\prod} \big( \mathscr{C}_2 \big)_i  \big( \mathscr{C}_2 \big)_j  
  \text{, } \end{align*}
 
 \begin{align*} \big( \mathcal{P}_5 \big)_8 \equiv A_3 {\tiny \big( u \big)  } \frac{\partial A_3 {\tiny\big( u^{\prime} \big)} }{\partial u^{\prime}}     \cdot \underset{1 \leq k \leq n-3}{\sum} \frac{\partial \big( \mathscr{C}_2\big)_k}{\partial u} \cdot \underset{1 \leq i \neq j \leq n-3}{\prod} \big( \mathscr{C}_2 \big)_i  \big( \mathscr{C}_2 \big)_j  
    \text{. }
\end{align*} } 
 
\end{itemize}

\subsubsection{Sixth Poisson bracket}

\begin{itemize}

\item[$\bullet$] \textit{Sixth Poisson bracket}. One has,

{\small \begin{align*}
  \mathcal{P}_6 \approx 
\big( \mathcal{P}_6 \big)_1  +   \big( \mathcal{P}_6 \big)_2  -     \big( \mathcal{P}_6 \big)_3            
     -  \big( \mathcal{P}_6 \big)_4   - \big( \mathcal{P}_6 \big)_5     -    \big( \mathcal{P}_6 \big)_6     -   \big( \mathcal{P}_6 \big)_7   -   \big( \mathcal{P}_6 \big)_8 -  \big( \mathcal{P}_6 \big)_9 
          -     \big( \mathcal{P}_6 \big)_{10}          \text{, } 
\end{align*} }

\noindent where,

{\small \begin{align*}    \big( \mathcal{P}_6 \big)_1  \equiv       \frac{  A_3 \big( u \big) B_3 \big(u^{\prime} \big)  
 }{u - u^{\prime}}     \cdot                       {\underset{m,n^{\prime} : m + n^{\prime} = n -3}{\sum}}   \big( \mathrm{sin} \big( 2 \eta \big) \big)^{n^{\prime}-1} \cdot   \underset{1 \leq j \leq n^{\prime}}{\underset{1 \leq i \leq m}{\prod}} \big( \mathscr{C}_2 \big)_i \big( \mathscr{C}_1 \big)_j                        \text{, }      \end{align*}

   \begin{align*}    \big( \mathcal{P}_6 \big)_2  \equiv       \frac{    B_3 \big( u \big) A_3 \big( u^{\prime} \big) }{u - u^{\prime}}      \cdot                    {\underset{m,n^{\prime} : m + n^{\prime} = n -3}{\sum}}  \big( \mathrm{sin} \big( 2 \eta \big) \big)^{n^{\prime}-1}  \cdot  \underset{1 \leq j \leq n^{\prime}}{\underset{1 \leq i \leq m}{\prod}} \big( \mathscr{C}_2 \big)_i \big( \mathscr{C}_1 \big)_j                 \text{, }      \end{align*}

   \begin{align*}   \big( \mathcal{P}_6 \big)_3  \equiv         2 B_3 \big( u \big) \frac{\partial A_3 \big( u \big)}{\partial u }     \big( \mathrm{sin} \big( 2 \eta \big) \big)^{n^{\prime}-1}   \cdot     \underset{1 \leq k \leq m}{\sum}   \frac{\partial \big( \mathscr{C}_2 \big)_k }{\partial u^{\prime}}   \cdot   \underset{1 \leq i \neq j \leq m}{\underset{1 \leq j \leq n^{\prime}}{ \prod}}  \big( \mathscr{C}_1 \big)_j  \big( \mathscr{C}_2 \big)_j        \text{, }      \end{align*}

   \begin{align*}  \big( \mathcal{P}_6 \big)_4   \equiv          2 B_3 \big( u \big) \frac{\partial B_3 \big( u \big)}{\partial u}   \big( \mathrm{sin} \big( 2 \eta \big) \big)^{n^{\prime}-1}   \cdot     \underset{1 \leq k \leq m}{\sum}            \frac{\partial \big( \mathscr{C}_2 \big)_k }{\partial u^{\prime}}  \cdot   \underset{1 \leq i \neq j \leq m}{\underset{1 \leq j \leq n^{\prime}}{ \prod}}  \big( \mathscr{C}_1 \big)_j  \big( \mathscr{C}_2 \big)_j        \text{, }      \end{align*}

   \begin{align*}  \big( \mathcal{P}_6 \big)_5  \equiv      2 A_3 \big( u \big) \frac{\partial A_3 \big( u \big)}{\partial u }  \big( \mathrm{sin} \big( 2 \eta \big) \big)^{n^{\prime}-1}   \cdot     \underset{1 \leq k \leq m}{\sum}          \frac{\partial \big( \mathscr{C}_2 \big)_k }{\partial u^{\prime}}   \cdot    \underset{1 \leq i \neq j \leq m}{\underset{1 \leq j \leq n^{\prime}}{ \prod}}  \big( \mathscr{C}_1 \big)_j  \big( \mathscr{C}_2 \big)_j       \text{, } \end{align*}

   \begin{align*}   \big( \mathcal{P}_6 \big)_6 \equiv   2 A_3 \big( u \big) \frac{\partial B_3 \big( u \big)}{\partial u}   \big( \mathrm{sin} \big( 2 \eta \big) \big)^{n^{\prime}-1}      \cdot    {\underset{1 \leq k \leq m}{\sum} }  \frac{\partial \big( \mathscr{C}_2 \big)_k }{\partial u}   \cdot   \underset{1 \leq i \neq j \leq m}{\underset{1 \leq j \leq n^{\prime}}{ \prod}}  \big( \mathscr{C}_1 \big)_j  \big( \mathscr{C}_2 \big)_j          \text{, }  \end{align*}

   \begin{align*}  \big( \mathcal{P}_6 \big)_7 \equiv     2 B_3 \big( u \big) \frac{\partial A_3 \big( u^{\prime} \big) }{\partial u^{\prime}}   \big( \mathrm{sin} \big( 2 \eta \big) \big)^{n^{\prime}-1} \cdot    {\underset{1 \leq k \leq m}{\sum} }  \frac{\partial \big( \mathscr{C}_2 \big)_k }{\partial u}   \cdot \underset{1 \leq i \neq j \leq m}{\underset{1 \leq j \leq n^{\prime}}{ \prod}}  \big( \mathscr{C}_1 \big)_j  \big( \mathscr{C}_2 \big)_j      \text{, } \end{align*}

   \begin{align*}  \big( \mathcal{P}_6 \big)_8 \equiv     2 B_3 \big( u \big) \frac{\partial B_3 \big( u^{\prime} \big) }{\partial u^{\prime}}   \big( \mathrm{sin} \big( 2 \eta \big) \big)^{n^{\prime}-1}   \cdot     {\underset{1 \leq k \leq m}{\sum} }  \frac{\partial \big( \mathscr{C}_2 \big)_k }{\partial u}   \cdot   \underset{1 \leq i \neq j \leq m}{\underset{1 \leq j \leq n^{\prime}}{ \prod}}  \big( \mathscr{C}_1 \big)_j  \big( \mathscr{C}_2 \big)_j               \text{, } \end{align*}

 \begin{align*}   
   \big( \mathcal{P}_6 \big)_9 \equiv        2 A_3 \big( u \big) \frac{\partial A_3 \big( u^{\prime} \big)}{\partial u^{\prime}}
 \big( \mathrm{sin} \big( 2 \eta \big) \big)^{n^{\prime}-1}      \cdot      {\underset{1 \leq k \leq m}{\sum} }  \frac{\partial \big( \mathscr{C}_2 \big)_k }{\partial u}  \cdot   \underset{1 \leq i \neq j \leq m}{\underset{1 \leq j \leq n^{\prime}}{ \prod}}  \big( \mathscr{C}_1 \big)_j  \big( \mathscr{C}_2 \big)_j                  \text{, } \end{align*}

 \begin{align*} \big( \mathcal{P}_6 \big)_{10} \equiv       2 A_3 \big( u \big) \frac{\partial B_3 \big( u^{\prime} \big) }{\partial u^{\prime}}      \big( \mathrm{sin} \big( 2 \eta \big) \big)^{n^{\prime}-1}    \cdot     {\underset{1 \leq k \leq m}{\sum} }  \frac{\partial \big( \mathscr{C}_2 \big)_k }{\partial u}  \cdot   \underset{1 \leq i \neq j \leq m}{\underset{1 \leq j \leq n^{\prime}}{ \prod}}  \big( \mathscr{C}_1 \big)_j  \big( \mathscr{C}_2 \big)_j     \text{. }  \end{align*} } 
 
\end{itemize}

\subsubsection{Seventh Poisson bracket}

\begin{itemize}

\item[$\bullet$] \textit{Seventh Poisson bracket}. One has,

{\small \begin{align*}
\mathcal{P}_7  \approx  \big( \mathcal{P}_7 \big)_1    +   \big( \mathcal{P}_7 \big)_2  +    \big( \mathcal{P}_7 \big)_3            -     \big( \mathcal{P}_7 \big)_4          +      \big( \mathcal{P}_7 \big)_5     -      \big( \mathcal{P}_7 \big)_6 
  -      \big( \mathcal{P}_7 \big)_7       +  \big( \mathcal{P}_7 \big)_8      +     \big( \mathcal{P}_7 \big)_9   +  \big( \mathcal{P}_7 \big)_{10}  \\   +   \big( \mathcal{P}_7 \big)_{11}   +   \big( \mathcal{P}_7 \big)_{12}        -      \big( \mathcal{P}_7 \big)_{13}              -   \big( \mathcal{P}_7 \big)_{14}     -    \big( \mathcal{P}_7 \big)_{15}     +        \big( \mathcal{P}_7 \big)_{16} \text{, } 
\end{align*} }

\noindent where,

{\small \begin{align*}
  \big( \mathcal{P}_7 \big)_{1}   \equiv    \frac{A_3 \big( u \big) B_3 \big( u^{\prime} \big)}{u - u^{\prime}}   \big( \mathrm{sin} \big( 2 \eta \big) \big)^{n-3}   \mathscr{C}_1 
           \text{,}    \end{align*}
 
 \begin{align*} \big( \mathcal{P}_7 \big)_{2}  \equiv  \frac{B_3 \big( u \big) A_3 \big( u^{\prime} \big)}{ u - u^{\prime} }   \big( \mathrm{sin} \big( 2 \eta \big) \big)^{n-3}   \mathscr{C}_1                    \text{,}    \end{align*}
 
 \begin{align*}   \big( \mathcal{P}_7 \big)_{3} \equiv          B_3 \big( u \big) \frac{\partial B_3 \big( u^{\prime} \big) }{\partial u}      \mathscr{C}_1  \cdot {\underset{m,n^{\prime} : m + n^{\prime} = n -3}{\sum}}  \big( \mathrm{sin} \big( 2 \eta \big) \big)^{n^{\prime}-1}   \cdot  \underset{1 \leq i \leq m}{\sum}    \frac{\partial}{\partial u}       \underset{1 \leq i \leq m}{\prod}       \big( \mathscr{C}_2 \big)_i                      \text{,}  \end{align*}
 
 \begin{align*}   \big( \mathcal{P}_7 \big)_{4}  \equiv        B_3 \big( u \big) \frac{\partial B_3 \big( u^{\prime} \big) }{\partial u^{\prime}}       \mathscr{C}_1    \cdot {\underset{m,n^{\prime} : m + n^{\prime} = n -3}{\sum}}  \big( \mathrm{sin} \big( 2 \eta \big) \big)^{n^{\prime}-1}   \cdot \underset{1 \leq i \leq m}{\sum}    \frac{\partial}{\partial u}       \underset{1 \leq i \leq m}{\prod}      \big( \mathscr{C}_2 \big)_i               \text{, }   \end{align*}
 
 \begin{align*}  \big( \mathcal{P}_7 \big)_{5}  \equiv         \frac{\partial B_3 \big( u^{\prime} \big)}{\partial u}       \mathscr{C}_1   \cdot {\underset{m,n^{\prime} : m + n^{\prime} = n -3}{\sum}}  \big( \mathrm{sin} \big( 2 \eta \big) \big)^{n^{\prime}-1}   \cdot \underset{1 \leq i \leq m}{\sum}          \underset{1 \leq i \leq m}{\prod}  \frac{\partial}{\partial u}      \big( \mathscr{C}_2 \big)_i             \text{,}   \end{align*}
 
 \begin{align*}  \big( \mathcal{P}_6 \big)_{5}   \equiv                       \frac{\partial B_3 \big( u^{\prime} \big)}{\partial u}    \mathscr{C}_1 \cdot   {\underset{m,n^{\prime} : m + n^{\prime} = n -3}{\sum}}  \big( \mathrm{sin} \big( 2 \eta \big) \big)^{n^{\prime}-1}  \cdot \underset{1 \leq i \leq m}{\sum}           \underset{1 \leq i \leq m}{\prod}   \frac{\partial}{\partial u}      \big( \mathscr{C}_2 \big)_i                \text{,}   \end{align*}
 
 \begin{align*}   \big( \mathcal{P}_7 \big)_{6}  \equiv               \frac{B_3 \big( u^{\prime} \big)}{\partial u}       \mathscr{C}_1 \cdot   {\underset{m,n^{\prime} : m + n^{\prime} = n -3}{\sum}}  \big( \mathrm{sin} \big( 2 \eta \big) \big)^{n^{\prime}-1}  \cdot  \underset{1 \leq i \leq m}{\sum}          \underset{1 \leq i \leq m}{\prod} \frac{\partial}{\partial u}      \big( \mathscr{C}_2 \big)_i         \text{,}    \end{align*}
 
 \begin{align*}  \big( \mathcal{P}_7 \big)_{7}  \equiv   B_3 \big( u \big) \frac{\partial A_3 \big( u^{\prime} \big)}{\partial u}    \mathscr{C}_1 \cdot  {\underset{m,n^{\prime} : m + n^{\prime} = n -3}{\sum}}  \big( \mathrm{sin} \big( 2 \eta \big) \big)^{n^{\prime}-1} \cdot   \underset{1 \leq i \leq m}{\sum} \bigg[  \frac{\partial}{\partial u} \underset{1 \leq i \leq m}{\prod}   \big( \mathscr{C}_2 \big)_i   \bigg]       \text{,}    \end{align*}
 
 \begin{align*}  \big( \mathcal{P}_7 \big)_{8}   \equiv          B_3 \big( u \big) \frac{\partial A_3 \big( u^{\prime} \big)}{\partial u}    \mathscr{C}_1  \cdot   {\underset{m,n^{\prime} : m + n^{\prime} = n -3}{\sum}}  \big( \mathrm{sin} \big( 2 \eta \big) \big)^{n^{\prime}-1}  \cdot   \underset{1 \leq i \leq m}{\sum} \bigg[     \frac{\partial}{\partial u} \underset{1 \leq i \leq m}{\prod}   \big( \mathscr{C}_2 \big)_i   \bigg]                    \text{,}    \end{align*}
 
 \begin{align*}      \big( \mathcal{P}_7 \big)_{9}        \equiv              B_3 \big( u \big) \frac{\partial A_3 \big( u^{\prime} \big)}{\partial u}   \mathscr{C}_1 \cdot   {\underset{m,n^{\prime} : m + n^{\prime} = n -3}{\sum}}  \big( \mathrm{sin} \big( 2 \eta \big) \big)^{n^{\prime}-1}  \cdot      \underset{1 \leq i \leq m}{\sum} \bigg[ \frac{\partial}{\partial u} \underset{1 \leq i \leq m}{\prod}   \big( \mathscr{C}_2 \big)_i   \bigg]   \text{,}    \end{align*}
 
 \begin{align*}   \big( \mathcal{P}_7 \big)_{10} \equiv               B_3 \big( u \big) \frac{\partial A_3 \big( u^{\prime} \big)}{\partial u^{\prime}}   \mathscr{C}_1  \cdot  {\underset{m,n^{\prime} : m + n^{\prime} = n -3}{\sum}}  \big( \mathrm{sin} \big( 2 \eta \big) \big)^{n^{\prime}-1}  \cdot   \underset{1 \leq i \leq m}{\sum} \bigg[  \frac{\partial}{\partial u} \underset{1 \leq i \leq m}{\prod}   \big( \mathscr{C}_2 \big)_i   \bigg]        
          \text{,}  \end{align*}
 
 \begin{align*}   \big( \mathcal{P}_7 \big)_{11}  \equiv     A_3 \big( u \big) \frac{\partial A_3 \big( u^{\prime} \big)}{\partial u^{\prime}} \mathscr{C}_1  \cdot  {\underset{m,n^{\prime} : m + n^{\prime} = n -3}{\sum}}  \big( \mathrm{sin} \big( 2 \eta \big) \big)^{n^{\prime}-1}  \cdot \underset{1 \leq i \leq m}{\sum}          \bigg[    \frac{\partial}{\partial u }   \underset{1 \leq i \leq m}{\prod}  \big( \mathscr{C}_2 \big)_i     \bigg]     \text{,}  \end{align*}
 
 \begin{align*}   \big( \mathcal{P}_7 \big)_{12} \equiv   A_3 \big( u \big) \frac{\partial A_3 \big( u^{\prime} \big)}{\partial u}   \mathscr{C}_1 \cdot   {\underset{m,n^{\prime} : m + n^{\prime} = n -3}{\sum}}  \big( \mathrm{sin} \big( 2 \eta \big) \big)^{n^{\prime}-1}  \cdot   \underset{1 \leq i \leq m}{\sum} \bigg[  \frac{\partial}{\partial u} \underset{1 \leq i \leq m}{\prod}   \big( \mathscr{C}_2 \big)_i   \bigg]                    \text{,} \end{align*}
 
 \begin{align*}    \big( \mathcal{P}_7 \big)_{13}   \equiv   \frac{\partial A_3 \big( u^{\prime}\big) }{\partial u}                               \mathscr{C}_1  \cdot  {\underset{m,n^{\prime} : m + n^{\prime} = n -3}{\sum}}  \big( \mathrm{sin} \big( 2 \eta \big) \big)^{n^{\prime}-1} \cdot   \underset{1 \leq i \leq m}{\sum} \bigg[  \frac{\partial}{\partial u} \underset{1 \leq i \leq m}{\prod}   \big( \mathscr{C}_2 \big)_i   \bigg]       \text{,}  \end{align*}
 
 \begin{align*}  
         \big( \mathcal{P}_7 \big)_{14}  \equiv     A_3 \big( u \big) \frac{\partial B_3 \big( u^{\prime}\big)}{\partial u^{\prime}}   \mathscr{C}_1 \cdot {\underset{m,n^{\prime} : m + n^{\prime} = n -3}{\sum}}  \big( \mathrm{sin} \big( 2 \eta \big) \big)^{n^{\prime}-1}  \cdot    \underset{1 \leq i \leq m}{\sum} \bigg[  \frac{\partial}{\partial u} \underset{1 \leq i \leq m}{\prod}   \big( \mathscr{C}_2 \big)_i   \bigg] \text{,} \end{align*}
 
 \begin{align*}   \big( \mathcal{P}_7 \big)_{15}  \equiv     A_3 \big( u \big)          \frac{\partial B_3 \big( u^{\prime}\big)}{\partial u^{\prime}}           \mathscr{C}_1  \cdot  {\underset{m,n^{\prime} : m + n^{\prime} = n -3}{\sum}}  \big( \mathrm{sin} \big( 2 \eta \big) \big)^{n^{\prime}-1} \cdot    \underset{1 \leq i \leq m}{\sum} \bigg[  \frac{\partial}{\partial u} \underset{1 \leq i \leq m}{\prod}   \big( \mathscr{C}_2 \big)_i   \bigg]   \text{,} \end{align*}
 
 \begin{align*}  \big( \mathcal{P}_7 \big)_{16}    \equiv    \frac{\partial B_3 \big(u^{\prime} \big)}{\partial u}             \mathscr{C}_1   \cdot  {\underset{m,n^{\prime} : m + n^{\prime} = n -3}{\sum}}  \big( \mathrm{sin} \big( 2 \eta \big) \big)^{n^{\prime}-1} \cdot    \underset{1 \leq i \leq m}{\sum} \bigg[  \frac{\partial}{\partial u} \underset{1 \leq i \leq m}{\prod}   \big( \mathscr{C}_2 \big)_i   \bigg]    \text{. }
\end{align*}   }

\end{itemize}

\subsubsection{Eighth Poisson bracket}

\begin{itemize}

\item[$\bullet$] \textit{Eighth Poisson bracket}. One has,

{\small \begin{align*}
   \mathcal{P}_8  \approx     -   \big( \mathcal{P}_8 \big)_1      -    \big( \mathcal{P}_8 \big)_2     -    \big( \mathcal{P}_8 \big)_3          -    \big( \mathcal{P}_8 \big)_4                -    \big( \mathcal{P}_8 \big)_5         -       \big( \mathcal{P}_8 \big)_6           -    \big( \mathcal{P}_8 \big)_7        
   -      \big( \mathcal{P}_8 \big)_8      +  \big( \mathcal{P}_8 \big)_9            +   \big( \mathcal{P}_8 \big)_{10}            +       \big( \mathcal{P}_8 \big)_{11} \\  -        \big( \mathcal{P}_8 \big)_{12}     -    \big( \mathcal{P}_8 \big)_{13}         \text{,}     \end{align*} }

\noindent where, 

{\small \begin{align*}
          \big( \mathcal{P}_8 \big)_1    \equiv    \frac{A_3 \big( u \big) B_3 \big( u^{\prime} \big)   \mathscr{C}_1   }{ u - u^{\prime} }   \cdot  {\underset{m,n^{\prime} : m + n^{\prime} = n -3}{\sum}} \big( \mathrm{sin} \big( 2 \eta \big) \big)^{n^{\prime}-1}  \cdot   \underset{1 \leq j \leq n-3}{\underset{1 \leq i \leq m}{\prod}} \big( \mathscr{C}_2 \big)_i \big( \mathscr{C}_2 \big)_j  \text{, } \end{align*}
  
  \begin{align*}   \big( \mathcal{P}_8 \big)_2     \equiv  \frac{B_3 \big( u \big) A_3 \big( u^{\prime} \big) 
 \mathscr{C}_1 }{u-u^{\prime}}  \cdot   {\underset{m,n^{\prime} : m + n^{\prime} = n -3}{\sum}} \big( \mathrm{sin} \big( 2 \eta \big) \big)^{n^{\prime}-1}   \cdot   \underset{1 \leq j \leq n-3}{\underset{1 \leq i \leq m}{\prod}} \big( \mathscr{C}_2 \big)_i \big( \mathscr{C}_2 \big)_j   \text{, }  \end{align*}
  
  \begin{align*}   \big( \mathcal{P}_8 \big)_3    \equiv      2      B_3 \big( u^{\prime} \big)    \frac{\partial B_3 \big( u \big)}{\partial u}   \mathscr{C}_1 \cdot {\underset{m,n^{\prime} : m + n^{\prime} = n -3}{\sum}} \big( \mathrm{sin} \big( 2 \eta \big) \big)^{n^{\prime}-1} \cdot      \underset{1 \leq k \leq n-3}{\sum} \frac{\partial \big( \mathscr{C}_2 \big)_k}{\partial u^{\prime}}  \cdot   \underset{1 \leq j \leq n-3}{\underset{1 \leq i \leq m}{\prod}} \big( \mathscr{C}_2 \big)_i \big( \mathscr{C}_2 \big)_j      \text{, } \end{align*}
  
  \begin{align*}   \big( \mathcal{P}_8 \big)_4     \equiv 2    B_3 \big( u^{\prime} \big)      \frac{\partial B_3 \big( u \big)}{\partial u^{\prime}}    \mathscr{C}_1 \cdot {\underset{m,n^{\prime} : m + n^{\prime} = n -3}{\sum}} \big( \mathrm{sin} \big( 2 \eta \big) \big)^{n^{\prime}-1} \cdot    \underset{1 \leq k \leq n-3}{\sum} \frac{\partial \big( \mathscr{C}_2 \big)_k}{\partial u^{\prime}} \cdot    \underset{1 \leq j \leq n-3}{\underset{1 \leq i \leq m}{\prod}} \big( \mathscr{C}_2 \big)_i \big( \mathscr{C}_2 \big)_j    \text{, } \end{align*}
  
  \begin{align*}     \big( \mathcal{P}_8 \big)_5         \equiv                 A_3 \big( u^{\prime} \big)  \frac{\partial A_3 \big( u \big) }{\partial u} \mathscr{C}_1    \cdot {\underset{m,n^{\prime} : m + n^{\prime} = n -3}{\sum}} \big( \mathrm{sin} \big( 2 \eta \big) \big)^{n^{\prime}-1} \cdot  \underset{1 \leq k \leq n-3}{\sum} \frac{\partial \big( \mathscr{C}_2 \big)_k}{\partial u^{\prime}}  \cdot  \underset{1 \leq j \leq n-3}{\underset{1 \leq i \leq m}{\prod}}  \big( \mathscr{C}_2 \big)_i \big( \mathscr{C}_2 \big)_j                 \text{, }  \end{align*}
  
  \begin{align*}   \big( \mathcal{P}_8 \big)_6     \equiv        A_3 \big( u^{\prime} \big) \frac{\partial A_3 \big( u \big) }{\partial u^{\prime}} \mathscr{C}_1 \cdot {\underset{m,n^{\prime} : m + n^{\prime} = n -3}{\sum}} \big( \mathrm{sin} \big( 2 \eta \big) \big)^{n^{\prime}-1} \cdot  \underset{1 \leq k \leq n-3}{\sum}             \frac{\partial \big( \mathscr{C}_2 \big)_k}{\partial u} \cdot  \underset{1 \leq j \leq n-3}{\underset{1 \leq i \leq m}{\prod}}  \big( \mathscr{C}_2 \big)_i \big( \mathscr{C}_2 \big)_j          \text{, }  \end{align*}
  
  \begin{align*}   \big( \mathcal{P}_8 \big)_7     \equiv  A_3 \big( u^{\prime} \big) \frac{\partial B_3 \big( u^{\prime} \big)}{\partial u^{\prime}} \mathscr{C}_1  \cdot {\underset{m,n^{\prime} : m + n^{\prime} = n -3}{\sum}} \big( \mathrm{sin} \big( 2 \eta \big) \big)^{n^{\prime}-1}  \cdot  \underset{1 \leq k \leq n-3}{\sum}  \frac{\big( \mathscr{C}_2 \big)_k}{\partial u^{\prime}} \cdot  \underset{1 \leq j \leq n-3}{\underset{1 \leq i \leq m}{\prod}}  \big( \mathscr{C}_2 \big)_i \big( \mathscr{C}_2 \big)_j   \text{, }  \end{align*}
  
  \begin{align*}   \big( \mathcal{P}_8 \big)_8     \equiv  A_3 \big( u^{\prime} \big) \frac{\partial B_3 \big( u^{\prime} \big)}{\partial u^{\prime}} \mathscr{C}_1 \cdot 
 {\underset{m,n^{\prime} : m + n^{\prime} = n -3}{\sum}} \big( \mathrm{sin} \big( 2 \eta \big) \big)^{n^{\prime}-1} \cdot      \underset{1 \leq k \leq n-3}{\sum}  \frac{\big( \mathscr{C}_2 \big)_k}{\partial u^{\prime}}   \cdot  \underset{1 \leq j \leq n-3}{\underset{1 \leq i \leq m}{\prod}}  \big( \mathscr{C}_2 \big)_i \big( \mathscr{C}_2 \big)_j   \text{, }  \end{align*}
  
  \begin{align*}   \big( \mathcal{P}_8 \big)_9     \equiv    A_3 \big( u^{\prime} \big) \frac{\partial B_3 \big( u^{\prime} \big)}{\partial u}  \mathscr{C}_1  \cdot {\underset{m,n^{\prime} : m + n^{\prime} = n -3}{\sum}} \big( \mathrm{sin} \big( 2 \eta \big) \big)^{n^{\prime}-1} \cdot      \underset{1 \leq k \leq n-3}{\sum} \frac{\partial \big( \mathscr{C}_2 \big)_k}{\partial u^{\prime}}  \cdot  \underset{1 \leq j \leq n-3}{\underset{1 \leq i \leq m}{\prod}}  \big( \mathscr{C}_2 \big)_i \big( \mathscr{C}_2 \big)_j     \text{, }   \end{align*}
  
  \begin{align*}    \big( \mathcal{P}_8 \big)_{10}     \equiv    \frac{\partial B_3 \big( u^{\prime} \big)}{\partial u}  \cdot {\underset{m,n^{\prime} : m + n^{\prime} = n -3}{\sum}} \big( \mathrm{sin} \big( 2 \eta \big) \big)^{n^{\prime}-1} \cdot  \underset{1 \leq k \leq n-3}{\sum} \frac{\partial \big( \mathscr{C}_2 \big)_k}{\partial u^{\prime}} \cdot  \underset{1 \leq j \leq n-3}{\underset{1 \leq i \leq m}{\prod}}  \big( \mathscr{C}_2 \big)_i \big( \mathscr{C}_2 \big)_j                  \text{, } \end{align*}
  
  \begin{align*}  \big( \mathcal{P}_8 \big)_{11}  \equiv                \frac{\partial A_3 \big( u^{\prime} \big)}{\partial u}  \cdot   {\underset{m,n^{\prime} : m + n^{\prime} = n -3}{\sum}} \big( \mathrm{sin} \big( 2 \eta \big) \big)^{n^{\prime}-1} \cdot  \underset{1 \leq k \leq n-3}{\sum} \frac{\partial \big( \mathscr{C}_2 \big)_k}{\partial u^{\prime}} \cdot \underset{1 \leq j \leq n-3}{\underset{1 \leq i \leq m}{\prod}}  \big( \mathscr{C}_2 \big)_i \big( \mathscr{C}_2 \big)_j                                                         \text{, }  \end{align*}
  
  \begin{align*}  \big( \mathcal{P}_8 \big)_{12}  \equiv    \frac{\partial A_3 \big( u^{\prime} \big)}{\partial u^{\prime}} \cdot  {\underset{m,n^{\prime} : m + n^{\prime} = n -3}{\sum}}  \big( \mathrm{sin} \big( 2 \eta \big) \big)^{n^{\prime}-1}  \cdot  \underset{1 \leq k \leq n-3}{\sum} \frac{\partial \big( \mathscr{C}_2 \big)_k}{\partial u^{\prime}} \cdot   \underset{1 \leq j \leq n-3}{\underset{1 \leq i \leq m}{\prod}}  \big( \mathscr{C}_2 \big)_i \big( \mathscr{C}_2 \big)_j     \text{, } \end{align*}
  
  \begin{align*} \big( \mathcal{P}_8 \big)_{13}  \equiv \frac{\partial B_3 \big( u^{\prime} \big)}{\partial u^{\prime}} \cdot   {\underset{m,n^{\prime} : m + n^{\prime} = n -3}{\sum}}  \big( \mathrm{sin} \big( 2 \eta \big) \big)^{n^{\prime}-1}  \cdot  \underset{1 \leq k \leq n-3}{\sum} \frac{\partial \big( \mathscr{C}_2 \big)_k}{\partial u^{\prime}} \cdot  \underset{1 \leq j \leq n-3}{\underset{1 \leq i \leq m}{\prod}}  \big( \mathscr{C}_2 \big)_i \big( \mathscr{C}_2 \big)_j    \text{. }
\end{align*} }

\end{itemize}

\subsubsection{Ninth Poisson bracket}

\begin{itemize}

\item[$\bullet$] \textit{Ninth Poisson bracket}. One has,

{\small \begin{align*}            \mathcal{P}_9   \approx     - \big( \mathcal{P}_9 \big)_1    - \big( \mathcal{P}_9 \big)_2 
   -     \big( \mathcal{P}_9 \big)_3     -   \big( \mathcal{P}_9 \big)_4   
  +  \big( \mathcal{P}_9 \big)_5        + \big( \mathcal{P}_9 \big)_6  -   \big( \mathcal{P}_9 \big)_7    - \big( \mathcal{P}_9 \big)_8     -  \big( \mathcal{P}_9 \big)_{10}    \text{, }
\end{align*} }

\noindent where, 

{\small \begin{align*}
   \big( \mathcal{P}_9 \big)_1   \equiv \frac{  \mathscr{C}_1 \mathscr{C}_2  B_3 \big( u^{\prime} \big) A_3 \big( u \big)}{u^{\prime} - u }   \cdot  {\underset{n, m^{\prime}: m + n^{\prime} = n-3}{\sum} } \big( \mathrm{sin} \big( 2 \eta \big) \big)^{n^{\prime}-1}                \text{, }  \end{align*}
  
  \begin{align*} \big( \mathcal{P}_9 \big)_2  \equiv  \frac{  \mathscr{C}_1 \mathscr{C}_2  A_3 \big(u^{\prime} \big) B_3 \big( u \big)}{u^{\prime}- u }  \cdot  {\underset{n, m^{\prime}: m + n^{\prime} = n-3}{\sum} } \big( \mathrm{sin} \big( 2 \eta \big) \big)^{n^{\prime}-1}                  \text{, }  \end{align*}
  
  \begin{align*}  \big( \mathcal{P}_9 \big)_3  \equiv    \frac{      2  \mathscr{C}^2_2   B_3 \big( u \big) 
 B_3 \big( u^{\prime} \big) A_3 \big( u \big)}{u^{\prime} - u }   \cdot 
 {\underset{n, m^{\prime}: m + n^{\prime} = n-3}{\sum} }    \big( \mathrm{sin} \big( 2 \eta \big) \big)^{2n^{\prime}-2}        \text{, } \end{align*}
  
  \begin{align*}  \big( \mathcal{P}_9 \big)_4  \equiv              \frac{  2 \mathscr{C}^2_2  A_3 \big( u \big)   A_3 \big( u^{\prime} \big) B_3 \big( u \big) }{u^{\prime} - u }    \cdot 
{\underset{n, m^{\prime}: m + n^{\prime} = n-3}{\sum} }     \big( \mathrm{sin} \big( 2 \eta \big) \big)^{2n^{\prime}-2}          \text{, }  \end{align*}
  
  \begin{align*}  \big( \mathcal{P}_9 \big)_5  \equiv      2 \frac{\partial B_3 \big( u^{\prime} \big)}{\partial u^{\prime}}    \bigg[      \underset{1 \leq j \leq m}{\sum}      \bigg[ \frac{\partial}{\partial u}                 \bigg[ \underset{1 \leq i \leq m}{\prod} \mathrm{sin} \big( u^{\prime} - v_{n-i} \pm \eta \sigma^z_{n-j} \big)  \text{ }    \bigg] \text{ } \bigg]    \text{ } \bigg]                 \text{, }  \end{align*}
  
  \begin{align*}  \big( \mathcal{P}_9 \big)_6   \equiv           2 \frac{\partial A_3 \big( u^{\prime} \big)}{\partial u^{\prime}}           \cdot  \underset{1 \leq j \leq m}{\sum}   \frac{\partial}{\partial u}                 \bigg[ \underset{1 \leq i \leq m}{\prod} \mathrm{sin} \big( u^{\prime} - v_{n-i} \pm \eta \sigma^z_{n-j} \big)  \text{ }    \bigg]       \text{, }  \end{align*}
  
  \begin{align*}   \big( \mathcal{P}_9 \big)_7   \equiv                        \frac{\partial A_3 \big( u^{\prime} \big)}{\partial u}   \cdot          \underset{1 \leq j \leq m}{\sum}      \bigg[ \frac{\partial}{\partial u^{\prime}}                 \bigg[ \underset{1 \leq i \leq m}{\prod} \mathrm{sin} \big( u^{\prime} - v_{n-i} \pm \eta \sigma^z_{n-j} \big)  \text{ }    \bigg]            \text{, }  \end{align*}
  
  \begin{align*}        \big( \mathcal{P}_9 \big)_8   \equiv          \frac{\partial B_3 \big(u^{\prime} \big)}{\partial u}    \cdot                \underset{1 \leq j \leq m}{\sum}      \bigg[ \frac{\partial}{\partial u^{\prime}}                 \bigg[ \underset{1 \leq i \leq m}{\prod} \mathrm{sin} \big( u^{\prime} - v_{n-i} \pm \eta \sigma^z_{n-j} \big)  \text{ }    \bigg]        \text{, }  \end{align*}
  
  \begin{align*}  
 \big( \mathcal{P}_9 \big)_9   \equiv    \frac{\partial A_3 \big( u^{\prime} \big)}{\partial u} \cdot   \underset{1 \leq j \leq m}{\sum}    \bigg[ \frac{\partial}{\partial u^{\prime}} \underset{1 \leq i \leq m}{\prod}  \mathrm{sin} \big( u^{\prime} - v_{n-i} \pm \eta \sigma^z_{n-j} \big) \bigg]             \text{, }  \end{align*}
  
  \begin{align*}   \big( \mathcal{P}_9 \big)_{10}  \equiv \frac{\partial B_3 \big( u^{\prime} \big)}{\partial u} \bigg[  \underset{1 \leq j \leq m}{\sum}    \bigg[ \frac{\partial}{\partial u^{\prime}} \bigg[ \underset{1 \leq i \leq m}{\prod}  \mathrm{sin} \big( u^{\prime} - v_{n-i} \pm \eta \sigma^z_{n-j} \big) \bigg] \text{ }  \bigg]  \text{. }
\end{align*}

 }

\end{itemize}

\subsection{Arguments for Lemma 1}

\noindent \textbf{Lemma} \textit{1} (\textit{collecting terms from the product of two by two L-operators}). The first entry of the $L$ operator product,

{\small \begin{align*}
     \overset{3}{\underset{i=0}{\prod}}   L_{\alpha , N - i  } {\tiny \big( \lambda_{\alpha} , v_{N-i} \big) }     \text{, } 
\end{align*} }

\noindent has an expansion of the form,

{\small \begin{align*}
      A_3 {\tiny\big( \lambda_{\alpha} \big) \equiv         }    \prod_{0 \leq i \leq 3} \mathrm{sin} {\tiny\big( \lambda_{\alpha} - v_{N-i} + \eta \sigma^z_{N-j} \big) }  +  \big( \mathrm{sin} {\tiny\big( 2 \eta \big) } \big)^2 \bigg\{           \underset{i \equiv 1, +}{\underset{i \equiv 0 , -}{\prod_{0 \leq i \leq 1}}}      \sigma^{-,+}_{N-i}    \text{ } \underset{2\leq i \leq 3}{\prod}   \mathrm{sin}  {\tiny\big( \lambda_{\alpha} - v_{N-i} + \eta \sigma^z_{N-j} \big) } \end{align*}
 
 \begin{align*}  
      +       \underset{i \equiv 3 , +}{\underset{i \equiv 2 , -}{\underset{2 \leq i \leq 3}{\prod} }}    \sigma^{-,+}_{N-i} \cdot    \underset{{0 \leq i \leq 1}}{\prod}  \mathrm{sin} {\tiny\big( \lambda_{\alpha} - v_{N-i} + \eta \sigma^z_{N-j} \big) }  +    {\tiny\big( \mathrm{sin} \big( 2 \eta \big)}^{-1}     \underset{i \equiv 3, +}{\underset{i \equiv 1, -}{\underset{i \mathrm{\text{ } odd \text{ }}: \text{ } 1 \leq i \leq 3}{\prod}} }\mathrm{sin} {\tiny\big( 2 \eta \big)}  \sigma^{-}_{N-i} \end{align*}
 
 \begin{align*}   \cdot 
  \underset{i \equiv 2, - \eta}{\underset{i \equiv 0 , + \eta}{\underset{i \text{ } \mathrm{even \text{ }} 
 :\text{ }  0 \leq i \leq 2}{\prod}}} \mathrm{sin} {\tiny\big( \lambda_{\alpha} - v_{N-i} \pm  \eta \sigma^z_{N-j} \big) }   +   \underset{i \equiv 2, +}{\underset{i \equiv 1, -}{\prod_{1 \leq i \leq 2}}}   \sigma^{-,+}_{N-i}    \end{align*}
 
 \begin{align*}      \cdot  \underset{i \mathrm{ \text{ } odd}: \text{ } 1 \leq i \leq 3}{\prod}     \mathrm{sin} {\tiny\big( \lambda_{\alpha} - v_{N-i} + \eta \sigma^z_{N-j} \big) }  
 +  \underset{ i \equiv 2 , +}{\underset{i \equiv 0 ,-}{\underset{i \text{ } \mathrm{even}\text{ } : \text{ } 0 \leq i \leq 2}{\prod}  }  }           \sigma^{-,+}_{N-i} \end{align*}
 
 \begin{align*}   \cdot \underset{i\text{ }  \mathrm{odd}\text{ } : \text{ } 1 \leq i \leq 3}{\prod}   \mathrm{sin} {\tiny\big( \lambda_{\alpha} - v_{N-i} - \eta \sigma^z_{N-j} \big) }    + 
 \mathrm{sin} {\tiny\big( 2 \eta \big)}   \cdot  \underset{i \equiv 3 , +}{\underset{i \equiv 2 , -}{\underset{i \equiv 1 , +}{\underset{i \equiv 0 , -}{\underset{0 \leq i \leq 3}{\prod}}  }} }\sigma^{-,+}_{N-i}   +       \underset{i \equiv 3 , +}{\underset{i \equiv 1 , -}{\underset{i \text{ } \mathrm{odd} \text{ } :\text{ } 1 \leq i \leq 3}{\prod}  }}    \sigma^{-,+}_{N-i}  \end{align*}
 
 \begin{align*}     \cdot        \underset{1 \leq i \leq 2}{\prod}        \mathrm{sin} {\tiny\big( \lambda_{\alpha} - v_{N-i} - \eta \sigma^z_{N-j} \big)  }          
 \bigg\}      \text{, } 
\end{align*} }

\noindent while the first entry of the L-operator product,

{\small \begin{align*}
    \prod_{i=0}^2 L_{\alpha , N-i} {\tiny \big( \lambda_{\alpha}, v_{N-i} \big) }  \text{, }
\end{align*} } 

\noindent has an expansion of the form,

{\small \begin{align*}
  A_2 {\tiny\big( \lambda_{\alpha} \big)} \equiv 
     \prod_{0 \leq i \leq 2} \mathrm{sin} {\tiny\big( \lambda_{\alpha} - v_{N-i} + \eta \sigma^z_{N-j} \big) } 
\end{align*}

 \begin{align*} + \big( \mathrm{sin} {\tiny\big( 2 \eta \big)}  \big)^2    \cdot \underset{i \equiv 1, +}{\underset{i \equiv 0 , -}{\prod_{0 \leq i \leq 1}}}      \sigma^{-,+}_{N-i}    \cdot  \mathrm{sin}  {\tiny \big( \lambda_{\alpha} - v_{N-2} + \eta \sigma^z_{N-2} \big)}  +  \big( \mathrm{sin} {\tiny\big( 2 \eta \big)}  \big)^2 \end{align*}

 \begin{align*}  \times   \underset{i \equiv 2, +}{\underset{i \equiv 1, -}{\prod_{1 \leq i \leq 2}}}   \sigma^{-,+}_{N-i}   \cdot     \mathrm{sin} {\tiny\big( \lambda_{\alpha} - v_N + \eta \sigma^z_N \big) }    +       \big( \mathrm{sin} {\tiny\big( 2 \eta \big)}  \big)^2   \cdot \underset{i \equiv 2, +}{\underset{i \equiv 0,-}{\prod_{\mathrm{even \text{ } } i : \text{ } 0 \leq i \leq 2}}}        \sigma^{-,+}_{N-i}  \end{align*}

 \begin{align*}   \times\mathrm{sin} {\tiny\big( \lambda_{\alpha} - v_{N-1} - \eta \sigma^z_{N-1} \big)}          
\text{. } 
\end{align*} } 

 \bigskip 

\noindent \textit{Proof of Lemma 1}. We collect terms from the first two terms in the product of $L$ operators,

{\small \[
\begin{bmatrix}
      \mathrm{sin} {\tiny  \big( \lambda_{\alpha} - v_N + \eta \sigma^z_{N} \big)    }              &        \mathrm{sin} {\tiny  \big( 2 \eta \big)  } \sigma^{-}_{N}                \\
         \mathrm{sin} {\tiny  \big( 2 \eta \big) }  \sigma^{+}_N       & \mathrm{sin} {\tiny  \big( \lambda_{\alpha} - v_N - \eta \sigma^z_N \big) } 
  \end{bmatrix} \begin{bmatrix}
          \mathrm{sin} {\tiny  \big( \lambda_{\alpha} - v_{N-1} + \eta \sigma^z_{N-1} \big) }               &       \mathrm{sin} {\tiny  \big( 2 \eta \big) } \sigma^{-}_{N-1}              \\
            \mathrm{sin} {\tiny  \big( 2 \eta \big) } \sigma^{+}_{N-1}        &       \mathrm{sin} {\tiny \big( \lambda_{\alpha} - v_{N-1} - \eta \sigma^z_{N-1} \big) }
  \end{bmatrix} \text{, } \] } 

\noindent from which one obtains a resultant matrix of the form,

{\small \[  \begin{bmatrix}
           \textbf{1}       &        \textbf{2}                \\
           \textbf{3}     &     \textbf{4} 
  \end{bmatrix} \text{, } 
\] }

\noindent from the following expression for the first and second entries,

{\small \begin{align*}
   \textbf{1}^0 \equiv \textbf{1} \equiv   \mathrm{sin} {\tiny  \big( \lambda_{\alpha} - v_N + \eta \sigma^z_{N} \big) } \text{ } \mathrm{sin} {\tiny  \big( \lambda_{\alpha} - v_{N-1} + \eta \sigma^z_{N-1}  \big) }  + \mathrm{sin} {\tiny \big( 2 \eta \big) }\sigma^{-}_N \mathrm{sin} {\tiny \big( 2 \eta \big)}  \sigma^{+}_{N-1} \text{, }   \end{align*}

 \begin{align*} \textbf{2}^0 \equiv  \textbf{2} \equiv  \mathrm{sin} {\tiny \big( \lambda_{\alpha} - v_N + \eta \sigma^z_N \big) } \mathrm{sin} {\tiny \big( 2 \eta \big)}  \sigma^{-}_{N-1} + \mathrm{sin} {\tiny \big( 2 \eta \big)}  \sigma^{-}_N \mathrm{sin} {\tiny  \big( \lambda_{\alpha} - v_{N-1} - \eta \sigma^z_{N-1} \big)    }      \text{. } 
\end{align*} }

\noindent implies that the expression for $A {\tiny  \big( \lambda_{\alpha} \big)}$, after repeating the computation for the two by two matrix multiplication $n$ times, can be obtained by grouping together terms under the $\big( \mathrm{sin} {\tiny \big( 2 \eta \big) }\big)^2$, in which,

{\small \begin{align*}
\text{ }  \underset{i \equiv 1, +}{\underset{i \equiv 0 , -}{\prod_{0 \leq i \leq 1}}}      \sigma^{-,+}_{N-i} \cdot  \underset{2\leq i \leq 3}{\prod}   \mathrm{sin}  {\tiny  \big( \lambda_{\alpha} - v_{N-i} + \eta \sigma^z_{N-j} \big) }+      \underset{i \equiv 3 , +}{\underset{i \equiv 2 , -}{\underset{2 \leq i \leq 3}{\prod} }}    \sigma^{-,+}_{N-i}  \cdot   \underset{{0 \leq i \leq 1}}{\prod}  \mathrm{sin} {\tiny  \big( \lambda_{\alpha} - v_{N-i} + \eta \sigma^z_{N-j} \big) }        \text{, } 
\end{align*} }

\noindent corresponding to a first group of terms,

{\small \begin{align*}
       \big( \mathrm{sin} {\tiny \big( 2 \eta \big) } \big)^{-1}      \cdot \underset{i \equiv 3, +}{\underset{i \equiv 1, -}{\underset{i \mathrm{\text{ } odd \text{ }}: \text{ } 1 \leq i \leq 3}{\prod}} }\mathrm{sin} {\tiny \big( 2 \eta \big) } \sigma^{-}_{N-i}   \cdot  
  \underset{i \equiv 2, - \eta}{\underset{i \equiv 0 , + \eta}{\underset{i \text{ } \mathrm{even \text{ }} 
 :\text{ }  0 \leq i \leq 2}{\prod}}} \mathrm{sin} {\tiny \big( \lambda_{\alpha} - v_{N-i} \pm  \eta \sigma^z_{N-j} \big) }   \text{, } 
\end{align*} }

\noindent corresponding to a second group of terms,

{\small \begin{align*}
    \underset{i \equiv 2, +}{\underset{i \equiv 1, -}{\prod_{1 \leq i \leq 2}}}   \sigma^{-,+}_{N-i}        \cdot \underset{i \mathrm{ \text{ } odd}: \text{ } 1 \leq i \leq 3}{\prod}     \mathrm{sin} {\tiny \big( \lambda_{\alpha} - v_{N-i} + \eta \sigma^z_{N-j} \big) }   \text{, } 
\end{align*} }

\noindent corresponding to a third group of terms,

{\small \begin{align*}
\underset{ i \equiv 2 , +}{\underset{i \equiv 0 ,-}{\underset{i \text{ } \mathrm{even}\text{ } : \text{ } 0 \leq i \leq 2}{\prod}  }  }           \sigma^{-,+}_{N-i} \cdot  \underset{i\text{ }  \mathrm{odd}\text{ } : \text{ } 1 \leq i \leq 3}{\prod}   \mathrm{sin} \big( \lambda_{\alpha} - v_{N-i} - \eta \sigma^z_{N-j} \big)     \text{, } 
\end{align*} }

\noindent corresponding to a fourth group of terms, and,

{\small \begin{align*}
       \mathrm{sin} \big( 2 \eta \big) \cdot    \underset{i \equiv 3, +}{\underset{i\equiv 2, -}{\underset{i \equiv 1, +}{\underset{i \equiv 0 ,-}{\prod_{0 \leq i \leq 3}}} } }\sigma^{-,+}_{N-i}    \text{, } 
\end{align*} }

\noindent corresponding to a fifth group of terms, and,

{\small \begin{align*}
     \underset{i \equiv 3 ,+}{{\underset{i \equiv 1 , -}{ \underset{i \text{ } \mathrm{odd} \text{ } :\text{ } 1 \leq i \leq 3}{\prod  } } }} \sigma^{-,+}_{N-i}     \cdot         \underset{1 \leq i \leq 2}{\prod}        \mathrm{sin} {\tiny  \big( \lambda_{\alpha} - v_{N-i} - \eta \sigma^z_{N-j} \big)  }         \text{, }
\end{align*} }

\noindent corresponding to a sixth group of terms, hence implying the desired form for the first coefficient,

{\small \begin{align*}
      A_3 {\tiny  \big( \lambda_{\alpha} \big)  } \equiv            \prod_{0 \leq i \leq 3} \mathrm{sin} {\tiny  \big( \lambda_{\alpha} - v_{N-i} + \eta \sigma^z_{N-j} \big) }   +  \big( \mathrm{sin} {\tiny \big( 2 \eta \big) } \big)^2 \bigg\{           \underset{i \equiv 1, +}{\underset{i \equiv 0 , -}{\prod_{0 \leq i \leq 1}}}      \sigma^{-,+}_{N-i} \end{align*}

 \begin{align*} 
      \times    \text{ } \underset{2\leq i \leq 3}{\prod}   \mathrm{sin}  {\tiny \big( \lambda_{\alpha} - v_{N-i} + \eta \sigma^z_{N-j} \big) } +       \underset{i \equiv 3 , +}{\underset{i \equiv 2 , -}{\underset{2 \leq i \leq 3}{\prod} }}    \sigma^{-,+}_{N-i}  \cdot   \underset{{0 \leq i \leq 1}}{\prod}  \mathrm{sin} {\tiny \big( \lambda_{\alpha} - v_{N-i} + \eta \sigma^z_{N-j} \big) }  
\end{align*}

 \begin{align*}   +   \big( \mathrm{sin} {\tiny  \big( 2 \eta \big) } \big)^{-1}     \cdot \underset{i \equiv 3, +}{\underset{i \equiv 1, -}{\underset{i \mathrm{\text{ } odd \text{ }}: \text{ } 1 \leq i \leq 3}{\prod}} }\mathrm{sin} {\tiny  \big( 2 \eta \big) } \sigma^{-}_{N-i}  \cdot 
  \underset{i \equiv 2, - \eta}{\underset{i \equiv 0 , + \eta}{\underset{i \text{ } \mathrm{even \text{ }} 
 :\text{ }  0 \leq i \leq 2}{\prod}}} \mathrm{sin} {\tiny  \big( \lambda_{\alpha} - v_{N-i} \pm  \eta \sigma^z_{N-j} \big) }   \end{align*}

 \begin{align*}   +    \underset{i \equiv 2, +}{\underset{i \equiv 1, -}{\prod_{1 \leq i \leq 2}}}   \sigma^{-,+}_{N-i}        \cdot \underset{i \mathrm{ \text{ } odd}: \text{ } 1 \leq i \leq 3}{\prod}     \mathrm{sin} {\tiny  \big( \lambda_{\alpha} - v_{N-i} + \eta \sigma^z_{N-j} \big)  }  \end{align*}  \begin{align*}  +    \underset{ i \equiv 2 , +}{\underset{i \equiv 0 ,-}{\underset{i \text{ } \mathrm{even}\text{ } : \text{ } 0 \leq i \leq 2}{\prod}  }  }           \sigma^{-,+}_{N-i} \cdot   \underset{i\text{ }  \mathrm{odd}\text{ } : \text{ } 1 \leq i \leq 3}{\prod}   \mathrm{sin} {\tiny  \big( \lambda_{\alpha} - v_{N-i} - \eta \sigma^z_{N-j} \big) }    \end{align*}

 \begin{align*}  +   \mathrm{sin} {\tiny  \big( 2 \eta \big)}   \cdot \underset{i \equiv 3 , +}{\underset{i \equiv 2 , -}{\underset{i \equiv 1 , +}{\underset{i \equiv 0 , -}{\underset{0 \leq i \leq 3}{\prod}}  }} }\sigma^{-,+}_{N-i}    + 
        \underset{i \equiv 3 , +}{\underset{i \equiv 1 , -}{\underset{i \text{ } \mathrm{odd} \text{ } :\text{ } 1 \leq i \leq 3}{\prod}  }}    \sigma^{-,+}_{N-i}     \cdot        \underset{1 \leq i \leq 2}{\prod}        \mathrm{sin} {\tiny  \big( \lambda_{\alpha} - v_{N-i} - \eta \sigma^z_{N-j} \big) }  
 \bigg\}     \text{, } 
\end{align*} }

\noindent from which we conclude the argument. \boxed{}

\subsection{Arguments for Lemma 2}

\noindent \textbf{Lemma} \textit{2} (\textit{collecting terms from the product of two by two L-operators}). The second entry of the $L$ operator product,

{\small \begin{align*}
     \overset{3}{\underset{i=0}{\prod}}   L_{\alpha , N - i  } {\tiny\big( \lambda_{\alpha} , v_{N-i} \big)   }      \text{, } 
\end{align*} }

\noindent has an expansion of the form,

{\small \begin{align*}
  B_3 {\tiny\big( \lambda_{\alpha} \big)}  \equiv   \big( \mathrm{sin} {\tiny\big( 2 \eta \big)} \big) \text{ } \sigma^{-}_{N-3} \cdot   \underset{0 \leq i \leq 2}{\prod}   \mathrm{sin} {\tiny\big( \lambda_{\alpha} - v_{N-i} + \eta \sigma^z_{N-j} \big)   }       +   \big( \mathrm{sin} {\tiny\big( 2 \eta \big)} \big)^3 \bigg\{   \big( \mathrm{sin} {\tiny\big( 2 \eta \big) }\big)^{-1} \end{align*}

 \begin{align*}    \times     \bigg[    \underset{i \equiv 1 , +}{\underset{i \equiv 0 , -}{\underset{0 \leq i \leq 1}{\prod}} }\sigma^{-,+}_{N-j}                \cdot    \underset{i \equiv 3 ,-}{ \underset{i \equiv 2 ,+}{\underset{2 \leq i \leq 3}{\prod } }}  \mathrm{sin} {\tiny\big( \lambda_{\alpha} - v_{N-i} \pm \eta \sigma^z_{N-j} \big) }     +  \underset{i \equiv 2 , +}{\underset{i \equiv 1 , -}{\underset{1 \leq i \leq 2}{\prod}} }\sigma^{-,+}_{N-i}   \end{align*}

 \begin{align*}   
  \cdot           \underset{i \equiv 3 ,+}{ \underset{ i \equiv 1 , -}{\underset{\mathrm{odd}\text{ } i \text{ } : \text{ }  1 \leq i \leq 3}{ \prod}} }           \mathrm{sin} {\tiny\big( \lambda_{\alpha}- v_{N-i} \pm \eta \sigma^z_{N-j} \big) } \end{align*}

 \begin{align*}    +   \sigma^{-}_{N-2} \bigg[      \underset{0 \leq i \leq 1}{ \prod}    \mathrm{sin} {\tiny\big( \lambda_{\alpha} - v_{N-i} + \eta \sigma^z_{N-j}      \big)  }                +     \underset{ i \equiv 3 , -}{\underset{i \equiv 1 , +}{\underset{i \equiv 0 , \text{ } \mathrm{odd} \text{ } i \text{ } : \text{ } 1 \leq i \leq 3}{ \prod   }}}  \mathrm{sin} {\tiny\big( \lambda_{\alpha} - v_{N-i} \pm \eta \sigma^z_{N-j} \big) }                             \bigg]  \end{align*}

 \begin{align*}     
  +  \underset{ 1 \leq i \leq 2}{\prod}       \mathrm{sin} {\tiny\big( \lambda_{\alpha} - v_{N-i} - \eta \sigma^z_{N-j}    \big) }   +  \sigma^{-}_N \mathrm{sin} {\tiny\big( \lambda_{\alpha} - v_{N-2} - \eta \sigma^z_{N-2} \big)} \cdot    \underset{1 \leq i \leq 3}{\prod}      \mathrm{sin} {\tiny\big( \lambda_{\alpha} - v_{N-i} - \eta \sigma^z_{N-j} \big)}   \end{align*}

 \begin{align*}      +  \big( 1 + \sigma^{-}_{N-1} \big)  \bigg[ \underset{i \equiv 2 , -}{\underset{i \equiv  0 , +}{\underset{\mathrm{even}\text{ } i \text{ } : \text{ }0 \leq i \leq 2}{\prod} }  }\mathrm{sin} {\tiny\big( \lambda_{\alpha} - v_{N-i} \pm \eta \sigma^z_{N-j} \big)}  \end{align*}

 \begin{align*}    +   \underset{i \equiv 2 , -}{\underset{i \equiv 0 , +}{{\underset{\mathrm{even}\text{ } i \text{ } : \text{ }0 \leq i \leq 2}{\prod} }}}\mathrm{sin} {\tiny\big( \lambda_{\alpha} - v_{N-i} \pm \eta \sigma^z_{N-j} \big)}  \bigg]   \text{ } \end{align*}

 \begin{align*}  +    \mathrm{sin} {\tiny\big( 2 \eta \big)}  \bigg[     \underset{i \equiv 2 , -}{\underset{i \equiv 1 , +}{\underset{i \equiv 0 ,-}{ \underset{0 \leq i \leq 2}{\prod}  }  }  }    \sigma^{-,+}_{N-i}   +       \sigma^{-}_{N-2} \cdot  \underset{\mathrm{even} \text{ } i \text{ } : \text{ } 0 \leq i \leq 2}{\prod}  \sigma^{-,+}_{N-i} \cdot  \underset{i \equiv 3 , +}{\underset{i \equiv 1, -}{\underset{\mathrm{odd} \text{ } i \text{ } : 1 \leq i \leq 3}{\prod}} }  \mathrm{sin} {\tiny\big( \lambda_{\alpha} - v_N \pm \eta \sigma^z_{N-j} \big)  }  \end{align*}

 \begin{align*}   +           \underset{i \equiv 1 , +}{\underset{i \equiv 0 , 2 , -}{\underset{0 \leq i \leq 2}{\prod}   }}      \sigma^{-,+}_{N-i}                   
 +       \underset{i \equiv 3 , - \eta}{\underset{ i \equiv 0 , 1 , + \eta}{\underset{0 \leq i \leq 1 , i \equiv 3}{\prod} } }            \mathrm{sin} {\tiny\big( \lambda_{\alpha} - v_{N-i}  \pm \eta \sigma^z_{N-j}  \big)}  \end{align*}

 \begin{align*}   + \underset{i \equiv 1 , +}{\underset{i \equiv 0,2,-}{\underset{0 \leq i \leq 2}{\prod} }}   \sigma^{-,+}_{N-i}   +      \underset{i \equiv 3 , - \eta}{\underset{i \equiv 0 , 1 + \eta}{\underset{0 \leq i \leq 1 , i \equiv 3}{\prod}}   } \mathrm{sin} {\tiny\big(        \lambda_{\alpha} - v_{N-i} \pm \eta \sigma^z_{N-j}              \big) }        +\sigma^{-}_{N-2} \cdot            \underset{\mathrm{even} \text{ } i \text{ } i : \text{ } 0 \leq i \leq 2 }{\prod}  \sigma^{-,+}_{N-i}                  \end{align*}

 \begin{align*}    + 
 \underset{i \equiv 2, -}{\underset{i \equiv 1, +}{\underset{i \equiv 0 , -}{ \underset{0 \leq i \leq 2}{\prod}  } }}   \sigma^{-,+}_{N-i}      \bigg]     + 
 \underset{i \equiv 1 , -}{\underset{i \equiv 0 ,+}{\underset{0 \leq i \leq 1}{\prod} }  }   \sigma^{-,+}_{N-i}    \cdot \mathrm{sin} {\tiny\big( \lambda_{\alpha} - v_{N-2} + \eta \sigma^z_{N-2} \big)}  \end{align*}

 \begin{align*}   + \underset{i \equiv 2 , +}{\underset{i \equiv 1 ,-}{\underset{1 \leq i \leq 2}{\prod} }  }  \sigma^{-,+}_{N-i} \cdot  \mathrm{sin} {\tiny\big( \lambda_{\alpha} - v_N + \eta \sigma^z_N \big) }  +  \underset{i \equiv 2 , +}{\underset{i \equiv 0 ,-}{\underset{\mathrm{even} \text{ } i \text{ } : \text{ } 0 \leq i \leq 2}{ \prod}  }   } \sigma^{-,+}_{N-i}  \cdot \mathrm{sin} {\tiny\big( \lambda_{\alpha} - v_N + \eta \sigma^z_N \big) }    \bigg\}   \text{, } 
 \end{align*} }

 \noindent while the second entry of the L-operator product,

{\small \begin{align*}
    \prod_{i=0}^2 L_{\alpha , N-i} {\tiny\big( \lambda_{\alpha}, v_{N-i} \big)}  \text{, }
\end{align*} } 

\noindent has an expansion of the form,

{\small \begin{align*}
    B_2 {\tiny\big( \lambda_{\alpha } \big)}  \equiv         \big(   \mathrm{sin} {\tiny\big( 2 \eta \big)}  \sigma^{-}_{N-2} \big) \cdot  \underset{{0 \leq i \leq 1}}{\prod}  \mathrm{sin} {\tiny\big( \lambda_{\alpha} - v_{N-i} + \eta \sigma^z_{N-j} \big)}    +  \big( \mathrm{sin} {\tiny\big( 2 \eta \big)} \big)^3 \cdot \underset{i\equiv 2, -}{\underset{i \equiv 1, +}{\underset{i \equiv 0 ,-}{\prod_{0 \leq i \leq 2}}} }\sigma^{-,+}_{N-i}  +    \big(   \mathrm{sin} {\tiny\big( 2 \eta \big) \sigma^{-}_{N-1}   \big)} \end{align*}

 \begin{align*} 
    \times  \underset{i \equiv 2, - \eta}{\underset{i \equiv 0 , + \eta}{\underset{\mathrm{even \text{ }} i :\text{ }  0 \leq i \leq 2}{\prod}}} \mathrm{sin} {\tiny\big( \lambda_{\alpha} - v_{N-i} \pm  \eta \sigma^z_{N-j} \big) }     +    \big( \mathrm{sin} {\tiny\big( 2 \eta \big) } \sigma^{-}_N \big) \cdot  \underset{1 \leq i \leq 2}{\prod}   \mathrm{sin} {\tiny\big( \lambda_{\alpha} - v_{N-i} - \eta \sigma^z_{N-j} \big)}        \text{. }
\end{align*} }

\bigskip

\noindent \textit{Proof of Lemma 2}. We collect terms from the first two terms in the product of $L$ operators,

\[
\begin{bmatrix}
      \mathrm{sin} {\tiny  \big( \lambda_{\alpha} - v_N + \eta \sigma^z_{N} \big)    }               &        \mathrm{sin} {\tiny  \big( 2 \eta \big) }  \sigma^{-}_{N}                \\
         \mathrm{sin} {\tiny  \big( 2 \eta \big) }  \sigma^{+}_N       & \mathrm{sin} {\tiny  \big( \lambda_{\alpha} - v_N - \eta \sigma^z_N \big) } 
  \end{bmatrix} \begin{bmatrix}
          \mathrm{sin} {\tiny  \big( \lambda_{\alpha} - v_{N-1} + \eta \sigma^z_{N-1} \big)  }              &       \mathrm{sin} {\tiny  \big( 2 \eta \big)  } \sigma^{-}_{N-1}              \\
            \mathrm{sin} {\tiny  \big( 2 \eta \big) }  \sigma^{+}_{N-1}        &       \mathrm{sin} {\tiny  \big( \lambda_{\alpha} - v_{N-1} - \eta \sigma^z_{N-1} \big) } 
  \end{bmatrix} \text{, } \]

\noindent from which one obtains a resultant matrix of the form,

\[  \begin{bmatrix}
           \textbf{1}       &        \textbf{2}                \\
           \textbf{3}     &     \textbf{4} 
  \end{bmatrix}  \text{, } 
\]

\noindent as rearrangements for collecting like terms under the prefactor $\big( \mathrm{sin} \big( \eta \big) \big)^3$ gives,

{\small \begin{align*}
 \big( \mathrm{sin} {\tiny  \big( 2 \eta \big) }  \big) \text{ } \sigma^{-}_{N-3} \cdot     \underset{0 \leq i \leq 2}{\prod}   \mathrm{sin} {\tiny  \big( \lambda_{\alpha} - v_{N-i} + \eta \sigma^z_{N-j} \big)  }        +  \big( \mathrm{sin} {\tiny \big( 2 \eta \big) } \big)^3 \bigg\{    \underset{i \equiv 1, +}{\underset{i \equiv 0 , +}{\underset{0 \leq i \leq 1}{\prod}  } }     \sigma^{-,+}_{N-i}  \cdot 
 \mathrm{sin} {\tiny \big( \lambda_{\alpha} - v_{N-2} + \eta \sigma^z_{N-2} \big) }  \end{align*}

 \begin{align*}  +     \underset{i \equiv 2, +}{\underset{i \equiv 1, -}{\underset{1 \leq i \leq 2}{\prod} } }    \sigma^{-,+}_{N-i} \cdot  \mathrm{sin} {\tiny  \big( \lambda_{\alpha} - v_N + \eta \sigma^z_N \big)  } 
  +    \underset{i \equiv 2, +}{\underset{i \equiv 0 ,-}{\underset{\mathrm{even \text{ } }i\text{ }  :\text{ }  0 \leq i \leq 2}{\prod} } } \sigma^{-,+}_{N-i}    \cdot  \mathrm{sin} {\tiny \big( \lambda_{\alpha }   - v_N + \eta \sigma^z_N \big) }  \end{align*}

 \begin{align*}   +   \big( \mathrm{sin} {\tiny \big( 2 \eta \big) } \big)^{-1}   \sigma^{-}_{N-2} \cdot       \underset{0 \leq i \leq 1}{\prod}    \mathrm{sin} {\tiny \big( \lambda_{\alpha} - v_{N-i} + \eta \sigma^z_{N-j} \big) }   \end{align*}

 \begin{align*}    
 + \big( \mathrm{sin} {\tiny  \big( 2 \eta \big) } \big)   \cdot  \underset{i \equiv 2, -}{\underset{i \equiv 1, +}{\underset{i \equiv 0 , -}{ \underset{0 \leq i \leq 2}{\prod}  } }}   \sigma^{-,+}_{N-i}               +  \big( \mathrm{sin} {\tiny  \big( 2 \eta \big) } \big)^{-1} \cdot    \underset{i \equiv 2, -\eta}{\underset{i \equiv 0 , + \eta }{\underset{\mathrm{even}\text{ } i \text{ } : \text{ } 0 \leq i \leq 2}{\prod} } }  \mathrm{sin} {\tiny  \big( \lambda_{\alpha} - v_{N-i} \pm \eta \sigma^z_{N-j} \big) } \end{align*}

 \begin{align*} +  \big( \mathrm{sin} {\tiny  \big( 2 \eta \big) } \big)^{-1} \cdot  \underset{1 \leq i \leq 2}{\prod}           \mathrm{sin}  {\tiny  \big(  \lambda_{\alpha}    - v_{N-i} - \eta \sigma^z_{N-j} \big) }   + \big( \mathrm{sin} {\tiny  \big( 2 \eta \big)  } \big)^{-1} \cdot \underset{i \equiv 1 , +}{\underset{i \equiv 0 ,-}{\underset{0 \leq i \leq 1}{\prod} }  }  \sigma^{-,+}_{N-i }        \end{align*}

 \begin{align*} \times 
       \underset{i \equiv 3 , \eta -}{\underset{i \equiv 2 , \eta +}{\underset{2 \leq i \leq 3}{\prod}}}     \mathrm{sin} {\tiny  \big( \lambda_{\alpha} - v_{N-i} \pm \eta \sigma^{z}_{N-j}       \big)    }                    + \big( \mathrm{sin} {\tiny  \big( 2 \eta \big)  } \big)^{-1}    \cdot       \underset{i \equiv 2 , +}{\underset{ i \equiv 1 , -}{ \underset{1 \leq i \leq 2}{\prod}}}       \sigma^{-,+}_{N-i}           \cdot       \underset{ i \equiv 3 , +}{\underset{i \equiv 1, -}{\underset{\mathrm{odd} \text{ } i : \text{ } 1 \leq i \leq 3}{\prod}  }}          \mathrm{sin} {\tiny  \big( \lambda_{\alpha} - v_N \pm \eta \sigma^z_{N-j}  \big)   }    \end{align*}

 \begin{align*}   +   \big( \mathrm{sin} {\tiny \big( 2 \eta \big)  } \big)   \text{ }   \sigma^{-}_{N-2}  \cdot  \underset{\mathrm{even}\text{ }  i \text{ } : \text{ } 0 \leq i \leq 2}{\prod}              \sigma^{-,+}_{N-i}                         \cdot         \underset{ i \equiv 3 , +}{\underset{i \equiv 1, -}{\underset{\mathrm{odd} \text{ } i : \text{ } 1 \leq i \leq 3}{\prod}  }}          \mathrm{sin} {\tiny \big( \lambda_{\alpha} - v_N \pm \eta \sigma^z_{N-j}  \big)  }          \end{align*}

 \begin{align*}    +      \big( \mathrm{sin} {\tiny  \big( 2 \eta \big) } \big)^{-1}              \sigma^{-}_{N-2} \cdot     \underset{i \equiv 3 ,-}{\underset{i \equiv 1 , +}{\underset{i \equiv 0, +}{\underset{i \equiv 0 \text{ } ,\text{ } \mathrm{odd}\text{ } i : \text{ } 1 \leq i \leq 3}{\prod} }  }  }         \mathrm{sin} {\tiny  \big( \lambda_{\alpha} - v_{N-1} \pm \eta \sigma^{z}_{N-j} \big)  }         + \big( \mathrm{sin} {\tiny \big( 2 \eta \big) }  \big) \text{ } \end{align*}

 \begin{align*} \times    \underset{i \equiv 3 , - \eta}{\underset{ i \equiv 0 , 1 , + \eta}{\underset{0 \leq i \leq 1 , i \equiv 3}{\prod} } }            \mathrm{sin} {\tiny \big( \lambda_{\alpha} - v_{N-i}  \pm \eta \sigma^z_{N-j}  \big)}       +      \big( \mathrm{sin} {\tiny  \big( 2 \eta \big) }  \big) \cdot              \underset{i \equiv 1 , +}{\underset{i \equiv 0 , 2 , -}{\underset{0 \leq i \leq 2}{\prod}   }}      \sigma^{-,+}_{N-i}                     + \big( \mathrm{sin} {\tiny \big( 2 \eta \big) }\big)^{-1} \text{ } \sigma^{-}_{N-1} \text{ } \end{align*}

 \begin{align*}  \times   \underset{i \equiv 2 , -\eta}{\underset{i \equiv 0 , + \eta}{\underset{\mathrm{even}\text{ } :\text{ } 0 \leq i \leq 2 , i \equiv 3 }{\prod}  }   }      \mathrm{sin} {\tiny  \big( \lambda_{\alpha} - v_{N-i} \pm \eta \sigma^z_{N-j}      \big)  }       +  \big( \mathrm{sin} {\tiny \big( 2 \eta \big)}  \big)^{-1} \sigma^{-}_N \text{ }    \mathrm{sin} {\tiny  \big( \lambda_{\alpha} - v_{N-2} - \eta \sigma^z_{N-2} \big)   }         \text{ } \end{align*}

 \begin{align*} \times   \underset{1 \leq i \leq 3}{\prod}   \mathrm{sin} {\tiny  \big( \lambda_{\alpha} - v_{N-i} - \eta \sigma^z_{N-j}           \big)  }                                           \bigg\}            \text{. } 
\end{align*} }

\noindent Altogether, grouping terms under similar lines similar to the computations provided above yields the desired representation for $B_3 {\tiny \big( \lambda_{\alpha} \big) }$, from which we conclude the argument. \boxed{}

 \subsection{Arguments for Lemma 3}

\noindent \textbf{Lemma} \textit{3} (\textit{collecting terms from the product of two by two L-operators}). The third entry of the $L$ operator product,

{\small \begin{align*}
     \overset{3}{\underset{i=0}{\prod}}   L_{\alpha , N - i  } {\tiny\big( \lambda_{\alpha} , v_{N-i} \big)  }    \text{, } 
\end{align*} }

\noindent has an expansion of the form,

{\small \begin{align*}
   C_3 {\tiny\big( \lambda_{\alpha} \big) } \equiv   \big( \mathrm{sin} {\tiny\big( 2 \eta \big) }\big)^3   \mathrm{sin} {\tiny\big( \lambda_{\alpha} - v_{N-n} - \eta \sigma^z_{N-n} \big)}  \cdot           \underset{0 \leq i \leq 3}{\prod}    \mathrm{sin} {\tiny\big( \lambda_{\alpha} - v_{N-i} - \eta \sigma^z_{N-j}  \big)  }      +   \big( \mathrm{sin} {\tiny\big( 2 \eta \big) } \big)^2   \bigg\{     \sigma^{+}_{N-i}   \bigg[ \mathrm{sin} {\tiny \big(   \lambda_{\alpha} - v_{N-i}} \end{align*}
 
 \begin{align*} {\tiny + \eta \sigma^z_{N-i}     \big)  }   \cdot                 \underset{i \equiv 2 , -}{\underset{i \equiv 0 , +}{\underset{\mathrm{even \text{ } } i \text{ } : \text{ }  0 \leq i \leq 3}{\prod  }}  }   \sigma^{-,+}_{N-i}     
   +   \mathrm{sin} {\tiny\big( \lambda_{\alpha} - v_{N-i} - \eta \sigma^z_{N-i} \big)} \cdot   \underset{i \equiv 3 , +}{\underset{i \equiv 1 , -}{\underset{\mathrm{odd \text{ } } i \text{ } : 1 \leq i \leq 3}{\prod}}}             \sigma^{-,+}_{N-i}                 \bigg]  
   +  \mathrm{sin} {\tiny \big( \lambda_{\alpha} - v_N - \eta \sigma^z_N \big)}   \end{align*}

 \begin{align*}  \times  \underset{i \equiv 3 , +}{\underset{i \equiv 2 , -}{\underset{i \equiv 1 , +}{\underset{1 \leq i \leq 3}{\prod}  }}  }\sigma^{-,+}_{N-i}      +   \sigma^{+}_{N-(n-3)}     \cdot      \underset{0 \leq i \leq 2}{\prod}    \mathrm{sin} {\tiny\big( \lambda_{\alpha} - v_{N-i} - \eta \sigma^z_{N-j} \big)  }                 \bigg\}   \end{align*}

 \begin{align*}  + \mathrm{sin} {\tiny\big( 2 \eta \big)} \bigg\{        \sigma^{+}_{N} \cdot \underset{1 \leq i \leq 
        3    }{\prod}           \mathrm{sin} {\tiny\big( \lambda_{\alpha} - v_{N-i} + \eta \sigma^z_{N-j} \big) }            +          \mathrm{sin} {\tiny\big( \lambda_{\alpha} - v_{N-n} + \eta \sigma^z_{N-n} \big)  }\end{align*}

 \begin{align*}    \times  \bigg[         \underset{i \equiv 2 , +}{\underset{i \equiv 0,-}{\underset{\mathrm{even \text{ } } i\text{ }  : \text{ } 0 \leq i \leq 3}{\prod} }  }         \mathrm{sin} {\tiny\big( \lambda_{\alpha} - v_{N-i} \pm \eta \sigma^z_{N-j} \big)  }       +       \sigma^{+}_{N-1}  \bigg]  \text{ }        \underset{0 \leq i \leq 3}{\prod} \mathrm{sin} {\tiny\big( \lambda_{\alpha} - v_{N-i} - \eta \sigma^z_{N-j} \big) }                                \bigg\}             \text{ }  \end{align*}

 \begin{align*}  + 
        \big( \mathrm{sin} \big( 2 \eta \big) \big)^3  \mathrm{sin} {\tiny\big( \lambda_{\alpha} - v_{N-3} - \eta \sigma^z_{N-3} \big) }  \cdot          \underset{0 \leq i \leq 1}{\prod}    \mathrm{sin} {\tiny\big( \lambda_{\alpha} - v_{N-i} - \eta \sigma^z_{N-j}  \big)   }       \text{, } 
\end{align*} } 

\noindent while the third entry of the L-operator product,

{\small \begin{align*}
    \prod_{i=0}^2 L_{\alpha , N-i} {\tiny\big( \lambda_{\alpha}, v_{N-i} \big) } \text{, }
\end{align*} } 

\noindent has an expansion of the form,

{\small \begin{align*}
C_2 {\tiny\big( \lambda_{\alpha} \big)}  \equiv            \big( \mathrm{sin} {\tiny\big( 2 \eta \big)} \sigma^{+}_N \big)  \cdot  \underset{1 \leq i \leq 2}{\prod} \mathrm{sin} {\tiny\big( \lambda_{\alpha} - v_{N-i} + \eta \sigma^z_{N-j} \big)}  +   \big( \mathrm{sin} {\tiny\big( 2 \eta \big)} \big) \cdot   \underset{i \equiv 2 , + \eta}{\underset{i \equiv 0 , - \eta} {\underset{\mathrm{even \text{ } } i : \text{ } 0 \leq i \leq 2}{\prod} }}           \mathrm{sin} {\tiny\big( \lambda_{\alpha} - v_{N-i}  \pm \eta \sigma^z_{N-j} \big)}   \end{align*}

 \begin{align*}     +  \big( \mathrm{sin} {\tiny\big( 2 \eta \big)} \big)^3  \underset{i \equiv 1}{\prod}       \big( \sigma^{-}_{N-i} \big)^2 \sigma^{+}_{N-i-1}         +   \big( \mathrm{sin} {\tiny\big( 2 \eta \big)} \sigma^{+}_{N-2}    \big)    \cdot     \underset{0 \leq i \leq 1}{\prod}  \mathrm{sin} {\tiny\big( \lambda_{\alpha} - v_{N-i} - \eta \sigma^z_{N-j} \big)  }    \text{. } 
\end{align*} }

\bigskip 

\noindent \textit{Proof of Lemma 3}. We collect terms from the first two terms in the product of $L$ operators,

{\small \[
\begin{bmatrix}
      \mathrm{sin} {\tiny  \big( \lambda_{\alpha} - v_N + \eta \sigma^z_{N} \big) }                 &        \mathrm{sin} {\tiny  \big( 2 \eta \big) } \sigma^{-}_{N}                \\
         \mathrm{sin} {\tiny  \big( 2 \eta \big)  } \sigma^{+}_N       & \mathrm{sin} {\tiny  \big( \lambda_{\alpha} - v_N - \eta \sigma^z_N \big) } 
  \end{bmatrix} \begin{bmatrix}
          \mathrm{sin} {\tiny  \big( \lambda_{\alpha} - v_{N-1} + \eta \sigma^z_{N-1} \big)  }              &       \mathrm{sin} {\tiny  \big( 2 \eta \big) }  \sigma^{-}_{N-1}              \\
            \mathrm{sin} {\tiny  \big( 2 \eta \big) }  \sigma^{+}_{N-1}        &       \mathrm{sin} {\tiny  \big( \lambda_{\alpha} - v_{N-1} - \eta \sigma^z_{N-1} \big)  } 
  \end{bmatrix} \text{, } \] } 

\noindent from which one obtains a resultant matrix of the form,

{\small \[  \begin{bmatrix}
           \textbf{1}       &        \textbf{2}                \\
           \textbf{3}     &     \textbf{4} 
  \end{bmatrix}  \text{, } 
\] }

\noindent which has the following expression for the third, and fourth, entries, with, 

{\small \begin{align*}
 \textbf{3}^0 \equiv \textbf{3} \equiv    \mathrm{sin} {\tiny  \big( 2 \eta \big) }  \sigma^{+}_N \mathrm{sin} {\tiny  \big( \lambda_{\alpha} - v_{N-1} + \eta \sigma^z_{N-1} \big) }  + \mathrm{sin} {\tiny  \big( \lambda_{\alpha} - v_N - \eta \sigma^z_N \big)  } \mathrm{sin} {\tiny  \big( 2 \eta \big) }  \sigma^{+}_{N-1}  \text{, } \end{align*}
 
 \begin{align*} \textbf{4}^0 \equiv \textbf{4} \equiv   \mathrm{sin} {\tiny  \big( 2 \eta \big) } \sigma^{-}_{N-1} \mathrm{sin} {\tiny  \big( 2 \eta \big) }  \sigma^{-}_{N-1} + \mathrm{sin} {\tiny  \big( \lambda_{\alpha} - v_N - \eta \sigma^z_N \big)  } \mathrm{sin} {\tiny  \big( \lambda_{\alpha} - v_{N-1} - \eta \sigma^z_{N-1} \big)   }     \text{, }  
\end{align*}  } 

\noindent From the product of operators,

{\small \[  \begin{bmatrix}
           \textbf{1}       &        \textbf{2}                \\
           \textbf{3}     &     \textbf{4} 
  \end{bmatrix} \begin{bmatrix}
      \mathrm{sin} {\tiny  \big( \lambda_{\alpha} - v_{N-2} + \eta \sigma^z_{N-2} \big)   }                &        \mathrm{sin} {\tiny \big( 2 \eta \big)}  \sigma^{-}_{N-2}                \\
         \mathrm{sin} {\tiny  \big( 2 \eta \big) }  \sigma^{+}_{N-2}       & \mathrm{sin} {\tiny  \big( \lambda_{\alpha} - v_{N-2} - \eta \sigma^z_{N-2} \big) } 
  \end{bmatrix} \] \[ \cdots \times \begin{bmatrix}
          \mathrm{sin} {\tiny  \big( \lambda_{\alpha} - v_{1} + \eta \sigma^z_{1} \big)      }           &       \mathrm{sin} \big( 2 \eta \big) \sigma^{-}_{1}              \\
            \mathrm{sin} {\tiny  \big( 2 \eta \big) }  \sigma^{+}_{1}        &       \mathrm{sin} {\tiny  \big( \lambda_{\alpha} - v_{1} - \eta \sigma^z_{1} \big)  } 
  \end{bmatrix}   \text{, } 
\]  }

\noindent takes the desired form,

{\small \begin{align*}
   C_3 {\tiny  \big( \lambda_{\alpha} \big)  }  \equiv   \big( \mathrm{sin} {\tiny  \big( 2 \eta \big) }  \big)^3   \mathrm{sin} {\tiny  \big( \lambda_{\alpha} - v_{N-n} - \eta \sigma^z_{N-n} \big) }  \cdot         \underset{0 \leq i \leq 3}{\prod}    \mathrm{sin} {\tiny  \big( \lambda_{\alpha} - v_{N-i} - \eta \sigma^z_{N-j}  \big)  }     \end{align*}

 \begin{align*} 
   +   \big( \mathrm{sin} {\tiny  \big( 2 \eta \big) }  \big)^2  \bigg\{      \sigma^{+}_{N-i}   \bigg[ \mathrm{sin} {\tiny  \big(   \lambda_{\alpha} - v_{N-i} + \eta \sigma^z_{N-j}     \big)   }       \cdot  \underset{i \equiv 2 , -}{\underset{i \equiv 0 , +}{\underset{\mathrm{even \text{ } } i \text{ } : \text{ }  0 \leq i \leq 3}{\prod  }}  }   \sigma^{-,+}_{N-i}    \end{align*}

 \begin{align*} 
   +  \mathrm{sin} {\tiny  \big( \lambda_{\alpha} - v_{N-i} - \eta \sigma^z_{N-i} \big) }  \cdot   \underset{i \equiv 3 , +}{\underset{i \equiv 1 , -}{\underset{\mathrm{odd \text{ } } i \text{ } : 1 \leq i \leq 3}{\prod}}}             \sigma^{-,+}_{N-i}                      \bigg] \end{align*}

 \begin{align*} 
      + \mathrm{sin} {\tiny  \big( \lambda_{\alpha} - v_N - \eta \sigma^z_N \big) } \cdot \underset{i \equiv 3 , +}{\underset{i \equiv 2 , -}{\underset{i \equiv 1 , +}{\underset{1 \leq i \leq 3}{\prod}  }}  }\sigma^{-,+}_{N-i}     +   \sigma^{+}_{N-(n-3)}     \cdot     \underset{0 \leq i \leq 2}{\prod}    \mathrm{sin} {\tiny  \big( \lambda_{\alpha} - v_{N-i} - \eta \sigma^z_{N-j} \big)  }             \bigg\}    \end{align*}

 \begin{align*} +  \mathrm{sin} {\tiny \big( 2 \eta \big) } \bigg\{         \sigma^{+}_{N} \cdot  \underset{1 \leq i \leq 
        3    }{\prod}           \mathrm{sin} {\tiny  \big( \lambda_{\alpha} - v_{N-i} + \eta \sigma^z_{N-j} \big)   }       +  \mathrm{sin} {\tiny  \big( \lambda_{\alpha} - v_{N-n} + \eta \sigma^z_{N-n} \big) }  \end{align*}

 \begin{align*} \times     \underset{i \equiv 2 , +}{\underset{i \equiv 0,-}{\underset{\mathrm{even \text{ } } i\text{ }  : \text{ } 0 \leq i \leq 3}{\prod} }  }         \mathrm{sin} {\tiny  \big( \lambda_{\alpha} - v_{N-i} \pm \eta \sigma^z_{N-j} \big)   }         + \sigma^{+}_{N-1}  \cdot        \underset{0 \leq i \leq 3}{\prod} \mathrm{sin} {\tiny  \big( \lambda_{\alpha} - v_{N-i} - \eta \sigma^z_{N-j} \big)     }                                              \bigg\}             \text{ } \end{align*}

 \begin{align*} +   \big( \mathrm{sin} {\tiny  \big( 2 \eta \big)  } \big)^3  \mathrm{sin} {\tiny  \big( \lambda_{\alpha} - v_{N-3} - \eta \sigma^z_{N-3} \big) } \cdot         \underset{0 \leq i \leq 1}{\prod}    \mathrm{sin} {\tiny  \big( \lambda_{\alpha} - v_{N-i} - \eta \sigma^z_{N-j}  \big) }       \text{, } 
\end{align*} }

\noindent from which we conclude the argument. \boxed{}

 \subsection{Arguments for Lemma 4}

\noindent \textbf{Lemma} \textit{4} (\textit{collecting terms from the product of two by two L-operators}). The fourth entry of the $L$ operator product,

\begin{align*}
     \overset{3}{\underset{i=0}{\prod}}   L_{\alpha , N - i  } {\tiny\big( \lambda_{\alpha} , v_{N-i} \big) }      \text{, } 
\end{align*}

\noindent has an expansion of the form,

\begin{align*}
  D_3 {\tiny\big( \lambda_{\alpha} \big)} \equiv            \underset{0 \leq i \leq 3}{\prod}    \mathrm{sin} {\tiny\big( \lambda_{\alpha} - v_{N-i} - \eta \sigma^z_{N-j} \big)}  +    \big(       \mathrm{sin} {\tiny\big( 2 \eta \big)}           \big)^2  \bigg\{     \underset{i \equiv 2 , -}{\underset{i \equiv 0 , +}{\underset{\mathrm{even \text{ } } i: \text{ }  0 \leq i \leq 2}{\prod}  } }    \sigma^{-,+}_{N-i}      \end{align*}

 \begin{align*} \times   \underset{i \equiv 3, -}{\underset{i \equiv 1 , +}{\underset{\mathrm{ odd} \text{ } i \text{ } : \text{ } 1 \leq i \leq 3}{\prod}}}\mathrm{sin} {\tiny\big(   \lambda_{\alpha} - v_{N-i} \pm \eta \sigma^z_{N-j}     \big)  }    + 
 \underset{i \equiv 2 , -}{\underset{i \equiv 1 , +}{\underset{1 \leq i \leq 2}{\prod}}   }    \sigma^{-,+}_{N-i}  \cdot   \underset{i \equiv 0, i \equiv 3}{\prod}     \mathrm{sin} {\tiny\big(    \lambda_{\alpha} - v_N - \eta \sigma^z_N  \big)   }      \end{align*}

 \begin{align*}    +   \underset{i \equiv 1}{\prod}   \sigma^{-}_{N-i} \cdot   \mathrm{sin} {\tiny\big( \lambda_{\alpha} - v_{N-2} - \eta \sigma^z_{N-2} \big)}           \bigg\}   
  +           \big( \mathrm{sin} {\tiny \big( 2 \eta \big)} \big)^3 \cdot  \underset{i \equiv 1}{\prod}       \big( \sigma^{-}_{N-i}  \big)^2 \cdot    \sigma^{+}_{N-i-1}  \end{align*}

 \begin{align*} +     \big( \mathrm{sin} {\tiny\big( 2 \eta \big)} \big)^2         \bigg[      \text{ }              \sigma^{+}_N \sigma^{+}_{N-3}   \cdot  \mathrm{sin} {\tiny\big( \lambda_{\alpha} - v_{N-i} + \eta \sigma^z_{N-j} \big)}        + 
  \sigma^{-}_{N-3} \cdot \underset{i \equiv 2 , + \eta}{\underset{i \equiv 0 , - \eta}{\underset{\mathrm{even \text{ } } i : \text{ } 0 \leq i \leq 2}{\prod} }}           \mathrm{sin} {\tiny\big( \lambda_{\alpha} - v_{N-i} \pm \eta \sigma^z_{N-j} \big)} \end{align*}

  \begin{align*} + \sigma^{+}_{N-2} \sigma^{-}_{N-3}  \cdot \underset{0 \leq i \leq 1}{\prod}  \mathrm{sin} {\tiny\big( \lambda_{\alpha} - v_{N-i} - \eta \sigma^z_{N-j} \big)}                                   \bigg]      \text{,
}
\end{align*}

\noindent while the fourth entry of the L-operator product,

\begin{align*}
    \prod_{i=0}^2 L_{\alpha , N-i} {\tiny\big( \lambda_{\alpha}, v_{N-i} \big)}  \text{, }
\end{align*}

\noindent has an expansion of the form,

{\small \begin{align*}
  D_2 {\tiny \big( \lambda_{\alpha} \big)}  \equiv        \big( \mathrm{sin} {\tiny \big( 2 \eta \big)}  \big)^2 \cdot   \underset{i \equiv 2 , -}{\underset{i \equiv 0 , +}{\underset{\mathrm{even \text{ } } i: \text{ }  0 \leq i \leq 2}{\prod}  } }    \sigma^{-,+}_{N-i}  \cdot \mathrm{sin} {\tiny\big(   \lambda_{\alpha} - v_{N-1} + \eta \sigma^z_{N-1}     \big) }                      + \big( \mathrm{sin} {\tiny\big( 2 \eta \big)}  \big)^2 \cdot \underset{i \equiv 2 , -}{\underset{i \equiv 1 , +}{\underset{1 \leq i \leq 2}{\prod}}   }    \sigma^{-,+}_{N-i}   \text{ }   \end{align*}

  \begin{align*} \times  \mathrm{sin} {\tiny\big(    \lambda_{\alpha} - v_N - \eta \sigma^z_N  \big) }        +           \big( \mathrm{sin} {\tiny\big( 2 \eta \big) }\big)^2 \cdot  \underset{i \equiv 1}{\prod}   \sigma^{-}_{N-i} \cdot    \mathrm{sin} {\tiny\big( \lambda_{\alpha} - v_{N-2} - \eta \sigma^z_{N-2} \big) }    \end{align*}

  \begin{align*}   +       \underset{0 \leq i \leq 2}{\prod}    \mathrm{sin} {\tiny\big( \lambda_{\alpha} - v_{N-i} - \eta \sigma^z_{N-j} \big) }   \text{. } 
\end{align*} }

\bigskip

\noindent With expressions for $A_3 {\tiny\big( \lambda_{\alpha} \big)}$, $B_3 {\tiny\big( \lambda_{\alpha} \big)}$, $C_3 {\tiny\big( \lambda_{\alpha} \big)}$ and $D_3 {\tiny\big( \lambda_{\alpha} \big)}$, below we obtain other expressions for,

\[  \begin{bmatrix}
           A_{n} {\tiny\big( \lambda_{\alpha} \big)    }   &    B_n {\tiny\big( \lambda_{\alpha} \big)}                     \\
           C_n {\tiny\big( \lambda_{\alpha} \big)}     &   D_n {\tiny\big( \lambda_{\alpha} \big)   } 
  \end{bmatrix}  \text{, } 
\]

\noindent in terms of,

\[  \begin{bmatrix}
             \textbf{1}^n   &             \textbf{2}^n      \\
           \textbf{3}^n   &  \textbf{4}^n    
  \end{bmatrix}  \text{. } 
\]

\bigskip

\noindent \textit{Proof of Lemma 4}. We collect terms from the first two terms in the product of $L$ operators,

\[
\begin{bmatrix}
      \mathrm{sin} {\tiny \big( \lambda_{\alpha} - v_N + \eta \sigma^z_{N} \big) }                   &        \mathrm{sin} {\tiny \big( 2 \eta \big) } \sigma^{-}_{N}                \\
         \mathrm{sin} {\tiny  \big( 2 \eta \big) } \sigma^{+}_N       & \mathrm{sin} {\tiny  \big( \lambda_{\alpha} - v_N - \eta \sigma^z_N \big) } 
  \end{bmatrix} \begin{bmatrix}
          \mathrm{sin} {\tiny  \big( \lambda_{\alpha} - v_{N-1} + \eta \sigma^z_{N-1} \big)  }               &       \mathrm{sin} {\tiny  \big( 2 \eta \big)  } \sigma^{-}_{N-1}              \\
            \mathrm{sin} {\tiny  \big( 2 \eta \big) }  \sigma^{+}_{N-1}        &       \mathrm{sin} {\tiny  \big( \lambda_{\alpha} - v_{N-1} - \eta \sigma^z_{N-1} \big) } 
  \end{bmatrix}  \text{, }\]

\noindent from which one obtains a resultant matrix of the form,

\[  \begin{bmatrix}
           \textbf{1}       &        \textbf{2}                \\
           \textbf{3}     &     \textbf{4} 
  \end{bmatrix}   \text{, } 
\]

\noindent from which the desired expression for the final entry of the monodromy matrix, $D_3 {\tiny  \big( \lambda_{\alpha} \big) } $, takes the form,

{\small \begin{align*}
  D_3 {\tiny \big( \lambda_{\alpha} \big) }  \equiv            \underset{0 \leq i \leq 3}{\prod}    \mathrm{sin} {\tiny \big( \lambda_{\alpha} - v_{N-i} - \eta \sigma^z_{N-j} \big)}   +  \big(       \mathrm{sin} {\tiny \big( 2 \eta \big)  }         \big)^2  \bigg\{    \underset{i \equiv 2 , -}{\underset{i \equiv 0 , +}{\underset{\mathrm{even \text{ } } i: \text{ }  0 \leq i \leq 2}{\prod}  } }    \sigma^{-,+}_{N-i}       \end{align*}
 
 \begin{align*}  \times   \text{ } \underset{i \equiv 3, -}{\underset{i \equiv 1 , +}{\underset{\mathrm{ odd} \text{ } i \text{ } : \text{ } 1 \leq i \leq 3}{\prod}}}\mathrm{sin} {\tiny  \big(   \lambda_{\alpha} - v_{N-i}   \pm \eta \sigma^z_{N-j}     \big)    }          +     \text{ } \underset{i \equiv 2 , -}{\underset{i \equiv 1 , +}{\underset{1 \leq i \leq 2}{\prod}}   }    \sigma^{-,+}_{N-i}   \cdot  \text{ }   \underset{i \equiv 0, i \equiv 3}{\prod}     \mathrm{sin} {\tiny  \big(    \lambda_{\alpha} - v_N - \eta \sigma^z_N  \big) }     +       \underset{i \equiv 1}{\prod}   \sigma^{-}_{N-i} \end{align*}
 
 \begin{align*}  \times   \underset{2 \leq i \leq 3}{\prod} \mathrm{sin}   {\tiny  \big( \lambda_{\alpha} - v_{N-2} - \eta \sigma^z_{N-2} \big) }         \bigg\}    + 
  \big( \mathrm{sin} {\tiny  \big( 2 \eta \big) }  \big)^3   \cdot  \underset{i \equiv 1}{\prod}       \big( \sigma^{-}_{N-i}  \big)^2   \cdot \sigma^{+}_{N-i-1} +  \big( \mathrm{sin} {\tiny  \big( 2 \eta \big) }  \big)^2         \bigg\{                    \sigma^{+}_N \sigma^{+}_{N-3}  \end{align*}
 
 \begin{align*}  \times    \underset{1 \leq i \leq 2}{\prod} \mathrm{sin} {\tiny  \big( \lambda_{\alpha} - v_{N-i}  + \eta \sigma^z_{N-j} \big) }  
 + \sigma^{-}_{N-3}  \cdot \underset{i \equiv 2 , + \eta}{\underset{i \equiv 0 , - \eta}{\underset{\mathrm{even \text{ } } i : \text{ } 0 \leq i \leq 2}{\prod} }}           \mathrm{sin} {\tiny  \big( \lambda_{\alpha} - v_{N-i} \pm \eta \sigma^z_{N-j} \big) }   + 
 \sigma^{+}_{N-2} \sigma^{-}_{N-3} \end{align*}
 
 \begin{align*}  \times   \text{ } \underset{0 \leq i \leq 1}{\prod}  \mathrm{sin} {\tiny  \big( \lambda_{\alpha} - v_{N-i}   - \eta \sigma^z_{N-j} \big) }                                    \bigg\}   \text{, 
}
\end{align*} }

\noindent from which we conclude the argument. \boxed{}

\subsection{Arguments for Lemma 5}

\noindent \textbf{Lemma} \textit{5} (\textit{iteratively obtaining the entries of the n th L-operator from the entries of the third L-operator}). The first entry of the $n$ th L-operator can be expressed in terms of, 

\[
\left\{\!\begin{array}{ll@{}>{{}}l}    \boxed{A_n {\tiny\big( \lambda_{\alpha} \big) }} =   \bigg[      A_3 {\tiny\big( \lambda_{\alpha} \big) }+ B_3 {\tiny\big( \lambda_{\alpha} \big)} \bigg]   \big( \mathrm{sin} {\tiny\big( 2 \eta \big)}  \big)^{n-3} \bigg[  \text{ }     \underset{1 \leq i \leq n-3}{\prod}     \sigma^{-,+}_{n-i}   \bigg]   +  \bigg[  A_3 {\tiny\big( \lambda_{\alpha} \big)}  + B_3 {\tiny\big( \lambda_{\alpha } \big)} \bigg]     \end{array}\right.
\]

\[
\left\{\!\begin{array}{ll@{}>{{}}l}     \cdot  \underset{1 \leq i \leq n-3}{\prod}   \mathrm{sin} {\tiny\big( \lambda_{\alpha} - v_{n-i} +    \eta \sigma^z_{n-j}     \big)}   +   \bigg[ A_3 {\tiny\big( \lambda_{\alpha}\big)  } +  B_3 {\tiny\big( \lambda_{\alpha} \big)} \bigg] \\  \times  \bigg\{  \text{ }  \underset{m,n^{\prime}: m+n^{\prime} = n-3}{\sum} \bigg[    \text{ }   \bigg[  \text{ }   \underset{1 \leq i \leq m}{\prod} \mathrm{sin} {\tiny\big( \lambda_{\alpha} - v_{n-i} \pm \eta \sigma^z_{n-j} \big) }  \bigg]   \cdot  \big( \mathrm{sin} {\tiny\big( 2 \eta \big)} \big)^{n^{\prime}-1}  \cdot    \underset{1 \leq j \leq n^{\prime}}{ \prod}  \sigma^{-,+}_{n-j}             \bigg]       \bigg\}                                    , \\ \\ \\   \boxed{B_n {\tiny\big( \lambda_{\alpha} \big) }} =     \bigg[      A_3 {\tiny\big( \lambda_{\alpha} \big)} + B_3 {\tiny\big( \lambda_{\alpha} \big)} \bigg]   \big( \mathrm{sin} {\tiny\big( 2 \eta \big)} \big)^{n-4} \bigg[ \text{ }     \underset{2 \leq i \leq n-3}{\prod}     \sigma^{-,+}_{n-i}   \bigg] 
 +  \bigg[  A_3 {\tiny\big( \lambda_{\alpha} \big) } + B_3 {\tiny\big( \lambda_{\alpha } \big) }\bigg] 
\\ \cdot \underset{2 \leq i \leq n-3}{\prod}   \mathrm{sin} {\tiny\big( \lambda_{\alpha} - v_{n-i} +    \eta \sigma^z_{n-j}     \big)} +   \bigg[ A_3 {\tiny\big( \lambda_{\alpha}\big)}  +  B_3 {\tiny\big( \lambda_{\alpha} \big)}  \bigg] \\  \times  \bigg\{  \text{ }   \underset{m,n^{\prime}: m+n^{\prime} = n-3}{\sum}   \bigg[    \text{ }   \bigg[ \text{ }   \underset{2 \leq i \leq m}{\prod} \mathrm{sin} {\tiny \big( \lambda_{\alpha} - v_{n-i} \pm \eta \sigma^z_{n-j} \big)  }  \bigg] \cdot \big( \mathrm{sin} {\tiny\big( 2 \eta \big) }\big)^{n^{\prime}-1} \cdot    \underset{2 \leq j \leq n^{\prime}}{ \prod}  \sigma^{-,+}_{n-j}     \text{ }      \bigg]     \bigg\}        , \\ \\  \\    \boxed{C_n {\tiny\big( \lambda_{\alpha} \big)}}  =   \bigg[      C_3 {\tiny\big( \lambda_{\alpha} \big) } + D_3 {\tiny\big( \lambda_{\alpha} \big) }\bigg]   \big( \mathrm{sin} {\tiny\big( 2 \eta \big)} \big)^{n-3} \bigg[ \text{ }     \underset{1 \leq i \leq n-3}{\prod}     \sigma^{-,+}_{n-i}   \bigg]  +  \bigg[  C_3 {\tiny\big( \lambda_{\alpha} \big)}  + D_3 {\tiny\big( \lambda_{\alpha } \big)} \bigg]  \\  \cdot  \underset{1 \leq i \leq n-3}{\prod}   \mathrm{sin} {\tiny\big( \lambda_{\alpha} - v_{n-i} +    \eta \sigma^z_{n-j}     \big)}    +   \bigg[ C_3 {\tiny\big( \lambda_{\alpha}\big) } +  D_3 {\tiny\big( \lambda_{\alpha} \big)} \bigg] \\ \cdot \bigg\{  \text{ }  \underset{m,n^{\prime}: m+n^{\prime} = n-3}{\sum}   \bigg[   \text{ }    \bigg[ \text{ }   \underset{1 \leq i \leq m}{\prod} \mathrm{sin} {\tiny\big( \lambda_{\alpha} - v_{n-i} \pm \eta \sigma^z_{n-j} \big)}    \bigg] \cdot \big( \mathrm{sin} {\tiny\big( 2 \eta \big)} \big)^{n^{\prime}-1}  \cdot  \underset{1 \leq j \leq n^{\prime}}{ \prod}  \sigma^{-,+}_{n-j}   \text{ }         \bigg]   \bigg\}              , \\ \\ \\  \boxed{D_n {\tiny\big( \lambda_{\alpha} \big)}}  =     \bigg[      C_3 {\tiny\big( \lambda_{\alpha} \big) }+ D_3 {\tiny\big( \lambda_{\alpha} \big)} \bigg]   \big( \mathrm{sin} {\tiny\big( 2 \eta \big) }\big)^{n-4}  \bigg[ \text{ }     \underset{2 \leq i \leq n-3}{\prod}     \sigma^{-,+}_{n-i}   \bigg] 
 + \bigg[  C_3 {\tiny\big( \lambda_{\alpha} \big) } + D_3 {\tiny\big( \lambda_{\alpha } \big)} \bigg] \\   \cdot \underset{2 \leq i \leq n-3}{\prod}   \mathrm{sin} {\tiny\big( \lambda_{\alpha} - v_{n-i} +    \eta \sigma^z_{n-j}     \big)}     +   \bigg[ C_3 {\tiny\big( \lambda_{\alpha}\big)}  +  D_3 {\tiny\big( \lambda_{\alpha} \big)} \bigg] \\  \times   \bigg\{ \text{ }   \underset{m,n^{\prime}: m+n^{\prime} = n-3}{\sum}   \bigg[  \text{ }     \bigg[ \text{ }   \underset{2 \leq i \leq m}{\prod} \mathrm{sin} {\tiny\big( \lambda_{\alpha} - v_{n-i} \pm \eta \sigma^z_{n-j} \big) }  \bigg] \cdot  \big( \mathrm{sin} {\tiny\big( 2 \eta \big)} \big)^{n^{\prime}-1} \cdot   \underset{2 \leq j \leq n^{\prime}}{ \prod}  \sigma^{-,+}_{n-j}     \bigg]        \bigg\}   . 
 \end{array}\right.
\]

\bigskip

\noindent With the above expressions, 

{\small \[  \begin{bmatrix}
           A_{n} {\tiny\big( \lambda_{\alpha} \big)    }   &    B_n {\tiny\big( \lambda_{\alpha} \big) }                    \\
           C_n {\tiny\big( \lambda_{\alpha} \big) }    &   D_n {\tiny\big( \lambda_{\alpha} \big)   } 
  \end{bmatrix} \equiv  \begin{bmatrix}
             \textbf{1}^n   &             \textbf{2}^n      \\
           \textbf{3}^n   &  \textbf{4}^n    
  \end{bmatrix} \equiv    \prod_{1 \leq i \leq n} \begin{bmatrix}
             \textbf{1}^i   &             \textbf{2}^i      \\
           \textbf{3}^i   &  \textbf{4}^i    
  \end{bmatrix} 
  \text{, } 
\] }

\noindent in terms of,

{\small \[      \begin{bmatrix}
 \bigg\{      A_3 {\tiny\big( \lambda_{\alpha} \big)} + B_3 {\tiny\big( \lambda_{\alpha} \big) }\bigg\} \cdot  \bigg\{   \big( \mathrm{sin} {\tiny\big( 2 \eta \big)} \big)^{n-3} \mathscr{A}_1  + \mathscr{A}_2 +  \mathscr{A}_3 \bigg\}  &  \bigg\{      A_3 {\tiny\big( \lambda_{\alpha} \big)} + B_3 {\tiny\big( \lambda_{\alpha} \big) }  \bigg\} \cdot  \bigg\{  \big( \mathrm{sin} {\tiny\big( 2 \eta \big) }\big)^{n-3} \mathscr{B}_1 + \mathscr{B}_2 + \mathscr{B}_3  \bigg\} \\   \\ \bigg\{         C_3 {\tiny\big( \lambda_{\alpha} \big)} + D_3 {\tiny\big( \lambda_{\alpha} \big)   }    \bigg\} \cdot   \bigg\{     \big( \mathrm{sin} {\tiny\big( 2 \eta \big) }\big)^{n-3} \mathscr{C}_1 + \mathscr{C}_2 + \mathscr{C}_3  \bigg\}  & \bigg\{    C_3 {\tiny\big( \lambda_{\alpha} \big) } + D_3 {\tiny\big( \lambda_{\alpha} \big)  }  \bigg\} \cdot  
 \bigg\{     \big( \mathrm{sin} {\tiny\big( 2 \eta \big)}  \big)^{n-3}\mathscr{D}_1 + \mathscr{D}_2 + \mathscr{D}_3   \bigg\}  \end{bmatrix}       \text{, }  \\ 
\] } 

\noindent for,

{\small \[
\left\{\!\begin{array}{ll@{}>{{}}l} 
   \mathscr{A}_1 \equiv  \mathscr{C}_1 \equiv     \underset{1 \leq i \leq n-3}{\prod} \big( \mathscr{C}_1 \big)_i  \equiv  \text{ }     \underset{1 \leq i \leq n-3}{\prod}     \sigma^{-,+}_{n-i}   \text{, }     \end{array}\right.
\]

\[
\left\{\!\begin{array}{ll@{}>{{}}l}     \mathscr{A}_2 \equiv    \mathscr{C}_2 \equiv \underset{1 \leq i \leq n-3}{\prod}  \big( \mathscr{A}_2 \big)_i   \equiv \underset{1 \leq i \leq n-3}{\prod}  \big( \mathscr{C}_2 \big)_i  \equiv           \underset{1 \leq i \leq n-3}{\prod}   \mathrm{sin} {\tiny\big( \lambda_{\alpha} - v_{n-i} +    \eta \sigma^z_{n-j}     \big)  }     \text{, } \\ \\   \mathscr{A}_3  \equiv   \mathscr{C}_3 \equiv \underset{m,n^{\prime}: m+n^{\prime} = n-3}{\sum} \bigg[ \underset{1 \leq j \leq n^{\prime}}{\underset{1 \leq i \leq m}{\prod}}  \big( \mathscr{C}_3 \big)_{i,j} \bigg]  \equiv   \text{ }  \underset{m,n^{\prime}: m+n^{\prime} = n-3}{\sum} \bigg\{    \bigg[ \text{ }   \underset{1 \leq i \leq m}{\prod} \mathrm{sin} {\tiny\big( \lambda_{\alpha} - v_{n-i} \pm \eta \sigma^z_{n-j} \big)}    \bigg] \\ \times  \big( \mathrm{sin} {\tiny\big( 2 \eta \big) } \big)^{n^{\prime}-1}  \bigg[ \text{ }   \underset{1 \leq j \leq n^{\prime}}{ \prod}  \sigma^{-,+}_{n-j}     \bigg]         \bigg\}      \text{, } 
   \end{array}\right.
\]  

\noindent corresponding to the first and third entries,

{\small \[
\left\{\!\begin{array}{ll@{}>{{}}l}  
   \mathscr{B}_1 \equiv  \mathscr{D}_1 \equiv   \underset{1 \leq i \leq n-3}{\prod}  \big( \mathscr{D}_1 \big)_i \equiv  \text{ }     \underset{2 \leq i \leq n-3}{\prod}     \sigma^{-,+}_{n-i} \text{, } \\ \\  \mathscr{B}_2 \equiv  \mathscr{D}_2 \equiv \underset{2 \leq i \leq n-3}{\prod}  \big( \mathscr{B}_2 \big)_i \equiv \underset{2 \leq i \leq n-3}{\prod} \big( \mathscr{D}_2 \big)_i  \equiv               \underset{2 \leq i \leq n-3}{\prod}   \mathrm{sin} {\tiny\big( \lambda_{\alpha} - v_{n-i} +    \eta \sigma^z_{n-j}     \big) }   \text{, }  \\ \\  \mathscr{B}_3  \equiv  \mathscr{D}_3 \equiv       \underset{m,n^{\prime}: m+n^{\prime} = n-3}{\sum}  \bigg[ \underset{ 2\leq j \leq n^{\prime}}{\underset{2 \leq i \leq m}{\prod}} \big( \mathscr{D}_3 \big)_{i,j}   \bigg]     \equiv    \text{ }   \underset{m,n^{\prime}: m+n^{\prime} = n-3}{\sum}  \bigg\{       \bigg[ \text{ }   \underset{2 \leq i \leq m}{\prod} \mathrm{sin} {\tiny\big( \lambda_{\alpha} - v_{n-i} \pm \eta \sigma^z_{n-j} \big)}   \bigg] \text{ } \\  \times \big( \mathrm{sin} {\tiny\big( 2 \eta \big)}  \big)^{n^{\prime}-1}   \bigg[ \text{ }   \underset{2 \leq j \leq n^{\prime}}{ \prod}  \sigma^{-,+}_{n-j}     \bigg]          \bigg\}     \text{, } 
   \end{array}\right.
\]  

\noindent corresponding to the second and fourth entries for $A_n {\tiny\big( \lambda_{\alpha} \big)}$, $B_n {\tiny\big( \lambda_{\alpha} \big)}$, $C_n {\tiny\big( \lambda_{\alpha} \big)}$ and $D_n {\tiny\big( \lambda_{\alpha} \big)}$.

\bigskip

\noindent \textit{Proof of Lemma 5}. Recall, from previous expansions of L-operators provided in the results above, that, 

{\small \[
\prod_{0 \leq i \leq 3} \begin{bmatrix}
\textbf{1}^i & \textbf{2}^i \\ \textbf{3}^i & \textbf{4}^i
\end{bmatrix} \text{, } \]} 

\noindent has the expansion,

{\small \[
\begin{bmatrix}
A_3 {\tiny\big( \lambda_{\alpha} \big)} \textbf{1}^3 + B_3 {\tiny\big( \lambda_{\alpha} \big) }\textbf{3}^3  & A_3 {\tiny\big( \lambda_{\alpha} \big)} \textbf{2}^3 + B_3 {\tiny\big( \lambda_{\alpha} \big)} \textbf{4}^3 \\  \\ C_3 {\tiny\big( \lambda_{\alpha} \big) }\textbf{1}^3 + D_3 {\tiny\big( \lambda_{\alpha} \big)} \textbf{3}^3 & C_3 {\tiny\big( \lambda_{\alpha} \big)} \textbf{2}^3 + D_3 {\tiny\big( \lambda_{\alpha} \big)}  \textbf{4}^3 
\end{bmatrix} \text{. }
\] }

\noindent We demonstrate that such an expansion as the one provided above holds from the following computatons. 
By direct computation, we begin by computing the product of L-operators,

\[ \begin{bmatrix}
             \textbf{1}^{0}   &             \textbf{2}^{0}      \\
           \textbf{3}^{0}   &  \textbf{4}^{0}    
  \end{bmatrix} \begin{bmatrix}
             \textbf{1}^{1}   &             \textbf{2}^{1}      \\
           \textbf{3}^{1}   &  \textbf{4}^{1}    
  \end{bmatrix} \begin{bmatrix}
             \textbf{1}^{2}   &             \textbf{2}^{2}      \\
           \textbf{3}^{2}   &  \textbf{4}^{2}    
  \end{bmatrix} \begin{bmatrix} \textbf{1}^3 & \textbf{2}^3 \\ \textbf{3}^3 & \textbf{4}^3
  \end{bmatrix} \text{, } \]

\noindent after which we compute the product of three arbitrary L-operators, 

\[ \begin{bmatrix}
             \textbf{1}^{i-3}   &             \textbf{2}^{i-3}      \\
           \textbf{3}^{i-3}   &  \textbf{4}^{i-3}    
  \end{bmatrix}    \begin{bmatrix}
             \textbf{1}^{i-2}   &             \textbf{2}^{i-2}      \\
           \textbf{3}^{i-2}   &  \textbf{4}^{i-2}    
  \end{bmatrix}    \begin{bmatrix}
             \textbf{1}^{i-1}   &             \textbf{2}^{i-1}      \\
           \textbf{3}^{i-1}   &  \textbf{4}^{i-1}    
  \end{bmatrix}     \text{, } 
\]

\noindent from the $n$ fold product,

\[ \prod_{1 \leq i \leq n} \begin{bmatrix}
             \textbf{1}^i   &             \textbf{2}^i      \\
           \textbf{3}^i   &  \textbf{4}^i    
  \end{bmatrix}       \text{. } 
\]

\noindent For the first product of two by two L-operators, for $i \equiv 0$, $i \equiv 1$, $i \equiv 2$, and $i \equiv 3$, by brute force one has,

\[ \begin{bmatrix}
             \textbf{1}^{0}\textbf{1}^1 + \textbf{2}^0 \textbf{3}^1    &             \textbf{1}^0 \textbf{2}^1 + \textbf{2}^0 \textbf{4}^1       \\
           \textbf{1}^1 \textbf{3}^0 + \textbf{3}^1 \textbf{4}^0  &    \textbf{2}^1 \textbf{3}^0  + \textbf{4}^0 \textbf{4}^1   
  \end{bmatrix} \begin{bmatrix}
             \textbf{1}^{2}   &             \textbf{2}^{2}      \\
           \textbf{3}^{2}   &  \textbf{4}^{2}    
  \end{bmatrix} \] \begin{align*}
 \end{align*} \[ \times  \begin{bmatrix} \textbf{1}^3 & \textbf{2}^3 \\ \textbf{3}^3 & \textbf{4}^3
  \end{bmatrix} \equiv 
\begin{bmatrix}
 {\tiny  \big( } \textbf{1}^{0}\textbf{1}^1 + \textbf{2}^0 \textbf{3}^1 {\tiny \big)  } \textbf{1}^2 + {\tiny \big( }  \textbf{1}^0 \textbf{2}^1 + \textbf{2}^0 \textbf{4}^1 {\tiny \big) }  \textbf{3}^2  &         {\tiny  \big( }  \textbf{1}^{0}\textbf{1}^1 + \textbf{2}^0 \textbf{3}^1 {\tiny \big)  } \textbf{2}^2 + {\tiny \big(  }  \textbf{1}^0 \textbf{2}^1 + \textbf{2}^0 \textbf{4}^1 {\tiny \big) } \textbf{4}^2 \\ {\tiny  \big(  } \textbf{1}^1 \textbf{3}^0 + \textbf{3}^1 \textbf{4}^0  {\tiny  \big) } \textbf{1}^2 + {\tiny  \big( } \textbf{2}^1 \textbf{3}^0 + \textbf{4}^0 \textbf{4}^1 {\tiny \big) } \textbf{3}^2   & {\tiny \big(} \textbf{1}^1 \textbf{3}^0 + \textbf{3}^1 \textbf{4}^0   {\tiny \big)} \textbf{2}^2 + {\tiny \big( }\textbf{2}^1 \textbf{3}^0 + \textbf{4}^0 \textbf{4}^1 {\tiny \big) } \textbf{4}^2 
\end{bmatrix}  \begin{bmatrix} \textbf{1}^3 & \textbf{2}^3 \\ \textbf{3}^3 & \textbf{4}^3
  \end{bmatrix}  \text{. } 
\]

\noindent From previous arguments given in \textbf{Lemmas} \textit{1}-\textit{4}, the fact that,

{\small \[
\begin{bmatrix}
  A_3 {\tiny  \big( \lambda_{\alpha} \big) }   & B_3 {\tiny \big( \lambda_{\alpha} \big)  } \\ C_3 {\tiny  \big( \lambda_{\alpha} \big) } & D_3 {\tiny \big( \lambda_{\alpha} \big) } 
\end{bmatrix} \text{, }
\]

\[ = 
\begin{bmatrix}
 {\tiny  \big( } \textbf{1}^{0}\textbf{1}^1 + \textbf{2}^0 \textbf{3}^1 {\tiny \big)}  \textbf{1}^2 + {\tiny \big( }  \textbf{1}^0 \textbf{2}^1 + \textbf{2}^0 \textbf{4}^1 {\tiny  \big) } \textbf{3}^2  &          {\tiny \big( }  \textbf{1}^{0}\textbf{1}^1 + \textbf{2}^0 \textbf{3}^1 {\tiny \big) } \textbf{2}^2 + {\tiny \big(  }  \textbf{1}^0 \textbf{2}^1 + \textbf{2}^0 \textbf{4}^1 {\tiny \big) }  \textbf{4}^2 \\ {\tiny  \big( }  \textbf{1}^1 \textbf{3}^0 + \textbf{3}^1 \textbf{4}^0  {\tiny  \big) } \textbf{1}^2 + {\tiny \big( }  \textbf{2}^1 \textbf{3}^0 + \textbf{4}^0 \textbf{4}^1 {\tiny \big)} \textbf{3}^2   & {\tiny \big(}  \textbf{1}^1 \textbf{3}^0 + \textbf{3}^1 \textbf{4}^0   {\tiny  \big) }\textbf{2}^2 + {\tiny \big( }  \textbf{2}^1 \textbf{3}^0 + \textbf{4}^0 \textbf{4}^1 {\tiny \big)}  \textbf{4}^2 
\end{bmatrix}  \text{. } 
\] }

\noindent In terms of $A_3 \big( \lambda_{\alpha} \big)$, $B_3 \big( \lambda_{\alpha} \big)$, $C_3 \big( \lambda_{\alpha} \big)$ and $D_3 \big( \lambda_{\alpha} \big)$, performing the final multiplication of L-operators implies,

\[
\begin{bmatrix}
  A_3 {\tiny  \big( \lambda_{\alpha} \big)  } & B_3 {\tiny \big( \lambda_{\alpha} \big) } \\ C_3 {\tiny \big( \lambda_{\alpha} \big) }  & D_3 {\tiny \big( \lambda_{\alpha} \big) } 
\end{bmatrix} \begin{bmatrix} \textbf{1}^3 & \textbf{2}^3 \\ \textbf{3}^3 & \textbf{4}^3
  \end{bmatrix} = 
\begin{bmatrix}
A_3 {\tiny  \big( \lambda_{\alpha} \big) } \cdot \textbf{1}^3 + B_3 {\tiny  \big( \lambda_{\alpha} \big) } \cdot  \textbf{3}^3  & A_3 {\tiny  \big( \lambda_{\alpha} \big) } \cdot  \textbf{2}^3 + B_3 {\tiny  \big( \lambda_{\alpha} \big)} \cdot  \textbf{4}^3  \\ C_3 {\tiny  \big( \lambda_{\alpha} \big) } \cdot \textbf{1}^3 + D_3 {\tiny  \big( \lambda_{\alpha} \big) } \cdot  \textbf{3}^3 & C_3 {\tiny  \big( \lambda_{\alpha} \big) } \cdot  \textbf{2}^3 + D_3 {\tiny  \big( \lambda_{\alpha} \big) } \cdot  \textbf{4}^3 
\end{bmatrix} \text{. } 
\]

\noindent Hence,

\[
\prod_{0 \leq i \leq 3} \begin{bmatrix}
\textbf{1}^i & \textbf{2}^i \\ \textbf{3}^i & \textbf{4}^i
\end{bmatrix} \equiv \begin{bmatrix}
A_3 {\tiny  \big( \lambda_{\alpha} \big)} \cdot  \textbf{1}^3 + B_3 {\tiny  \big( \lambda_{\alpha} \big)} \cdot  \textbf{3}^3  & A_3 {\tiny  \big( \lambda_{\alpha} \big) } \cdot  \textbf{2}^3 + B_3 {\tiny  \big( \lambda_{\alpha} \big)} \cdot  \textbf{4}^3  \\ C_3 {\tiny  \big( \lambda_{\alpha} \big)} \cdot  \textbf{1}^3 + D_3 {\tiny  \big( \lambda_{\alpha} \big)  } \cdot \textbf{3}^3 & C_3 {\tiny  \big( \lambda_{\alpha} \big) } \cdot  \textbf{2}^3 + D_3 {\tiny  \big( \lambda_{\alpha} \big) } \cdot \textbf{4}^3 
\end{bmatrix}  \text{. } 
\]

\noindent With such an expression, we make use of the following relations,

{\small \[
\left\{\!\begin{array}{ll@{}>{{}}l}   
     A_3 {\tiny  \big( \lambda_{\alpha} \big) } = A_2 {\tiny  \big( \lambda_{\alpha} \big) }  \textbf{1}^2   +    B_2 {\tiny  \big( \lambda_{\alpha} \big) } \textbf{3}^2 \equiv  A_2 {\tiny  \big( \lambda_{\alpha} \big)}  \mathrm{sin} {\tiny  \big( \lambda_{\alpha} - v_2 + \eta \sigma^z_2 \big) }   +    B_2 \big( \lambda_{\alpha} \big) \mathrm{sin} {\tiny \big( 2 \eta \big) }  \sigma^{+}_2  \text{, } \\   \vdots  \\   D_3 {\tiny  \big( \lambda_{\alpha} \big) }  = C_2 {\tiny  \big( \lambda_{\alpha} \big)  } \textbf{2}^2 + D_2 {\tiny  \big( \lambda_{\alpha} \big) }  \textbf{4}^2  \equiv  C_2 {\tiny  \big( \lambda_{\alpha} \big) }  \mathrm{sin} {\tiny  \big( 2 \eta \big)  } \sigma^{-}_2  + D_2 {\tiny  \big( \lambda_{\alpha} \big)   }    \mathrm{sin} {\tiny  \big( \lambda_{\alpha} - v_2 - \eta \sigma^z_{2} \big)   }   \text{. } 
   \end{array}\right.
\]    } 

\noindent To demonstrate the following arguments for \textbf{Lemma} \textit{6} can be used to deduce that \textbf{Lemma} \textit{5} holds, we perform the following computations.

\bigskip

\noindent Observe, for the first entry of the $n$ th L-operator,

  \[\prod_{1 \leq i \leq n} \begin{bmatrix}
             \textbf{1}^i   &             \textbf{2}^i      \\
           \textbf{3}^i   &  \textbf{4}^i    
  \end{bmatrix}  \text{, } \]

\noindent that,

{\small \begin{align*}
   A_n {\tiny  \big( \lambda_{\alpha} \big) } = A_{n-1} {\tiny  \big( \lambda_{\alpha} \big) }   \textbf{1}^{n-1}   +    B_{n-1} {\tiny  \big( \lambda_{\alpha} \big) }  \textbf{3}^{n-1} \equiv  A_{n-1} {\tiny  \big( \lambda_{\alpha} \big)}  \mathrm{sin} {\tiny  \big( \lambda_{\alpha} - v_{n-1} + \eta \sigma^z_{n-1} \big) }   +    B_{n-1} {\tiny \big( \lambda_{\alpha} \big) }   \mathrm{sin} {\tiny \big( 2 \eta \big)}  \sigma^{+}_{n-1}    \text{. } \tag{1} 
\end{align*} }

\noindent From the equality above, making the substitution for $A_{n-1} {\tiny  \big( \lambda_{\alpha} \big)}$ in terms of $A_{n-2} {\tiny \big( \lambda_{\alpha} \big)}$,  implies,

{\small \begin{align*}
 A_{n-1} {\tiny  \big( \lambda_{\alpha} \big) } =  A_{n-2} {\tiny  \big( \lambda_{\alpha} \big) }   \textbf{1}^{n-2}   +    B_{n-2} {\tiny \big( \lambda_{\alpha} \big)}  \textbf{3}^{n-2} \equiv  A_{n-2} {\tiny  \big( \lambda_{\alpha} \big) }  \mathrm{sin} {\tiny  \big( \lambda_{\alpha} - v_{n-2} + \eta \sigma^z_{n-2} \big)  }   +    B_{n-2} {\tiny  \big( \lambda_{\alpha} \big) }  \mathrm{sin} {\tiny  \big( 2 \eta \big) } \sigma^{+}_{n-2}    \text{, } 
\end{align*} }

\noindent from which, along the lines of computations performed in the manipulations above, rearrangements from $(1^{*}-2)$ of terms implies,

{\small \begin{align*}
            A_{n-4} {\tiny  \big( \lambda_{\alpha} \big)  } \big( \mathrm{sin} {\tiny  \big( 2 \eta \big)  } \big)^2 \mathrm{sin} {\tiny  \big( \lambda_{\alpha} - v_{n-2} - \eta \sigma^z_{n-2} \big) }  \cdot \underset{3 \leq i \leq 4}{\prod} \sigma^{-}_{n-i}   +  B_{n-4} \big( \lambda_{\alpha} \big)  \big( \mathrm{sin} {\tiny  \big( 2 \eta \big)  } \big)^2 \end{align*}
 
 \begin{align*} \times  \mathrm{sin} {\tiny  \big( \lambda_{\alpha} - v_{n-2} - \eta \sigma^z_{n-2} \big)   }      \cdot     \underset{3 \leq i \leq 4}{\prod }\sigma^{-,+}_{n-i}   +  A_{n-4} {\tiny  \big( \lambda_{\alpha} \big) } \mathrm{sin} {\tiny  \big( 2 \eta \big) }  \sigma^{-}_{n-4}  \cdot  \underset{2 \leq i \leq 3}{\prod}        \mathrm{sin} {\tiny  \big( \lambda_{\alpha} - v_{n-i} - \eta \sigma^z_{n-j} \big)  }  \end{align*}
 
 \begin{align*}   + B_{n-4} {\tiny \big( \lambda_{\alpha} \big)  }    \cdot   \underset{2 \leq i \leq 4}{\prod}        \mathrm{sin} {\tiny  \big( \lambda_{\alpha}  - v_{n-i} - \eta \sigma^z_{n-j} \big) }    \text{, } 
\end{align*} }

\noindent from which we conclude the that the four desired product expansions hold.

\bigskip 

\noindent To make use of the expressions in the matrix product above, we perform additional computations below, in which, 

{\small \[    \begin{bmatrix}
             \textbf{1}^{i-2}   &             \textbf{2}^{i-2}      \\
           \textbf{3}^{i-2}   &  \textbf{4}^{i-2}    
  \end{bmatrix}    \begin{bmatrix}
             \textbf{1}^{i-1}   &             \textbf{2}^{i-1}      \\
           \textbf{3}^{i-1}   &  \textbf{4}^{i-1}    
  \end{bmatrix}  \equiv  \begin{bmatrix}
      \textbf{1}^{i-2} \textbf{1}^{i-1} + \textbf{2}^{i-2} \textbf{3}^{i-1}          &   \textbf{1}^{i-2} \textbf{2}^{i-1 } + \textbf{2}^{i-2} \textbf{4}^{i-1}                \\ \textbf{3}^{i-2} \textbf{1}^{i-1} + \textbf{4}^{i-2} \textbf{3}^{i-1}
              &   \textbf{3}^{i-2} \textbf{2}^{i-1} + \textbf{4}^{i-1} \textbf{4}^{i-2}  
  \end{bmatrix}    \text{. } 
\] }

\noindent Substituting the expression for the product above with the $i-3$ th L-operator implies,

{\small \[ \begin{bmatrix}
             \textbf{1}^{i-3}   &             \textbf{2}^{i-3}      \\
           \textbf{3}^{i-3}   &  \textbf{4}^{i-3}    
  \end{bmatrix}   \begin{bmatrix}
      \textbf{1}^{i-2} \textbf{1}^{i-1} + \textbf{2}^{i-2} \textbf{3}^{i-1}          &   \textbf{1}^{i-2} \textbf{2}^{i-1 } + \textbf{2}^{i-2} \textbf{4}^{i-1}                \\ \textbf{3}^{i-2} \textbf{1}^{i-1} + \textbf{4}^{i-2} \textbf{3}^{i-1}
              &   \textbf{3}^{i-2} \textbf{2}^{i-1} + \textbf{4}^{i-1} \textbf{4}^{i-2}  
  \end{bmatrix} \text{, } \]  \begin{align*}
 \end{align*} \[  = \bigg[  \begin{smallmatrix}
    \prod_{1 \leq j \leq 3} \textbf{1}^{i-j} + \textbf{1}^{i-3} \textbf{2}^{i-2} \textbf{3}^{i-1} + \textbf{1}^{i-1} \textbf{2}^{i-3} \textbf{3}^{i-2} + \textbf{2}^{i-3} \textbf{3}^{i-1} \textbf{4}^{i-2}    &    \mathcal{E}_1   \\ \textbf{3}^{i-3}  \prod_{1 \leq j \leq 2} \textbf{1}^{i-j}    + \textbf{2}^{i-2} \prod_{j \text{ } \mathrm{odd} \text{ } : 1 \leq j \leq 3}  \textbf{3}^{i-j} + \textbf{1}^{i-1} \textbf{3}^{i-2} \textbf{4}^{i-3} + \textbf{3}^{i-1} \prod_{2 \leq j \leq 3} \textbf{4}^{i-j} & \mathcal{E}_2
  \end{smallmatrix}  \bigg]  \text{, }
\] } 

\noindent for,

{\small \begin{align*}
  \mathcal{E}_1 \equiv \prod_{2 \leq j \leq 3} \textbf{1}^{i-j} \textbf{2}^{i-1} + \textbf{1}^{i-3} \textbf{2}^{i-2} \textbf{4}^{i-1} + \prod_{j \text{ } \mathrm{odd} \text{ } : 1 \leq j \leq 3} \textbf{2}^{i-j} \textbf{3}^{i-2} + \textbf{2}^{i-3} \cdot  \prod_{1 \leq j \leq 2} \textbf{4}^{i-j}    \text{, } \end{align*}
 
 \begin{align*} \mathcal{E}_2 \equiv  \textbf{1}^{i-2} \textbf{2}^{i-1} \textbf{3}^{i-3} + \textbf{2}^{i-2} \textbf{3}^{i-3} \textbf{4}^{i-1} + \textbf{2}^{i-1} \textbf{3}^{i-2} \textbf{4}^{i-3} + \prod_{1 \leq j \leq 3} \textbf{4}^{i-j}  \text{, }
\end{align*} } 

\noindent from the fact that, each entry of the matrix above is respectively given by,

{\small \begin{align*}
\textbf{1}^{i-3} \bigg[  \prod_{1 \leq j \leq 2} \textbf{1}^{i-j} + \textbf{2}^{i-2} \textbf{3}^{i-1} \bigg]  + \textbf{2}^{i-3} \bigg[  \textbf{3}^{i-2} \textbf{1}^{i-1} + \textbf{4}^{i-2} \textbf{3}^{i-1} \bigg]         \text{, } \\  \textbf{1}^{i-3} \bigg[ \textbf{1}^{i-2} \textbf{2}^{i-1} + \textbf{2}^{i-2} \textbf{4}^{i-1} \bigg]   + \textbf{2}^{i-3} \bigg[    \textbf{3}^{i-2} \textbf{2}^{i-1} + \prod_{1 \leq j \leq 2} \textbf{4}^{i-j}     \bigg]  \text{, } \\  \textbf{3}^{i-3} \bigg[ \prod_{1 \leq j \leq 2} \textbf{1}^{i-j} + \textbf{2}^{i-2} \textbf{3}^{i-1} \bigg]  + \textbf{4}^{i-3} \bigg[  \textbf{3}^{i-2} \textbf{1}^{i-1} + \textbf{4}^{i-2} \textbf{3}^{i-1} \bigg]      \text{, } \\   \textbf{3}^{i-3} \bigg[  \textbf{1}^{i-2} \textbf{2}^{i-1} + \textbf{2}^{i-2} \textbf{4}^{i-1} \bigg]  + \textbf{4}^{i-3} \bigg[  \textbf{3}^{i-2} \textbf{2}^{i-1} + \prod_{1 \leq j \leq 2} \textbf{4}^{i-j} \bigg]         \text{. } 
\end{align*}} 

\noindent Iterating the computation, by making use of the expression above, implies that the matrix product,

{\small \[ \bigg[ \begin{smallmatrix}
    \prod_{1 \leq j \leq 3} \textbf{1}^{i-j} + \textbf{1}^{i-3} \textbf{2}^{i-2} \textbf{3}^{i-1} + \textbf{1}^{i-1} \textbf{2}^{i-3} \textbf{3}^{i-2} + \textbf{2}^{i-3} \textbf{3}^{i-1} \textbf{4}^{i-2}    &  ( \mathcal{E}_1 )^{\prime}  \\ \textbf{3}^{i-3}  \prod_{1 \leq j \leq 2} \textbf{1}^{i-j}    + \textbf{2}^{i-2} \prod_{j \text{ } \mathrm{odd} \text{ } : 1 \leq j \leq 3}  \textbf{3}^{i-j} + \textbf{1}^{i-1} \textbf{3}^{i-2} \textbf{4}^{i-3} + \textbf{3}^{i-1} \prod_{2 \leq j \leq 3} \textbf{4}^{i-j} &  ( \mathcal{E}_2 )^{\prime} 
  \end{smallmatrix}  \bigg]  
\begin{bmatrix}
   \textbf{1}^{i-1}   &   \textbf{2}^{i-1}  \\   \textbf{3}^{i-1}    & \textbf{4}^{i-1}
  \end{bmatrix} \text{, } 
\]} 

\noindent for,

{\small \begin{align*}
     \big( \mathcal{E}_1 \big)^{\prime}  \equiv  \prod_{2 \leq j \leq 3} \textbf{1}^{i-j} \textbf{2}^{i-1} + \textbf{1}^{i-3} \textbf{2}^{i-2} \textbf{4}^{i-1} + \prod_{j \text{ } \mathrm{odd} \text{ } : 1 \leq j \leq 3} \textbf{2}^{i-j} \textbf{3}^{i-2} + \textbf{2}^{i-3} \cdot  \prod_{1 \leq j \leq 2} \textbf{4}^{i-j}   \text{, } \end{align*}

 \begin{align*} \big( \mathcal{E}_2 \big)^{\prime } \equiv   \textbf{1}^{i-2} \textbf{2}^{i-1} \textbf{3}^{i-3} + \textbf{2}^{i-2} \textbf{3}^{i-3} \textbf{4}^{i-1} + \textbf{2}^{i-1} \textbf{3}^{i-2} \textbf{4}^{i-3} + \prod_{1 \leq j \leq 3} \textbf{4}^{i-j}  \text{, }
\end{align*} }

\noindent is equivalent to the matrix with entries that are respectively given by,

{\small \[
\text{First entry} \equiv  \text{ } 
\left\{\!\begin{array}{ll@{}>{{}}l}         \textbf{1}^{i-1} \big(    \prod_{1 \leq j \leq 3} \textbf{1}^{i-j} + \textbf{1}^{i-3} \textbf{2}^{i-2} \textbf{3}^{i-1} + \textbf{1}^{i-1} \textbf{2}^{i-3} \textbf{3}^{i-2} + \textbf{2}^{i-3} \textbf{3}^{i-1} \textbf{4}^{i-2}    \big)    \\ + 
 \textbf{3}^{i-1} \big(     \prod_{2 \leq j \leq 3} \textbf{1}^{i-j} \textbf{2}^{i-1} + \textbf{1}^{i-3} \textbf{2}^{i-2} \textbf{4}^{i-1} + \prod_{j \text{ } \mathrm{odd} \text{ } : 1 \leq j \leq 3} \textbf{2}^{i-j} \textbf{3}^{i-2} + \textbf{2}^{i-3}\\ \times  \prod_{1 \leq j \leq 2} \textbf{4}^{i-j}    \big)    
\end{array}\right.  \]
\begin{align*}
 \end{align*}  
\[ = 
\left\{\!\begin{array}{ll@{}>{{}}l}         \big( \textbf{1}^{i-1} \big)^2    \prod_{2 \leq j \leq 3} \textbf{1}^{i-j} + \prod_{\mathrm{odd}\text{ } j \text{ } : 1 \leq j \leq 3} \textbf{1}^{i-j}  \textbf{2}^{i-2} \textbf{3}^{i-1} + \big( \textbf{1}^{i-1} \big)^2  \textbf{2}^{i-3} \textbf{3}^{i-2} + \textbf{1}^{i-1} \textbf{2}^{i-3} \textbf{3}^{i-1} \textbf{4}^{i-2}  \\ + 
  \prod_{2 \leq j \leq 3} \textbf{1}^{i-j} \textbf{2}^{i-1} \textbf{3}^{i-1}  + \textbf{1}^{i-3} \textbf{2}^{i-2}   \textbf{3}^{i-1} \textbf{4}^{i-1} + \prod_{j \text{ } \mathrm{odd} \text{ } : 1 \leq j \leq 3} \textbf{2}^{i-j} \prod_{1 \leq j \leq 2}  \textbf{3}^{i-j} + \textbf{2}^{i-3} \big(  \textbf{3}^{i-1} \big)^2   \textbf{3}^{i-2}  \\ \times  \prod_{1 \leq j \leq 2} \textbf{4}^{i-j}        
\end{array}\right. \text{, } 
\]

 } 

\noindent with similar expressions holding for the remaining three entries of the product representation for the L-operator. Altogether, the expressions obtained for the entries above imply, 

{\small \[ \begin{bmatrix}
             \textbf{1}^{i-3}   &             \textbf{2}^{i-3}      \\
           \textbf{3}^{i-3}   &  \textbf{4}^{i-3}    
  \end{bmatrix}    \begin{bmatrix}
             \textbf{1}^{i-2}   &             \textbf{2}^{i-2}      \\
           \textbf{3}^{i-2}   &  \textbf{4}^{i-2}    
  \end{bmatrix}    \begin{bmatrix}
             \textbf{1}^{i-1}   &             \textbf{2}^{i-1}      \\
           \textbf{3}^{i-1}   &  \textbf{4}^{i-1}    
  \end{bmatrix} \equiv \begin{bmatrix}
   \text{First entry}     & \text{Second entry} \\ \text{Third entry} & \text{Fourth entry}
  \end{bmatrix}
\text{, } 
\] } 

\noindent from which the desired form for the product of $n$ L-operators,

{\small \[ \prod_{1 \leq i \leq n} \begin{bmatrix}
             \textbf{1}^i   &             \textbf{2}^i      \\
           \textbf{3}^i   &  \textbf{4}^i    
  \end{bmatrix}       \text{, } 
\]}

\noindent would take the desired form, in which,

{\small \[\begin{bmatrix}
   \textit{First entry}     & \textit{Second entry} \\ \textit{Third entry} & \textit{Fourth entry}
  \end{bmatrix} = \begin{bmatrix}
 \bigg\{       A_{i-3} {\tiny\big( \lambda_{\alpha} \big)} + B_{i-3} {\tiny\big( \lambda_{\alpha} \big)} \bigg\} \cdot \bigg\{   \big( \mathrm{sin} {\tiny\big( 2 \eta \big)} \big)^{n-(i-3)} \mathscr{A}^{\prime}_1  + \mathscr{A}^{\prime}_2 +  \mathscr{A}^{\prime}_3 \bigg\}  &  \big( \mathcal{E}_1 \big)^{\prime\prime} \\   \\ \bigg\{         C_{i-3} {\tiny\big( \lambda_{\alpha} \big) } + D_{i-3} {\tiny\big( \lambda_{\alpha} \big) }      \bigg\}  \cdot \bigg\{  \big( \mathrm{sin} {\tiny\big( 2 \eta \big)} \big)^{n-(i-3)} \mathscr{C}^{\prime}_1 + \mathscr{C}^{\prime}_2 + \mathscr{C}^{\prime}_3  \bigg\}  &   \big( \mathcal{E}_2 \big)^{\prime\prime} \end{bmatrix} \text{, }
 \] }

 \noindent where,

 {\small \begin{align*}
   \big( \mathcal{E}_1 \big)^{\prime\prime} \equiv  \bigg\{       A_{i-3} {\tiny\big( \lambda_{\alpha} \big)} + B_{i-3} {\tiny\big( \lambda_{\alpha} \big)} \bigg\} \cdot   \bigg\{    \big( \mathrm{sin} {\tiny\big( 2 \eta \big)} \big)^{n-(i-4)} \mathscr{B}^{\prime}_1  + \mathscr{B}^{\prime}_2 +  \mathscr{B}^{\prime}_3 \bigg\}    \text{, } \\    \big( \mathcal{E}_2 \big)^{\prime\prime} \equiv  \bigg\{       C_{i-3} {\tiny\big( \lambda_{\alpha} \big)} + D_{i-3} {\tiny\big( \lambda_{\alpha} \big)} \bigg\} \cdot   \bigg\{    \big( \mathrm{sin} {\tiny\big( 2 \eta \big)} \big)^{n-(i-4)} \mathscr{D}^{\prime}_1  + \mathscr{D}^{\prime}_2 +  \mathscr{D}^{\prime}_3 \bigg\}    \text{, }
 \end{align*} } 

\noindent for,

{\small \begin{align*}
  \mathscr{A}^{\prime}_1 \equiv  \mathscr{C}^{\prime}_1 \equiv \underset{1 \leq i \leq n - (i-3 )}{\prod} \sigma^{-,+}_{n-i} \text{, } \end{align*}

  \begin{align*}  \mathscr{A}^{\prime}_2 \equiv \mathscr{C}^{\prime}_2 \equiv \underset{1 \leq i \leq n - (i-3)}{\prod}    \mathrm{sin} {\tiny\big( \lambda_{\alpha} - v_{n-i} + \eta \sigma^z_{n-i} \big) }       \text{, } \end{align*}

  \begin{align*}           \mathscr{A}^{\prime}_3 \equiv \mathscr{C}^{\prime}_3 \equiv   \underset{m,n^{\prime} : m + n^{\prime} = n-3}{\sum}  \bigg\{     \underset{1 \leq i \leq m}{\prod} \mathrm{sin} {\tiny\big( \lambda_{\alpha} - v_{n-i} \pm \eta \sigma^z_{n-i} \big) }   \cdot   \big( \mathrm{sin} {\tiny\big( 2 \eta \big) } \big)^{n^{\prime}-1}  \cdot   \underset{1 \leq j \leq n^{\prime}}{ \prod}  \sigma^{-,+}_{n-j}          \bigg\}       \text{, } 
\end{align*} }

\noindent and,

{\small \begin{align*}
    \mathscr{B}^{\prime}_1 \equiv  \mathscr{D}^{\prime}_1 \equiv    \text{ }     \underset{2 \leq i \leq n-(i-3)}{\prod}     \sigma^{-,+}_{n-i}  \text{, } \end{align*}

  \begin{align*}    \mathscr{B}^{\prime}_2 \equiv  \mathscr{D}^{\prime}_2 \equiv               \underset{2 \leq i \leq n-(i-3)}{\prod}   \mathrm{sin} {\tiny\big( \lambda_{\alpha} - v_{n-i} +    \eta \sigma^z_{n-i}     \big)  }     \text{, } \end{align*}

  \begin{align*}  \mathscr{B}^{\prime}_3  \equiv  \mathscr{D}^{\prime}_3 \equiv    \text{ }  \underset{m,n^{\prime} : m + n^{\prime} = n-3}{\sum}  \bigg\{       \underset{2 \leq i \leq m}{\prod} \mathrm{sin} {\tiny\big( \lambda_{\alpha} - v_{n-i} \pm \eta \sigma^z_{n-i} \big) }  \cdot \big( \mathrm{sin} {\tiny\big( 2 \eta \big)} \big)^{n^{\prime}-1} \cdot  \underset{2 \leq j \leq n^{\prime}}{ \prod}  \sigma^{-,+}_{n-j}        \bigg\}     \text{. } 
\end{align*} } 

\noindent Iterating the previous arguments for computing the entries of each L-operator, from the product,

{\small \[\begin{bmatrix}
 \bigg\{       A_{i-3} {\tiny\big( \lambda_{\alpha} \big)} + B_{i-3} {\tiny\big( \lambda_{\alpha} \big)} \bigg\} \cdot   \bigg\{    \big( \mathrm{sin} {\tiny\big( 2 \eta \big)} \big)^{n-(i-3)} \mathscr{A}^{\prime}_1  + \mathscr{A}^{\prime}_2 +  \mathscr{A}^{\prime}_3 \bigg\}   &  \big( \mathcal{E}_1 \big)^{\prime\prime\prime } \\   \\ \bigg\{         C_{i-3} {\tiny\big( \lambda_{\alpha} \big)} + D_{i-3} {\tiny\big( \lambda_{\alpha} \big) }       \bigg\} \cdot   \bigg\{   \big( \mathrm{sin} {\tiny\big( 2 \eta \big)} \big)^{n-(i-3)} \mathscr{C}^{\prime}_1 + \mathscr{C}^{\prime}_2 + \mathscr{C}^{\prime}_3  \bigg\}   &  \big( \mathcal{E}_2 \big)^{\prime\prime\prime }  \end{bmatrix} 
  \cdot   \prod_{i \leq i^{\prime} \leq n}    \begin{bmatrix}
             \textbf{1}^{i^{\prime}}   &             \textbf{2}^{i^{\prime}}      \\
           \textbf{3}^{i^{\prime}}   &  \textbf{4}^{i^{\prime}}    
  \end{bmatrix}   \text{, } 
 \] }

 \noindent for,

{\small \begin{align*}
   \big( \mathcal{E}_1 \big)^{\prime\prime\prime } \equiv         \bigg\{       A_{i-3} {\tiny\big( \lambda_{\alpha} \big)} + B_{i-3} {\tiny\big( \lambda_{\alpha} \big)} \bigg\} \cdot   \bigg\{    \big( \mathrm{sin} {\tiny\big( 2 \eta \big) } \big)^{n-(i-4)} \mathscr{B}^{\prime}_1  + \mathscr{B}^{\prime}_2 +  \mathscr{B}^{\prime}_3 \bigg\}      \text{, } \end{align*}

  \begin{align*}    \big( \mathcal{E}_2 \big)^{\prime\prime\prime } \equiv  \bigg\{       C_{i-3} {\tiny\big( \lambda_{\alpha} \big)}  + D_{i-3} {\tiny\big( \lambda_{\alpha} \big) }\bigg\} \cdot   \bigg\{   \big( \mathrm{sin} {\tiny\big( 2 \eta \big)} \big)^{n-(i-4)} \mathscr{D}^{\prime}_1  + \mathscr{D}^{\prime}_2 +  \mathscr{D}^{\prime}_3 \bigg\}  
      \text{, }
 \end{align*} } 

\noindent yields similar expressions for ${\tiny\big\{\mathscr{A}_i\big\}_{1 \leq i \leq 3}}$, ${\tiny\big\{ \mathscr{B}_i \big\}_{1 \leq i \leq 3}}$, ${\tiny\big\{ \mathscr{C}_i \big\}_{1 \leq i \leq 3}}$ and ${\tiny\big\{ \mathscr{D}_i \big\}_{1 \leq i \leq 3}}$, from ${\tiny\big\{\mathscr{A}^{\prime}_i\big\}_{1 \leq i \leq 3}}$, ${\tiny\big\{ \mathscr{B}^{\prime}_i \big\}_{1 \leq i \leq 3}}$, ${\tiny\big\{ \mathscr{C}^{\prime}_i \big\}_{1 \leq i \leq 3}}$ and ${\tiny\big\{ \mathscr{D}^{\prime}_i \big\}_{1 \leq i \leq 3}}$, from which we conclude the argument. \boxed{}

\subsection{Argument for Lemma BE 1}

\noindent \textit{Proof of Lemma BE 1}. Observe, from,

 {\small \begin{align*} \frac{\partial}{\partial v_k} \bigg[  \mathrm{sinh} \big( \frac{\eta}{2} + i \alpha_j - v_k \big)   \bigg]  \text{, } \end{align*} }

\noindent that one can write,

{\small \begin{align*}
\bigg\{   \frac{\mathrm{sinh} \big( \frac{\eta}{2} + i \alpha_j - v_k \big)}{\mathrm{sinh} \big( \frac{\eta}{2} + i \alpha_j - v_k \big)}    \cdot   \frac{\mathrm{cosh} \big( \frac{\eta}{2} + i \alpha_j - v_k \big)}{\mathrm{cosh} \big( \frac{\eta}{2} + i \alpha_j - v_k \big)}   \bigg\}  \mathrm{cosh} \big( \frac{\eta}{2} + i \alpha_j - v_k \big)   =       \mathrm{coth} \big( \frac{\eta}{2} + i \alpha_j - v_k \big)    {\mathrm{sinh} \big( \frac{\eta}{2} + i \alpha_j - v_k \big)}   \text{. } 
\end{align*} }

\noindent Also, observe, from,

{\small \begin{align*}
         \frac{\partial}{\partial v_k}    \bigg[   \mathrm{sinh} \big( \frac{\eta}{2} + i \alpha_j +  v_k \big)       \bigg]  \text{, }
\end{align*} } 

\noindent that one can write,

{\small \begin{align*}
     \bigg\{  \frac{\mathrm{sinh} \big( \frac{\eta}{2} + i \alpha_j + v_k \big)}{\mathrm{sinh} \big( \frac{\eta}{2} + i \alpha_j + v_k \big)}    \cdot    \frac{\mathrm{cosh} \big( \frac{\eta}{2} + i \alpha_j + v_k \big)}{\mathrm{cosh} \big( \frac{\eta}{2} + i \alpha_j + v_k \big)}   \bigg\}  \mathrm{cosh} \big( \frac{\eta}{2} + i \alpha_j + v_k \big)      =      \mathrm{coth} \big( \frac{\eta}{2} + i \alpha_j + v_k \big)    {\mathrm{sinh} \big( \frac{\eta}{2} + i \alpha_j + v_k \big)}       \text{, } 
\end{align*} }

\noindent implying,

{\small \begin{align*}
   \mathscr{U}^j_1 =    \mathrm{coth} \big( \frac{\eta}{2} + i \alpha_j - v_k \big)                  \text{, } 
\end{align*} }

\noindent and,

{\small \begin{align*}
  \mathscr{U}^j_2 =    \mathrm{coth} \big( \frac{\eta}{2} + i \alpha_j + v_k \big)             \text{. } 
\end{align*} }

\noindent For the remaining two functions, similarly observe, from,

{\small \begin{align*}
     \frac{\partial}{\partial \alpha_m} \bigg[   \mathrm{sinh} \big( i {\tiny \big( \alpha_j - \alpha_m \big)}  + \eta \big)          \bigg]  \text{, } 
\end{align*} } 

\noindent that one can write,

{\small \begin{align*}
   -  \bigg\{          \frac{\mathrm{sinh} \big(       i {\tiny  \big( \alpha_j - \alpha_m \big) }  + \eta      \big) }{\mathrm{sinh} \big(   i {\tiny  \big( \alpha_j - \alpha_m \big) }  + \eta  
 \big) }         \cdot      \frac{\mathrm{cosh} \big(   i {\tiny  \big( \alpha_j - \alpha_m \big) }  + \eta   \big) }{\mathrm{cosh} \big(  i {\tiny  \big( \alpha_j - \alpha_m \big) }  + \eta   \big) }                \bigg\}   \mathrm{cosh}  \big( i {\tiny  \big( \alpha_j - \alpha_m \big) } + \eta \big)  =  -        \mathrm{coth} \big(    i {\tiny  \big( \alpha_j - \alpha_m \big)} + \eta   \big)  \end{align*}

  \begin{align*} \times \mathrm{sinh} \big( i {\tiny  \big( \alpha_j }  {\tiny  - \alpha_m \big) } + \eta \big)   \text{. } 
\end{align*} }

\noindent Also, observe from,

{\small \begin{align*}
 \frac{\partial}{\partial \alpha_m} \big(   \mathrm{sinh} \big( i {\tiny  \big( \alpha_j - \alpha_m \big) }  - \eta \big)    \big)  \text{, } 
\end{align*} } 

\noindent that one can write,

{\small \begin{align*}
  -    \bigg\{  \frac{\mathrm{sinh} \big(   i {\tiny \big( \alpha_j - \alpha_m \big)} - \eta        \big) }{\mathrm{sinh} \big(  i {\tiny \big( \alpha_j - \alpha_m \big)} - \eta     \big) } \cdot  \frac{\mathrm{cosh} \big(  i {\tiny \big( \alpha_j - \alpha_m \big)} - \eta    \big) }{\mathrm{cosh} \big(  i {\tiny \big( \alpha_j - \alpha_m \big)}  - \eta    \big) } \bigg\}   
   \mathrm{cosh} \big( i {\tiny \big( \alpha_j - \alpha_m \big) }- \eta \big) =  -    \mathrm{coth} \big( i {\tiny \big( \alpha_j - \alpha_m \big) }- \eta \big)         \end{align*}

  \begin{align*} \times             \mathrm{sinh} \big( i {\tiny \big( \alpha_j}   {\tiny - \alpha_m \big) }  - \eta \big)   \text{, } 
\end{align*}  }

\noindent implying,

{\small \begin{align*}
   \mathscr{U}^j_3  =   - \mathrm{coth} \big(    i {\tiny  \big( \alpha_j - \alpha_m \big)  } + \eta   \big)     \text{. } 
\end{align*} }

\noindent and,

{\small \begin{align*}
   \mathscr{U}^j_4 =   \mathrm{coth} \big(    i {\tiny  \big( \alpha_j - \alpha_m \big)  } - \eta   \big)  \text{. } 
\end{align*} }

\noindent One can directly obtain the functions $\mathscr{U}^j_1 , \mathscr{U}^j_2, \mathscr{U}^j_3$ and $\mathscr{U}^j_4$ for any $j$, which take the form as unions over $\textbf{N}$ provided at the end of the statement for \textbf{Lemma} \textit{BE 1}, from which we conclude the argument. \boxed{}

 \subsection{Arguments for Lemma BE 2}

\noindent \textit{Proof of Lemma BE 2}. Observe,

{\small \begin{align*}
   \overset{N}{\underset{k=1}{\prod}}    \frac{\partial F_1 \big( \frac{1}{2} , \alpha_j , - v_k \big) }{\partial F_2 \big( \frac{1}{2} , - \alpha_j , v_k \big) } \overset{(\textbf{Lemma} \text{ } \textit{BE 1})}{=} \overset{N}{\underset{k=1}{\prod}} \frac{\mathscr{U}^j_1 \big( \frac{\eta}{2} , \alpha_j , - v_k  \big) F_1  \big( \frac{1}{2} , \alpha_j , - v_k \big)        }{ \mathscr{U}^j_2 \big( \frac{\eta}{2} , - \alpha_j , v_k \big)    F_2  \big( \frac{1}{2} , - \alpha_j , v_k \big)  }  \text{, } \tag{*}
\end{align*} }

\noindent from dividing,

{\small \begin{align*}
   \overset{N}{\underset{k=1}{\prod}}    \frac{\partial F_1}{\partial v_k} \big( \frac{1}{2} , \alpha_j , - v_k \big) \overset{(\textbf{Lemma} \text{ } \textit{BE 1})}{=} \overset{N}{\underset{k=1}{\prod}} {\mathscr{U}^j_1 \big( \frac{\eta}{2} , \alpha_j , - v_k  \big) F_1  \big( \frac{1}{2} , \alpha_j , - v_k \big)        } \text{, } 
\end{align*} }

\noindent with,

{\small \begin{align*}
   \overset{N}{\underset{k=1}{\prod}}    \frac{\partial F_2}{\partial v_k}   \big( \frac{1}{2} , - \alpha_j , v_k \big) \overset{(\textbf{Lemma} \text{ } \textit{BE 1})}{=} \overset{N}{\underset{k=1}{\prod}} { \mathscr{U}^j_2 \big( \frac{\eta}{2} , - \alpha_j , v_k \big)    F_2  \big( \frac{1}{2} , - \alpha_j , v_k \big)  }  \text{, } 
\end{align*} }

\noindent for the left hand side of the Bethe equations, and,

{\small \begin{align*}
    \overset{n}{\underset{m=1 , m \neq j }{\prod} }   
       \frac{\partial F_3 \big(   1 , \alpha_j - \alpha_m , 0 \big) }{\partial F_4  \big( - 1 , \alpha_j - \alpha_m , 0  \big) } \overset{(\textbf{Lemma} \text{ } \textit{BE 1})}{=}    \overset{n}{\underset{m=1 , m \neq j }{\prod} }   
       \frac{ \mathscr{U}^j_3 \big(   1 , \alpha_j - \alpha_m , 0    \big)  F_3  \big( 1 , \alpha_j - \alpha_m , 0   \big)  }{\mathscr{U}^j_4 \big(   - 1 , \alpha_j - \alpha_m , 0       \big)       F_4 \big( - 1, \alpha_j - \alpha_m  , 0  \big)        }     \text{, } \tag{**}
\end{align*} }

\noindent from dividing,

{\small \begin{align*}
 \overset{n}{\underset{m=1 }{\prod} }   
      \frac{\partial F_3}{\partial \alpha_m } \big(   1 , \alpha_j - \alpha_m , 0 \big) \overset{(\textbf{Lemma} \text{ } \textit{BE 1})}{=}   \overset{n}{\underset{m=1   }{\prod} }   \mathscr{U}^j_3 \big(   1 , \alpha_j - \alpha_m , 0    \big) \mathrm{sinh} \big( i {\tiny  \big( \alpha_j - \alpha_m \big) }  + \eta \big)\text{, } 
\end{align*} }

\noindent with,

{\small \begin{align*}
   \overset{n}{\underset{m=1 , m \neq j }{\prod} }   
      \frac{\partial F_4}{\partial \alpha_m } \big(   1 , \alpha_j - \alpha_m , 0 \big) \overset{(\textbf{Lemma} \text{ } \textit{BE 1})}{=}    \overset{n}{\underset{m=1 , m \neq j }{\prod} }  \mathscr{U}^j_4 \big(   - 1 , \alpha_j - \alpha_m , 0       \big)       \mathrm{sinh} \big( i {\tiny  \big( \alpha_j - \alpha_m \big) }  - \eta \big) \text{, } 
\end{align*} }

\noindent and multiplying the quotient with,

{\small \begin{align*}
   \mathrm{exp} {\tiny  \big( 2 H N \big) }  \text{, } 
\end{align*} }

\noindent for the right hand side of the Bethe equations. Altogether,

{\small \begin{align*}
 (*) = (**)   \text{, } 
\end{align*} } 

\noindent from which we conclude the argument. \boxed{}

\subsection{Arguments for Lemma BE 3}

\noindent \textit{Proof of Lemma BE 3}. By the definition of hyperbolic sine, write,

{\small \begin{align*}
  \frac{F_1  \big( \frac{1}{2} , \alpha_j , - v_k \big)}{  F_2  \big( \frac{1}{2} , - \alpha_j , v_k \big) } = \frac{\mathrm{sinh} \big(       \frac{\eta}{2} + i \alpha_j - v_k  \big)}{\mathrm{sinh} \big( \frac{\eta}{2} - i \alpha_j + v_k  \big) } =  \frac{\frac{1}{2}}{\frac{1}{2}} \bigg[ \frac{-\mathrm{exp} \big( - {\tiny \big( \frac{\eta}{2} + i \alpha_j - v_k \big) } 
 \big) + \mathrm{exp} \big( \frac{\eta}{2} + i \alpha_j - v_k \big) }{ - \mathrm{exp} \big( - {\tiny  \big( \frac{\eta}{2} + i \alpha_j - v_k \big) } 
 \big) + \mathrm{exp} \big( \frac{\eta}{2} + i \alpha_j - v_k \big)} \bigg]   \text{, } 
\end{align*} }

\noindent which is equivalent to,

{\small \begin{align*}
 \frac{-\mathrm{exp} \big( - {\tiny \big( \frac{\eta}{2} + i \alpha_j - v_k \big)}  \big) }{- \mathrm{exp} \big( - \big( \frac{\eta}{2} + i \alpha_j - v_k \big) 
 \big) + \mathrm{exp} \big( \frac{\eta}{2} + i \alpha_j - v_k \big)} + \frac{\mathrm{exp} \big( \frac{\eta}{2} + i \alpha_j - v_k \big) }{-\mathrm{exp} \big( - {\tiny \big( \frac{\eta}{2} + i \alpha_j - v_k \big) } 
 \big) + \mathrm{exp} \big( \frac{\eta}{2} + i \alpha_j - v_k \big)}    \text{, } 
\end{align*} }

\noindent after having separated exponentials in the numerator of the expression, which is in turn also equivalent to,

{\small \begin{align*}
   \frac{\mathrm{exp} \big( - {\tiny  \big( \frac{\eta}{2} + i \alpha_j - v_k \big)}  \big) }{  \mathrm{sinh} \big( \frac{\eta}{2} + i \alpha_j - v_k \big)    \big( \mathrm{exp} \big( - \big( \frac{\eta}{2} + i \alpha_j - v_k \big) 
 \big) - \mathrm{exp} \big( \frac{\eta}{2} + i \alpha_j - v_k \big) \big) }  \cdot   F_1 \big( \frac{1}{2} , \alpha_j , - v_k \big) \end{align*}

 \begin{align*} +   \frac{\mathrm{exp} \big( \frac{\eta}{2} + i \alpha_j - v_k \big) }{ \mathrm{sinh} \big( \frac{\eta}{2} - i \alpha_j + v_k \big)  \big(  - \mathrm{exp} \big( - {\tiny \big( \frac{\eta}{2} + i \alpha_j - v_k \big)}  
 \big) + \mathrm{exp} \big( \frac{\eta}{2} + i \alpha_j - v_k \big) \big) } \cdot  F_2 \big( \frac{1}{2} , - \alpha_j , v_k \big)     \text{, } 
\end{align*} }

\noindent in the basis spanned by $F_1$ and $F_2$. On the other hand, again by the definition of hyperbolic sine, write,

{\small \begin{align*}
  \frac{F_3 \big( 1 , \alpha_j - \alpha_m , 0 \big) }{F_4 \big( -1 , \alpha_j - \alpha_m , 0 \big) } = \frac{\mathrm{sinh} \big(     i {\tiny  \big( \alpha_j - \alpha_m \big)}  + \eta     \big) }{\mathrm{sinh} \big(  i {\tiny  \big( \alpha_j - \alpha_m \big)}  - \eta          \big) } = \frac{\frac{1}{2}}{\frac{1}{2}} \bigg[     \frac{\mathrm{exp} \big( i {\tiny \big( \alpha_j - \alpha_m \big) + \eta \big) } - \mathrm{exp} \big( \big( i {\tiny \big( \alpha_m - \alpha_j \big)} - \eta \big) }{\mathrm{exp} \big( i {\tiny \big( \alpha_j - \alpha_m \big)} - \eta \big) - \mathrm{exp} \big( i {\tiny \big( \alpha_m - \alpha_j \big)}  + \eta \big)  }      \bigg] \text{, } 
\end{align*} } 

\noindent which is equivalent to,

{\small \begin{align*}
    \frac{\mathrm{exp} \big( i {\tiny  \big( \alpha_j - \alpha_m \big)}  + \eta \big) }{\mathrm{exp} \big( i {\tiny  \big( \alpha_j - \alpha_m \big) }  - \eta \big) - \mathrm{exp} \big( i {\tiny \big( \alpha_m - \alpha_j \big)}  + \eta \big) } - \frac{\mathrm{exp} \big( i {\tiny \big( \alpha_m - \alpha_j \big)} - \eta \big) }{\mathrm{exp} \big( i {\tiny \big( \alpha_j - \alpha_m \big)} - \eta \big) - \mathrm{exp} \big( i {\tiny \big( \alpha_m - \alpha_j \big)} + \eta \big) }       \text{, } 
\end{align*} }

\noindent after having separated exponentials in the numerator of the expression, which in turn is also equivalent to,

{\small \begin{align*}
 \frac{\mathrm{exp} \big( i {\tiny  \big( \alpha_j - \alpha_m \big) } + \eta \big) }{ \mathrm{sinh} \big( i {\tiny \big( \alpha_j - \alpha_m \big)}  + \eta \big)  \big( \mathrm{exp} \big( i {\tiny \big( \alpha_j - \alpha_m \big)} - \eta \big) - \mathrm{exp} \big( i {\tiny \big( \alpha_m - \alpha_j \big)} + \eta \big) \big)  } \cdot   F_3 {\tiny \big( 1 , \alpha_j - \alpha_m , 0 \big) } \end{align*}

 \begin{align*}   -         \frac{\mathrm{exp} \big( i {\tiny \big( \alpha_m - \alpha_j \big) } - \eta \big) }{ \mathrm{sinh} \big( i {\tiny \big( \alpha_j - \alpha_m \big) } - \eta \big)  \big( \mathrm{exp} \big( i {\tiny \big( \alpha_j - \alpha_m \big)} - \eta \big) - \mathrm{exp} \big( i {\tiny \big( \alpha_m - \alpha_j \big) } + \eta \big) \big) }      \cdot F_4 {\tiny  \big( - 1 , \alpha_j - \alpha_m , 0 \big)}   \text{, } 
\end{align*} }

\noindent for the basis spanned by $F_3$ and $F_4$. Taking $\mathscr{S}_1, \mathscr{S}_2, \mathscr{S}_3$, and $\mathscr{S}_4$ from the quantities above provides the desired form of the linear combination for $F_{1,2}$ and $F_{3,4}$, from which we conclude the argument. \boxed{}

\subsection{Arguments for Lemma BE 4}

\noindent \textit{Proof of Lemma BE 4}. It suffices to determine one basis, $\big\{ F^j_1 , F^j_2 \big\}$, and another basis, $\big\{ F^j_3 , F^j_4 \big\}$, for each $j$. The computation from the previous result can be repeated a countably many number of times for any solution to the Bethe equations, hence yielding the desired span from which we conclude the argument. \boxed{}

\subsection{First Poisson bracket, $\mathcal{P}_1$, for $ \big\{ A {\tiny\big( u \big)} , A {\tiny\big( u^{\prime} \big) }\big\} $}

\noindent The first term, $\mathcal{P}_1$, approximately equals,

{\small \begin{align*}
     \big[ \big( \mathrm{sin}  {\tiny\big( 2 \eta \big) }\big)^{n-3} 
\mathscr{C}_1  \big]^2  \bigg[      \frac{A_3 {\tiny \big( u \big) }B_3 {\tiny\big( u^{\prime} \big)}}{ u - u^{\prime} }       -    \frac{A_3 {\tiny\big( u^{\prime} \big)}  B_3 {\tiny\big( u \big)}}{u^{\prime} - u}      \bigg]    \text{. }
\end{align*} }

\subsection{Second Poisson bracket, $\mathcal{P}_2$, for $ \big\{ A {\tiny\big( u \big)} , A {\tiny\big( u^{\prime} \big) } \big\} $}

   \noindent The second term, $\mathcal{P}_2$, approximately equals,

{\small \begin{align*}
       - \big[ \big( \mathrm{sin } {\tiny\big( 2 \eta \big) } \big)^{n-3} \mathscr{C}_1    \big]          \bigg[ \frac{A_3 {\tiny\big( u \big)} B_3 {\tiny\big( u^{\prime} \big)} }{u - u^{\prime}}   -    \frac{B_3 {\tiny\big( u \big)} A_3 {\tiny\big( u^{\prime} \big)} }{u^{\prime} - u }   \bigg]           \text{. }
\end{align*} }

\subsection{Third Poisson bracket, $\mathcal{P}_3$, for $ \big\{ A {\tiny\big( u \big)} , A {\tiny\big( u^{\prime} \big)} \big\} $}

\noindent The third term approximately equals,

{\small \begin{align*}
  \mathcal{P}_3 \approx    \bigg[   {\underset{m,n^{\prime} : m + n^{\prime} = n -3}{\sum}}    \big( \mathrm{sin} {\tiny\big( 2 \eta \big) } \big)^{n^{\prime}-1}   \bigg]  \bigg[  \frac{\partial  \big[ \underset{1 \leq i \leq m}{\prod} \big( \mathscr{C}_2 \big)_i  \big] }{\partial u} \frac{\partial B_3 {\tiny\big( u \big) }   }{\partial u^{\prime}}        + \frac{\partial   \big[ \underset{1 \leq i \leq m}{\prod} \big( \mathscr{C}_2 \big)_i  \big]  }{\partial u} \frac{\partial A_3 {\tiny\big( u \big)}  }{\partial u^{\prime}} \bigg]  \end{align*}

  \begin{align*}  + \bigg[    {\underset{m,n^{\prime} : m + n^{\prime} = n -3}{\sum}}   \big( \mathrm{sin} {\tiny\big( 2 \eta \big) } \big)^{n^{\prime}-1}      \bigg]\bigg[ \frac{\partial \big[      \underset{1 \leq i \leq m}{\prod} \big( \mathscr{C}_2 \big)_i     \big] }{\partial u^{\prime}}  \frac{\partial B_3 {\tiny\big( u \big)} }{\partial u}  + \frac{\partial \big[      \underset{1 \leq i \leq m}{\prod} \big( \mathscr{C}_2 \big)_i     \big]}{\partial u^{\prime}}    \frac{\partial A_3 {\tiny\big( u \big) } }{\partial u}   \bigg]  \end{align*}

  \begin{align*} 
 -   
  \bigg[   {\underset{1\leq j \leq  n^{\prime}}{\underset{1 \leq i \leq m}{\sum} }}   \big( \mathscr{C}_3 \big)_{i,j}             \bigg] \bigg[  \big( \mathrm{sin} {\tiny\big( 2 \eta \big)}  \big)^{n-3}    \frac{B_3 {\tiny\big( u \big)} A_3 {\tiny\big( u^{\prime} \big)} }{u - u^{\prime}}     +   \big( \mathrm{sin} {\tiny \big( 2 \eta \big) } \big)^{n-3}       
   \frac{A_3 {\tiny\big( u \big) }B_3 {\tiny\big( u^{\prime} \big)} }{u - u^{\prime}}      \bigg]   \text{. }
\end{align*} }

\subsection{Fourth Poisson bracket, $\mathcal{P}_4$, for $ \big\{ A {\tiny\big( u \big)} , A {\tiny\big( u^{\prime} \big) }\big\} $}

\noindent The fourth term, $\mathcal{P}_4$, approximately equals,

\begin{align*}
     - \big[ \big( \mathrm{sin} {\tiny\big( 2 \eta \big) }\big)^{n-3}              \big( \mathscr{C}_1 \big)_i           \big]  \bigg[     \frac{A_3 {\tiny\big( u \big) } B_3 {\tiny\big( u^{\prime} \big) }}{u-u^{\prime}}       +  \frac{B_3 {\tiny\big( u \big) } A_3 {\tiny\big( u^{\prime} \big)}  }{u-u^{\prime}} \bigg]   +     \underset{1 \leq i \leq n-3}{\sum} \mathscr{A}_2 \bigg[ 
     \text{ }         \bigg[   A_3 {\tiny\big( u \big) }
 \frac{\partial   B_3 {\tiny\big( u^{\prime} \big)} }{\partial u^{\prime}}              \frac{\partial \big( \mathscr{C}_2 \big)_i}{\partial u}       \end{align*}

  \begin{align*}   +  B_3 {\tiny\big( u \big) }\frac{\partial B_3 {\tiny\big( u^{\prime} \big) }}{\partial u}   \frac{\partial \big( \mathscr{C}_2 \big)_i}{\partial u^{\prime}  }       \bigg] 
 +    \bigg[   A_3 {\tiny\big( u \big) }  \frac{\partial A_3 {\tiny\big( u^{\prime} \big) }}{\partial u^{\prime}}                 \frac{\partial  \big( \mathscr{C}_2 \big)_i }{\partial u}    + B_3 {\tiny\big( u \big) }\frac{\partial A_3 {\tiny\big( u^{\prime} \big) } }{\partial u}  \frac{\partial \big( \mathscr{C}_2 \big)_i}{\partial u^{\prime}  }      \bigg]  \text{ }           \bigg]      \text{. }
\end{align*}

\subsection{Fifth Poisson bracket, $\mathcal{P}_5$, for $ \big\{ A {\tiny\big( u \big)} , A {\tiny\big( u^{\prime} \big) }\big\} $ }

\noindent The fifth term approximately equals,

{\small \begin{align*}
  \mathcal{P}_5  \approx   -   \frac{B_3 {\tiny\big(u^{\prime} \big)}  A_3 {\tiny\big( u \big)  }  B_3 {\tiny\big( u \big) }   \mathscr{C}_2  }{u^{\prime} - u} \cdot \frac{\partial  B_3 {\tiny\big( u^{\prime} \big) } }{\partial u^{\prime}}  \cdot        \underset{1 \leq i \leq n-3}{\sum}          \frac{\partial \big( \mathscr{C}_2 \big)_i}{\partial u} \cdot  \underset{1 \leq j  \leq n-3}{\prod} \big( \mathscr{C}_2 \big)_j  \end{align*}

  \begin{align*} -   \frac{B_3 {\tiny\big(u^{\prime} \big) } A_3 {\tiny\big( u \big) }  A_3 {\tiny\big( u \big)}  \mathscr{C}_2  }{u^{\prime} - u}  \cdot \frac{\partial  B_3 {\tiny\big( u^{\prime} \big)} }{\partial u^{\prime}} \cdot 
     \underset{1 \leq k \leq n-3}{\sum}          \frac{\partial \big( \mathscr{C}_2 \big)_k}{\partial u}  \cdot \underset{1 \leq j  \leq n-3}{\prod} \big( \mathscr{C}_2 \big)_j      \end{align*}

  \begin{align*}    -     
         \frac{B_3 {\tiny\big( u \big) }A_3 {\tiny\big( u^{\prime} \big) }  B_3 {\tiny\big( u \big) } \mathscr{C}_2 }{u - u^{\prime}}  \cdot \frac{\partial  B_3 {\tiny\big( u^{\prime} \big) } }{\partial u^{\prime}} \cdot    \underset{1 \leq k \leq n-3}{\sum}          \frac{\partial \big( \mathscr{C}_2 \big)_k}{\partial u} \cdot \underset{1 \leq j  \leq n-3}{\prod} \big( \mathscr{C}_2 \big)_j  \end{align*}

  \begin{align*}   -        \frac{B_3 {\tiny\big( u \big) } A_3 {\tiny\big( u^{\prime} \big)}  A_3 {\tiny\big( u \big) }  \mathscr{C}_2 }{u - u^{\prime}}  \frac{\partial  B_3 {\tiny\big( u^{\prime} \big) } }{\partial u^{\prime}}   \cdot       \underset{1 \leq k \leq n-3}{\sum}          \frac{\partial \big( \mathscr{C}_2 \big)_k}{\partial u}  \cdot\underset{1 \leq j \leq n-3}{\prod} \big( \mathscr{C}_2 \big)_j     \end{align*}

  \begin{align*}    -   B_3 {\tiny\big( u \big)  }   \frac{\partial  A_3 {\tiny\big( u^{\prime} \big) }}{\partial u^{\prime}}    \cdot           \underset{1 \leq k \leq n-3}{\sum}  \frac{\partial \big( \mathscr{C}_2 \big)_k}{\partial u}  \cdot \underset{1 \leq i \neq j \leq n-3}{\prod} \big( \mathscr{C}_2 \big)_i  \big( \mathscr{C}_2 \big)_j  
  \end{align*}

  \begin{align*} +  B_3 {\tiny\big( u \big)}  \frac{\partial A_3 {\tiny\big( u^{\prime} \big)} }{\partial u}     \cdot   \underset{1 \leq k \leq n-3}{\sum}   \frac{\partial \big( \mathscr{C}_2 \big)_k }{\partial u}      \cdot \underset{1 \leq i \neq j \leq n-3}{\prod} \big( \mathscr{C}_2 \big)_i  \big( \mathscr{C}_2 \big)_j  
     \end{align*}

  \begin{align*}     - A_3 {\tiny\big( u \big) } \frac{\partial A_3 {\tiny\big( u^{\prime} \big)} }{\partial u}      \cdot \underset{1 \leq k \leq n-3}{\sum} \frac{\partial \big( \mathscr{C}_2\big)_k}{\partial u}   \cdot\underset{1 \leq i \neq j \leq n-3}{\prod} \big( \mathscr{C}_2 \big)_i  \big( \mathscr{C}_2 \big)_j  
    \end{align*}

  \begin{align*}   - A_3 {\tiny\big( u \big) } \frac{\partial A_3 {\tiny\big( u^{\prime} \big)} }{\partial u^{\prime}} \cdot\underset{1 \leq k \leq n-3}{\sum} \frac{\partial \big( \mathscr{C}_2\big)_k}{\partial u}   \cdot \underset{1 \leq i \neq j \leq n-3}{\prod} \big( \mathscr{C}_2 \big)_i  \big( \mathscr{C}_2 \big)_j  
      \text{. }
\end{align*} }

\subsection{Sixth Poisson bracket, $\mathcal{P}_6$, for $ \big\{ A {\tiny\big( u \big) }, A {\tiny\big( u^{\prime} \big) } \big\} $}

\noindent The sixth term, $\mathcal{P}_6$, approximately equals,

{\small \begin{align*}
  \mathcal{P}_6 \approx 
  \frac{  A_3 {\tiny\big( u \big) } B_3 {\tiny\big(u^{\prime} \big) }  
 }{u - u^{\prime}}       \cdot                       {\underset{m,n^{\prime} : m + n^{\prime} = n -3}{\sum}}   \big( \mathrm{sin} {\tiny\big( 2 \eta \big)}  \big)^{n^{\prime}-1}  \cdot  \underset{1 \leq j \leq n^{\prime}}{\underset{1 \leq i \leq m}{\prod}} \big( \mathscr{C}_2 \big)_i \big( \mathscr{C}_1 \big)_j          \end{align*}

  \begin{align*} + \frac{    B_3 {\tiny\big( u \big)} A_3 {\tiny\big( u^{\prime} \big)} }{u - u^{\prime}}      \cdot                        {\underset{m,n^{\prime} : m + n^{\prime} = n -3}{\sum}}  \big( \mathrm{sin} {\tiny\big( 2 \eta \big)}  \big)^{n^{\prime}-1}  \cdot    \underset{1 \leq j \leq n^{\prime}}{\underset{1 \leq i \leq m}{\prod}} \big( \mathscr{C}_2 \big)_i \big( \mathscr{C}_1 \big)_j            \end{align*}

  \begin{align*}  - 
         2 B_3 {\tiny\big( u \big) } \frac{\partial A_3{\tiny \big( u \big)} }{\partial u }     \big( \mathrm{sin} {\tiny\big( 2 \eta \big)} \big)^{n^{\prime}-1}     \cdot    \underset{1 \leq k \leq m}{\sum}   \frac{\partial \big( \mathscr{C}_2 \big)_k }{\partial u^{\prime}}   \cdot\text{ }   \underset{1 \leq i \neq j \leq m}{\underset{1 \leq j \leq n^{\prime}}{ \prod}}  \big( \mathscr{C}_1 \big)_j  \big( \mathscr{C}_2 \big)_j              \end{align*}

  \begin{align*}
     - 2 B_3 {\tiny\big( u \big) } \frac{\partial B_3 {\tiny\big( u \big) } }{\partial u}   \big( \mathrm{sin} {\tiny\big( 2 \eta \big)}  \big)^{n^{\prime}-1}    \cdot   \underset{1 \leq k \leq m}{\sum}            \frac{\partial \big( \mathscr{C}_2 \big)_k }{\partial u^{\prime}}  \cdot \text{ }   \underset{1 \leq i \neq j \leq m}{\underset{1 \leq j \leq n^{\prime}}{ \prod}}  \big( \mathscr{C}_1 \big)_j  \big( \mathscr{C}_2 \big)_j \end{align*}

  \begin{align*}  - 2 A_3 {\tiny\big( u \big) }\frac{\partial A_3 {\tiny\big( u \big)} }{\partial u }  \big( \mathrm{sin} {\tiny\big( 2 \eta \big) }\big)^{n^{\prime}-1}   \cdot   \underset{1 \leq k \leq m}{\sum}          \frac{\partial \big( \mathscr{C}_2 \big)_k }{\partial u^{\prime}}   \cdot \text{ }   \underset{1 \leq i \neq j \leq m}{\underset{1 \leq j \leq n^{\prime}}{ \prod}}  \big( \mathscr{C}_1 \big)_j  \big( \mathscr{C}_2 \big)_j   \end{align*}

  \begin{align*}   - 2 A_3 {\tiny\big( u \big) } \frac{\partial B_3 {\tiny\big( u \big) } }{\partial u}   \big( \mathrm{sin} {\tiny\big( 2 \eta \big) } \big)^{n^{\prime}-1}   \cdot  {\underset{1 \leq k \leq m}{\sum} }  \frac{\partial \big( \mathscr{C}_2 \big)_k }{\partial u}  \cdot  \text{ }   \underset{1 \leq i \neq j \leq m}{\underset{1 \leq j \leq n^{\prime}}{ \prod}}  \big( \mathscr{C}_1 \big)_j  \big( \mathscr{C}_2 \big)_j          \end{align*}

  \begin{align*}  - 2 B_3 {\tiny\big( u \big) }\frac{\partial A_3 {\tiny\big( u^{\prime} \big)} }{\partial u^{\prime}}   \big( \mathrm{sin} {\tiny\big( 2 \eta \big) } \big)^{n^{\prime}-1}    \cdot {\underset{1 \leq k \leq m}{\sum} }  \frac{\partial \big( \mathscr{C}_2 \big)_k }{\partial u}   \cdot  \text{ }   \underset{1 \leq i \neq j \leq m}{\underset{1 \leq j \leq n^{\prime}}{ \prod}}  \big( \mathscr{C}_1 \big)_j  \big( \mathscr{C}_2 \big)_j     \end{align*}

  \begin{align*} -  2 B_3 {\tiny\big( u \big)} \frac{\partial B_3 {\tiny\big( u^{\prime} \big)} }{\partial u^{\prime}}   \big( \mathrm{sin} {\tiny\big( 2 \eta \big) } \big)^{n^{\prime}-1}     \cdot   {\underset{1 \leq k \leq m}{\sum} }  \frac{\partial \big( \mathscr{C}_2 \big)_k }{\partial u}   \cdot \text{ }   \underset{1 \leq i \neq j \leq m}{\underset{1 \leq j \leq n^{\prime}}{ \prod}}  \big( \mathscr{C}_1 \big)_j  \big( \mathscr{C}_2 \big)_j   \end{align*}

  \begin{align*} - 2 A_3 {\tiny\big( u \big) } \frac{\partial A_3 {\tiny\big( u^{\prime} \big)} }{\partial u^{\prime}}  \big( \mathrm{sin} {\tiny \big( 2 \eta \big)  } \big)^{n^{\prime}-1}     \cdot    {\underset{1 \leq k \leq m}{\sum} }  \frac{\partial \big( \mathscr{C}_2 \big)_k }{\partial u}  \cdot  \text{ }   \underset{1 \leq i \neq j \leq m}{\underset{1 \leq j \leq n^{\prime}}{ \prod}}  \big( \mathscr{C}_1 \big)_j  \big( \mathscr{C}_2 \big)_j     \end{align*}

  \begin{align*}
          -      2 A_3 {\tiny\big( u \big) }\frac{\partial B_3 {\tiny\big( u^{\prime} \big) } }{\partial u^{\prime}}       \big( \mathrm{sin} {\tiny\big( 2 \eta \big) } \big)^{n^{\prime}-1}  \cdot     {\underset{1 \leq k \leq m}{\sum} }  \frac{\partial \big( \mathscr{C}_2 \big)_k }{\partial u}   \cdot\text{ }   \underset{1 \leq i \neq j \leq m}{\underset{1 \leq j \leq n^{\prime}}{ \prod}}  \big( \mathscr{C}_1 \big)_j  \big( \mathscr{C}_2 \big)_j              \text{. }
\end{align*} }

\subsection{Seventh Poisson bracket, $\mathcal{P}_7$, for $ \big\{ A {\tiny\big( u \big)} , A {\tiny\big( u^{\prime} \big)}  \big\} $}

\noindent The seventh term, $\mathcal{P}_7$, approximately equals,

{\small \begin{align*}
\mathcal{P}_7  \approx    \frac{A_3 {\tiny \big( u \big) } B_3 {\tiny \big( u^{\prime} \big)} }{u - u^{\prime}}   \cdot \big( \mathrm{sin} {\tiny\big( 2 \eta \big) } \big)^{n-3}   \mathscr{C}_1 
  + \frac{B_3 {\tiny\big( u \big) } A_3 {\tiny\big( u^{\prime} \big) } }{ u - u^{\prime} }  \cdot  \big( \mathrm{sin} {\tiny \big( 2 \eta \big) }  \big)^{n-3}   \mathscr{C}_1     \end{align*}

  \begin{align*} +               B_3 {\tiny\big( u \big) } \frac{\partial B_3 {\tiny\big( u^{\prime} \big)}  }{\partial u}      \mathscr{C}_1 \cdot {\underset{m,n^{\prime} : m + n^{\prime} = n -3}{\sum}}  \big( \mathrm{sin} {\tiny\big( 2 \eta \big) } \big)^{n^{\prime}-1} \cdot \underset{1 \leq i \leq m}{\sum}   \frac{\partial}{\partial u}       \underset{1 \leq i \leq m}{\prod}       \big( \mathscr{C}_2 \big)_i \end{align*}

  \begin{align*}  -         B_3 \big( u \big) \frac{\partial B_3 {\tiny\big( u^{\prime} \big) } }{\partial u^{\prime}}      \mathscr{C}_1 \cdot  {\underset{m,n^{\prime} : m + n^{\prime} = n -3}{\sum}}  \big( \mathrm{sin} {\tiny\big( 2 \eta \big)}  \big)^{n^{\prime}-1} \cdot \underset{1 \leq i \leq m}{\sum}     \frac{\partial}{\partial u}       \underset{1 \leq i \leq m}{\prod}      \big( \mathscr{C}_2 \big)_i       \end{align*}

  \begin{align*}     +   \frac{\partial B_3 {\tiny\big( u^{\prime} \big)} }{\partial u}        \mathscr{C}_1   \cdot {\underset{m,n^{\prime} : m + n^{\prime} = n -3}{\sum}}  \big( \mathrm{sin} {\tiny\big( 2 \eta \big) } \big)^{n^{\prime}-1} \cdot\underset{1 \leq i \leq m}{\sum}          \underset{1 \leq i \leq m}{\prod}   \frac{\partial}{\partial u}      \big( \mathscr{C}_2 \big)_i            \end{align*}

  \begin{align*}  -     \frac{\partial B_3 {\tiny\big( u^{\prime} \big)} }{\partial u}       \mathscr{C}_1 \cdot {\underset{m,n^{\prime} : m + n^{\prime} = n -3}{\sum}}  \big( \mathrm{sin} {\tiny\big( 2 \eta \big)  } \big)^{n^{\prime}-1} \cdot\underset{1 \leq i \leq m}{\sum}       \underset{1 \leq i \leq m}{\prod}  \frac{\partial}{\partial u}      \big( \mathscr{C}_2 \big)_i      \end{align*}

  \begin{align*}       -        \frac{B_3 {\tiny\big( u^{\prime} \big) } }{\partial u}                \mathscr{C}_1 \cdot   {\underset{m,n^{\prime} : m + n^{\prime} = n -3}{\sum}}  \big( \mathrm{sin} {\tiny \big( 2 \eta \big) } \big)^{n^{\prime}-1}  \cdot \underset{1 \leq i \leq m}{\sum}          \underset{1 \leq i \leq m}{\prod}  \bigg[ \frac{\partial}{\partial u}      \big( \mathscr{C}_2 \big)_i             \end{align*}

  \begin{align*}    +  B_3 {\tiny \big( u \big) } \frac{\partial A_3 {\tiny \big( u^{\prime} \big) } }{\partial u}   \mathscr{C}_1 \cdot {\underset{m,n^{\prime} : m + n^{\prime} = n -3}{\sum}}  \big( \mathrm{sin} {\tiny \big( 2 \eta \big) } \big)^{n^{\prime}-1} \cdot    \underset{1 \leq i \leq m}{\sum}   \frac{\partial}{\partial u} \underset{1 \leq i \leq m}{\prod}   \big( \mathscr{C}_2 \big)_i  \end{align*}

  \begin{align*} +   B_3 {\tiny\big( u \big) } \frac{\partial A_3 {\tiny \big( u^{\prime} \big)} }{\partial u}   \mathscr{C}_1 \cdot {\underset{m,n^{\prime} : m + n^{\prime} = n -3}{\sum}}  \big( \mathrm{sin} {\tiny \big( 2 \eta \big) } \big)^{n^{\prime}-1} \cdot \underset{1 \leq i \leq m}{\sum}    \frac{\partial}{\partial u} \underset{1 \leq i \leq m}{\prod}   \big( \mathscr{C}_2 \big)_i     \end{align*}

  \begin{align*}  +  B_3 {\tiny \big( u \big) }  \frac{\partial A_3 {\tiny \big( u^{\prime} \big) } }{\partial u}   \mathscr{C}_1 \cdot {\underset{m,n^{\prime} : m + n^{\prime} = n -3}{\sum}}  \big( \mathrm{sin} \big( 2 \eta \big) \big)^{n^{\prime}-1}  \cdot  \underset{1 \leq i \leq m}{\sum} \frac{\partial}{\partial u} \underset{1 \leq i \leq m}{\prod}   \big( \mathscr{C}_2 \big)_i \end{align*}

  \begin{align*}+   B_3 {\tiny \big( u \big)  } \frac{\partial A_3 {\tiny \big( u^{\prime} \big) } }{\partial u^{\prime}}      \mathscr{C}_1 \cdot {\underset{m,n^{\prime} : m + n^{\prime} = n -3}{\sum}}  \big( \mathrm{sin} {\tiny \big( 2 \eta \big) } \big)^{n^{\prime}-1}  \cdot    \underset{1 \leq i \leq m}{\sum}  \frac{\partial}{\partial u} \underset{1 \leq i \leq m}{\prod}   \big( \mathscr{C}_2 \big)_i     \end{align*}

  \begin{align*}
      +  A_3 {\tiny \big( u \big) } \frac{\partial A_3 {\tiny \big( u^{\prime} \big) } }{\partial u^{\prime}} \mathscr{C}_1  \cdot{\underset{m,n^{\prime} : m + n^{\prime} = n -3}{\sum}}  \big( \mathrm{sin} {\tiny \big( 2 \eta \big) } \big)^{n^{\prime}-1}  \cdot \underset{1 \leq i \leq m}{\sum}        \frac{\partial}{\partial u }   \underset{1 \leq i \leq m}{\prod}  \big( \mathscr{C}_2 \big)_i         \end{align*}

  \begin{align*}  -  A_3 {\tiny \big( u \big) } \frac{\partial A_3 {\tiny\big( u^{\prime} \big) } }{\partial u}    \mathscr{C}_1 \cdot  {\underset{m,n^{\prime} : m + n^{\prime} = n -3}{\sum}}  \big( \mathrm{sin} {\tiny \big( 2 \eta \big)  } \big)^{n^{\prime}-1}  \cdot    \underset{1 \leq i \leq m}{\sum}   \frac{\partial}{\partial u} \underset{1 \leq i \leq m}{\prod}   \big( \mathscr{C}_2 \big)_i       \end{align*}

  \begin{align*}       -    \frac{\partial A_3 {\tiny \big( u^{\prime}\big)}  }{\partial u}                                \mathscr{C}_1  \cdot  {\underset{m,n^{\prime} : m + n^{\prime} = n -3}{\sum}}  \big( \mathrm{sin} {\tiny \big( 2 \eta \big) } \big)^{n^{\prime}-1}  \cdot  \underset{1 \leq i \leq m}{\sum}  \frac{\partial}{\partial u} \underset{1 \leq i \leq m}{\prod}   \big( \mathscr{C}_2 \big)_i         \end{align*}

  \begin{align*}   -    A_3 {\tiny \big( u \big) } \frac{\partial B_3 {\tiny \big( u^{\prime}\big)} }{\partial u^{\prime}} \mathscr{C}_1  \cdot {\underset{m,n^{\prime} : m + n^{\prime} = n -3}{\sum}}  \big( \mathrm{sin} {\tiny \big( 2 \eta \big) } \big)^{n^{\prime}-1}  \cdot  \underset{1 \leq i \leq m}{\sum}  \frac{\partial}{\partial u} \underset{1 \leq i \leq m}{\prod}   \big( \mathscr{C}_2 \big)_i    \end{align*}

  \begin{align*}-    A_3 {\tiny \big( u \big) }          \frac{\partial B_3 {\tiny \big( u^{\prime}\big) } }{\partial u^{\prime}}           \mathscr{C}_1 \cdot {\underset{m,n^{\prime} : m + n^{\prime} = n -3}{\sum}}  \big( \mathrm{sin} {\tiny \big( 2 \eta \big)  } \big)^{n^{\prime}-1}  \cdot  \underset{1 \leq i \leq m}{\sum}  \frac{\partial}{\partial u} \underset{1 \leq i \leq m}{\prod}   \big( \mathscr{C}_2 \big)_i     \end{align*}

  \begin{align*} +             \frac{\partial B_3 {\tiny \big(u^{\prime} \big) } }{\partial u}             \mathscr{C}_1 \cdot {\underset{m,n^{\prime} : m + n^{\prime} = n -3}{\sum}}  \big( \mathrm{sin} {\tiny \big( 2 \eta \big) } \big)^{n^{\prime}-1}  \cdot   \underset{1 \leq i \leq m}{\sum}   \frac{\partial}{\partial u} \underset{1 \leq i \leq m}{\prod}   \big( \mathscr{C}_2 \big)_i      \text{. }
\end{align*} }

\subsection{Eighth Poisson bracket, $\mathcal{P}_8$, for $ \big\{ A {\tiny \big( u \big) }  , A {\tiny \big( u^{\prime} \big) }  \big\} $}

\noindent The eighth term, $\mathcal{P}_8$, approximately equals,

 {\small \begin{align*}
   \mathcal{P}_8  \approx     -       \frac{A_3 {\tiny\big( u \big)} B_3 {\tiny\big( u^{\prime} \big) }  \mathscr{C}_1   }{ u - u^{\prime} }    \cdot  {\underset{m,n^{\prime} : m + n^{\prime} = n -3}{\sum}} \big( \mathrm{sin} {\tiny\big( 2 \eta \big) } \big)^{n^{\prime}-1} \cdot \underset{1 \leq j \leq n-3}{\underset{1 \leq i \leq m}{\prod}} \big( \mathscr{C}_2 \big)_i \big( \mathscr{C}_2 \big)_j   \end{align*}

  \begin{align*} -  \frac{B_3 {\tiny\big( u \big) }A_3 {\tiny\big( u^{\prime} \big) }
 \mathscr{C}_1 }{u-u^{\prime}} \cdot  {\underset{m,n^{\prime} : m + n^{\prime} = n -3}{\sum}} \big( \mathrm{sin} {\tiny \big( 2 \eta \big) } \big)^{n^{\prime}-1} \cdot   \text{ }   \underset{1 \leq j \leq n-3}{\underset{1 \leq i \leq m}{\prod}} \big( \mathscr{C}_2 \big)_i \big( \mathscr{C}_2 \big)_j      \end{align*}

  \begin{align*}-  2      B_3 {\tiny\big( u^{\prime} \big)   } \frac{\partial B_3 {\tiny\big( u \big) } }{\partial u}   \mathscr{C}_1 \cdot {\underset{m,n^{\prime} : m + n^{\prime} = n -3}{\sum}} \big( \mathrm{sin} {\tiny\big( 2 \eta \big)}  \big)^{n^{\prime}-1}  \cdot      \underset{1 \leq k \leq n-3}{\sum} \frac{\partial \big( \mathscr{C}_2 \big)_k}{\partial u^{\prime}}  \cdot   \text{ }   \underset{1 \leq j \leq n-3}{\underset{1 \leq i \leq m}{\prod}} \big( \mathscr{C}_2 \big)_i \big( \mathscr{C}_2 \big)_j      \end{align*}

  \begin{align*}     - 2    B_3 {\tiny\big( u^{\prime} \big)  }    \frac{\partial B_3 {\tiny\big( u \big)}}{\partial u^{\prime}}    \mathscr{C}_1  \cdot {\underset{m,n^{\prime} : m + n^{\prime} = n -3}{\sum}} \big( \mathrm{sin} {\tiny\big( 2 \eta \big)}  \big)^{n^{\prime}-1} \cdot        \underset{1 \leq k \leq n-3}{\sum} \frac{\partial \big( \mathscr{C}_2 \big)_k}{\partial u^{\prime}} \cdot   \text{ }   \underset{1 \leq j \leq n-3}{\underset{1 \leq i \leq m}{\prod}} \big( \mathscr{C}_2 \big)_i \big( \mathscr{C}_2 \big)_j             \end{align*}

  \begin{align*}      -   A_3 {\tiny\big( u^{\prime} \big)}  \frac{\partial A_3 {\tiny\big( u \big)}  }{\partial u} \mathscr{C}_1   \cdot {\underset{m,n^{\prime} : m + n^{\prime} = n -3}{\sum}} \big( \mathrm{sin} {\tiny\big( 2 \eta \big)} \big)^{n^{\prime}-1} \cdot      \underset{1 \leq k \leq n-3}{\sum} \frac{\partial \big( \mathscr{C}_2 \big)_k}{\partial u^{\prime}}   \cdot  \underset{1 \leq j \leq n-3}{\underset{1 \leq i \leq m}{\prod}}  \big( \mathscr{C}_2 \big)_i \big( \mathscr{C}_2 \big)_j     \end{align*}

  \begin{align*}    -   A_3 {\tiny\big( u^{\prime} \big)}  \frac{\partial A_3 {\tiny\big( u \big)}  }{\partial u^{\prime}} \mathscr{C}_1 \cdot {\underset{m,n^{\prime} : m + n^{\prime} = n -3}{\sum}} \big( \mathrm{sin} {\tiny\big( 2 \eta \big) } \big)^{n^{\prime}-1} \cdot     \underset{1 \leq k \leq n-3}{\sum}             \frac{\partial \big( \mathscr{C}_2 \big)_k}{\partial u} \cdot  \underset{1 \leq j \leq n-3}{\underset{1 \leq i \leq m}{\prod}}  \big( \mathscr{C}_2 \big)_i \big( \mathscr{C}_2 \big)_j             \end{align*}

  \begin{align*}     -     A_3 {\tiny\big( u^{\prime} \big) } \frac{\partial B_3 {\tiny\big( u^{\prime} \big) } }{\partial u^{\prime}} \mathscr{C}_1  \cdot  {\underset{m,n^{\prime} : m + n^{\prime} = n -3}{\sum}} \big( \mathrm{sin} {\tiny\big( 2 \eta \big)}  \big)^{n^{\prime}-1}  \cdot  \underset{1 \leq k \leq n-3}{\sum}  \frac{\big( \mathscr{C}_2 \big)_k}{\partial u^{\prime}}  \cdot  \underset{1 \leq j \leq n-3}{\underset{1 \leq i \leq m}{\prod}}  \big( \mathscr{C}_2 \big)_i \big( \mathscr{C}_2 \big)_j      \end{align*}

  \begin{align*}
   -       A_3 {\tiny\big( u^{\prime} \big)} \frac{\partial B_3 {\tiny\big( u^{\prime} \big)} }{\partial u^{\prime}} \mathscr{C}_1 \cdot 
 {\underset{m,n^{\prime} : m + n^{\prime} = n -3}{\sum}} \big( \mathrm{sin} {\tiny\big( 2 \eta \big)} \big)^{n^{\prime}-1} \cdot        \underset{1 \leq k \leq n-3}{\sum}  \frac{\big( \mathscr{C}_2 \big)_k}{\partial u^{\prime}} \cdot  \underset{1 \leq j \leq n-3}{\underset{1 \leq i \leq m}{\prod}}  \big( \mathscr{C}_2 \big)_i \big( \mathscr{C}_2 \big)_j  \end{align*}

  \begin{align*} +   A_3 {\tiny\big( u^{\prime} \big)} \frac{\partial B_3 {\tiny\big( u^{\prime} \big)}}{\partial u}  \mathscr{C}_1 \cdot  {\underset{m,n^{\prime} : m + n^{\prime} = n -3}{\sum}} \big( \mathrm{sin} {\tiny\big( 2 \eta \big)} \big)^{n^{\prime}-1} \cdot       \underset{1 \leq k \leq n-3}{\sum} \frac{\partial \big( \mathscr{C}_2 \big)_k}{\partial u^{\prime}}  \cdot   \underset{1 \leq j \leq n-3}{\underset{1 \leq i \leq m}{\prod}}  \big( \mathscr{C}_2 \big)_i \big( \mathscr{C}_2 \big)_j  \end{align*}

  \begin{align*}            +      \frac{\partial B_3 {\tiny\big( u^{\prime} \big)}}{\partial u}  \cdot  {\underset{m,n^{\prime} : m + n^{\prime} = n -3}{\sum}} \big( \mathrm{sin} {\tiny\big( 2 \eta \big)}  \big)^{n^{\prime}-1} \cdot   \underset{1 \leq k \leq n-3}{\sum} \frac{\partial \big( \mathscr{C}_2 \big)_k}{\partial u^{\prime}} \cdot  \underset{1 \leq j \leq n-3}{\underset{1 \leq i \leq m}{\prod}}  \big( \mathscr{C}_2 \big)_i \big( \mathscr{C}_2 \big)_j         \end{align*}

  \begin{align*} +  \frac{\partial A_3 {\tiny\big( u^{\prime} \big)} }{\partial u}  \cdot {\underset{m,n^{\prime} : m + n^{\prime} = n -3}{\sum}} \big( \mathrm{sin} {\tiny\big( 2 \eta \big)}  \big)^{n^{\prime}-1} \cdot  \underset{1 \leq k \leq n-3}{\sum} \frac{\partial \big( \mathscr{C}_2 \big)_k}{\partial u^{\prime}}  \cdot  \underset{1 \leq j \leq n-3}{\underset{1 \leq i \leq m}{\prod}}  \big( \mathscr{C}_2 \big)_i \big( \mathscr{C}_2 \big)_j             \end{align*}

  \begin{align*} -     \frac{\partial A_3 {\tiny\big( u^{\prime} \big)} }{\partial u^{\prime}} \cdot  {\underset{m,n^{\prime} : m + n^{\prime} = n -3}{\sum}}  \big( \mathrm{sin} {\tiny\big( 2 \eta \big) }  \big)^{n^{\prime}-1}  \cdot \underset{1 \leq k \leq n-3}{\sum} \frac{\partial \big( \mathscr{C}_2 \big)_k}{\partial u^{\prime}} \cdot  \underset{1 \leq j \leq n-3}{\underset{1 \leq i \leq m}{\prod}}  \big( \mathscr{C}_2 \big)_i \big( \mathscr{C}_2 \big)_j       \end{align*}

  \begin{align*} -     \frac{\partial B_3 {\tiny\big( u^{\prime} \big)} }{\partial u^{\prime}} \cdot   {\underset{m,n^{\prime} : m + n^{\prime} = n -3}{\sum}}  \big( \mathrm{sin} {\tiny \big( 2 \eta \big) }  \big)^{n^{\prime}-1} \cdot   \underset{1 \leq k \leq n-3}{\sum} \frac{\partial \big( \mathscr{C}_2 \big)_k}{\partial u^{\prime}}   \cdot \underset{1 \leq j \leq n-3}{\underset{1 \leq i \leq m}{\prod}}  \big( \mathscr{C}_2 \big)_i \big( \mathscr{C}_2 \big)_j          \text{. }     \end{align*} }

\subsection{Ninth Poisson bracket, $\mathcal{P}_9$, for $ \big\{ A {\tiny \big( u \big) }  , A {\tiny \big( u^{\prime} \big)}  \big\} $}

\noindent The ninth term approximately equals,

{\small \begin{align*}            \mathcal{P}_9   \approx     - \frac{  \mathscr{C}_1 \mathscr{C}_2  B_3 {\tiny\big( u^{\prime} \big)} A_3 {\tiny\big( u \big)}}{u^{\prime} - u }   \cdot   {\underset{n, m^{\prime}: m + n^{\prime} = n-3}{\sum} } \big( \mathrm{sin} {\tiny\big( 2 \eta \big) } \big)^{n^{\prime}-1}                \end{align*}

  \begin{align*}
-  \frac{  \mathscr{C}_1 \mathscr{C}_2  A_3 {\tiny\big(u^{\prime} \big)}  B_3 {\tiny\big( u \big)} }{u^{\prime}- u }  \cdot  {\underset{n, m^{\prime}: m + n^{\prime} = n-3}{\sum} } \big( \mathrm{sin} {\tiny\big( 2 \eta \big)}  \big)^{n^{\prime}-1}                   
\end{align*}

  \begin{align*} -   \frac{      2  \mathscr{C}^2_2   B_3 {\tiny\big( u \big) }
 B_3 {\tiny\big( u^{\prime} \big) }A_3 {\tiny\big( u \big)} }{u^{\prime} - u }    \cdot 
 {\underset{n, m^{\prime}: m + n^{\prime} = n-3}{\sum} }    \big( \mathrm{sin} {\tiny\big( 2 \eta \big)}  \big)^{2n^{\prime}-2}         \end{align*}

  \begin{align*} -  \frac{  2 \mathscr{C}^2_2  A_3 {\tiny\big( u \big) }  A_3 {\tiny\big( u^{\prime} \big)} B_3 {\tiny\big( u \big) } }{u^{\prime} - u }    \cdot 
{\underset{n, m^{\prime}: m + n^{\prime} = n-3}{\sum} }     \big( \mathrm{sin} {\tiny\big( 2 \eta \big)}  \big)^{2n^{\prime}-2}     \end{align*}

  \begin{align*}
  + 2 \frac{\partial A_3 {\tiny\big( u^{\prime} \big)} }{\partial u^{\prime}}        \cdot \underset{1 \leq j \leq m}{\sum} \frac{\partial}{\partial u}                 \bigg[ \underset{1 \leq i \leq m}{\prod} \mathrm{sin} {\tiny\big( u^{\prime} - v_{n-i} \pm \eta \sigma^z_{n-j} \big) }  \text{ }    \bigg]    \end{align*}

  \begin{align*}        + 2 \frac{\partial B_3 {\tiny\big( u^{\prime} \big)} }{\partial u^{\prime}}   \cdot   \underset{1 \leq j \leq m}{\sum}  \frac{\partial}{\partial u}                 \bigg[ \underset{1 \leq i \leq m}{\prod} \mathrm{sin} {\tiny\big( u^{\prime} - v_{n-i} \pm \eta \sigma^z_{n-j} \big) }  \text{ }    \bigg]     \end{align*}

  \begin{align*} - \frac{\partial A_3 {\tiny\big( u^{\prime} \big)} }{\partial u}   \cdot               \underset{1 \leq j \leq m}{\sum}       \frac{\partial}{\partial u^{\prime}}                 \bigg[ \underset{1 \leq i \leq m}{\prod} \mathrm{sin} {\tiny\big( u^{\prime} - v_{n-i} \pm \eta \sigma^z_{n-j} \big)}   \text{ }    \bigg]       \end{align*}

  \begin{align*} - \frac{\partial B_3 {\tiny\big(u^{\prime} \big)} }{\partial u}   \cdot               \underset{1 \leq j \leq m}{\sum}     \frac{\partial}{\partial u^{\prime}}                 \bigg[ \underset{1 \leq i \leq m}{\prod} \mathrm{sin} {\tiny\big( u^{\prime} - v_{n-i} \pm \eta \sigma^z_{n-j} \big) } \text{ }    \bigg] \end{align*}

  \begin{align*}    -   \frac{\partial A_3 {\tiny\big( u^{\prime} \big)}}{\partial u} \cdot  \underset{1 \leq j \leq m}{\sum} \frac{\partial}{\partial u^{\prime}} \bigg[ \underset{1 \leq i \leq m}{\prod}  \mathrm{sin} {\tiny\big( u^{\prime} - v_{n-i} \pm \eta \sigma^z_{n-j} \big)}  \bigg] \end{align*}

  \begin{align*} - \frac{\partial B_3 {\tiny\big( u^{\prime} \big)} }{\partial u} \cdot   \underset{1 \leq j \leq m}{\sum}   \frac{\partial}{\partial u^{\prime}} \bigg[ \underset{1 \leq i \leq m}{\prod}  \mathrm{sin} {\tiny\big( u^{\prime} - v_{n-i} \pm \eta \sigma^z_{n-j} \big)}  \bigg]   \text{. }
\end{align*} }

\subsection{Binomial expansion for the integral against the density of Bethe roots}

\noindent By direct computation,

 {\small
 
\begin{align*}
   \int_C   \bigg\{            \bigg( \frac{\sqrt{2}}{n!} \bigg)^4   \bigg(         \frac{   \frac{\eta}{2} - u + i \alpha - iu  }{ \frac{\eta}{2} - u + i \alpha - iu    }      \times \frac{\frac{\eta}{2} + i \alpha}{\frac{\eta}{2} - i \alpha }                     \bigg)^n                  \bigg\}    \text{ } \mathrm{d} \alpha  =     \bigg( \frac{\sqrt{2}}{n!} \bigg)^4   \cdot  \int_C   \bigg\{          \bigg(         \frac{   \frac{\eta}{2} - u + i \alpha - iu  }{ \frac{\eta}{2} - u + i \alpha - iu    }      \times \frac{\frac{\eta}{2} + i \alpha}{\frac{\eta}{2} - i \alpha }                     \bigg)^n                  \bigg\}    \text{ } \mathrm{d} \alpha   \\    \\               =     \bigg( \frac{\sqrt{2}}{n!} \bigg)^4   \cdot  \int_C   \bigg\{                      \frac{\big( \frac{\eta}{2} \big)^2}{ \big( \frac{\eta}{2} - u + i \alpha - iu \big) \big(   \frac{\eta}{2} - i \alpha   \big)   }    +   \frac{\frac{\eta}{2}  \cdot i \alpha }{ \big( \frac{\eta}{2} - u + i \alpha - iu \big) \big(   \frac{\eta}{2} - i \alpha   \big)   }     -   \frac{u \cdot \frac{\eta}{2}  }{ \big( \frac{\eta}{2} - u + i \alpha - iu \big) \big(   \frac{\eta}{2} - i \alpha   \big)   }    \\ \\   - \frac{ u \cdot i \alpha}{ \big( \frac{\eta}{2} - u + i \alpha - iu \big) \big(   \frac{\eta}{2} - i \alpha   \big)}    + \frac{i \alpha \cdot \frac{\eta}{2}}{ \big( \frac{\eta}{2} - u + i \alpha - iu \big) \big(   \frac{\eta}{2} - i \alpha   \big)}           + \frac{\big( i \alpha \big)^2}{\big( \frac{\eta}{2} - u + i \alpha - iu \big) \big(   \frac{\eta}{2} - i \alpha   \big)} \\ \\ - \frac{iu \cdot \frac{\eta}{2}}{\big( \frac{\eta}{2} - u + i \alpha - iu \big) \big(   \frac{\eta}{2} - i \alpha   \big)}    + \frac{ u \cdot \alpha }{\big( \frac{\eta}{2} - u + i \alpha - iu \big) \big(   \frac{\eta}{2} - i \alpha   \big)}     \bigg\}^n     \text{ } \mathrm{d} \alpha    \\    
\end{align*}

\begin{align*}
  =       \bigg( \frac{\sqrt{2}}{n!} \bigg)^4   \cdot  \int_C      \bigg\{   \underset{0 \leq k \leq n}{\sum}        \binom{n}{k}      \bigg[      \frac{\big( \frac{\eta}{2} \big)^2}{ \big( \frac{\eta}{2} - u + i \alpha - iu \big) \big(   \frac{\eta}{2} - i \alpha   \big)   }           \bigg]^{n-k} \cdot \bigg[   \frac{\frac{\eta}{2}  \cdot i \alpha }{ \big( \frac{\eta}{2} - u + i \alpha - iu \big) \big(   \frac{\eta}{2} - i \alpha   \big)   }    \end{align*}

\begin{align*}  -   \frac{u \cdot \frac{\eta}{2}  }{ \big( \frac{\eta}{2} - u + i \alpha - iu \big) \big(   \frac{\eta}{2} - i \alpha   \big)   }               - \frac{ u \cdot i \alpha}{ \big( \frac{\eta}{2} - u + i \alpha - iu \big) \big(   \frac{\eta}{2} - i \alpha   \big)}    + \frac{i \alpha \cdot \frac{\eta}{2}}{ \big( \frac{\eta}{2} - u + i \alpha - iu \big) \big(   \frac{\eta}{2} - i \alpha   \big)}         \end{align*}

\begin{align*}   + \frac{\big( i \alpha \big)^2}{\big( \frac{\eta}{2} - u + i \alpha - iu \big) \big(   \frac{\eta}{2} - i \alpha   \big)}                    - \frac{iu \cdot \frac{\eta}{2}}{\big( \frac{\eta}{2} - u + i \alpha - iu \big) \big(   \frac{\eta}{2} - i \alpha   \big)}    + \frac{ u \cdot \alpha }{\big( \frac{\eta}{2} - u + i \alpha - iu \big) \big(   \frac{\eta}{2} - i \alpha   \big)}           \bigg]^k               \bigg\}     \text{ } \mathrm{d} \alpha                         \\ 
\end{align*}

\begin{align*}
=      \bigg( \frac{\sqrt{2}}{n!} \bigg)^4   \cdot  \int_C      \bigg\{   \underset{0 \leq k \leq n}{\sum}        \binom{n}{k}      \bigg[      \frac{\big( \frac{\eta}{2} \big)^2}{ \big( \frac{\eta}{2} - u + i \alpha - iu \big) \big(   \frac{\eta}{2} - i \alpha   \big)   }           \bigg]^{n-k}   \cdot \bigg\{                 \underset{0 \leq k^{\prime} \leq k}{\sum}              \binom{k}{k^{\prime}}  \bigg[       \frac{\frac{\eta}{2}  \cdot i \alpha }{ \big( \frac{\eta}{2} - u + i \alpha - iu \big) \big(   \frac{\eta}{2} - i \alpha   \big)   }      \bigg]^{k-k^{\prime}} \end{align*}

\begin{align*} \cdot \bigg[      -   \frac{u \cdot \frac{\eta}{2}  }{ \big( \frac{\eta}{2} - u + i \alpha - iu \big) \big(   \frac{\eta}{2} - i \alpha   \big)   }               - \frac{ u \cdot i \alpha}{ \big( \frac{\eta}{2} - u + i \alpha - iu \big) \big(   \frac{\eta}{2} - i \alpha   \big)}    + \frac{i \alpha \cdot \frac{\eta}{2}}{ \big( \frac{\eta}{2} - u + i \alpha - iu \big) \big(   \frac{\eta}{2} - i \alpha   \big)}         \end{align*}

\begin{align*}   + \frac{\big( i \alpha \big)^2}{\big( \frac{\eta}{2} - u + i \alpha - iu \big) \big(   \frac{\eta}{2} - i \alpha   \big)}                    - \frac{iu \cdot \frac{\eta}{2}}{\big( \frac{\eta}{2} - u + i \alpha - iu \big) \big(   \frac{\eta}{2} - i \alpha   \big)}    + \frac{ u \cdot \alpha }{\big( \frac{\eta}{2} - u + i \alpha - iu \big) \big(   \frac{\eta}{2} - i \alpha   \big)}          \bigg]^{k^{\prime}}                     \bigg\} \bigg\}   \text{ } \mathrm{d} \alpha                        \\ 
\end{align*}

\begin{align*}
=      \bigg( \frac{\sqrt{2}}{n!} \bigg)^4   \cdot  \int_C      \bigg\{   \underset{0 \leq k \leq n}{\sum}        \binom{n}{k}      \bigg[      \frac{\big( \frac{\eta}{2} \big)^2}{ \big( \frac{\eta}{2} - u + i \alpha - iu \big) \big(   \frac{\eta}{2} - i \alpha   \big)   }           \bigg]^{n-k}   \cdot \bigg\{                 \underset{0 \leq k^{\prime} \leq k}{\sum}              \binom{k}{k^{\prime}}  \bigg[       \frac{\frac{\eta}{2}  \cdot i \alpha }{ \big( \frac{\eta}{2} - u + i \alpha - iu \big) \big(   \frac{\eta}{2} - i \alpha   \big)   }      \bigg]^{k-k^{\prime}} \end{align*}

\begin{align*}  \cdot \bigg\{   \underset{0 \leq k^{\prime\prime} \leq k^{\prime}}{\sum}        \binom{k^{\prime}}{k^{\prime\prime}}   \bigg[                  -   \frac{u \cdot \frac{\eta}{2}  }{ \big( \frac{\eta}{2} - u + i \alpha - iu \big) \big(   \frac{\eta}{2} - i \alpha   \big)   }      \bigg]^{k^{\prime} - k^{\prime\prime}} \cdot \bigg[                 - \frac{ u \cdot i \alpha}{ \big( \frac{\eta}{2} - u + i \alpha - iu \big) \big(   \frac{\eta}{2} - i \alpha   \big)} \end{align*}

\begin{align*}  + \frac{i \alpha \cdot \frac{\eta}{2}}{ \big( \frac{\eta}{2} - u + i \alpha - iu \big) \big(   \frac{\eta}{2} - i \alpha   \big)}       + \frac{\big( i \alpha \big)^2}{\big( \frac{\eta}{2} - u + i \alpha - iu \big) \big(   \frac{\eta}{2} - i \alpha   \big)}                    - \frac{iu \cdot \frac{\eta}{2}}{\big( \frac{\eta}{2} - u + i \alpha - iu \big) \big(   \frac{\eta}{2} - i \alpha   \big)}  \end{align*}

\begin{align*}   + \frac{ u \cdot \alpha }{\big( \frac{\eta}{2} - u + i \alpha - iu \big) \big(   \frac{\eta}{2} - i \alpha   \big)}           \bigg]^{k^{\prime\prime}}       \bigg\}                   \bigg\} \bigg\}   \text{ } \mathrm{d} \alpha                        \\ 
\end{align*}

\begin{align*}
=      \bigg( \frac{\sqrt{2}}{n!} \bigg)^4   \cdot  \int_C      \bigg\{   \underset{0 \leq k \leq n}{\sum}        \binom{n}{k}      \bigg[      \frac{\big( \frac{\eta}{2} \big)^2}{ \big( \frac{\eta}{2} - u + i \alpha - iu \big) \big(   \frac{\eta}{2} - i \alpha   \big)   }           \bigg]^{n-k}   \cdot \bigg\{                 \underset{0 \leq k^{\prime} \leq k}{\sum}              \binom{k}{k^{\prime}}  \bigg[       \frac{\frac{\eta}{2}  \cdot i \alpha }{ \big( \frac{\eta}{2} - u + i \alpha - iu \big) \big(   \frac{\eta}{2} - i \alpha   \big)   }      \bigg]^{k-k^{\prime}} \end{align*}

\begin{align*} \cdot \bigg\{   \underset{0 \leq k^{\prime\prime} \leq k^{\prime}}{\sum}        \binom{k^{\prime}}{k^{\prime\prime}}   \bigg[                  -   \frac{u \cdot \frac{\eta}{2}  }{ \big( \frac{\eta}{2} - u + i \alpha - iu \big) \big(   \frac{\eta}{2} - i \alpha   \big)   }      \bigg]^{k^{\prime} - k^{\prime\prime}} \cdot \bigg\{      \underset{0 \leq k^{\prime\prime\prime} \leq k^{\prime\prime}}{\sum}   \binom{k^{\prime\prime}}{k^{\prime\prime\prime}}    \bigg[   - \frac{ u \cdot i \alpha}{ \big( \frac{\eta}{2} - u + i \alpha - iu \big) \big(   \frac{\eta}{2} - i \alpha   \big)}   \bigg]^{k^{\prime\prime} - k^{\prime\prime\prime}} \end{align*}

\begin{align*} \cdot \bigg[  \frac{i \alpha \cdot \frac{\eta}{2}}{ \big( \frac{\eta}{2} - u + i \alpha - iu \big) \big(   \frac{\eta}{2} - i \alpha   \big)}                                                                                           + \frac{\big( i \alpha \big)^2}{\big( \frac{\eta}{2} - u + i \alpha - iu \big) \big(   \frac{\eta}{2} - i \alpha   \big)}                    - \frac{iu \cdot \frac{\eta}{2}}{\big( \frac{\eta}{2} - u + i \alpha - iu \big) \big(   \frac{\eta}{2} - i \alpha   \big)}    \\ + \frac{ u \cdot \alpha }{\big( \frac{\eta}{2} - u + i \alpha - iu \big) \big(   \frac{\eta}{2} - i \alpha   \big)}         \bigg]^{k^{\prime\prime\prime}}  \bigg\}       \bigg\}                   \bigg\} \bigg\}   \text{ } \mathrm{d} \alpha                        \\ 
\end{align*}

\begin{align*}
=      \bigg( \frac{\sqrt{2}}{n!} \bigg)^4   \cdot  \int_C      \bigg\{   \underset{0 \leq k \leq n}{\sum}        \binom{n}{k}      \bigg[      \frac{\big( \frac{\eta}{2} \big)^2}{ \big( \frac{\eta}{2} - u + i \alpha - iu \big) \big(   \frac{\eta}{2} - i \alpha   \big)   }           \bigg]^{n-k}   \cdot \bigg\{                 \underset{0 \leq k^{\prime} \leq k}{\sum}              \binom{k}{k^{\prime}}  \bigg[       \frac{\frac{\eta}{2}  \cdot i \alpha }{ \big( \frac{\eta}{2} - u + i \alpha - iu \big) \big(   \frac{\eta}{2} - i \alpha   \big)   }      \bigg]^{k-k^{\prime}} \end{align*}

\begin{align*}  \cdot \bigg\{   \underset{0 \leq k^{\prime\prime} \leq k^{\prime}}{\sum}        \binom{k^{\prime}}{k^{\prime\prime}}   \bigg[                  -   \frac{u \cdot \frac{\eta}{2}  }{ \big( \frac{\eta}{2} - u + i \alpha - iu \big) \big(   \frac{\eta}{2} - i \alpha   \big)   }      \bigg]^{k^{\prime} - k^{\prime\prime}} \cdot \bigg\{      \underset{0 \leq k^{\prime\prime\prime} \leq k^{\prime\prime}}{\sum}   \binom{k^{\prime\prime}}{k^{\prime\prime\prime}}    \bigg[   - \frac{ u \cdot i \alpha}{ \big( \frac{\eta}{2} - u + i \alpha - iu \big) \big(   \frac{\eta}{2} - i \alpha   \big)}   \bigg]^{k^{\prime\prime} - k^{\prime\prime\prime}} \end{align*}

\begin{align*} \cdot \bigg\{     \underset{0 \leq k^{\prime\prime\prime\prime} \leq k^{\prime\prime\prime}}{\sum}       \binom{k^{\prime\prime\prime}}{k^{\prime\prime\prime\prime}}        \bigg[    \frac{i \alpha \cdot \frac{\eta}{2}}{ \big( \frac{\eta}{2} - u + i \alpha - iu \big) \big(   \frac{\eta}{2} - i \alpha   \big)}                   \bigg]^{k^{\prime\prime\prime\prime} - k^{\prime\prime\prime}} \cdot \bigg[      \frac{\big( i \alpha \big)^2}{\big( \frac{\eta}{2} - u + i \alpha - iu \big) \big(   \frac{\eta}{2} - i \alpha   \big)}               \end{align*}

\begin{align*}  - \frac{iu \cdot \frac{\eta}{2}}{\big( \frac{\eta}{2} - u + i \alpha - iu \big) \big(   \frac{\eta}{2} - i \alpha   \big)}    + \frac{ u \cdot \alpha }{\big( \frac{\eta}{2} - u + i \alpha - iu \big) \big(   \frac{\eta}{2} - i \alpha   \big)}                    \bigg]^{k^{\prime\prime\prime\prime}}   \bigg\}                     \bigg\}       \bigg\}                   \bigg\} \bigg\}   \text{ } \mathrm{d} \alpha                        \\ 
\end{align*}

\begin{align*}
=      \bigg( \frac{\sqrt{2}}{n!} \bigg)^4   \cdot  \int_C      \bigg\{   \underset{0 \leq k \leq n}{\sum}        \binom{n}{k}      \bigg[      \frac{\big( \frac{\eta}{2} \big)^2}{ \big( \frac{\eta}{2} - u + i \alpha - iu \big) \big(   \frac{\eta}{2} - i \alpha   \big)   }           \bigg]^{n-k}   \cdot \bigg\{                 \underset{0 \leq k^{\prime} \leq k}{\sum}              \binom{k}{k^{\prime}}  \bigg[       \frac{\frac{\eta}{2}  \cdot i \alpha }{ \big( \frac{\eta}{2} - u + i \alpha - iu \big) \big(   \frac{\eta}{2} - i \alpha   \big)   }      \bigg]^{k-k^{\prime}} \end{align*}

\begin{align*} \cdot \bigg\{   \underset{0 \leq k^{\prime\prime} \leq k^{\prime}}{\sum}        \binom{k^{\prime}}{k^{\prime\prime}}   \bigg[                  -   \frac{u \cdot \frac{\eta}{2}  }{ \big( \frac{\eta}{2} - u + i \alpha - iu \big) \big(   \frac{\eta}{2} - i \alpha   \big)   }      \bigg]^{k^{\prime} - k^{\prime\prime}} \cdot \bigg\{      \underset{0 \leq k^{\prime\prime\prime} \leq k^{\prime\prime}}{\sum}   \binom{k^{\prime\prime}}{k^{\prime\prime\prime}}    \bigg[   - \frac{ u \cdot i \alpha}{ \big( \frac{\eta}{2} - u + i \alpha - iu \big) \big(   \frac{\eta}{2} - i \alpha   \big)}   \bigg]^{k^{\prime\prime} - k^{\prime\prime\prime}} \end{align*}

\begin{align*} \cdot \bigg\{     \underset{0 \leq k^{\prime\prime\prime\prime} \leq k^{\prime\prime\prime}}{\sum}       \binom{k^{\prime\prime\prime}}{k^{\prime\prime\prime\prime}}        \bigg[    \frac{i \alpha \cdot \frac{\eta}{2}}{ \big( \frac{\eta}{2} - u + i \alpha - iu \big) \big(   \frac{\eta}{2} - i \alpha   \big)}                   \bigg]^{k^{\prime\prime\prime\prime} - k^{\prime\prime\prime}} \cdot \bigg\{                          \underset{0 \leq k^{\prime\prime\prime\prime\prime} \leq k^{\prime\prime\prime\prime} }{\sum}      \binom{k^{\prime\prime\prime\prime}}{k^{\prime\prime\prime\prime\prime}}  \end{align*}

\begin{align*} \times        \bigg[     \frac{\big( i \alpha \big)^2}{\big( \frac{\eta}{2} - u + i \alpha - iu \big) \big(   \frac{\eta}{2} - i \alpha   \big)}                \bigg]^{k^{\prime\prime\prime\prime} - k^{\prime\prime\prime\prime\prime} }   \cdot \bigg[           - \frac{iu \cdot \frac{\eta}{2}}{\big( \frac{\eta}{2} - u + i \alpha - iu \big) \big(   \frac{\eta}{2} - i \alpha   \big)}  \end{align*}

\begin{align*}    + \frac{ u \cdot \alpha }{\big( \frac{\eta}{2} - u + i \alpha - iu \big) \big(   \frac{\eta}{2} - i \alpha   \big)}                      \bigg]^{k^{\prime\prime\prime\prime\prime} }         \bigg\}   \bigg\}                     \bigg\}       \bigg\}                   \bigg\} \bigg\}   \text{ } \mathrm{d} \alpha                        \\ 
\end{align*}

\begin{align*}
=      \bigg( \frac{\sqrt{2}}{n!} \bigg)^4   \cdot  \int_C      \bigg\{   \underset{0 \leq k \leq n}{\sum}        \binom{n}{k}      \bigg[      \frac{\big( \frac{\eta}{2} \big)^2}{ \big( \frac{\eta}{2} - u + i \alpha - iu \big) \big(   \frac{\eta}{2} - i \alpha   \big)   }           \bigg]^{n-k}   \cdot \bigg\{                 \underset{0 \leq k^{\prime} \leq k}{\sum}              \binom{k}{k^{\prime}}  \bigg[       \frac{\frac{\eta}{2}  \cdot i \alpha }{ \big( \frac{\eta}{2} - u + i \alpha - iu \big) \big(   \frac{\eta}{2} - i \alpha   \big)   }      \bigg]^{k-k^{\prime}} \end{align*}

\begin{align*} \cdot \bigg\{   \underset{0 \leq k^{\prime\prime} \leq k^{\prime}}{\sum}        \binom{k^{\prime}}{k^{\prime\prime}}   \bigg[                  -   \frac{u \cdot \frac{\eta}{2}  }{ \big( \frac{\eta}{2} - u + i \alpha - iu \big) \big(   \frac{\eta}{2} - i \alpha   \big)   }      \bigg]^{k^{\prime} - k^{\prime\prime}} \cdot \bigg\{      \underset{0 \leq k^{\prime\prime\prime} \leq k^{\prime\prime}}{\sum}   \binom{k^{\prime\prime}}{k^{\prime\prime\prime}}    \bigg[   - \frac{ u \cdot i \alpha}{ \big( \frac{\eta}{2} - u + i \alpha - iu \big) \big(   \frac{\eta}{2} - i \alpha   \big)}   \bigg]^{k^{\prime\prime} - k^{\prime\prime\prime}} \end{align*}

\begin{align*} \cdot \bigg\{     \underset{0 \leq k^{\prime\prime\prime\prime} \leq k^{\prime\prime\prime}}{\sum}       \binom{k^{\prime\prime\prime}}{k^{\prime\prime\prime\prime}}        \bigg[    \frac{i \alpha \cdot \frac{\eta}{2}}{ \big( \frac{\eta}{2} - u + i \alpha - iu \big) \big(   \frac{\eta}{2} - i \alpha   \big)}                   \bigg]^{k^{\prime\prime\prime\prime} - k^{\prime\prime\prime}} \cdot \bigg\{                          \underset{0 \leq k^{\prime\prime\prime\prime\prime} \leq k^{\prime\prime\prime\prime} }{\sum}      \binom{k^{\prime\prime\prime\prime}}{k^{\prime\prime\prime\prime\prime}}  \end{align*}

\begin{align*} \times        \bigg[     \frac{\big( i \alpha \big)^2}{\big( \frac{\eta}{2} - u + i \alpha - iu \big) \big(   \frac{\eta}{2} - i \alpha   \big)}                \bigg]^{k^{\prime\prime\prime\prime} - k^{\prime\prime\prime\prime\prime} }   \cdot \bigg\{          \underset{0 \leq   k^{\prime\prime\prime\prime\prime\prime} \leq k^{\prime\prime\prime\prime\prime} }{\sum}        \binom{k^{\prime\prime\prime\prime\prime}}{k^{\prime\prime\prime\prime\prime\prime}}    \bigg[               - \frac{iu \cdot \frac{\eta}{2}}{\big( \frac{\eta}{2} - u + i \alpha - iu \big) \big(   \frac{\eta}{2} - i \alpha   \big)} \bigg]^{k^{\prime\prime\prime\prime\prime}-k^{\prime\prime\prime\prime\prime\prime}}        \end{align*}

\begin{align*}   \cdot     \bigg[  \frac{ u \cdot \alpha }{\big( \frac{\eta}{2} - u + i \alpha - iu \big) \big(   \frac{\eta}{2} - i \alpha   \big)}       \bigg]^{k^{\prime\prime\prime\prime\prime\prime}}   \bigg\}       \bigg\}   \bigg\}                     \bigg\}       \bigg\}                   \bigg\} \bigg\}   \text{ } \mathrm{d} \alpha   . 
\end{align*}

 }

\section{Declarations}

\subsection{Ethics approval and consent to participate}

The author consents to participate in the peer review process.

\subsection{Consent for publication}

The author consents to submit the following work for publication.

\subsection{Availability of data and materials}

Not applicable

\subsection{Competing interests}

Not applicable

\subsection{Funding}

Not applicable

\subsection{Author's contributions}

PR wrote the manuscript, and performed rounds of editing for submitting the work.

\nocite{*}
\bibliography{sn-bibliography}

\end{document}